\documentclass[superscriptaddress,aps,pra,nofootinbib,notitlepage,10pt]{revtex4-1}
\usepackage{graphicx}
\usepackage{amsmath}
\usepackage{amssymb}
\usepackage{amsthm}
\usepackage[version=3]{mhchem}
\usepackage{bm}
\usepackage{amsfonts}
\usepackage{comment}
\usepackage[colorlinks]{hyperref}
\hypersetup{
	pdfstartview={FitH},
	pdfnewwindow=true,
	colorlinks=true,
	linkcolor=blue,
	citecolor=blue,
	filecolor=blue,
	urlcolor=blue}
\usepackage[caption=false]{subfig}
\usepackage[all]{hypcap}
\usepackage{tikz}
\usepackage{verbatim}
\usepackage{subfig}
\usetikzlibrary{arrows}
\usepackage{units}
\usepackage{multirow}
\usepackage{soul}
\usepackage{float}
\makeatletter
\let\newfloat\newfloat@ltx
\makeatother
\usepackage{algorithm}
\usepackage{algpseudocode}
\usepackage{url}
\usepackage{cleveref}
\usepackage{adjustbox}
\usepackage{cancel}
\usepackage{textgreek}
\newcommand{\insertrev}[1]{{\textcolor{black}{#1}}}

\newcommand{\eq}[1]{Eq.~\hyperref[eq:#1]{(\ref*{eq:#1})}}
\renewcommand{\sec}[1]{\hyperref[sec:#1]{Section~\ref*{sec:#1}}}
\newcommand{\app}[1]{\hyperref[app:#1]{Appendix~\ref*{app:#1}}}
\newcommand{\tab}[1]{\hyperref[tab:#1]{Table~\ref*{tab:#1}}}
\newcommand{\fig}[1]{\hyperref[fig:#1]{Figure~\ref*{fig:#1}}}
\newcommand{\figa}[2]{\hyperref[fig:#1]{Figure~\ref*{fig:#1}#2}}
\newcommand{\figx}[2]{\hyperref[fig:#1]{Figure~\ref*{fig:#1}(#2)}}
\newcommand{\thm}[1]{\hyperref[thm:#1]{Theorem~\ref*{thm:#1}}}
\newcommand{\lem}[1]{\hyperref[lem:#1]{Lemma~\ref*{lem:#1}}}
\newcommand{\cor}[1]{\hyperref[cor:#1]{Corollary~\ref*{cor:#1}}}
\newcommand{\defn}[1]{\hyperref[def:#1]{Definition~\ref*{def:#1}}}
\newcommand{\alg}[1]{\hyperref[alg:#1]{Algorithm~\ref*{alg:#1}}}

\newcommand{\iter}{\mathcal{I}}

\newcommand{\zetabits}{\aleph}
\newcommand{\rotbits}{\beth}
\newcommand{\rankone}{L}
\newcommand{\ranktwo}{\Xi}
\newcommand{\sparsity}{S}
\newcommand{\rotprec}{b_r}

\newcommand{\sel}{\textsc{select}}
\newcommand{\prep}{\textsc{prepare}}

\def\bra#1{\mathinner{\langle{#1}|}}
\def\ket#1{\mathinner{|{#1}\rangle}}
\newcommand{\braket}[2]{\langle #1|#2\rangle}
\newcommand{\proj}[1]{\ket{#1}\!\!\bra{#1}}

\newcommand{\be}{\begin{equation}}
\newcommand{\ee}{\end{equation}}
\newcommand{\ba}{\begin{eqnarray}}
\newcommand{\ea}{\end{eqnarray}}

\newcommand{\nn}{\nonumber \\}

\newcommand{\factor}{\eta} 

\newcommand{\Google}{\affiliation{Google Quantum AI, Venice, CA, USA}}
\newcommand{\Columbia}{\affiliation{Department of Chemistry, Columbia University, New York, NY, USA}}

\newcommand{\Macquarie}{\affiliation{Department of Physics and Astronomy, Macquarie University, Sydney, NSW, Australia}}
\newcommand{\UW}{\affiliation{Department of Physics, University of Washington, Seattle, WA, USA}}
\newcommand{\PNNL}{\affiliation{Pacific Northwest National Laboratory, Richland, WA, USA}}
\newcommand{\ketbra}[2]{|#1\rangle\!\langle #2|}

\newcommand{\wid}{m}
\newcommand{\dm}{d}
\newcommand{\chunk}{k}

\input{Qcircuit}

\makeatletter
\def\l@subsubsection#1#2{}
\makeatother

\begin{document}

\title{Even more efficient quantum computations of chemistry\texorpdfstring{\\}{} through tensor hypercontraction}

\date{\today}
\author{Joonho Lee}
\email{corresponding author: linusjoonho@gmail.com}
\thanks{these authors contributed equally.}
\Columbia
\author{Dominic W. Berry}
\email{corresponding author: dominic.berry@mq.edu.au}
\thanks{these authors contributed equally.}
\Macquarie
\author{Craig Gidney}
\Google
\author{William J.~Huggins}
\Google
\author{Jarrod R.~McClean}
\Google
\author{Nathan Wiebe}
\UW
\PNNL
\author{Ryan Babbush}
\email{corresponding author: ryanbabbush@gmail.com}
\Google

\begin{abstract}
We describe quantum circuits with only $\widetilde{\cal O}(N)$ Toffoli complexity that block encode the spectra of quantum chemistry Hamiltonians in a basis of $N$ arbitrary (e.g., molecular) orbitals. With ${\cal O}(\lambda / \epsilon)$ repetitions of these circuits one can use phase estimation to sample in the molecular eigenbasis, where $\lambda$ is the 1-norm of Hamiltonian coefficients and $\epsilon$ is the target precision. This is the lowest complexity that has been shown for quantum computations of chemistry within an arbitrary basis. Furthermore, up to logarithmic factors, this matches the scaling of the most efficient prior block encodings that can only work with orthogonal basis functions diagonalizing the Coloumb operator (e.g., the plane wave dual basis). Our key insight is to factorize the Hamiltonian using a method known as tensor hypercontraction (THC) and then to transform the Coulomb operator into an isospectral diagonal form with a non-orthogonal basis defined by the THC factors. We then use qubitization to simulate the non-orthogonal THC Hamiltonian, in a fashion that avoids most complications of the non-orthogonal basis. We also reanalyze and reduce the cost of several of the best prior algorithms for these simulations in order to facilitate a clear comparison to the present work. In addition to having lower asymptotic scaling spacetime volume, compilation of our algorithm for challenging finite-sized molecules such as FeMoCo reveals that our method requires the least fault-tolerant resources of any known approach. 
By laying out and optimizing the surface code resources required of our approach we show that FeMoCo can be simulated using about four million physical qubits and under four days of runtime, assuming $1\,${\textmu}s cycle times and physical gate error rates no worse than $0.1\%$.
\end{abstract}

\maketitle
\tableofcontents

\makeatletter
\let\toc@pre\relax
\let\toc@post\relax
\makeatother 

\listoftables

\section{Introduction}

\subsection{Background}

The quantum computation of quantum chemistry is commonly regarded as one of the most promising applications of quantum computers \cite{Bauer2020QuantumScience,Cao2019QuantumComputing,McArdle2020QuantumChemistry}. This is because there are many applications of quantum chemistry that pertain to the development of practical technologies and, for at least some of these applications, quantum algorithms appear to provide an exponential scaling advantage relative to the best known classical approaches \cite{Aspuru-Guzik2005}. Specifically, most algorithmic work in this area focuses on the construction of quantum circuits that precisely sample in the eigenbasis of the electronic Hamiltonian. This enables the very precise preparation of any electronic eigenstate that can be reasonably approximated with a classically tractable theory, such as the ground states of many molecules. The ability to arbitrarily refine the accuracy of an approximation to molecular eigenstates is valuable because high precision is often required to predict important properties of these systems, such as the rates of chemical reactions, excitation energies, barrier heights, and non-covalent molecular interaction energies \cite{mardirossian2017thirty}. The ``holy grail'' of quantum chemistry is to have a generally applicable electronic structure method that yields relative energies with errors less than 1.6 milliHartrees (so-called ``chemical accuracy''\cite{ruedenberg1973energy}), which is still a long-standing challenge in the field.

The most efficient rigorous approaches to this problem in quantum computing use the quantum phase estimation algorithm \cite{Kitaev1995B,Abrams1999} to sample in the eigenbasis of the molecular Hamiltonian by measuring the phase accumulated on an initial state under the application of a unitary operator with eigenvalues that are related to those of the electronic Hamiltonian. The electronic Hamiltonian in an arbitrary second-quantized basis can be expressed as
\begin{equation}
H = T + V \qquad \qquad T = \sum_{\sigma\in\{\uparrow,\downarrow\}} \sum_{p,q=1}^{N/2} T_{p q} a_{p,\sigma}^\dagger a_{q,\sigma}
\qquad \qquad 
V = \frac{1}{2}\sum_{\alpha,\beta\in\{\uparrow,\downarrow\}} \sum_{p,q,r,s=1}^{N/2}
V_{pqrs}
a_{p,\alpha}^\dagger a_{q,\alpha}
a_{r,\beta}^\dagger a_{s,\beta}
\label{eq:full}
\end{equation}
where $N$ is the number of spin-orbital basis functions used to discretize the Hamiltonian, $T_{pq}$ represents a matrix element of the effective one-body operator,  $V_{pqrs}$ is a matrix element of the two-body Coulomb operator, and $a_{p,\sigma}^\dagger$ and $a_{p,\sigma}$ are fermionic raising and lowering operators for the $p^{\rm th}$ single particle orbital with spin $\sigma$. Note that unlike the convention in \cite{Berry2019B}, we do not absorb the factor of ``1/2'' into the $V_{pqrs}$ (this is consistent with conventions of \cite{vonBurg2020} to avoid confusion).
$V_{pqrs}$ is the central tensor in this work, which is defined as
\begin{equation}
V_{pqrs} = 
\int\!\int \mathrm{d}\mathbf{r}_1\: \mathrm{d}\mathbf{r}_2
\frac{\phi_p(\mathbf r_1) \phi_q(\mathbf r_1) \phi_r(\mathbf r_2) \phi_s(\mathbf r_2)}{|\mathbf r_1 - \mathbf r_2|}
\end{equation}
where $\{\phi_i\}_{i=1}^{N/2}$ is a set of real-valued single particle spatial-orbital basis functions.
\insertrev{
We note that the matrix element $T_{pq}$ is defined as
\begin{equation}
T_{pq} = h_{pq} - \frac12 \sum_{r=1}^{N/2} V_{prrs}
\end{equation}
with $h_{pq}$ being a matrix element of the one-body operator that includes the kinetic energy operator and electron-nuclear Coulomb operator.}

While most work has focused on encoding the eigenspectra of this Hamiltonian in a unitary for phase estimation by synthesizing the time-evolution operator $e^{-i H t}$ \cite{Whitfield2010}, several recent papers (including this one) have demonstrated increased efficiency by instead synthesizing a qubitized quantum walk \cite{Low2016} with eigenvalues proportional to $e^{\pm i \arccos(H / \lambda)}$, where $\lambda$ is a parameter related to the norm of the Hamiltonian. By repeating this quantum walk a number of times scaling as ${\cal O}(\lambda / \epsilon)$ one is able to prepare eigenstates of $H$ and estimate the associated eigenvalue such that the error in the estimated eigenvalue is no greater than $\epsilon$  \cite{Poulin2017,Berry2018}.

Throughout this paper, all discussion of an error $\epsilon$ in the eigenspectra refers to error relative to the finite sized basis of $N$ spin-orbitals that we have used to discretize our system. 
The finite sized basis we choose also introduces a discretization error with respect to the representation of the system in the continuum limit. 
While this basis set discretization error can be asymptotically refined as $\epsilon = {\cal O}(1/N)$ for common choices of the single-particle basis such as plane waves or molecular orbitals, the precise constant factors in this scaling depend on the particular basis functions used (which also determine the coefficients $T_{pq}$ and $V_{pqrs}$ in \eq{full}). 
In the last few years, several papers have demonstrated quantum algorithms for chemistry with reduced complexity \cite{BabbushLow,BabbushSpectraB,Low2018,Kivlichan2019,McClean2020,vonBurg2020}. Some algorithms achieved a lower scaling by performing simulation in a basis \cite{BabbushLow,McClean2020,white2017hybrid} that diagonalizes the Coulomb operator $V$ such that $V_{pqrs} = 0$ unless $p=q$ and $r=s$. However, those representations typically require a significantly larger $N$ in order to model molecular systems within target accuracy of the continuum limit. While such representations might prove practical for molecules when used in first quantization \cite{Kassal2008,BabbushContinuum}, in second quantization the requirement of a significantly larger basis translates into needing significantly more qubits and thus those approaches seem impractical for simulating molecular systems (as opposed to, e.g., crystalline solid-state systems).

Given this, there is a need for efficient quantum algorithms that are compatible with arbitrary basis functions and not limited to those that diagonalize the Coulomb operator. Such quantum algorithms can exploit the molecular orbital basis directly and thereby significantly reduce the required number of basis functions for a target accuracy towards the continuum limit. There are only three prior papers that have fully determined the cost of performing such a quantum algorithm for chemistry within an error-correcting code \cite{Reiher2017,Berry2019B,vonBurg2020}.

The first of these papers by Reiher \emph{et al.}~\cite{Reiher2017} deployed a Trotter-based approach to study the quantum simulation of the FeMo cofactor of the Nitrogenase enzyme, also known as ``FeMoCo'' (${\rm Fe}_7 {\rm Mo} {\rm S}_9 {\rm C}$), a molecule that is important for understanding the mechanism of biological nitrogen fixation \cite{Beinert1997}. The algorithm used for that work had T gate complexity scaling as approximately ${\cal O}(N^2 S / \epsilon^{3/2})$ where $S$ is the Hamiltonian sparsity ($S = {\cal O}(N^4)$ in an arbitrary basis but with a sufficiently large, localized basis sometimes $S={\cal O}(N^2)$ \cite{McClean2014}). The work of Reiher \emph{et al.}~required more than $10^{14}$ T gates to simulate an active space model of FeMoCo. These papers focus on counting (and reducing) the required number of T gates (or Toffoli gates) because within practical error-correcting codes such as the surface code \cite{Fowler2012}, these gates require significantly more time to implement than any other gate and also require a very large number of physical qubits for their implementation. If implemented in the surface code using gates with $10^{-3}$ error rates, the most efficient protocols for implementing T gates require roughly 15 qubitseconds \cite{Litinski2018,Fowler2018} of spacetime volume. At those rates, just distilling the magic states needed for the Reiher \emph{et al.} FeMoCo calculation would require four million qubitdecades (e.g., one million qubits running for four decades or one billion qubits running for two weeks).

This work and the two papers \cite{Berry2019B,vonBurg2020} employ a qubitization based approach \cite{Low2016}, and improve considerably over the Trotter-based methods of \cite{Reiher2017}. The primary difference between these more recent algorithms is that each adapts qubitization to a different type of tensor factorization of the Coulomb operator. The first of these papers, and the first to suggest combining qubitization with tensor factorizations of the Coulomb operator was by Berry \emph{et al.}~\cite{Berry2019B}. That work presents two approaches (one with lower asymptotic complexity, and one with lower gate counts for molecules such as the FeMoCo active space). The approach with the lower finite sized gate counts is based on using qubitization to exploit sparsity structures in the electronic Hamiltonian. 
We discuss that method (referred to throughout as the ``sparse'' method) and its cost in \app{sparse} and show that in \cite{Berry2019B}, an error led us to overestimate the complexity.
We have also recalculated the number of nonzero entries that must be retained in the Hamiltonian, as well as making a number of minor improvements to the algorithm, leading to the sparse approach improving over the Toffoli complexity of the Reiher \emph{et al.}~result by roughly a factor of $1,\!100\times$ rather than the factor of $700\times$ reported in \cite{Berry2019B,vonBurg2020}. In terms of asymptotic complexity this approach has Toffoli complexity $\widetilde{\cal O}((N + \sqrt{\sparsity}) \lambda / \epsilon)$\footnote{Here and throughout the paper we use $\widetilde{\cal O}(\cdot)$ to indicate an asymptotic upper bound suppressing polylogarithmic factors in the scaling.} and space complexity $\widetilde{\cal O}(N + \sqrt{\sparsity})$, where $\lambda$ is a parameter related to the Hamiltonian norm (we discuss $\lambda$ further towards the end of the introduction).

The lower scaling method (but with slightly higher constant factors) presented by Berry \emph{et al.}~\cite{Berry2019B} combines qubitization with the first step of a tensor factorization originally discussed for quantum computing in \cite{Motta2018v2}. Here we refer to that representation as the ``single low rank'' factorization or simply ``SF''. The SF uses an eigendecomposition of the two-electron integral tensor, similar to the Cholesky decomposition~\cite{aquilante2011cholesky} or density fitting~\cite{Whitten1973,Baerends1973,Jafri1974} as commonly used in electronic structure literature. The single low rank factorization algorithm obtains Toffoli and space complexities of $\widetilde{\cal O}(N^{3/2} \lambda / \epsilon)$ and $\widetilde{\cal O}(N^{3/2})$, respectively. We discuss that method and its cost in \app{low_rank}.

The most recent paper by von Burg \emph{et al.}~\cite{vonBurg2020} adapts qubitization to a second tensor factorization that occurs on top of the SF used by Berry \emph{et al.} This second tensor factorization was also first described for quantum computing in \cite{Motta2018v2} and corresponds to a diagonalization of the squared one-body operators, which dates back to \cite{peng2017highly} in the electronic structure literature. We refer to this as the ``double low rank'' factorization or simply DF. We review the method and its cost in \app{double_low_rank}. Von Burg \emph{et al.}~claim that their DF algorithm gives more than an order of magnitude complexity improvement relative to the SF algorithm of Berry \emph{et al.}; however, the numbers they compare to are actually those for the sparse algorithm of Berry \emph{et al.}~(which is more efficient than the SF method for molecules they consider). 
Furthermore, for the more accurate FeMoCo Hamiltonian introduced by Li \emph{et al.}~\cite{Li2019}, the sparse algorithm requires half as many logical qubits and 2/3 the number of Toffoli gates as the DF algorithm (so in that case, the sparse algorithm is considerably cheaper).
The Toffoli complexity of the DF approach is  $\widetilde{\cal O}(N \lambda \sqrt{\ranktwo}/\epsilon)$ with space complexity $\widetilde{\cal O}(N\sqrt{\ranktwo})$ where $\ranktwo$ is the average rank of the second tensor factorization discussed in \cite{Motta2018v2}. In most cases (including the regimes that are usually of interest for small quantum computers) $\ranktwo$ will scale around ${\cal O}(N)$, giving similar scaling as the SF method. There is some evidence that when $N$ is growing towards the thermodynamic limit (the number of atoms is very large and increasing while fixing the ratio of spin-orbitals to atoms) $\ranktwo$ can scale as ${\cal O}(\log N)$ \cite{peng2017highly}; however, this is not the case in our numerics on hydrogen systems which reveal scaling of ${\cal O}(N)$.

There are several other commonalities between the algorithms of \cite{Berry2019B,vonBurg2020} and the ones developed here which are worth discussing before introducing the main techniques of this paper. In addition to combining qubitization \cite{Low2016} and tensor factorizations, all three works involve the use of unary iteration, QROM, coherent alias sampling, and qubitized phase estimation bounds from \cite{BabbushSpectraB} as well as a more advanced version of QROM with fanout first developed in \cite{Lowpreparation} and then optimized for use in our context \cite{Berry2019B} (where it is referred to as ``QROAM'' -- a portmanteau of QROM and QRAM). 
Using these tools one can often construct algorithms to realize qubitized quantum walks with gate complexity that scales as ${\cal O}(\sqrt{\Gamma})$ where one requires ${\cal O}(\sqrt{\Gamma})$ ancilla, and $\Gamma$ is the amount of information required to specify the Hamiltonian within a particular tensor factorization. Then, by performing phase estimation on the resultant qubitized quantum walks, one is able to sample in the Hamiltonian eigenbasis with Toffoli complexity $\widetilde{\cal O}(\sqrt{\Gamma} \lambda / \epsilon)$.

For the sparse algorithm of \cite{Berry2019B}, no tensor factorization is employed and thus the Hamiltonian (the same as in \eq{full}) contains a number of integrals scaling as $\Gamma = \widetilde{\cal O}(\sparsity)$. Thus, accounting for a complexity proportional to $N$ required to perform controlled operations, we end up with an algorithm having Toffoli complexity $\widetilde{\cal O}((N + \sqrt{\sparsity}) \lambda / \epsilon)$ and space complexity $\widetilde{\cal O}(N + \sqrt{\sparsity})$. For the SF algorithm of \cite{Berry2019B} we rely on the single low rank factorized Hamiltonian shown in \app{low_rank} which is specified with an amount of information scaling as $\Gamma = \widetilde{\cal O}(N^3)$. This leads to an algorithm with Toffoli complexity $\widetilde{\cal O}(N^{3/2} \lambda/\epsilon)$ and space complexity $\widetilde{\cal O}(N^{3/2})$. Finally, the DF algorithm of \cite{vonBurg2020} relies on the factorization shown in \app{double_low_rank} which compresses the Hamiltonian to $\Gamma = \widetilde{\cal O}(N^2 \ranktwo)$ pieces of information. Accordingly, this approach yields gate complexity $\widetilde{\cal O}(N \lambda \sqrt{\ranktwo}/\epsilon)$ and space complexity $\widetilde{\cal O}(N \sqrt{\ranktwo})$.

The last element of the complexity of these algorithms that we should discuss is the scaling of $\lambda$. The quantity $\lambda$ is a type of norm of the Hamiltonian that is being simulated, and depends on how it is expressed. For example, for the ``sparse'' algorithm of \cite{Berry2019B} the value of $\lambda$ is simply the sum of the absolute value of Hamiltonian coefficients (i.e., the 1-norm) appearing in \eq{full}.  Numerical studies reveal that $\lambda$ usually scales somewhere in between ${\cal O}(N)$ and ${\cal O}(N^3)$ depending on details of the particular system as well as the algorithm and how one is increasing $N$ (e.g., the scaling is lower if $N$ is growing because the number of basis functions per atom is fixed but the number of atoms is increasing and the scaling is higher if $N$ is growing because the number of atoms is fixed but the basis size is growing). The largest $\lambda$ tends to be that associated with the SF used in \cite{Berry2019B}, followed by the sparse representation used in \cite{Berry2019B}, with the smallest $\lambda$ of three approaches discussed so far being the $\lambda$ associated with the DF used in \cite{vonBurg2020}. In most contexts the main advantage of the DF algorithm relative to the SF algorithm is not a substantially smaller $\Gamma$ (or less complex quantum walk) but rather, a smaller $\lambda$. Thus, it is critical to consider how the algorithmic choices that one makes will affect the value of $\lambda$.

\subsection{Overview of results}

We now discuss the main results of this paper. We will present a new qubitization based algorithm that results from exploiting structure in the molecular Hamiltonian that emerges from a tensor factorization of the Coulomb operator known as tensor hypercontraction (THC) \cite{Hohenstein2012,Parrish2012,Hohenstein2012a}. THC is a very compact representation for the Hamiltonian which gives $\Gamma = \widetilde{\cal O}(N^2)$ regardless of how one increases $N$. It is relatively straightforward to apply qubitization directly to the THC representation and a method for that is presented in \app{standard_thc}. This results in an approach with Toffoli complexity $\widetilde{\cal O}(N \lambda / \epsilon)$ and space complexity $\widetilde{\cal O}(N)$. However, it turns out that directly applying qubitization to the THC representation is not particularly efficient because this causes $\lambda$ to become even larger than the single low rank $\lambda$ of \cite{Berry2019B} (which was already orders of magnitude larger than the double low rank $\lambda$ of \cite{vonBurg2020} for systems like FeMoCo). To remedy this, we discuss a different approach to utilizing the THC representation of the Hamiltonian.

We show that one can use the THC tensors to define a larger orbital basis with exactly the same resolution as the original basis while diagonalizing the Coulomb operator. This form of the Hamiltonian would then be amenable to simulation using the efficient strategies of \cite{BabbushSpectraB,Low2018,Kivlichan2019} except for the fact that the orbitals are non-orthogonal. However, by separately rotating into this non-orthogonal basis before operating on each tensor factor of the Hamiltonian we are able to avoid many of the complications that would usually arise from simulating a non-orthogonal operator. To achieve this, we use a strategy for storing and implementing non-orthogonal Givens rotation based orbital basis transformations \cite{kivlichan2018quantum} with rotation angles loaded from QROM that is similar to techniques developed for qubitizing orthogonal basis rotations in  \cite{vonBurg2020}. As these rotations add significant cost to the method, we also introduce a strategy for coarse graining the angles of the basis transformation that allows us to precisely control a tradeoff between the complexity of these rotations and the accuracy of the Hamiltonian (a similar technique would also reduce the cost of the algorithm by von Burg \emph{et al.}). Ultimately, our method does not require additional system qubits (beyond those for QROM and other minor ancillary functions) and results in asymptotic Toffoli complexity $\widetilde{\cal O}(N \lambda / \epsilon)$ and space complexity $\widetilde{\cal O}(N)$, while essentially matching the small $\lambda$ values obtained in the double low rank algorithm.

As can be seen in \tab{big_table} our algorithm improves over the spacetime volume of the approaches in \cite{Berry2019B,vonBurg2020} by a factor of about $\widetilde{\cal O}(N)$ in most contexts. In fact, the scaling is often even better than that due to the lower scaling of $\lambda$ associated with our representation, as can be seen from numerics on hydrogen systems in \tab{hydrogen_scalings}. Surprisingly, the $\lambda$ value associated with our algorithm sometimes scales even less than ${\cal O}(N^2)$, which is the scaling of the $\lambda$ for the lowest scaling qubitization approach requiring orthogonal basis functions that diagonalize the Coulomb operator \cite{BabbushSpectraB}.

\begin{table*}
\begin{tabular}{|c|c|c|c|c|}\hline
Year
& Reference
& Primary algorithmic innovation
& Space complexity
& Toffoli/T complexity\\
\hline
2005
& Aspuru-Guzik \emph{et al}.~\cite{Aspuru-Guzik2005}
& First algorithm (no compilation or bounds)
& ${\cal O}(N)$
& ${\cal O}(\textrm{poly}(N / \epsilon))$\\
2010
& Whitfield \emph{et al}.~\cite{Whitfield2010}
& First compilation (no Trotter bounds)
& ${\cal O}(N)$
& ${\cal O}(\textrm{poly}(N / \epsilon))$\\
2012
& Seeley \emph{et al}.~\cite{Seeley2012}
& Use of Bravyi-Kitaev transformation
& ${\cal O}(N)$
& ${\cal O}(\textrm{poly}(N / \epsilon))$\\
2013
& Wecker \emph{et al}.~\cite{Wecker2014B}
& First chemistry specific Trotter bounds
& ${\cal O}(N)$
& $\widetilde{\cal O}(N^{10} / \epsilon^{3/2})$\\
2013
& Toloui \emph{et al}.~\cite{Toloui2013}
& Use of first quantization
& ${\cal O}(\eta \log N)$
& $\widetilde{\cal O}(\eta^2 N^{8}/ \epsilon^{3/2})$\\
2014
& Hastings \emph{et al}.~\cite{Hastings2015}
& Better compilation and multi-resolution Trotter
& ${\cal O}(N)$
& $\widetilde{\cal O}(N^8 / \epsilon^{3/2})$\\
2014
& Poulin \emph{et al}.~\cite{Poulin2014}
& Tighter Trotter bounds and ordering
& ${\cal O}(N)$
& $\widetilde{\cal O}(N^6 / \epsilon^{3/2})$\\
2014
& McClean \emph{et al}.~\cite{McClean2014}
& Exploiting Hamiltonian sparsity with Trotter
& ${\cal O}(N)$
& $\widetilde{\cal O}(N^4 \sparsity / \epsilon^{3/2})$\\
2014
& Babbush \emph{et al}.~\cite{BabbushTrotter}
& Tighter system specific Trotter bounds
& ${\cal O}(N)$
& $\widetilde{\cal O}(N^2 \sparsity / \epsilon^{3/2})$\\
2015
& Babbush \emph{et al}.~ \cite{BabbushSparse1}
& Use of Taylor series (database method)
& ${\cal O}(N)$
& $\widetilde{\cal O}(N^4 \lambda_V / \epsilon)$\\
2015
& Babbush \emph{et al}.~ \cite{BabbushSparse1}
& Use of Taylor series (on-the-fly method)
& ${\cal O}(N)$
& $\widetilde{\cal O}(N^5 / \epsilon)$\\
2015
& Babbush \emph{et al}.~\cite{BabbushSparse2}
& Use of Taylor series with first quantization
& ${\cal O}(\eta \log N)$
& $\widetilde{\cal O}(\eta^2 N^3 / \epsilon)$\\
2016
& Reiher \emph{et al}.~\cite{Reiher2017}
& First T count and tighter Trotter bounds
& ${\cal O}(N)$
& $\widetilde{\cal O}(N^2 \sparsity  / \epsilon^{3/2})$\\
2018
& Motta \emph{et al.}~\cite{Motta2018v2}
& Use of low rank factorization with Trotter
& ${\cal O}(N)$
& $\widetilde{\cal O}(N^4 \ranktwo / \epsilon^{3/2})$\\
2018
& Campbell~\cite{Campbell2018B}
& Use of randomized compiling with Trotter
& ${\cal O}(N)$
& $\widetilde{\cal O}(\lambda_V^2 / \epsilon^2)$\\
2019
& Berry \emph{et al}.~\cite{Berry2019B}
& Use of qubitization (sparse method)
& $\widetilde{\cal O}(N + \sqrt{\sparsity} )$
& $\widetilde{\cal O}((N + \sqrt{\sparsity}) \lambda_V / \epsilon)$\\
2019
& Berry \emph{et al}.~\cite{Berry2019B}
& Use of qubitization (single factorization)
& $\widetilde{\cal O}(N^{3/2})$
& $\widetilde{\cal O}(N^{3/2} \lambda_{\rm SF} / \epsilon)$\\
2019
& Kivlichan \emph{et al}.~\cite{Kivlichan2019PhaseHamiltoniansB}
& Better randomized compiled phase estimation
& ${\cal O}(N)$
& $\widetilde{\cal O}(\lambda_V^2 / \epsilon^2)$\\
2020
& von Burg \emph{et al}.~\cite{vonBurg2020}
& Use of qubitization (double factorization)
& $\widetilde{\cal O}(N \sqrt{\ranktwo})$
& $\widetilde{\cal O}(N  \lambda_{\rm DF} \sqrt{\ranktwo}/ \epsilon)$\\
2020
& Present work
& Use of tensor hypercontraction
& $\widetilde{\cal O}(N)$
& $\widetilde{\cal O}(N \lambda_{\zeta} / \epsilon)$\\
\hline
\end{tabular}
\caption[Lowest asymptotic scaling quantum algorithms for chemistry]{\label{tab:big_table} History of the lowest asymptotic scaling quantum algorithms for simulating quantum chemistry in an arbitrary (e.g., molecular orbital) basis. $N$ is the number of arbitrary orbital basis functions, $\eta$ is the number of electrons (relevant only in first-quantized simulations) and $\epsilon$ is the target precision to which we estimate the Hamiltonian eigenvalues using phase estimation. $\sparsity$ is the sparsity of the electronic Hamiltonian; usually $\sparsity = {\cal O}(N^4)$ when using an arbitrary basis but sometimes the scaling can be lower. The $\lambda$ parameters (discussed extensively in this paper) are roughly the 1-norm of the Coulomb operator associated with the representation in which we simulate the system. In general, we expect that ${\cal O}(\lambda_\zeta) \leq {\cal O}(\lambda_{\rm DF}) \leq {\cal O}(\lambda_V) \leq {\cal O}(\lambda_{\rm SF})$ but the precise scaling is difficult to report since it depends on the specific molecule and how $N$ is growing. Roughly, the $\lambda$ values have scaling that is typically between ${\cal O}(N)$ and ${\cal O}(N^3)$. Here, $\ranktwo$ is the average rank of the second tensor factorization discussed in Motta \emph{et al.}~\cite{Motta2018v2}, which is also important for the complexity of the work of von Burg \emph{et al.}~\cite{vonBurg2020}. 
In general (for example, when scaling towards the continuum limit) we would expect that $\ranktwo = {\cal O}(N)$. 
But for large systems that are growing because we are adding more atoms while keeping the basis to atom ratio fixed, $\ranktwo$ can be smaller; for the hydrogen chains studied in this work we observe $\ranktwo = {\cal O}(N)$ up to 100 hydrogen atoms. 
See \tab{hydrogen_scalings} for a better sense of how algorithms that depend on $\sparsity$, $\lambda$ and $\ranktwo$ parameters scale for hydrogen benchmarks. The work of Motta \emph{et al.}~\cite{Motta2018v2} does not determine the scaling of the Trotter error for the Trotter steps compiled therein and so this table assumes (rather speculatively) that the Trotter error scaling for those Trotter steps is the same as the Trotter error scaling for the Trotter steps of Reiher \emph{et al.}~\cite{Reiher2017}. This table omits methods that require special basis functions or non-Galerkin representations. All such representations are less compact for molecules compared to molecular orbitals and thus require more qubits to reach the same level of accuracy. Notable examples include the grid bases used in \cite{Kassal2008,Kivlichan2016}, the discontinuous Galerkin techniques of \cite{McClean2020}, the plane waves required by \cite{BabbushContinuum,vonBurg2020,Low2018} and the basis sets diagonalizing the Coulomb operator required by \cite{BabbushLow,kivlichan2018quantum,Kivlichan2019}.}
\end{table*}

\begin{table*}[t]
\def\arraystretch{1.2}
\begin{tabular}{|c|c|c|c|c|}
\hline
\multirow{2}{*}{Algorithm}
& \multicolumn{2}{c|}{${\rm H}_4$ continuum limit} 
& \multicolumn{2}{c|}{hydrogen chain thermodynamic limit}\\
\cline{2-5}
    & space complexity
    & Toffoli complexity
    & space complexity
    & Toffoli complexity\\
\hline
Babbush \emph{et al.}~\cite{BabbushSparse1} (Taylor series database)
  & ${\cal O}(N)$
  & $\widetilde{\cal O}(N^{7.1}/\epsilon)$
  & ${\cal O}(N)$
  & $\widetilde{\cal O}(N^{5.3}/\epsilon)$\\
Campbell/Kivlichan \emph{et al.}~\cite{Campbell2018B,Kivlichan2019PhaseHamiltoniansB} (qDRIFT)
  & ${\cal O}(N)$
  & $\widetilde{\cal O}(N^{6.2}/\epsilon^2)$
  & ${\cal O}(N)$
  & $\widetilde{\cal O}(N^{2.5}/\epsilon^2)$\\
 Berry \emph{et al.} \cite{Berry2019B} (single factorization)
  & $\widetilde{\cal O}(N^{1.5})$
  & $\widetilde{\cal O}(N^{3.8}/\epsilon)$
  & $\widetilde{\cal O}(N^{1.5})$
  & $\widetilde{\cal O}(N^{4.5}/\epsilon)$\\
 Berry \emph{et al.} \cite{Berry2019B} (sparse) 
  & $\widetilde{\cal O}(N^{1.9})$
  & $\widetilde{\cal O}(N^{5.0}/\epsilon)$
  & $\widetilde{\cal O}(N)$
  & $\widetilde{\cal O}(N^{2.3}/\epsilon)$\\
 von Burg \emph{et al.} \cite{vonBurg2020} (double factorization)
  & $\widetilde{\cal O}(N^{1.5})$
  & $\widetilde{\cal O}(N^{3.8}/\epsilon)$
  & $\widetilde{\cal O}(N^{1.5})$
  & $\widetilde{\cal O}(N^{3.4}/\epsilon)$\\
 this work (tensor hypercontraction)
  & $\widetilde{\cal O}(N)$
  & $\widetilde{\cal O}(N^{3.1}/\epsilon)$
  & $\widetilde{\cal O}(N)$
  & $\widetilde{\cal O}(N^{2.1}/\epsilon)$\\
\hline
\end{tabular}
\caption[Scaling of $\lambda$-dependent algorithms for hydrogen systems]{\label{tab:hydrogen_scalings}
Empirical complexity of algorithms that have scalings obscured by $\lambda$ values in \tab{big_table} for two benchmark chemical series. $N$ is the number of spin-orbital basis functions and $\epsilon$ is the target precision to which we would aim to realize phase estimation. This table summarizes the findings of numerics discussed in \sec{numerics} which reveal the scaling with $N$ that one might observe in practice for these algorithms when the system size is growing towards either the continuum or thermodynamic limits. For the continuum limit we focus on ${\rm H}_4$ Hamiltonians with each hydrogen placed on the corners of a square plaquette with side length of 2.0 Bohr radii. We then increase the number of molecular orbitals used to represent the system and determine the scaling of the associated algorithms. For scalings towards the thermodynamic limit we fix the ratio of basis functions to atoms and increase the number of hydrogens in a one-dimensional hydrogen chain, with atom spacings again at 1.4 Bohr radii. For more information on these calculations, see \sec{numerics}.  The justification for the above scalings for qDRIFT in the case where $\epsilon$ is the root-mean-square error is given in \app{drift}.
}
\label{tab:h4hchain}
\end{table*}

In addition to having the best asymptotic scaling compared to prior approaches, our algorithm also outperforms the finite spacetime volume for all molecules studied here including hydrogen chains and FeMoCo. We study active space models of FeMoCo proposed by Reiher \emph{et al.}~\cite{Reiher2017} as well as by Li \emph{et al.}~\cite{Li2019}. 
The Reiher Hamiltonian was found to be qualitatively incorrect in describing the ground state of FeMoCo as
it does not capture the open-shell nature of the system~\cite{Li2019}. The Li Hamiltonian was then proposed and shown to capture the open-shell nature of the ground state properly~\cite{Li2019}. We focus on the FeMoCo Hamiltonians primarily because they are regarded as a standard benchmark for quantum computing. For the FeMoCo Hamiltonians of Reiher \emph{et al.}~and Li \emph{et al.}~we find a reduction in spacetime volume of about $3\times$ and $6\times$ (respectively) compared to \cite{vonBurg2020}. The results for FeMoCo are summarized in \tab{femoco_scalings}.

We also carefully analyze the surface code resources required to simulate the FeMoCo Hamiltonian of Reiher \emph{et al.} \cite{Reiher2017} using our THC approach. Rather than just focus on the cost of distillation, we fully lay out the surface code computation in spacetime and optimize resource usage. We determine that the computation could execute using approximately four million physical qubits and run in under four days, assuming surface code cycle times of 1 microsecond and gate error rates of about $0.1\%$. We find that if error rates were reduced to $0.01\%$ that the computation could complete using about one million physical qubits and under two days.

\begin{table*}[t]
\def\arraystretch{1.2}
\begin{tabular}{|c|c|c|c|c|}
\hline
\multirow{2}{*}{Algorithm}
& \multicolumn{2}{c|}{Reiher \emph{et al.}~FeMoCo \cite{Reiher2017}} 
& \multicolumn{2}{c|}{Li \emph{et al.}~FeMoCo \cite{Li2019}}\\
\cline{2-5}
    & logical qubits
    & Toffoli count
    & logical qubits
    & Toffoli count\\
\hline
Reiher \emph{et al.}~\cite{Reiher2017} (Trotter)
  & 111
  & $5.0 \times 10^{13}$
  & ---
  & ---\\
  \hline
Campbell/Kivlichan \emph{et al.}~\cite{Campbell2018B,Kivlichan2019PhaseHamiltoniansB} (qDRIFT) \eqref{eq:ToffqDRIFT}, \eqref{eq:ancqDrift}
  & 288
  & $5.2 \times 10^{27}$
  & 328
  & $1.8\times 10^{28}$\\
  qDRIFT with 95\% confidence interval \eqref{eq:CIest} & 270 & $1.9\times 10^{16}$ & 310 & $1.0\times 10^{16}$\\
 \hline
 Berry \emph{et al.} \cite{Berry2019B} (single factorization) \eqref{eq:lowranktof}, \eqref{eq:lowranklog}
 
 & 3,320
 & $9.5 \times 10^{10}$
 & 3,628
 & $1.2 \times 10^{11}$\\
 \hline
 Berry \emph{et al.} \cite{Berry2019B} (sparse) \eqref{eq:sparsetoffoli}, \eqref{eq:sparsequbits}
 & 2,190
 &$8.8\times 10^{10}$
 & 2,489
 &$4.4\times 10^{10}$
  \\
  \hline
 von Burg \emph{et al.} \cite{vonBurg2020} (double factorization) \eqref{eq:MStoffolis}, \eqref{eq:MSqubits}
  & 3,725
  & $1.0\times 10^{10}$
  & 6,404
  & $6.4 \times 10^{10}$\\
  \hline
 this work (tensor hypercontraction) \eqref{eq:THCtoffolis}, \eqref{eq:THCqubits} 
  & 2,142
  & $5.3\times 10^{9}$
  & 2,196
  & $3.2\times 10^{10}$\\
\hline
\end{tabular}
\caption[Resources required to simulate FeMoCo]{\label{tab:femoco_scalings}
Here we report the finite resources required for various recent algorithms to quantum phase estimate two FeMoCo active spaces to within chemical accuracy. The reason for focusing on two different active spaces is that several papers benchmarked their methods for the FeMoCo Hamiltonian of Reiher \emph{et al.}~\cite{Reiher2017}, but the work of Li \emph{et al.}~\cite{Li2019} later showed that there is almost no open-shell character in the ground state of the active space model by Reiher \emph{et al.} and proposed a slightly larger active space with important open-shell character.
We report the lowest known Toffoli and logical qubit counts of prior algorithms, consistent with our accounting of these costs in \app{sparse}, \app{low_rank}, \app{double_low_rank} and \app{drift}. 
Note that \app{sparse} and \app{low_rank} show how to reduce the Toffoli counts reported for the algorithms of Berry \emph{et al.}~by factors of roughly $13\times$ for the Reiher Hamiltonian and $8\times$ for the Li Hamiltonian (single factorization) and $3 \times$ for the Reiher Hamiltonian and $2 \times$ for the Li Hamiltonian (sparse), respectively, and we use these more optimized resource estimates here. Likewise, the resource estimates reported for von Burg \emph{et al.}~\cite{vonBurg2020} are slightly different from what is reported in their paper because we use a different criterion for determining the truncation thresholds (which, as we discuss in this paper, is a justified, but a tighter criterion than what is used in their work). There were some errors with the algorithm of von Burg \emph{et al.}~as presented in their work, which we correct here in \app{double_low_rank}. Finally, because the Trotter based methods from Reiher \emph{et al.}~are more naturally bottlenecked by T gates than by Toffoli gates, we report half the number of T gates that would be required since a Toffoli gate is roughly twice the cost of a T gate within the surface code \cite{Gidney2018pub}.
{We used 10 bits for state preparation in methods of Berry \emph{et al.}, von Burg \emph{et al.}, and tensor hypercontraction.
A total of 16 bits for the Reiher Hamiltonian and 20 bits for the Li Hamiltonian were used for rotations in tensor hypercontraction and double factorization.}
}
\end{table*}

Finally, in \app{drift} we also provide a detailed analysis of the error in phase estimation when combined with stochastic approximations such as qDRIFT~\cite{Campbell2018B}.  We find that while existing analyses naturally lead to reasonable region estimates for the estimated phase, the distribution of errors for the phase estimation procedure can have fat tails.  These tails manifest in \tab{femoco_scalings} in the form of impractically large numbers of Toffoli gates needed to perform phase estimation on both benchmark examples for FeMoCo.  Specifically, we find that the cost of performing qDRIFT unmodified is nearly $18$ orders of magnitude greater than the cost of performing the optimized form of qubitization that we consider for these benchmark molecules. If we only require that an estimate within an $\alpha$-confidence region is reported, then the costs can be reduced by twelve orders of magnitude.  
Alternatively, quantifying the performance according to the Hodges-Lehmann estimator provides a further order of magnitude improvement. This illustrates a poorly appreciated fact in the quantum simulation literature: for simulation techniques like qDRIFT that have high failure probability, we need to carefully specify the error metric used for the eigenphase yielded by the algorithm because some error metrics (such as the mean square error) can be much harder to minimize than others.

\subsection{Paper organization}

Our paper is organized as follows. In \sec{thc} we describe the THC factorization of the electronic Hamiltonian. We give some background on how the factorization can be obtained, and discuss how it has been shown to scale. We outline how one can directly apply qubitization to the THC Hamiltonian, although with large constant factors in the scaling. Then, we introduce an elegant use of the THC Hamiltonian that corresponds to a diagonal Coulomb operator in a larger non-orthogonal basis.

In
\sec{algorithm} we give a complete description of how qubitization can be combined with the novel (diagonal and non-orthogonal) THC representation to give the most efficient known quantum algorithm for simulating electronic structure in an arbitrary basis. We compile our approach all the way to Clifford + Toffoli gates and report the constant factors in the leading order scaling for both the total Toffoli complexity and total ancilla required.

In \sec{numerics} we analyze the finite resources required to perform the algorithm of \sec{algorithm} for the simulation of several real systems: FeMoCo and hydrogen chains of various sizes. These numerics demonstrate the effectiveness of the THC representation, help to elucidate certain aspects of the scaling of our approach, and allow us to compare the cost of our approach to prior methods. The second part of this section discusses the layout of the Reiher FeMoCo Hamiltonian simulation in the surface code. Our analysis considers the cost of routing, distillation, and analyzes how many physical qubits are required at all points in the computation. Finally, we reflect on the significance of these results and suggest future directions for research in \sec{conclusion}.

\app{sparse}, \app{low_rank} and \app{double_low_rank} contain extensive and self-contained descriptions of the prior methods from \cite{Berry2019B} and \cite{vonBurg2020}, adapted to the notation and conventions of this paper, and in some cases with improved bounds on the complexity of those methods. Furthermore, they also contain numerical details associated with each method that are not presented in the main text. \app{drift} covers the randomized compiled methods of \cite{Campbell2018B} which are not particularly related to the other methods here but also have scaling that depends on $\lambda$ so we are able to analyze the exact resources required by that approach with numerics that were already available to us. Then, in \app{standard_thc} we perform a detailed analysis of how one can directly use the THC representation without projecting into the non-orthgonal basis. This leads to an exceptionally simple-to-understand algorithm (more straightforward than the other approach of this paper or those of \cite{Berry2019B,vonBurg2020}) but with worse constant factors in the scaling compared to the primary approach of this paper due to a larger value of $\lambda$. The remaining two appendices discuss technical details pertaining to the implementation of important circuit primitives used throughout this work. In \app{contig} we discuss a technique and costings for computing contiguous registers. Finally, in \app{qrom} we describe and analyze the cost of modifying the QROM procedure \cite{Lowpreparation,Berry2019B} to output two registers at a time.

\section{Tensor Hypercontraction Representations for Quantum Simulation}
\label{sec:thc}

\subsection{The standard tensor hypercontraction representation}
\label{sec:standard_thc}

The tensor hypercontraction (THC) representation \cite{Hohenstein2012,Parrish2012,Hohenstein2012a} of the electronic Hamiltonian factorizes the Coulomb operator $V$ from \eq{full} as
\begin{equation}
\label{eq:thc}
V \approx G = \frac{1}{2} \sum_{\alpha, \beta \in \{\uparrow, \downarrow\}} \sum_{p,q,r,s=1}^{N/2} G_{pqrs} a^\dagger_{p,\alpha} a_{q,\alpha} a^\dagger_{r,\beta} a_{s,\beta}
\qquad \qquad
G_{pqrs} =
\sum_{\mu, \nu = 1}^{M} \chi_{p}^{(\mu)} \chi_{q}^{(\mu)} \zeta_{\mu\nu} \chi_{r}^{(\nu)}  \chi_{s}^{(\nu)} \end{equation}
where $\chi_p^{(\mu)}$ and $\zeta_{\mu\nu}= \zeta_{\nu\mu}$ are real scalars obtained via existing algorithms ({\it vide infra}) and empirical studies \cite{Hohenstein2012,Parrish2012,Hohenstein2012a,Parrish2013a,Hohenstein2013a,Hohenstein2013,benedikt2013tensor,Parrish2014,Lu2015,KokkilaSchumacher2015,Song2016,Lu2016,Hu2017,Song2017,Lu2017,Song2017a,hummel2017low,schutski2017tensor,Dong2018,Hu2018,Duchemin2019,lee2019systematically,A.Matthews2020ImprovedHypercontraction} suggest that the approximation $G$ encodes low energy eigenvalues of $V$ to within error $\epsilon_{\textsc{thc}}$ so long as
\begin{equation}
\label{eq:M_scaling}
 M = {\cal O}\left(N \,\textrm{polylog}\left(1/\epsilon_{\textsc{thc}}\right)\right).
\end{equation}
The $\epsilon_{\textsc{thc}}$ inside of \eq{M_scaling} is an error in the energy \emph{per atom}. Thus, if we instead intend $\epsilon_{\textsc{thc}}$ to represent a fixed additive error in the total energy then when $N$ is growing towards the thermodynamic limit (i.e., if we are fixing the ratio of basis functions to atoms and adding more atoms) $M = {\cal O}(N \,\textrm{polylog} (N/\epsilon_{\textsc{thc}}))$, but when $N$ is growing towards the continuum limit (i.e., we are fixing the number of atoms and adding more basis functions) $M = {\cal O}(N \,\textrm{polylog} (1/\epsilon_{\textsc{thc}}))$.

While the behavior that the THC rank $M$ should scale near linearly in $N$ has been observed in many contexts \cite{Hohenstein2012,Parrish2012,Hohenstein2012a,Parrish2013a,Hohenstein2013a,Hohenstein2013,benedikt2013tensor,Parrish2014,Lu2015,KokkilaSchumacher2015,Song2016,Lu2016,Hu2017,Song2017,Lu2017,Song2017a,hummel2017low,schutski2017tensor,Dong2018,Hu2018,Duchemin2019,lee2019systematically,A.Matthews2020ImprovedHypercontraction}, the most rigorous result establishing the poly-logarithmic dependence on $\epsilon_{\textsc{thc}}$ comes from \cite{A.Matthews2020ImprovedHypercontraction}. Specifically, the work of \cite{A.Matthews2020ImprovedHypercontraction} employs perturbation theory to empirically evaluate the scaling of the scaling of $M$ in \eq{thc}, ultimately concluding that (within second-order perturbation theory) $M = {\cal O}(N \,\textrm{polylog} (1/\epsilon_{\textsc{thc}}))$. 
The leading order constant for this scaling is both system and basis dependent. Nevertheless, generally speaking, 
to achieve 50 {\textmu}Hartree per atom (a widely accepted accuracy threshold for approximating $V$ to within chemical accuracy for molecular simulations), one needs at least as many THC basis functions as density-fitting basis functions when using ${\cal O}(N^4)$ algorithms \cite{schutski2017tensor} or slightly more with efficient ${\cal O}(N^3)$ algorithms~\cite{lee2019systematically}. With the ${\cal O}(N^4)$ algorithm in \cite{schutski2017tensor}, one needs $M$ equal to between $2N$ and $3N$ for a broad class of chemical problems to achieve chemical accuracy.

The structure of the tensor factorization in \eq{thc} is rather different from either the SF or the DF used in \cite{Berry2019B,vonBurg2020}. For instance, unlike with the low rank decompositions, it is unclear how one might combine the THC factorization with reduced scaling product formulas for time-evolution \cite{Motta2018v2} or better methods of performing energy measurements for variational algorithms \cite{Huggins2019EfficientComputersNew}. Similar to how the work of \cite{Berry2019B,vonBurg2020} combines qubitization with the tensor factorizations described in \app{low_rank} and \app{double_low_rank}, here we will discuss how the THC factorization leads to an advantage when combined with qubitization; however, our approach will require different qubitization oracles (and thus different algorithms) from those in \cite{Berry2019B,vonBurg2020}.

\subsection{Numerical computation of the tensor hypercontraction factorization}
\label{eq:numerical_thc}

The problem of obtaining the factorization of \eq{thc} is often numerically ill-conditioned and has been the subject of research for many years in quantum chemistry \cite{Hohenstein2012,Parrish2012,Hohenstein2012a,Parrish2013a,Hohenstein2013a,Hohenstein2013,benedikt2013tensor,Parrish2014,Lu2015,KokkilaSchumacher2015,Song2016,Lu2016,Hu2017,Song2017,Lu2017,Song2017a,hummel2017low,schutski2017tensor,Dong2018,Hu2018,Duchemin2019,lee2019systematically,A.Matthews2020ImprovedHypercontraction}. 
We write the THC factorization problem as an $L_2$ norm minimization problem where
we seek minimizers, $\chi$ and $\zeta$, for
\begin{equation}
    \mathcal L =  \left\| V-G \right\|_2 = \sum_{pqrs} \left| V_{pqrs} - G_{pqrs} \right|^2.
 \label{eq:lagrangian}
\end{equation}
This objective function $\mathcal L$ is quartic in $\chi$ and linear in $\zeta$.
This generally exhibits multiple minima as well as a flat optimization landscape,
which makes finding global minimizers challenging. 
We developed a strategy to cope
with numerical difficulties, which was found to be effective for systems considered in this work.
We provide details of the strategy below.
All numerical results in this paper obtained from the following protocol. The resulting THC tensors are available in \cite{zenodo}.

\subsubsection{Initial guess}

Due to the nonlinear nature of \eq{lagrangian}, good initial guesses are often critical in obtaining accurate THC factorizations.
We generate random guesses for $\chi$ and $\zeta$ when the real-space molecular orbital representation is unavailable.
This is the case for the FeMoCo Hamiltonians since only Hamiltonian matrix elements $V_{pqrs}$ are reported in literature.
In this work, we tried 20 random guesses and picked the best performing one in the end. 

On the other hand, when we have real-space molecular orbital representation available (which is the case for hydrogen systems and other chemical systems in general) 
we utilize the interpolative separable density fitting (ISDF) technique to generate an accurate initial guess for $\chi$ and $\zeta$.
The ISDF approach was originally proposed by Lu and Ying \cite{Lu2016}. In terms of the classical precomputation required, this approach is more efficient than the direct gradient descent approach. It is based on the intuition that the THC factorization is an interpolative decomposition of the electronic pair density in real-space:
\begin{equation}
\phi_p\left(\mathbf r\right) \phi_q\left(\mathbf r\right)
=
\sum_{\mu=1}^M \xi_\mu\left(\mathbf r\right) \phi_p\left(r_\mu\right) \phi_r\left(r_\mu\right)
\end{equation}
where $\phi_p(\mathbf r)$ is the $p^{\rm th}$ single-particle orbital represented on a grid $\{\mathbf r\}$,  $r_\mu$ denotes the interpolation points and $\xi_\mu(\mathbf  r)$ is called an interpolation vector which can be obtained via a simple least-squares fit.
Once $r_\mu$ and $\xi_\mu({\mathbf r})$ are found, the THC factorization is then obtained via
\begin{align}
\chi_p^{(\mu)} = \phi_p\left(r_\mu\right) \qquad \qquad 
\zeta_{\mu\nu} = \int \! \mathrm d {\mathbf r_1} \! \int \mathrm d  {\mathbf r_2} \frac{\xi_\mu( {\mathbf r_1}) \xi_\nu( {\mathbf r_2})}{|{\mathbf r_1} -  {\mathbf r_2}|}.
\end{align}
ISDF involves no nonlinear optimization and thereby it is numerically robust and provides straightforward initial guess for subsequent direct minimizations. Finding interpolation points can be also performed with a linear complexity \cite{lee2019systematically} via the centroid Voronoi tessellation (CVT) approach \cite{Dong2018}. Determining $\xi_\mu({\mathbf r})$ scales as $\mathcal O(N^3)$ and is ultimately the bottleneck of this algorithm. While this is more economical than also performing further optimization (see the next subsection), the resulting THC factorization is not as compact for given accuracy. Therefore, in this work, we use the ISDF/CVT approach to generate a good initial guess from which we perform further optimization.

\subsubsection{Optimization}

While ISDF initial guesses are quite accurate, random guesses are sometimes very far away from any local minima.
This is problematic when an ISDF initial guesse is unavailable, which is the case for FeMoCo.
When starting from random initial guesses, we found that a quasi-Newton method such as the limited memory Broyden-Fletcher-Goldfarb-Shanno algorithm with box constraints (L-BFGS-B) algorithm~\cite{2020SciPy-NMeth}
is quite robust in the warm-up stage.
We note that quasi-Newton methods were previously studied for use in obtaining THC factorizations in \cite{schutski2017tensor}. 
The necessary gradient of \eq{lagrangian} for BFGS is obtained via the automatic differentiation method in JAX \cite{jax2018github}.
Unfortunately, L-BFGS-B alone could not produce accurate factorization for all of the systems considered in this work.
Therefore, in some cases we also had to perform another optimization with a different solver. 

After an L-BFGS-B run, we employ a popular machine learning optimizer called AdaGrad \cite{duchi2011adaptive} as implemented in JAX to finalize the factorization \cite{jax2018github}.
We found that L-BFGS-B tends to get easily stuck in unwanted local minima and subsequent optimization via AdaGrad
helped to escape from a local minimum and improved the accuracy of our optimization by more than an order of magnitude.

\subsection{Diagonal Coulomb operators from projecting into the auxillary tensor hypercontraction basis}
\label{sec:diagonal_thc}

One can attempt to apply qubitization directly to the Hamiltonian representation of \eq{thc}. In fact, doing this is relatively straightforward and in \app{standard_thc} we describe an algorithm based on exactly that approach. We show that this results in an algorithm with Toffoli complexity $\widetilde{\cal O}(N \lambda_{\textsc{thc}} / \epsilon)$, which prima facie, is relatively low complexity. The method of \app{standard_thc} is also exceptionally simple as far as arbitrary basis quantum chemistry qubitizations go; thus, that approach might be valuable for pedagogical purposes. However, the problem with directly applying qubitization to this form of the THC Hamiltonian is that we end up with a $\lambda_{\textsc{thc}}$ defined as
\begin{equation}
\label{eq:thc_lambda}
\lambda_{\textsc{thc}} = \sum_{p,q,r,s=1}^{N/2} \sum_{\mu, \nu = 1}^{M} \left |\chi_{p}^{(\mu)} \chi_{q}^{(\mu)} \zeta_{\mu\nu} \chi_{r}^{(\nu)}  \chi_{s}^{(\nu)} \right | .
\end{equation}
Because the sum over $\mu$ and $\nu$ appears outside of the absolute value in this expression, $\lambda_{\textsc{thc}}$ is much larger than even $\lambda_V$, the 1-norm of the original Hamiltonian.
Note that $\lambda_V$ is already larger than $\lambda_{\rm DF}$ (the $\lambda$ value associated with the von Burg \emph{et al.}~\cite{vonBurg2020} algorithm), and so this larger $\lambda$ value should be avoided. To avoid this blow up in $\lambda$, the main approach of this paper will focus on simulating a different Hamiltonian that we derive from the THC representation.

As discussed in the introduction, a number of recent papers have demonstrated very high efficiency quantum algorithms for simulating the electronic Hamiltonian in a basis that diagonalizes the Coulomb operator \cite{BabbushLow,BabbushSpectraB,Low2018,Kivlichan2019,Childs2019BError}. The downside of these approaches is that the required orthogonal basis sets often require many more qubits in order to reach chemical accuracy for molecules, compared to arbitrary basis functions such as molecular orbitals which lead to the Coulomb operator in \eq{full}. On the contrary, the THC representation of the Coulomb operator in \eq{thc} may be more compact, but it is not diagonal. However, as we show below, it is possible to use the $\chi$ tensor from the THC factorization to define a larger ``auxillary'' basis in which the Coulomb operator is diagonal.

We can achieve this diagonal representation by first defining a non-unitary (and non-orthogonal) basis rotation corresponding to a projection of the original fermion ladder operators into a larger basis:
\begin{equation}
\label{eq:projection}
c^\dagger_{\mu, \sigma} = \sum_{p=1}^{N/2} \chi_{p}^{(\mu)} a^\dagger_{p, \sigma}
\qquad \qquad c_{\mu, \sigma} = \sum_{p=1}^{N/2} \chi_{p}^{(\mu)} a_{p, \sigma}
\end{equation}
where the creation / annihilation operators $c^\dagger_{\mu}$ / $c_\mu$ act on a larger space of $2 M$ spin-orbitals rather than $N$ spin-orbitals. Without loss of generality, we are taking $\chi^{(\mu)}$ to be a normalized vector for each $\mu$ (because constant factors can be absorbed into $\zeta_{\mu\nu}$).
Using this, we rewrite \eq{thc} as
\begin{align}
G &= \frac{1}{2} \sum_{\alpha, \beta \in \{\uparrow, \downarrow\}} 
\sum_{\mu, \nu = 1}^{M}
\left(\sum_{p=1}^{N/2}\chi_{p}^{(\mu)}a^\dagger_{p,\alpha}\right) 
\left(\sum_{q=1}^{N/2} \chi_{q}^{(\mu)} a_{q,\alpha} \right)
\left(\sum_{r=1}^{N/2} \chi_{r}^{(\nu)} a^\dagger_{r,\beta} \right)
\left(\sum_{s=1}^{N/2}  \chi_{s}^{(\nu)}   a_{s,\beta} \right)
 \zeta_{\mu\nu}
 \\
&=
\frac{1}{2} \sum_{\alpha, \beta \in \{\uparrow, \downarrow\}} 
\sum_{\mu, \nu = 1}^{M}
(c_{\mu,\alpha}^\dagger
c_{\mu,\alpha})
(c_{\nu,\beta}^\dagger
c_{\nu,\beta})
 \zeta_{\mu\nu} \, .\nonumber
\end{align}
This provides a diagonal form of the Coulomb operator in the expanded basis:
\begin{equation}
\label{eq:larger_basis}
G = \frac{1}{2} \sum_{\alpha, \beta \in \{\uparrow, \downarrow\}}  \sum_{\mu, \nu = 1}^{M} \zeta_{\mu\nu} n_{\mu, \alpha} n_{\nu, \beta}
\end{equation}
where $n_{\mu, \sigma} = c^\dagger_{\mu, \sigma} c_{\mu, \sigma}$ is the number operator in the larger basis. To our knowledge, we are the first to derive this representation of the Hamiltonian (it does not seem to appear in any prior literature even for classical electronic structure). Even though the basis size (and thus the number of qubits) has been increased by a factor that is roughly between 4 and 10 in most contexts, the appeal of highly efficient algorithms presented in \cite{BabbushLow,BabbushSpectraB,Low2018,Kivlichan2019,Childs2019BError} for diagonal Coulomb operators makes this representation interesting. Unfortunately, this larger basis is not orthogonal, which complicates our approach to directly simulating this Hamiltonian. For example, one cannot use methods such as those in \cite{BabbushSpectraB,Kivlichan2019} that were developed to simulate diagonal electronic Hamiltonians in orthogonal basis sets. Instead, we will pursue a qubitization based approach that involves rotating into the underlying non-orthogonal basis, one tensor factor at a time, avoiding complications that arise if the rotations were done globally.
The Hamiltonian in \eq{larger_basis} will be referred to as the non-orthogonal THC Hamiltonian. 

\subsection{Deriving the \texorpdfstring{$\lambda$}{lambda} value associated with the non-orthogonal THC Hamiltonian representation}

We will now give a very high level overview of the main approach of this paper, which involves qubitizing \eq{larger_basis}, and we will derive the $\lambda$ value associated with that representation. In \cite{vonBurg2020} it was shown that when qubitizing an operator in a different basis, the basis rotation does not need to rotate all the bases at once. Instead, because it is controlled by a register, it is sufficient to perform basis rotations independently for each number operator controlled by the register.
As shown in Eq.~(51) of \cite{vonBurg2020}, only $N/2$ Givens rotations are needed, instead of $\mathcal{O}(N^2)$ if all $N$ basis vectors were being rotated at once.
So in our case, we can control on $\mu$ to rotate the basis to that described by $n_{\nu,\alpha}$ using $N$ Givens rotations in the same way as shown in Eq.~(51) of \cite{vonBurg2020}.
Since $\chi^{(\mu)}$ is taken to be a normalized vector for each $\mu$, this is a valid rotation for each individual $\mu$.
It does not matter that the basis is not orthogonal, because the rotations are done individually for each $\mu$ separately, rather than jointly for all $\mu$ together.

It is also possible to apply other methods from \cite{vonBurg2020} associated with implementing these rotations.
\begin{enumerate}
\item Use a QROM to output the $N/2$ rotation angles controlled on the register $\mu$ which takes $M$ values.
\item Since the number operators can be represented by $(\openone-Z)/2$, we can take the identity parts out and combine them with the one-body terms, and the remaining two-body terms will have a $\lambda$-value that is divided by $4$.
\end{enumerate}
Note also that the procedure to rotate to and from the new basis must be done twice, once for the operators dependent on $\nu$ and again for those dependent on $\mu$.
The net result is that the complexity for a single step is $\widetilde{\mathcal{O}}(N)$, and the value of $\lambda$ has a contribution from two-body terms
\begin{equation}\label{eq:thirdlambda}
\lambda_\zeta = \frac 12 \sum_{\mu,\nu} |\zeta_{\mu\nu}|.
\end{equation}

To determine the $\lambda$ value more carefully, $G$ can be given as
\begin{equation}
G = \frac{1}{2} \sum_{\alpha, \beta \in \{\uparrow, \downarrow\}} 
\sum_{\mu, \nu = 1}^{M}
U_\mu^\dagger n_{1,\alpha} U_\mu U_\nu^\dagger n_{1,\beta} U_\nu
 \zeta_{\mu\nu} \, .
\end{equation}
In actually implementing this, we would use $\alpha$ or $\beta$ to control swaps between the qubits representing spin up and spin down orbitals, so $U_\mu$ would only need to act on $N/2$ qubits, but for simplicity this will not be shown explicitly here.
Then by taking $n_{1,\alpha}=(\openone-Z_{1,\alpha})/2$, we have
\begin{align}\label{eq:Gexp}
G &= \frac{1}{8} \sum_{\alpha, \beta \in \{\uparrow, \downarrow\}} 
\sum_{\mu, \nu = 1}^{M}
U_\mu^\dagger (\openone-Z_{1,\alpha}) U_\mu U_\nu^\dagger (\openone-Z_{1,\beta}) U_\nu
 \zeta_{\mu\nu} \\
 &= -\frac{1}{8} \sum_{\alpha, \beta \in \{\uparrow, \downarrow\}} 
\sum_{\mu, \nu = 1}^{M}
U_\mu^\dagger U_\mu U_\nu^\dagger U_\nu
 \zeta_{\mu\nu}
+
\frac{1}{8} \sum_{\alpha, \beta \in \{\uparrow, \downarrow\}} 
\sum_{\mu, \nu = 1}^{M}
U_\mu^\dagger (\openone-Z_{1,\alpha}) U_\mu U_\nu^\dagger  U_\nu
 \zeta_{\mu\nu}\nn
 & \quad +
 \frac{1}{8} \sum_{\alpha, \beta \in \{\uparrow, \downarrow\}} 
\sum_{\mu, \nu = 1}^{M}
U_\mu^\dagger U_\mu U_\nu^\dagger (\openone - Z_{1,\beta}) U_\nu
 \zeta_{\mu\nu} +
 \frac{1}{8} \sum_{\alpha, \beta \in \{\uparrow, \downarrow\}} 
\sum_{\mu, \nu = 1}^{M}
U_\mu^\dagger Z_{1,\alpha} U_\mu U_\nu^\dagger Z_{1,\beta} U_\nu
 \zeta_{\mu\nu}\nn
 &=-\frac{1}{2} \openone
\sum_{\mu, \nu = 1}^{M}
 \zeta_{\mu\nu}
+
\frac{1}{2} \sum_{\alpha \in \{\uparrow, \downarrow\}} 
\sum_{\mu, \nu = 1}^{M}
U_\mu^\dagger (\openone-Z_{1,\alpha}) U_\mu 
 \zeta_{\mu\nu}
+
 \frac{1}{8} \sum_{\alpha, \beta \in \{\uparrow, \downarrow\}} 
\sum_{\mu, \nu = 1}^{M}
U_\mu^\dagger Z_{1,\alpha} U_\mu U_\nu^\dagger Z_{1,\beta} U_\nu
 \zeta_{\mu\nu} \, . \nonumber
\end{align}
The first term corresponds to an overall energy shift that can be ignored.
The third term is the two-body term that gives the contribution to $\lambda$ given in \eq{thirdlambda}.
The middle term can be given as
\begin{align}
T^{(2\to 1)} &= \sum_{\alpha \in \{\uparrow, \downarrow\}} 
\sum_{\mu, \nu = 1}^{M}
U_\mu^\dagger n_{1,\alpha} U_\mu 
 \zeta_{\mu\nu} =  \sum_{\alpha \in \{\uparrow, \downarrow\}} 
\sum_{\mu, \nu = 1}^{M} \left(\sum_{p=1}^{N/2}\chi_{p}^{(\mu)}a^\dagger_{p,\alpha}\right) 
\left(\sum_{q=1}^{N/2} \chi_{q}^{(\mu)} a_{q,\alpha} \right)\zeta_{\mu\nu}\\
&= \sum_{\alpha \in \{\uparrow, \downarrow\}} \left( \sum_{\mu, \nu = 1}^{M} \chi_{p}^{(\mu)}\chi_{q}^{(\mu)}\zeta_{\mu\nu}\right)a^\dagger_{p,\alpha}a_{q,\alpha} \, . \nonumber
\end{align}
Now note that
\begin{align}
\sum_{r=1}^{N/2} G_{pqrr} &=
\sum_{\mu, \nu = 1}^{M} \chi_{p}^{(\mu)} \chi_{q}^{(\mu)} \zeta_{\mu\nu} \sum_{r=1}^{N/2} \chi_{r}^{(\nu)}  \chi_{r}^{(\nu)}= \sum_{\mu, \nu = 1}^{M} \chi_{p}^{(\mu)} \chi_{q}^{(\mu)} \zeta_{\mu\nu} \, .
\end{align}
For this term we can replace the approximation $G_{pqrr}$ with the exact $V_{pqrr}$, so the approximation is only used in the two-body term.
Therefore, combining $T^{(2\to 1)}$ with $T$ gives a new one-body operator
\begin{equation}
T' = T + T^{(2\to 1)} = \sum_{\sigma\in\{\uparrow,\downarrow\}} \sum_{p,q=1}^{N/2} T'_{p q} a_{p,\sigma}^\dagger a_{q,\sigma}
\qquad {\rm with} \qquad
T'_{pq} = T_{pq} + \sum_{r=1}^{N/2} V_{pqrr} \, .
\end{equation}
Then $T'$ can be diagonalized as
\begin{align}\label{eq:Texp}
T' = \sum_{\sigma\in\{\uparrow,\downarrow\}} \sum_{\ell=1}^{N/2} t_\ell  n_{\ell,\sigma} 
&= \sum_{\sigma\in\{\uparrow,\downarrow\}} \sum_{\ell=1}^{N/2} t_\ell U_{T,\ell}^\dagger n_{1,\sigma} U_{T,\ell}
=  \sum_{\ell=1}^{N/2} t_\ell \openone
- \frac 12\sum_{\sigma\in\{\uparrow,\downarrow\}} \sum_{\ell=1}^{N/2} t_\ell U_{T,\ell}^\dagger Z_{1,\sigma} U_{T,\ell}
\end{align}
where $t_\ell$ are eigenvalues of $T'_{p q}$, and $U_{T,p}$ are individual rotations for $T$ similar to the $U_{p}$ for $G$.
The first term is an overall energy shift, and the second term is the one-body term that contributes towards $\lambda$.
The net result of this is
\begin{equation}
\lambda = \sum_{\ell=1}^{N/2} |t_\ell| +  \frac 12 \sum_{\mu,\nu} |\zeta_{\mu\nu}| = {\cal O}\left(\lambda_\zeta\right).
\end{equation}
As we will discuss later on, $\lambda_\zeta$ actually scales even better than the $\lambda$ values associated with any prior algorithm in the literature, including the $\lambda_{\rm DF}$ associated with the doubled factorized algorithm of von Burg \emph{et al.}~\cite{vonBurg2020}.

\section{Qubitizing the Non-Orthogonal Tensor Hypercontraction Hamiltonian}
\label{sec:algorithm}

\subsection{Approach to qubitization}
\label{sec:qubitization}

Our approach to encoding the eigenspectra of the THC representation of the electronic Hamiltonian in a unitary for phase estimation will use the linear combination of unitaries (LCU) query model \cite{Childs2012}. Specifically, we will use qubitization \cite{Low2016} to block encode \cite{Gilyen2019QuantumArithmetics} the Hamiltonian eigenspectra as a Szegedy quantum walk \cite{Szegedy2004}. What all LCU methods have in common is that they involve simulating or block encoding the Hamiltonian from a representation where it can be accessed as a linear combination of unitaries:
\begin{equation}
\label{eq:lcu}
H = \sum_{\ell=1}^{L} \omega_\ell \, U_\ell
\end{equation}
where $U_\ell$ are unitary operators and the $\omega_\ell$ are scalars. 
LCU methods are defined in terms of queries to two oracle circuits that are commonly defined as
\begin{align}
\label{eq:oracles}
\textsc{select} \ket{\ell} \ket{\psi} \mapsto \ket{\ell} U_\ell \ket{\psi}
\qquad \qquad
\textsc{prepare} \ket{0}^{\otimes \log L} & \mapsto \sum_{\ell=1}^L \sqrt{\frac{\omega_\ell}{\lambda}} \ket{\ell} \equiv \ket{\cal L} \qquad \qquad \lambda = \sum_{\ell=1}^L \left| \omega_\ell \right |,
\end{align}
where $\ket{\psi}$ is the system register, $\ket{\ell}$ is an ancilla register which usually indexes the terms in \eq{lcu} in binary, and this is the general definition of $\lambda$.

Currently, the most practical fault-tolerant approaches for simulating quantum chemistry are based on qubitization \cite{Low2016,BabbushSpectraB,Berry2019B,vonBurg2020}. Following the analysis and techniques of \cite{BabbushSpectraB}, one can use phase estimation based on qubitized quantum walks to sample in the eigenbasis of a Hamiltonian with error in the sampled eigenvalue bounded from above by $\epsilon$ with Toffoli complexity scaling exactly as
\begin{equation}
\label{eq:lcu_cost}
\left\lceil\frac{\pi \lambda}{2\epsilon_{\textsc{pea}}}\right\rceil \left(C_S + C_P + C_{P^\dagger} + \log L  + {\cal O}\left(1\right)\right),
\end{equation}
where $C_S$ is the gate complexity of $\textsc{select}$, $C_P$ is the gate complexity of $\textsc{prepare}$, $C_{P^\dagger} \leq C_P$ is the gate complexity of uncomputing $\textsc{prepare}$, and $\epsilon_{\textsc{pea}}$ is the allowable error in the phase estimation. 

The multiplying factor of $\lceil\pi\lambda/(2\epsilon_{\textsc{pea}})\rceil$ corresponds to the number of repetitions of the LCU step used in the phase estimation.
In \cite{BabbushSpectraB} the number of repetitions was taken to be a power of $2$, because that makes the phase estimation particularly simple, with each qubit of the control registers controlling a number of repetitions that is a power of 2.
It is also possible to use a number of repetitions that is an arbitrary integer. This was assumed by von Burg \emph{et al.}~\cite{vonBurg2020}, although doing this requires a more sophisticated control which those authors do not show how to perform. We will explain how this can be accomplished. The general principle is to control each step using the unary iteration procedure introduced in \cite{BabbushSpectraB}.

To explain the procedure in more detail, one should combine the $\textsc{select}$ operation together with a reflection $R = 2 \proj{\cal L}\otimes \openone - \openone$ based on the $\textsc{prepare}$ operation, to create a step of a quantum walk ${\cal W}$ which has eigenvalues proportional to $e^{\pm i \arccos(E_n / \lambda)}$ where $H\ket{n}=E_n \ket{n}$  \cite{Poulin2017,Berry2018}.
In order for this procedure to work, the $\textsc{select}$ needs to be self-inverse.
It is possible to obtain the same complexity if there were controlled application of $\textsc{select}$ or $\textsc{select}^\dagger$, but we avoid that because it doubles the complexity, and instead will construct $\textsc{select}$ so it is self-inverse.

\begin{figure}[tbh]
\centering
\includegraphics[]{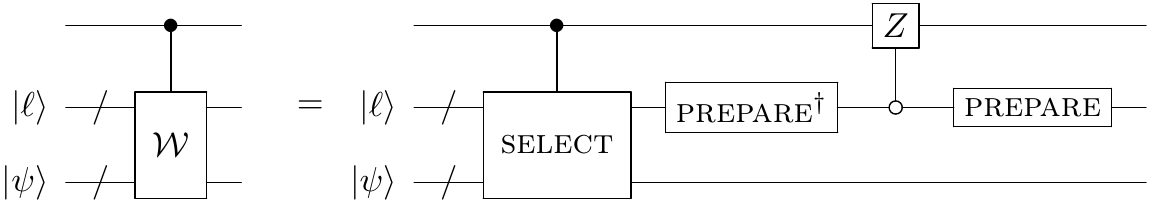}
  \caption{\label{fig:walk_circuit}
    A circuit realizing the qubitized quantum walk operator ${\cal W}$ controlled on an ancilla qubit. Note that the $Z$ gate with the 0-control is actually controlled on the zero state of the entire $\ket{\ell}$ register and not just a single qubit. Accordingly, implementation of that controlled-$Z$ has gate complexity $\log L+{\cal O}(1)$ where $\lceil\log L\rceil$ is the size of the $\ket{\ell}$ register.}
\end{figure}

\begin{figure}[tbh]
\centering
\includegraphics[scale=0.921]{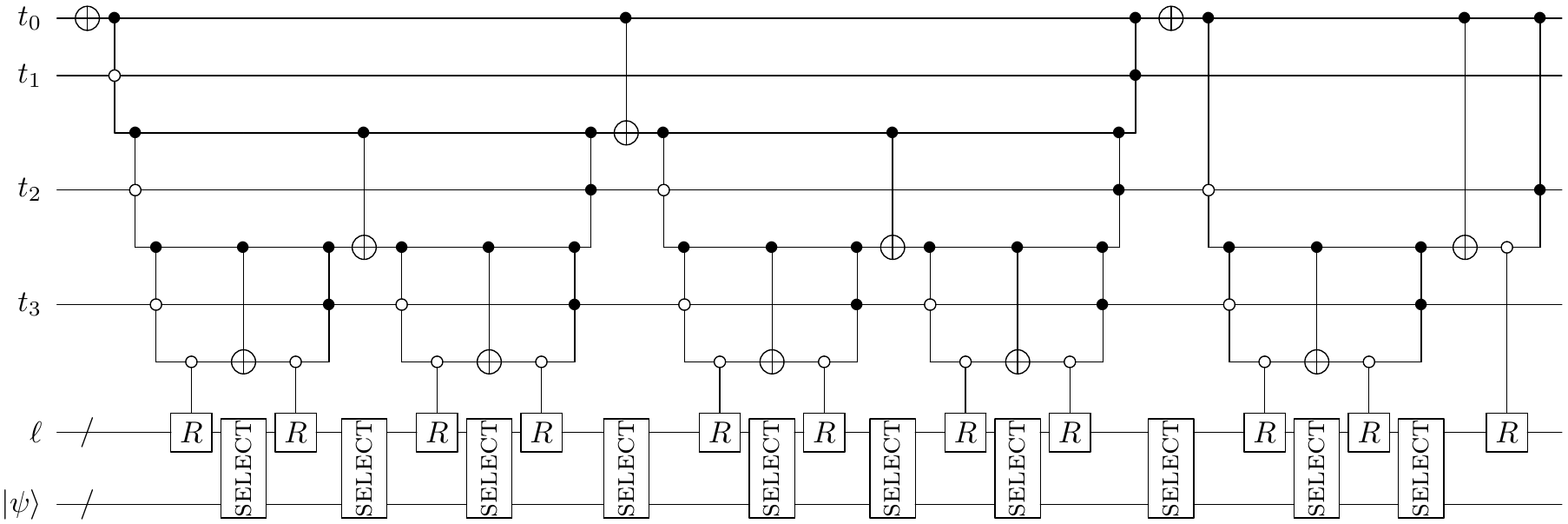}
\caption{\label{fig:controls}
 A unary iteration circuit for selecting ${\cal W}^{2t-10}$ for $t$ from $0$ to $10$. Here $R$ indicates the inverse preparation, reflection, and preparation. As in \fig{walk_circuit} only the reflection need be made controlled.}
 \end{figure}

Simplistically, one could make ${\cal W}$ controlled as shown in \fig{walk_circuit}, but for the purpose of obtaining the complexity shown in \eq{lcu_cost}, one needs to control application of ${\cal W}$ and its inverse.
Given that $\textsc{select}$ is self-inverse, one can obtain ${\cal W}^\dagger$ simply by performing the reflection before $\textsc{select}$ instead of after.
That means that one could control between ${\cal W}$ and ${\cal W}^\dagger$ by performing a controlled reflection before $\textsc{select}$ as well as after (and not making $\textsc{select}$ controlled). But, it is not necessary to perform two controlled reflections, only one.
To see why, consider the case where the control is selecting four applications of ${\cal W}$ with the rest of the operations being ${\cal W}^\dagger$.
Then we need to perform the sequence of operations
\begin{equation}
(RU)(RU)(RU)(RU)(UR)(UR)(UR)\ldots = (RU)(RU)(RU)(RU)\openone(UR)(UR)(UR)\ldots,
\end{equation}
where $U$ is being used for $\textsc{select}$ and $R$ is being used for the reflection (i.e.\ the combined inverse preparation, reflection about the zero state, and preparation).
We will only use $U$ or $R$ with this meaning in this equation and in \fig{controls}.
Here, after the fourth $U$, we simply perform the identity rather than a reflection.
This means that we \emph{always} perform the reflection in between each successive $\textsc{select}$, except when the number of times we have performed $\textsc{select}$ matches the number in the control register.
That means we only need one controlled reflection in between each $\textsc{select}$, which corresponds to removing the control on $\textsc{select}$ in \fig{walk_circuit}.

Next we explain how to address cases where $\lceil\pi \lambda/(2\epsilon_{\textsc{pea}})\rceil$ is not a power of 2.
To achieve that,
one can simply prepare the optimal control state as described in \cite{BabbushSpectraB} except using a superposition over a number of basis states that is smaller than a power of 2.
Then, instead of controlling on each successive qubit of this state, use the unary iteration procedure of \cite{BabbushSpectraB} to control the reflection.
The unary iteration procedure introduces a trivial additional cost of one Toffoli per step, and doubles the ancilla cost of the control register for the temporary ancillas for the unary iteration.
To show explicitly how this control is done, see the example in \fig{controls}.

In our algorithm there are four sources of error:
\begin{enumerate}
    \item Error due to measurement in phase estimation $\epsilon_{\textsc{pea}}$ which first appears in \eq{lcu_cost}.
    \item The approximation of the Hamiltonian coefficients as part of the coherent alias sampling procedure from \cite{BabbushSpectraB} that is used as part of our \prep\, strategy.
    \item The approximation of the individual Givens rotations needed for the basis rotations.
    \item The approximation of tensor hypercontraction $\epsilon_{\textsc{thc}}$ which first appears in \eq{M_scaling}.
\end{enumerate}
To bound the overall error, one could take these individual errors and simply add them together.
However, the sources of error 2 to 4 all contribute to error in the approximation of the Hamiltonian.
That is, they together give an approximate Hamiltonian, and one can simply estimate the error due to using this approximate Hamiltonian.

One approach would be to determine the root-mean-square sum of the errors of each of the coefficients in the Hamiltonian as is done in \cite{vonBurg2020}.
Based on classical simulations we found that this overestimates the error,
so we will instead perform classical calculations of the error in the energy to estimate the allowable truncations.
We will include all error from approximation of the Hamiltonian into $\epsilon_{\textsc{thc}}$, and require that
\begin{equation}
\label{eq:epsilon}
    \epsilon \geq 
    \epsilon_{\textsc{pea}} + \epsilon_{\textsc{thc}},
\end{equation}
where $\epsilon$ is our target accuracy, which we will take to be 0.0016 Hartree (chemical accuracy) for the resource estimates of this paper.
We will take $\epsilon_{\textsc{pea}}\le 0.001$ Hartree and $\epsilon_{\textsc{thc}}\le 0.0006$ Hartree.
The allowable error for phase estimation in \cite{vonBurg2020} was $0.0009$ Hartree, but we will recalculate the cost with $\epsilon_{\textsc{pea}}\le 0.001$ Hartree for the DF method of \cite{vonBurg2020} for a fair comparison.
We also recalculate the costs using the same parameters for the sparse and SF approaches.

Note that in quantum chemistry it is typical to require the differences in energy between two configurations to be determined to chemical accuracy, which na{\'i}vely would require accuracy of 0.0008 Hartree on each estimation of each energy.
That precision is expected to not be necessary, because the approximations made in the Hamiltonian for the two configurations have correlated errors.
For the phase estimation, the errors would add if the computation was performed independently for the two configurations, but it is possible to use the control register to control a forward evolution for one Hamiltonian on one target system, \emph{and} reverse evolution on a second target system with the Hamiltonian for the second configuration. This can be combined with the improved phase estimation techniques of \cite{Berry2018} which help to ensure one is projecting into the correct states at lower cost than the entire phase estimation. The phase estimation will then provide an estimate of the energy difference.
We therefore continue to assume that the accuracy required is overall 0.0016 Hartree, which also provides consistency with prior work.

We will now discuss how to implement and compile \prep\, and \sel\, for our algorithm.

\subsection{State preparation for the non-orthogonal tensor hypercontraction Hamiltonian}

The method to perform the \prep~step is based on expressing the Hamiltonian in a non-orthogonal basis as in \eq{larger_basis}. We can rewrite the Hamiltonian in the form
\begin{equation}
  H=  - \frac 12\sum_{\sigma\in\{\uparrow,\downarrow\}} \sum_{\ell=1}^{N/2} t_\ell U_{T,\ell}^\dagger Z_{1,\sigma} U_{T,\ell}
  +\frac{1}{8} \sum_{\alpha, \beta \in \{\uparrow, \downarrow\}} 
\sum_{\mu, \nu = 1}^{M}\zeta_{\mu\nu}
U_\mu^\dagger Z_{1,\alpha} U_\mu U_\nu^\dagger Z_{1,\beta} U_\nu
\end{equation}
where the first term is from $T'$ in \eq{Texp} and the second term is from $G$ in \eq{Gexp}.
The state we need is
\begin{equation}
\frac 1{\sqrt\lambda} \ket{+}\ket{+}\left[ \sum_{\ell=1}^{N/2} \sqrt{|t_\ell|} \ket{\ell} \ket{M+1}
+ \frac 1{\sqrt{2}} \sum_{\mu,\nu=1}^M \sqrt{|\zeta_{\mu\nu}|} \ket{\mu}\ket{\nu} \right] \, .
\end{equation}
Here the first and second registers give the superpositions over spins, and the third and fourth registers store $\mu$ and $\nu$.
The last register takes the value $M+1$ to flag that this is the first term in the Hamiltonian.
For simplicity we adopt the convention that these registers are labelled starting from 1, though 1 would be stored as binary 0 in the register.

\begin{figure}[tbh]
\centering
\includegraphics[scale=0.925]{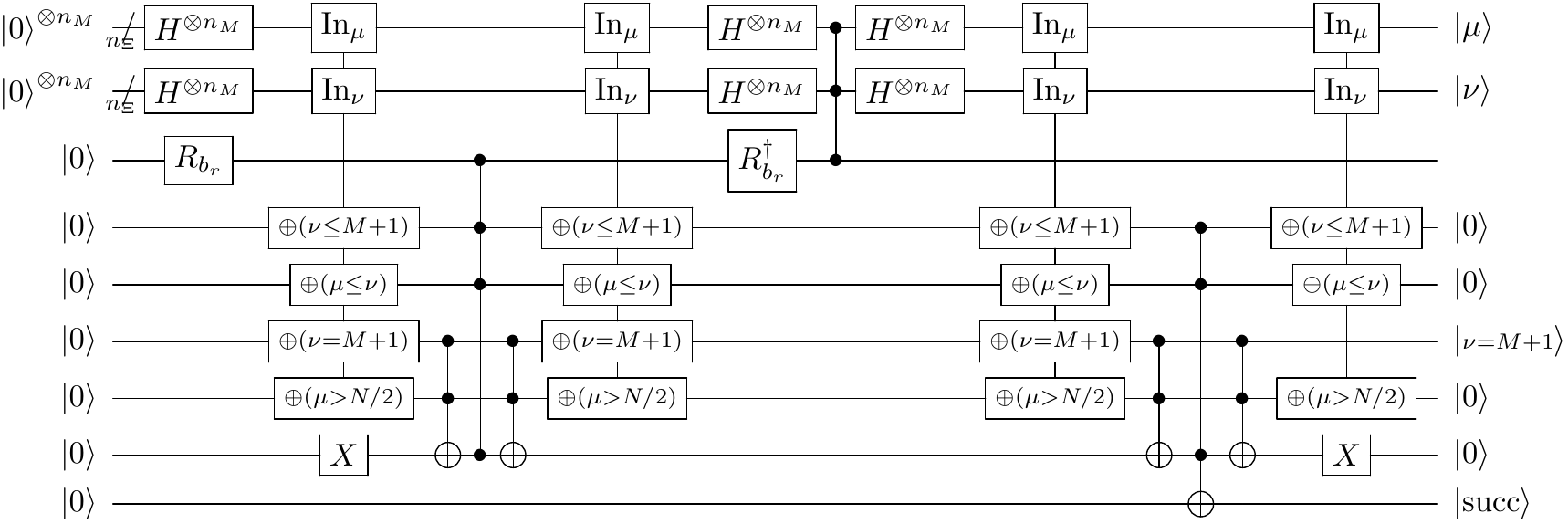}
\caption{The circuit for creating an equal superposition over the $\mu$ and $\nu$ registers, with $n_M=n_M=\lceil\log(M+1)\rceil$, and $R_{\rotprec}$ being a $Y$ rotation to $\rotprec$ bits of precision.
The state is prepared on the first two registers, the last register flags success of the entire procedure, and the fourth-last register records whether $\nu$ was equal to $M+1$.
We use $\ket{\mu}$ and $\ket{\nu}$ to label the outputs on the top two registers, though these are in a superposition state.}
\label{fig:eqprep}
\end{figure}

For the state preparation, we have a three step procedure where we first prepare an equal superposition over the $\mu$ and $\nu$ registers, then perform coherent alias sampling \cite{BabbushSpectraB}, then swap the $\mu$ and $\nu$ registers controlled by an ancilla in a $\ket{+}$ state.
This means we only need to initially prepare the range $\mu\le \nu$, and the number of coefficients needed in the state preparation is
\begin{equation}
d=N/2+M(M+1)/2 .
\end{equation}
In the first step we need to create an equal superposition over $\mu\le \nu\le M+1$ for registers 3 and 4, with $\mu\le N/2$ for $\nu=M+1$.
The procedure needed is depicted in \fig{eqprep}.

The method is to perform Hadamards on all qubits for these registers, then perform the inequality tests $\nu\le M+1$, $\mu\le \nu$, and $\mu\le N/2$ controlled on $\nu=M+1$. These inequality tests have cost $4(n_M-1)$, with $n_M=\lceil\log(M+1)\rceil$.
Rotate an ancilla qubit by adding a constant into a phase gradient register (as described in Appendix A of~\cite{Sanders2020CompilationOptimizationB}), to obtain an overall amplitude for success on this qubit and the three inequality tests approximately 1/2.
This needs $\rotprec-3$ Toffolis, using $\rotprec$ bits of precision, which can typically be taken to be about 7.
We can flag failure on an ancilla and perform the identity (instead of \sel) in the case of failure.
Then we need to reflect on five registers, which we do using three Toffolis.
Next, we can invert the inequality tests (which may be done without further Toffoli costs using the out-of-place adders of \cite{GidneyAdder}), 
and invert the rotation with cost $\rotprec-3$.
Then inverting the Hadamards and reflecting about zero has cost $2n_M-1$.
Then we perform the Hadamards and inequality tests again with cost $4n_M-3$.
Checking that the inequality tests are satisfied has cost 3, giving a total cost of  $10\lceil\log (M+1)\rceil+2\rotprec-9$ Toffolis.

\begin{figure}[tbh]
\centering
\includegraphics[scale=0.925]{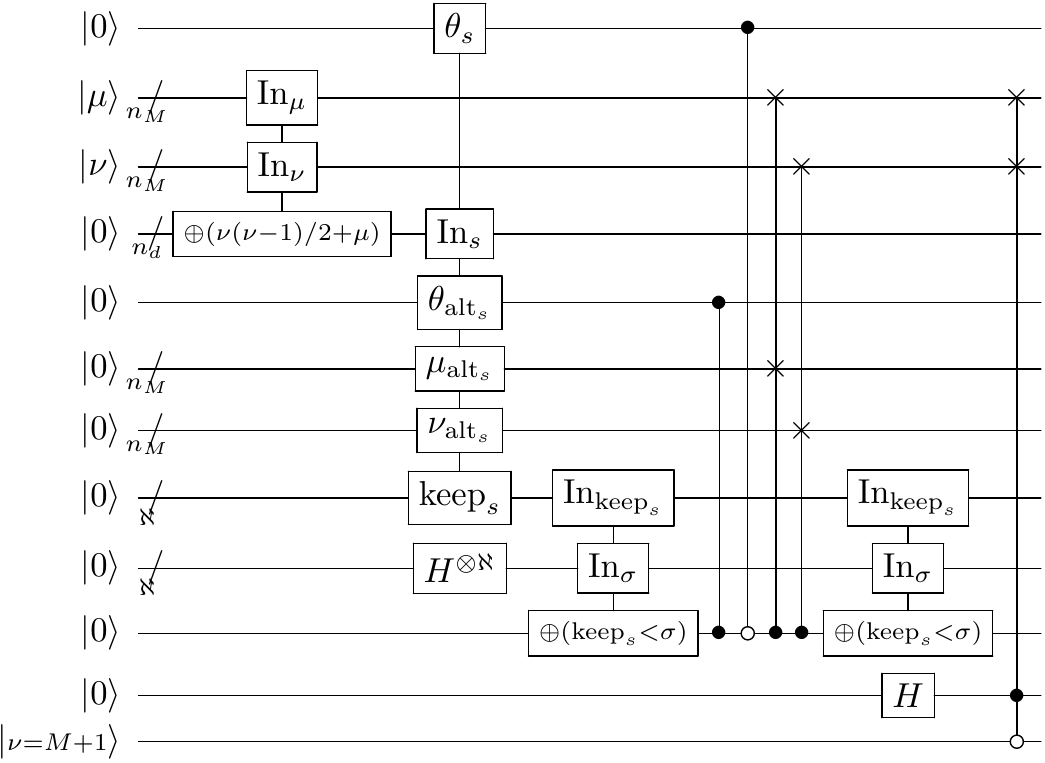}
    \caption{The state preparation after preparation of the equal superposition state in \fig{eqprep}. Here $n_d=\lceil\log d\rceil$, and we use $\ket{\mu}$ and $\ket{\nu}$ to label the input registers that are in a superposition. The modifications to this state preparation over that in \cite{BabbushSpectraB} are that a contiguous register is calculated in the beginning, and at the end the $\mu$ and $\nu$ registers are swapped controlled on a $\ket{+}$ state and $\nu$ not being equal to $M+1$, which is contained in a register output from the procedure for creating the equal superposition.
    Since we simply need to perform a $Z$ gate on the sign qubit $\theta$, instead of explicitly performing a controlled swap with the alt value, we can perform two controlled phase operations thereby eliminating one non-Clifford gate.
    These controlled phase gates need not be performed when inverting the state preparation.}
    \label{fig:stateprep}
\end{figure}

Once we have prepared the equal superposition over the $\mu$ and $\nu$ registers, we can then prepare the state as shown in \fig{stateprep}.
In more detail, the steps needed are as follows.
\begin{enumerate}
\item Use Hadamards to create the $\ket{+}$ states on the first two registers.
\item Create a new contiguous register from registers 3 and 4.
This contiguous register can be given as
\begin{equation}
\label{eq:contiguous}
s = \nu(\nu-1)/2 +\mu,
\end{equation}
where we are using the convention in this equation that numbering starts from 1.
Because we are using the convention that $\nu=M+1$ flags the first term where $\mu\le N/2$, the allowed range of values for $\mu$ and $\nu$ gives a contiguous range of values for $s$.
The complexity of computing $s$ is $n_M^2+n_M-1$, which is shown in \app{contig}.
\item Use the new contiguous register to output alternate values for registers 4 and 5, a sign qubit and an alternate sign qubit (to account for the signs in the linear combination of unitaries), and keep values.
The size of the output is then
\begin{equation}
m = 2n_M + 2 + \zetabits
\end{equation}
where $\zetabits$ is the number of bits for the keep register.
The cost of the QROM is then
\begin{equation}
\left\lceil\frac{d}{k_{s1}}\right\rceil + m(k_{s1}-1).
\end{equation}
\item Perform an inequality test between the keep register and a register in an equal superposition, with cost $\zetabits$.
\item Perform a swap controlled on the result of the inequality test, with cost $2n_M$.
We can simply perform controlled phase gates on the sign qubits instead of explicitly swapping them, which is why we simply have the cost of swapping the $\mu$ and $\nu$ registers with the alternate registers.
\item Controlled on a $\ket{+}$ state, \emph{and} the result of the test for $\nu=M+1$ from step 2, swap the $\mu$ and $\nu$ registers, with cost $n_M+1$.
Here the $+1$ is the Toffoli needed to perform the AND operation on two qubits for the control.
\end{enumerate}
In inverting the state preparation, the cost of the inverse QROM is reduced to 
\begin{equation}
\left\lceil\frac{d}{k_{s2}}\right\rceil + k_{s2},
\end{equation}
but the other costs are unchanged.
Adding all these costs together gives
\begin{equation}
C_P+C_{P^\dagger} = 28n_M+4\rotprec-18+2n_M^2+2\zetabits+\left\lceil\frac{d}{k_{s1}}\right\rceil + m(k_{s1}-1)+\left\lceil\frac{d}{k_{s2}}\right\rceil + k_{s2},
\end{equation}
with $m = 2n_M + 2 + \zetabits$, $n_M=\lceil \log (M+1)\rceil$.

The number of bits used for the keep register was $\zetabits=\lceil 2.5+\log(10\lambda/\epsilon)\rceil$ in \cite{vonBurg2020}.
The error in the keep probability is translated to that in the state in the following way.
The squared amplitudes of the state needed are $|t_\ell|/\lambda$ and $|\zeta_{\mu\nu}|/\lambda$ for $\mu\ne\nu$, but $|\zeta_{\mu\mu}|/2\lambda$ for $\mu=\nu$.
These amplitudes are compared to initial squared amplitudes in the equal superposition of $1/d$.
Taking $\zetabits$ as the number of bits for the keep probability is equivalent to discretizing the squared amplitudes to the nearest $1/(2^\zetabits d)$ (see Eq.~(35) of \cite{BabbushSpectraB}).
It would at first be expected that this would lead to an error no larger than $\lambda/(2^{\zetabits+1}d)$ in $|t_\ell|$ or $|\zeta_{\mu\nu}|$ for $\mu\ne\nu$, or $\lambda/(2^{\zetabits}d)$ for $|\zeta_{\mu\mu}|$, with a factor of $2$ reduction in the error because rounding gives error no larger than half the discretization size.
The subtlety is that if we round all the squared amplitudes, the result will no longer be normalized.
To maintain normalization it will be necessary to discretize some of the amplitudes in a different way than rounding, resulting in an error larger than $\lambda/(2^{\zetabits+1}d)$, but still no larger than $\lambda/(2^{\zetabits}d)$.
That is why the factor of 2 is not included in \cite{BabbushSpectraB}.
However, it is possible to make the typical errors smaller than $\lambda/(2^{\zetabits+1}d)$ by adjusting the threshold for rounding from $1/2$ to maintain normalization.
In practice only a very small adjustment is needed.

\subsection{Hamiltonian selection oracle for the non-orthogonal tensor hypercontraction Hamiltonian}

\begin{figure}[tbh]
\includegraphics[scale=0.925]{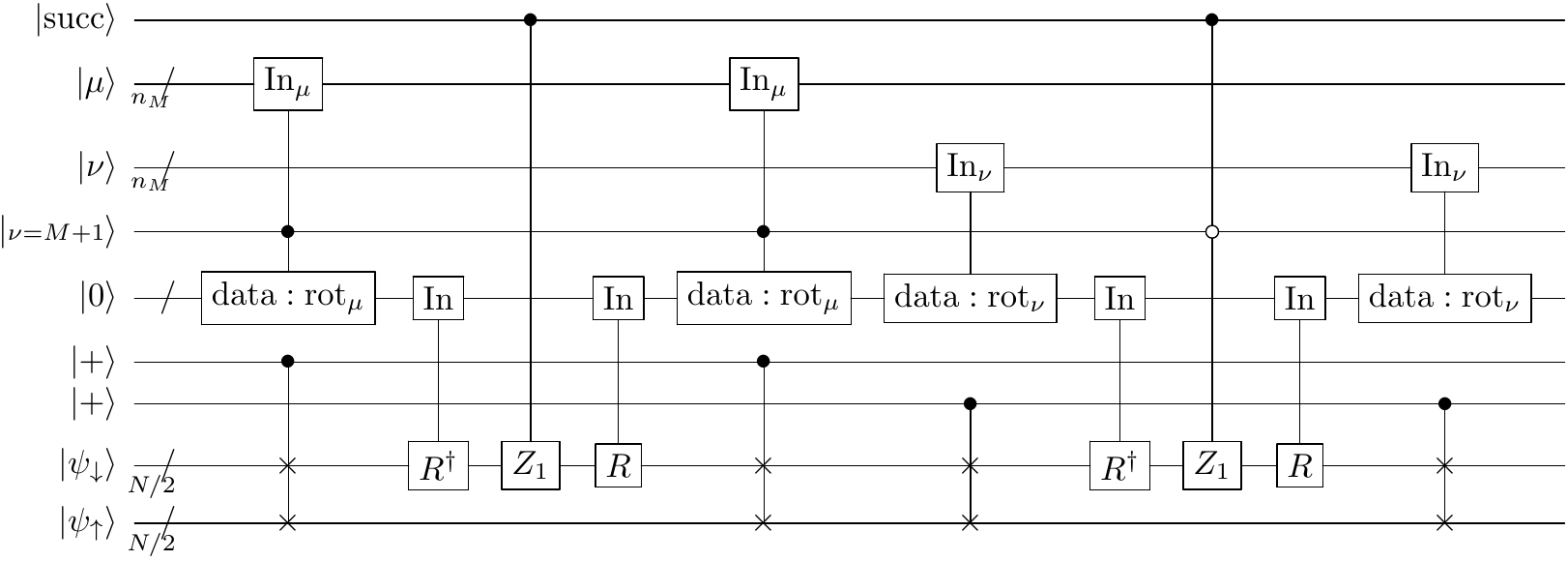}
    \caption{The circuit for performing the controlled operations (select) for the linear combination of unitaries.
    The registers from top to bottom are the success flag register, the $\mu$ and $\nu$ control registers, the qubit flagging that $\nu=M+1$, a blank ancilla to put the database of rotations in, two $\ket{+}$ states to provide the superposition over the operations on the spin up and down components of the state, and the qubits representing the spin down and up orbitals.
    The controlled $R$ is indicating the basis change obtained from the sequence of rotations with angles provided in the database, equivalent to $U_{\vec u,0}U_{\vec u,1}$ in \cite{vonBurg2020}, which is depicted in Eqs.\ (58) and (59) of the Supplementary Material of that work.
    The registers labelled $\ket{\psi_\downarrow}$ and $\ket{\psi_\uparrow}$ correspond to the qubits for the spin-down and spin-up orbitals.
    The operation $Z_1$ acts on only one of the qubits of the $\ket{\psi_\downarrow}$ register, similar to the circuit diagram labelled (57) in the Supplementary Material of \cite{vonBurg2020}.}
    \label{fig:select}
\end{figure}

For the \sel~operation, we have two steps.
In the first we need to perform the operation $U_\mu$ or $U_{T,\mu}$ for the two-electron or one-electron terms, respectively.
In the second we only need to perform $U_\nu$.
To perform these operations, we can use a QROM to generate a sequence of $N/2$ rotations, then perform the rotations using the method given in \cite{vonBurg2020}.
The costs are as follows.
\begin{enumerate}
\item There is a cost of $N/2$ to perform the swaps controlled on the spin qubit.
\item The dominant cost is the QROM.  The output size is large, so it is most efficient to perform the QROM as in \cite{BabbushSpectraB}.
In the first step (for $\mu$) we have $M+N/2$ sets of data to output which has complexity $M+N/2-2$.
Note that we need to output one set of rotations if the $\nu$ register is in the state $\ket{M+1}$, or a different set if the $\nu$ register is not in that state.
The value $\nu=M+1$ is still flagged in an ancilla.
In the second step (for $\nu$) we only need to generate the rotations for the two-electron terms, so the cost is $M-2$.
\item We can perform the rotations with cost $N(\rotbits-2)$ by adding the rotations into the phase gradient state.
Here $\rotbits$ is the number of bits of precision used for the rotation angles.
\item The $Z$ operation has no Toffoli cost.
It needs to be controlled on the qubit which is the result of testing whether $\nu=M+1$ for the case where we apply the operations for $\nu$.
\item We invert the rotations with cost $N(\rotbits-2)$.
\item To erase the QROM, in the case of $\mu$ we need to subdivide the bits into the more- and less-significant bits, where the more-significant bit includes that distinguishing the case $\nu=M+1$.
We have a cost of $\lceil M/k_{r1}\rceil$ for the case $\nu<M+1$, and a cost $\lceil N/2k_{r1}\rceil$ for the case $\nu=M+1$, then a cost $k_{r1}$ for the remaining $k_{r1}$ less significant bits, for a total of
\begin{equation}
\left\lceil \frac{M}{k_{r1}} \right\rceil +\left\lceil \frac{N}{2k_{r1}} \right\rceil+ k_{r1}.
\end{equation}
For the case of $\mu$, we only have to perform the QROM on $M$ possible values, giving a cost of
\begin{equation}
\left\lceil \frac{M}{k_{r2}} \right\rceil + k_{r2}.
\end{equation}
\item Perform the swaps controlled by the ancillas (for spin) again with cost $N/2$.
\end{enumerate}
This procedure needs to be performed twice, once for $\mu$ and once for $\nu$.
The complete procedure is shown in \fig{select}.

A subtlety is that the operation needs to be controlled by success of the state preparation, which can be achieved simply by making the $Z$ operation controlled, which is a Clifford gate, so does not add to the Toffoli cost.
Another subtlety is that for $\nu$, we perform no operation if $\nu=M+1$.
That can be achieved simply by making the $Z$ operation controlled by the qubit output by the result of the equality test as well, requiring just one more Toffoli gate.

\begin{figure}[tbh]
    \centering
\includegraphics[scale=0.925]{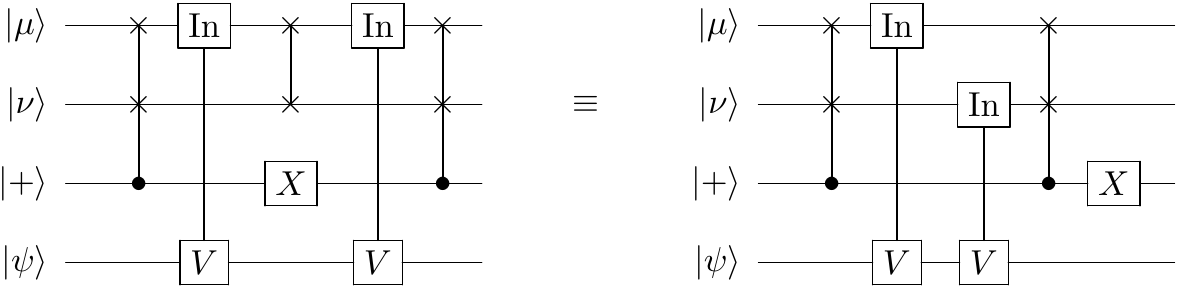}
    \caption{This shows that the procedure for generating the two-electron terms is self-inverse.
    The self-inverse controlled operations $V$ are equivalent to $V_\mu=U_\mu^\dagger Z_{1,\alpha} U_\mu$, or $V_\nu$, but we omit the dependence on $\mu$ and $\nu$ because it is in a superposition of being dependent on both.
    On the left is the form of the circuit as we described it in \eq{startsym} to \eq{endsym}, which is obviously self-inverse because of the symmetry of the circuit.
    On the right is a simplified form of the circuit.}
    \label{fig:selfinverse}
\end{figure}

A major subtlety is that the $\textsc{select}$ operation needs to be made self-inverse, and the operation as depicted is not self-inverse.
This problem can be averted by using a NOT operation on the register controlling the swap of the $\mu$ and $\nu$ registers.
To see why this is so, consider the prepared state (ignoring the one-electron term for simplicity)
\begin{equation}\label{eq:startsym}
\sum_{\mu\le \nu} \zeta'_{\mu\nu} \ket{\mu}\ket{\nu},
\end{equation}
where $\zeta'_{\mu\nu}$ are the amplitudes needed for the asymmetric state.
Introduce the ancilla in the $\ket{+}$ state and perform a controlled swap, giving
\begin{equation}
\frac 1{\sqrt{2}}\sum_{\mu\le \nu} \zeta'_{\mu\nu} \left( \ket{0}\ket{\mu}\ket{\nu}+\ket{1}\ket{\nu}\ket{\mu}\right).
\end{equation}
Then, denoting $V_\mu=U_\mu^\dagger Z_{1,\alpha} U_\mu$ and similarly for $\nu$, and performing the controlled operation on the target system in state $\ket\psi$, we have
\begin{equation}
\frac 1{\sqrt{2}}\sum_{\mu\le \nu} \zeta'_{\mu\nu} \left( \ket{0}\ket{\mu}\ket{\nu}V_\mu\ket\psi +\ket{1}\ket{\nu}\ket{\mu}V_\nu\ket\psi\right).
\end{equation}
Now performing an $X$ on the ancilla qubit and swapping the $\mu$ and $\nu$ registers, we have
\begin{equation}
\frac 1{\sqrt{2}}\sum_{\mu\le \nu} \zeta'_{\mu\nu} \left( \ket{0}\ket{\mu}\ket{\nu}V_\nu\ket\psi +\ket{1}\ket{\nu}\ket{\mu}V_\mu\ket\psi\right).
\end{equation}
Then, again performing the controlled operation on the target system, we have
\begin{equation}
\frac 1{\sqrt{2}}\sum_{\mu\le \nu} \zeta'_{\mu\nu} \left( \ket{0}\ket{\mu}\ket{\nu}V_\mu V_\nu\ket\psi +\ket{1}\ket{\nu}\ket{\mu}V_\nu V_\mu\ket\psi\right).
\end{equation}
Performing the controlled swap again gives
\begin{equation}
\frac 1{\sqrt{2}}\sum_{\mu\le \nu} \zeta'_{\mu\nu} \left( \ket{0}\ket{\mu}\ket{\nu}V_\mu V_\nu\ket\psi +\ket{1}\ket{\mu}\ket{\nu}V_\nu V_\mu\ket\psi\right).
\end{equation}
Finally, projecting onto $\ket +$ for the ancilla qubit gives
\begin{equation}\label{eq:endsym}
\frac 1{2}\sum_{\mu\le \nu} \zeta'_{\mu\nu} \ket{\mu}\ket{\nu}\left(V_\mu V_\nu+V_\nu V_\mu\right)\ket\psi.
\end{equation}
Thus, this procedure has given the desired sum of operations $V_\mu V_\nu+V_\nu V_\mu$ and is self-inverse.
As shown in \fig{selfinverse}, the form of the circuit we have just described can be simplified so that there are just controlled swaps performed on the $\mu$ and $\nu$ registers at the beginning and the end. Those controlled swaps correspond to the controlled swap at the end of \fig{stateprep}, with the controlled swap at the end corresponding to that done when inverting the state preparation.

\begin{figure}[tbh]
    \centering
\includegraphics[scale=0.925]{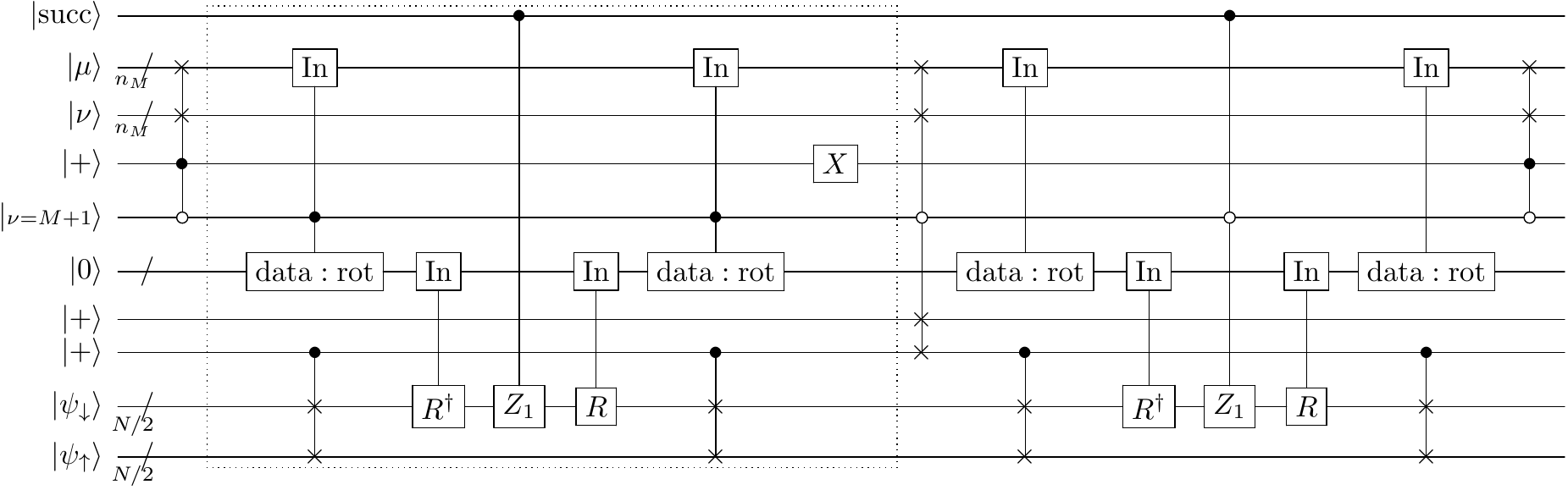}
    \caption{The circuit for performing the controlled operations (\textsc{select}) for the linear combination of unitaries in a form that is more clearly self-inverse.
    The registers from top to bottom are the success flag register, the $\mu$ and $\nu$ control registers, the qubit in the $\ket{+}$ state to control swapping the $\mu$ and $\nu$ registers, the qubit flagging that $\nu=M+1$, a blank ancilla to put the database of rotations in, two $\ket{+}$ states to provide the superposition over the operations on the spin up and down components of the state, and the qubits representing the spin down and up orbitals.
    The dotted box shows the only part that is applied when $\nu=M+1$.
    The controlled $R$ is the rotation of the Majorana basis, equivalent to $U_{\vec u,0}U_{\vec u,1}$ in \cite{vonBurg2020}.
    The registers labelled $\ket{\psi_\downarrow}$ and $\ket{\psi_\uparrow}$ correspond to the qubits for the spin-down and spin-up orbitals.
    The operation $Z_1$ acts on only one of the qubits of the $\ket{\psi_\downarrow}$ register, similar to the circuit diagram labelled (57) in the Supplementary Material of \cite{vonBurg2020}.}
    \label{fig:select1}
\end{figure}

To be more specific, for the complete \textsc{select} operation we can include the controlled swaps, and perform the operation as shown in \fig{select1}.
For that form of the \textsc{select} operation the controlled swap at the end of state preparation in \fig{stateprep} would not be performed, because it is being bundled into the \textsc{select} operation.
As can be seen from \fig{select1}, that form of the operation is more clearly self-inverse, though there is the subtlety that we have the qubit with $\nu=M+1$ flagging the one-electron component of the Hamiltonian.
In the case $\nu\ne M+1$, the circuit simplifies to a completely symmetric form.
In the case $\nu=M+1$, only the left half of the circuit (shown in the dotted box) is performed, which is itself clearly self-inverse due to symmetry.
The swap of the $\mu$ and $\nu$ registers is shown as controlled by the register flagging $\nu=M+1$ here, which would require more Toffoli gates.
However, this form is just used to illustrate that the circuit is self-inverse, and that controlled swap can be moved to the end and combined with the other controlled swap, resulting in no more Toffolis being needed.
There is one more Toffoli needed in order to apply the controlled swap between the two ancilla qubits in the $\ket{+}$ state which give the spin.
As a result, the total complexity for the \sel\ operation is only two Toffolis more (the second $Z$ operation must be made controlled) than described in the list above, and can be given as
\begin{equation}
C_S = 2M+4N\rotbits-\frac{11N}{2}+\left\lceil \frac{M}{k_{r1}} \right\rceil +\left\lceil \frac{N}{2k_{r1}} \right\rceil+ k_{r1}+\left\lceil \frac{M}{k_{r2}} \right\rceil + k_{r2}   -2.
\end{equation}

\subsection{Overall costs of the non-orthogonal tensor hypercontraction Hamiltonian simulation}

In order to construct the entire step of the quantum walk, we also need a reflection on the ancillas used for the control state. We need $2n_M+\zetabits+4$ qubits, which are as follows.
\begin{enumerate}
\item We need $2n_M$ for the $\mu$ and $\nu$ registers.
\item The equal superposition state for the coherent alias sampling needs $\zetabits$ qubits.
\item One qubit that controls the swap of the $\mu$ and $\nu$ registers.
\item One ancilla that is rotated to ensure the amplitude amplification for the equal superposition state preparation works.
\item Two qubits encoding the superposition over spins.
\end{enumerate}
The complexity needed to perform a reflection on this many qubits is $2n_M+\zetabits+2$.
Another cost needed for each step is to perform a step of the unary iteration on the control register.
This has a cost of another single Toffoli for each step.
The output of that unary iteration needs to be used to control the reflection about the ancilla as well, which adds another single Toffoli.
Another single Toffoli cost is the controlled swap of the spin registers in \fig{select1}.
Altogether that gives an additional cost of $2n_M+5$ for each step, in addition to the cost of the \prep\ and \sel\ operations.
The total complexity for a single step can then be given as
\begin{align}\label{eq:THCtoffolis}
C_S+C_P+C_{P^\dagger}+2n_M+\zetabits+4 
&=30n_M+4\rotprec-16+2n_M^2+3\zetabits+\left\lceil\frac{d}{k_{s1}}\right\rceil + m(k_{s1}-1)+\left\lceil\frac{d}{k_{s2}}\right\rceil + k_{s2} \nn
&\quad + 
2M+4N\rotbits-\frac{11N}{2}+\left\lceil \frac{M}{k_{r1}} \right\rceil +\left\lceil \frac{N}{2k_{r1}} \right\rceil+ k_{r1}+\left\lceil \frac{M}{k_{r2}} \right\rceil + k_{r2}\, ,
\end{align}
with $m = 2n_M + 2 + \zetabits$, $n_M=\lceil\log(M+1)\rceil$.
For the total cost, this needs to be multiplied by the number of iterations needed for the phase estimation
\begin{equation}
\iter = \lceil\pi\lambda/(2\epsilon_{\textsc{PEA}})\rceil .
\end{equation}

Next we consider the costs for the number of logical qubits.
\begin{enumerate}
\item The control register for the phase estimation needs $\lceil\log(\mathcal{I}+1) \rceil$ qubits.
The unary iteration to control on this register needs another $\lceil\log(\mathcal{I}+1) \rceil-1$ qubits.
\item $N$ qubits used for the system register.
\item There are $2n_M$ qubits used for the $\mu$ and $\nu$ registers.
\item There is an ancilla with $\zetabits$ bits that is placed in an equal superposition.
\item We need another qubit for the $\ket{+}$ state to control the swap of the $\mu$ and $\nu$ registers.
\item In preparing the equal superposition state we also need to perform a rotation on a single ancilla.
\item There are $2$ qubits used for the spin registers in the prepared state.
\item One qubit is used for flagging success of the inequality tests in the state preparation.
\item One qubit is used to flag if $\nu=M+1$.
\item There are $\rotbits$ qubits for the phase gradient state.
\item The contiguous register has size $\lceil\log d\rceil$.
\item The QROM has qubit cost (including the output) of $mk_{s1}+\lceil\log d/k_{s1}\rceil$.
Of these, $m(k_{s1}-1)+\lceil\log d/k_{s1}\rceil$ are temporary ancillas which can be reused later.
That is, $\lceil\log d/k_{s1}\rceil$ temporary ancillas for QROM, and $m(k_{s1}-1)$ are extra outputs that may be erased by measuring in the $X$ basis.
If the costs in the following steps are smaller, then we can ignore them and use the cost of the QROM here instead.
\item The size of the data output for the rotations is $\rotbits N/2$, and there will be $\lceil\log{M}\rceil$ temporary ancillas as well.
\item There are $\rotbits-2$ temporary qubits when adding into the phase gradient state for rotations, which will be larger than $\lceil\log{M}\rceil$ in the previous step for the systems of interest.
\item The ancilla costs for erasing the QROM can be ignored because they can reuse ancillas that were previously erased.
\end{enumerate}
This costing gives us a total number of logical qubits
\begin{equation}\label{eq:THCqubits}
2\lceil\log(\mathcal{I}+1) \rceil+N+2n_M+\rotbits+\lceil\log d\rceil+\zetabits+5+\max\left( mk_{s1}+\lceil\log d/k_{s1}\rceil ~,~m+\rotbits N/2 +\rotbits-2 \right).
\end{equation}

\subsection{Error metrics for approximate tensors}

The evaluation of Hamiltonian approximation errors, i.e.~$\epsilon_{\textsc{thc}}$, is critical for precisely estimating the resource requirements of any quantum simulation method based on tensor factorizations, including tensor hypercontraction and the low rank methods. Moreover, to fairly compare against other methods such as the single factorization (SF), double factorization (DF) and the sparse method, one should use a consistent error metric for all.
We will take the perspective that one should estimate the error in the {\it exact} ground state energy made by approximating the Hamiltonian matrix elements since preparing ground states is very often the goal.
Obviously, estimating the error in the exact ground state energy is not feasible for problems whose ground state computations are not classically tractable. We list below desired properties which a good error metric should satisfy:
\begin{enumerate}
\item An error metric should be classically tractable to evaluate so that computing it for a handful of medium-sized systems is relatively straightforward.
\item An error metric should be size-extensive and size-consistent. Size-extensivity and size-consistency are important properties of the exact ground state wavefunction and energy \cite{szabo1996modern,shavitt2009many}. Violating these properties often results in poor approximations of the exact ground state wavefunctions and therefore these are important properties to preserve. Size-extensivity asserts the asymptotic scaling of the energy to be linear in system size and size-consistency asserts the product separability of wavefunctions and the additivity of energies when two subsystems are infinitely far part. Therefore, any error metrics that attempt to bound the ground state energy should satisfy these two properties. 
\item An error metric should provide a reliable bound on the exact ground state energy error and also be well-correlated with the actual error in the energy. By well correlated, we mean that improving the error metric should lead to the improvement in the ground state energy error as well.
This is generally very difficult to meet and also verify because, in general, we do not know the exact ground state energy.
\end{enumerate}
We summarize existing error metrics here and discuss which one is most suitable for our purposes.
We refer to the approximate integral tensor as $\widetilde{V}$ and keep the discussion of error metrics as general as possible so that it can be applied to any form of approximation methods.

In von Burg \emph{et al.}~\cite{vonBurg2020}, two error metrics were introduced and advocated: (1) coherent error ($\epsilon_\text{co}$) and (2) incoherent error ($\epsilon_\text{in}$).
In fact, these are the 1-norm and 2-norm measures of the difference between the exact and approximate integral tensors; i.e.,
\begin{align}
\epsilon_\text{co}
\equiv
\sum_{pqrs}
\left|V_{pqrs} - \widetilde{V}_{pqrs}\right| , \qquad \qquad
\epsilon_\text{in}
\equiv
\sqrt{
\sum_{pqrs}
\left|V_{pqrs} - \widetilde{V}_{pqrs}\right|^2 
} .
\end{align}
The incoherent scheme provides a rigorous bound to the shift in the ground state energy but we find these bounds to be too loose. In examples considered here, they are not well correlated with the error in the ground state energy despite bounding that error. For instance, a slight improvement in $\epsilon_\text{in}$ can lead to a drastic improvement in the ground state energy estimate.
More importantly, both $\epsilon_\text{co}$ and $\epsilon_\text{in}$ scale up to quartically with system size (at the very least super-linearly in general) which grows far too quickly with system size and violates the size-extensivity criterion. Furthermore, $\epsilon_\text{in}$ is no longer size-consistent.
We believe that the violation of size-extensivity and size-consistency is what makes these error metrics often not well correlated with the error in the exact ground state energy.

Berry \emph{et al.}~\cite{Berry2019B} advocated error metrics based on classical quantum chemistry methods called the second-order M{\o}ller-Plesset perturbation theory (MP2) and 
configuration interaction with singles and doubles (CISD).
The MP2 correlation energy error metric can be thought of as a weighted signed difference between two tensors:
\begin{equation}
\epsilon_\text{MP2}
=
\sum_{pqrs} w_{pqrs} \left(V_{pqrs} - \widetilde{V}_{pqrs}\right) ,
\label{eq:mp2}
\end{equation}
where $w_{pqrs}$ is a positive weight that is defined as the orbital energy differences defined for each quartet $(p,q,r,s)$.
It can be rigorously shown that \eq{mp2} satisfies size-consistency and size-extensivity since the MP2 correlation energy satisfies these \cite{shavitt2009many}.
We note that the correlation energy in a finite basis is defined as the energy difference between an approximate method and a mean-field approach (Hartree-Fock) in the same basis\cite{szabo1996modern}.
Unfortunately, the MP2 correlation energy behaves quite erratically in many cases~\cite{lee2018regularized} so the error bound based on the MP2 correlation energy
may not be well correlated with the actual error in the exact ground state energy.
Since there are other classical approaches that go beyond MP2, one may consider other options as well. 
The CISD correlation energy error metric is the option that Berry \emph{et al.}~\cite{Berry2019B} chose. 
Unfortunately, the CISD correlation energy is neither size-extensive nor size-consistent.
Consequently, in the limit of infinite system size, the CISD correlation energy approaches zero.

In this work, we advocate the use of an error metric based on the correlation energy error in the coupled-cluster with singles, doubles, and perturbative triples (CCSD(T)) approach \cite{raghavachari1989fifth}.
CCSD(T) is so-called the ``gold standard'' method in classical simulations of quantum chemistry. 
While its quantitative accuracy on strongly correlated systems is often doubtful, it is still the best classical method that 
satisfies criteria (1) and (2) above. Whether it meets criterion (3) is again difficult to assess, but for some Hamiltonians for which CCSD(T) is qualitatively accurate, it is reasonable to expect that (3) is satisfied.
Furthermore, CCSD(T) contains the MP2 wavefunction contribution in it in the sense that some diagrams contained in CCSD(T) exactly correspond to those of MP2. 
Therefore, we think that measuring the errors made by approximating the Hamiltonian elements based on CCSD(T) would be representative of classical simulation methods. Ultimately, it is possible that the errors in the CCSD(T) correlation energy are not well correlated with the errors in the exact ground state energy. We leave more precise understanding of these error metrics for future study and focus on comparing different approaches based on the CCSD(T) correlation energy error. 
For strongly correlated Hamiltonians, one can consider high-spin states where single determinant wavefunctions provide a good description. In this case, the CCSD(T) energy must be well-correlated with the exact high-spin eigenstates. 
As long as the underlying integral approximation does not assume the underlying spin state (which is the case for all approximations considered here), this should facilitate a fair comparison between the methods discussed in this work.

In addition to the inherent factorization error,
it is useful to further account for the error made by the
approximation of the Hamiltonian coefficients ($\zeta$) as part of coherent alias sampling
and the individual qubit rotations via $\chi$. 
This is achieved by first making approximate representations for $\chi$ and $\zeta$ for fixed
numbers of bits to represent these tensors, $\rotbits$ and $\zetabits$, respectively.
That is, $\chi$ is represented by a sequence of rotations with angles $\{\theta_p^{(\mu)}\}$ and
these angles are approximated by the limited precision given by $\rotbits$.
More specifically,
a rotation vector, $\vec{\chi}^{(\mu)}$ is parametrized by a set of angles $\{\theta_p^{(\mu)}\}$ which are defined recursively \cite{vonBurg2020} as
\begin{equation}
\chi_p^{(\mu)}\left[\vec{\theta}^{(\mu)}\right] = \cos(2\theta_p^{(\mu)}) \Pi_{q<p} \sin(2\theta_q^{(\mu)}).
\end{equation}
Next, we define units for $\theta$ and $\zeta$,
\begin{align}
u_{\theta} = \frac{2\pi}{2^\rotbits}  \qquad \qquad
u_{\zeta}^o = \frac{\lambda_z}{d2^{\zetabits}} \qquad \qquad
u_{\zeta}^d = \frac{\lambda_z}{d2^{\zetabits-1}}
\end{align}
where $u_{\theta}$ is the unit for $\theta$,
$u_{\zeta}^o$ is the unit for off-diagonal elements of $\zeta$, and
$u_{\zeta}^d$ is the unit for diagonal elements of $\zeta$.
Finally, write approximate $\chi$ and $\zeta$ as $\tilde{\chi}$ and $\tilde{\zeta}$, respectively, by
\begin{align}
{\tilde\chi_p}^{(\mu)} = {\chi_p}^{(\mu)}\left[u_\theta\, \times \text{round}(\vec{\theta}^{(\mu)}/u_\theta)\right]
\qquad \qquad
\tilde{\zeta}_{\mu\nu} = 
\begin{cases}
u_{\zeta}^o \times \text{round}( \zeta_{\mu\nu} / u_{\zeta}^o + x) ~~\text{for} ~~\mu \ne \nu \\
u_{\zeta}^d \times \text{round}( \zeta_{\mu\mu} / u_{\zeta}^d + x) ~~\text{for} ~~\mu = \nu
\end{cases}
\end{align}
where $x$ is a small constant to ensure the normalization of $|\tilde{\zeta}|$.
We employed a sextic polynomial fit for $x\in[-1,1]$
to find the optimal $x$ that normalizes $|\tilde{\zeta}|$. The resulting $\lambda$ of $|\tilde{\zeta}|$ was normalized to 1 with a small error on the order of $10^{-6}$.
These can be used to build an approximate representation $\widetilde{G}$ for $G$.
In addition to evaluating the CCSD(T) correlation energy error using ${G}$, we
also evaluate the error using $\widetilde{G}$ to provide more precise resource estimates.

\section{Resource Estimates for Real Systems}
\label{sec:numerics}

\subsection{Resource estimates for simulating active space models of FeMoCo molecule}

Keeping with the tradition of the work by Reiher \emph{et al.}~\cite{Reiher2017}, Berry \emph{et al.}~\cite{Berry2019B} and von Burg \emph{et al.}~\cite{vonBurg2020}, here we will benchmark our methods on the problem of simulating the ground state of active space models of FeMoCo. While this is not the only quantum chemistry problem worth studying on a quantum computer, it is easier to compare to prior methods if we study the same molecule. We look forward to also deploying our method to some of the catalysis Hamiltonians studied by von Burg \emph{et al.}~\cite{vonBurg2020} as soon as they are made available by those authors. Unfortunately, the first of these papers Reiher \emph{et al.}~used an active space for FeMoCo that was later found to be problematic for the ground state simulation by Li \emph{et al.}~\cite{Li2019}. 
The ground state of the active space proposed by Reiher \emph{et al.}~does not capture the open-shell nature (i.e., strong correlation) of the FeMoCo model cluster and therefore the gold standard CCSD(T) calculation was found to be off by only 5 milliHartree from the near-exact density matrix renormalization group (DMRG) energy~\cite{Li2019}.
Therefore, in their work, Li \emph{et al.}~propose an alternative active space. Whereas the original Reiher Hamiltonian involved 108 qubits, the integrals by Li \emph{et al.}~involve 152 qubits and is thus a more challenging problem. The work by Berry \emph{et al.}~estimates the cost for both Hamiltonians but here we focus on assessing the resources that would be required to simulate the Li \emph{et al.}~Hamiltonian, as the more accurate active space. 

\begin{figure}[ht!]
\includegraphics[scale=0.45]{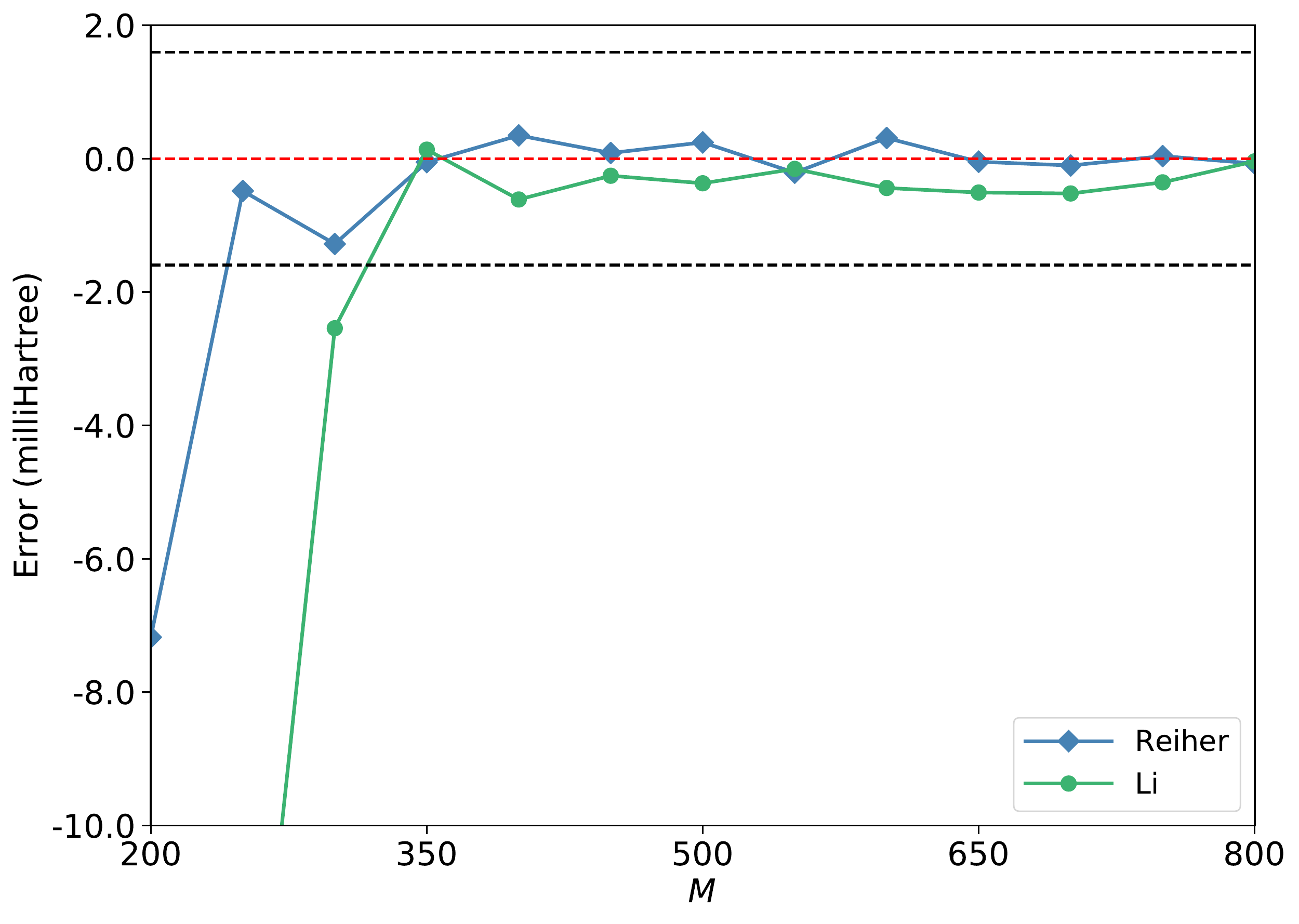}
\caption{\label{fig:femoco}
Error in the CCSD(T) correlation energy (milliHartree) for the Reiher \emph{et al.}~\cite{Reiher2017} and the Li \emph{et al.}~\cite{Li2019} Hamiltonian as a function of the THC rank $M$. Black dotted lines indicate chemical accuracy of 1.6 milliHartree. Chemical accuracy for both Hamiltonians is achieved for $M\ge350$.
Note that the energy improvement is not monotonic in $M$ in part due to difficulties associated with the non-linear optimization of THC factors.
}
\end{figure}

In \fig{femoco} we analyze the dependence of the shift in the ground state energy on the THC rank $M$. For the Reiher \emph{et al.}~Hamiltonian, as mentioned above, the gold standard CCSD(T) is quantitatively accurate so it is
quite sensible to assess the error of approximate integral tensors based on the error in the CCSD(T) correlation energy.
The ground state computation involves $N=108$ qubits and 54 electrons and the target spin state is $S=0$ (singlet).
We employ spin-restricted Hartree-Fock orbitals for this Hamiltonian.
As we can see, we achieve chemical accuracy for $M \ge 250$.

For the Li Hamiltonian, there are 152 spin orbitals and 113 electrons in this active space model. 
For this Hamiltonian, CCSD(T) is no longer quantitatively accurate for low-spin states (such as the ground state $S=3/2$) due to its inability to capture
the antiferromagnetically-coupled open-shell nature. However, if we consider a high-spin state such as $S=35/2$, CCSD(T) with spin-unrestricted Hartree-Fock orbitals
should be quantitatively accurate compared to the exact energy of the $S=35/2$ state.
Furthermore, there is no aspect in the THC factorization (or other methods compared in this work) specialized for a specific spin state.
Therefore, we believe that the CCSD(T) error metric for the $S=35/2$ is well correlated with the actual error made by integral approximations in the exact $S=3/2$ ground state energy.
As shown in \fig{femoco}, THC achieves chemical accuracy for $M\ge350$.

For both Hamiltonians, we assess
the CCSD(T) correlation energy error
with approximate rotation angles and $\zeta$. 
We found that 10 bits for state preparation and 16 bits for rotations were enough for the Reiher Hamiltonian,
whereas 10 bits for state preparation and 20 bits for rotations were needed for the Li Hamiltonian.
The resulting CCSD(T) correlation energy error, $\lambda$, Toffoli count, and logical qubit count are available
in \tab{femocothcR} for the Reiher Hamiltonian and \tab{femocothcL} for the Li Hamiltonian.
Given these data, we recommend $M=350$ for the Reiher Hamiltonian and $M=450$ for the Li Hamiltonian.

\begin{table*}
\begin{tabular}{|c|c|c|c|c|}
\hline
$M$ & CCSD(T) Error (m$E_h$) & $\lambda$ & Toffoli count & logical qubits \\ \hline
250 & -1.31 & 294.1 & 4.4$\times 10^9$ & 1115 \\ \hline
300 & -1.12 & 302.8 & 4.9$\times 10^9$ & 1183 \\ \hline
\color{blue}350 & \color{blue}-0.29 & \color{blue}306.3 & \color{blue}5.3$\times 10^9$ & \color{blue}2142 \\ \hline
400 & -0.18 & 315.1 & 5.6$\times 10^9$ & 2144 \\ \hline
450 & 0.13 & 327.9 & 6.1$\times 10^9$ & 2144 \\ \hline
500 & 0.03 & 339.2 & 6.6$\times 10^9$ & 2146 \\ \hline
550 & -0.07 & 343.0 & 7.1$\times 10^9$ & 2278 \\ \hline
600 & -0.10 & 347.8 & 7.6$\times 10^9$ & 2278 \\ \hline
650 & -0.29 & 361.4 & 8.2$\times 10^9$ & 2278 \\ \hline
700 & -0.10 & 365.1 & 8.7$\times 10^9$ & 2278 \\ \hline
750 & -0.09 & 373.6 & 9.3$\times 10^9$ & 4327 \\ \hline
800 & -0.04 & 380.2 & 9.7$\times 10^9$ & 4327 \\ \hline
\end{tabular}
\caption[Tensor hypercontraction overheads for Reiher FeMoCo Hamiltonian]{THC costing for the Reiher Hamiltonian with 10 bits for state preparation and 16 bits for rotations. For our analysis of this Hamiltonian, we use $M=350$, which is the highlighted entry.}
\label{tab:femocothcR}
\end{table*}
\begin{table*}
\begin{tabular}{|c|c|c|c|c|}
\hline
$M$ & CCSD(T) Error (m$E_h$) & $\lambda$ & Toffoli count & logical qubits \\ \hline
350 & 0.39 & 1279.0 & 3.2$\times 10^{10}$ & 2194 \\ \hline
400 & -1.14 & 1258.4 & 3.2$\times 10^{10}$ & 2196 \\ \hline
\color{blue}450 & \color{blue}-0.18 & \color{blue}1201.5 & \color{blue}3.2$\times 10^{10}$ & \color{blue}2196 \\ \hline
500 & -0.49 & 1214.9 & 3.3$\times 10^{10}$ & 2196 \\ \hline
550 & -0.08 & 1161.2 & 3.3$\times 10^{10}$ & 2328 \\ \hline
600 & -0.16 & 1140.8 & 3.4$\times 10^{10}$ & 2328 \\ \hline
650 & -0.09 & 1132.2 & 3.5$\times 10^{10}$ & 2328 \\ \hline
700 & -0.39 & 1119.8 & 3.6$\times 10^{10}$ & 2328 \\ \hline
750 & -0.23 & 1114.4 & 3.6$\times 10^{10}$ & 4377 \\ \hline
800 & -0.32 & 1123.7 & 3.8$\times 10^{10}$ & 4377 \\ \hline
\end{tabular}
\caption[Tensor hypercontraction overheads for Li FeMoCo Hamiltonian]{THC costing for the Li Hamiltonian with 10 bits for state preparation and 20 bits for rotations. For our analysis of this Hamiltonian, we use $M=450$, which is the highlighted entry.}
\label{tab:femocothcL}
\end{table*}

\subsection{Resource estimates and scaling analysis for hydrogen chain and \texorpdfstring{\ce{H4}}{H4} benchmarks}

In order to study the scaling of these methods in this section we analyze a chemical series consisting of hydrogen chains. We use the same chemical series as the one studied in \cite{Berry2019B} in order to facilitate a direct comparison with that work: one-dimensional hydrogen chains with a spacing of 1.4 Bohr. For the finite size scaling, we use the STO-6G minimal basis set and grow the system by adding more hydrogens. Denoting the number of hydrogens in a chain by $N_H$, we study the chemical series from $N_H = 10$ to $N_H = 100$. 
When estimating thermodynamic asymptotes, 
it is often convenient to fix the error per particle as opposed to the total error for every system size \cite{weigend2006accurate}. Indeed, the most widely used integral factorization, the resolution-of-the-identity approximation, typically yields
an error per atom to be 50-60 {\textmu}Hartrees \cite{weigend2006accurate}. We follow a similar standard in this section.

For the hydrogen chain calculations,
we used a threshold of $5\times10^{-5}$ for the sparse method, 
0.0015 for the single factorization method using the modified Cholesky factorization \cite{aquilante2010molcas} (which yields roughly 2 Cholesky vectors per H atom), and
0.01 for the double factorization method (see \eq{dfthresh}).
For THC, we fixed the THC rank to be 7 $\times N_H$.
These thresholds were enough to obtain the CCSD(T) correlation energy error per atom to be less than 50 {\textmu}Hartrees as shown in \tab{hchain}.

For \ce{H4}, we used the cc-pVTZ basis set \cite{Dunning1989} to obtain $V$
and
considered truncated Hamiltonians in the molecular orbital basis with varying number of orbitals from $N/2=$ 10 to 56. 
The number of Cholesky vectors used in the SF method was fixed to be 300 and the
THC rank was held at 576.
This was enough to maintain the accuracy of 1--10 {\textmu}Hartrees for $N/2$ greater than 35 in all methods.
With respect to the basis set size, 
the data size for state preparation scales
cubically with basis set size in the case of the SF method and quadratically with basis set size for THC \cite{lee2019systematically}. Therefore, we focus on the scaling of $\lambda$ for SF and THC methods.
We used a threshold of $5\times10^{-5}$ for the sparse method and
$1\times10^{-4}$ for the DF method.
On the other hand, the data size of sparse and DF methods varies with system so we obtain representation for each instance of the truncated Hamiltonian. 
This gives us the scaling of data size with respect to the basis set size.
In all systems, the sparse method $\lambda$ values were evaluated with localized orbitals using the Edmiston-Ruedenberg method \cite{edmiston1963localized} to exploit spatial locality. All other methods used canonical Hartree-Fock orbitals.

\begin{figure}[ht!]
\includegraphics[scale=0.55]{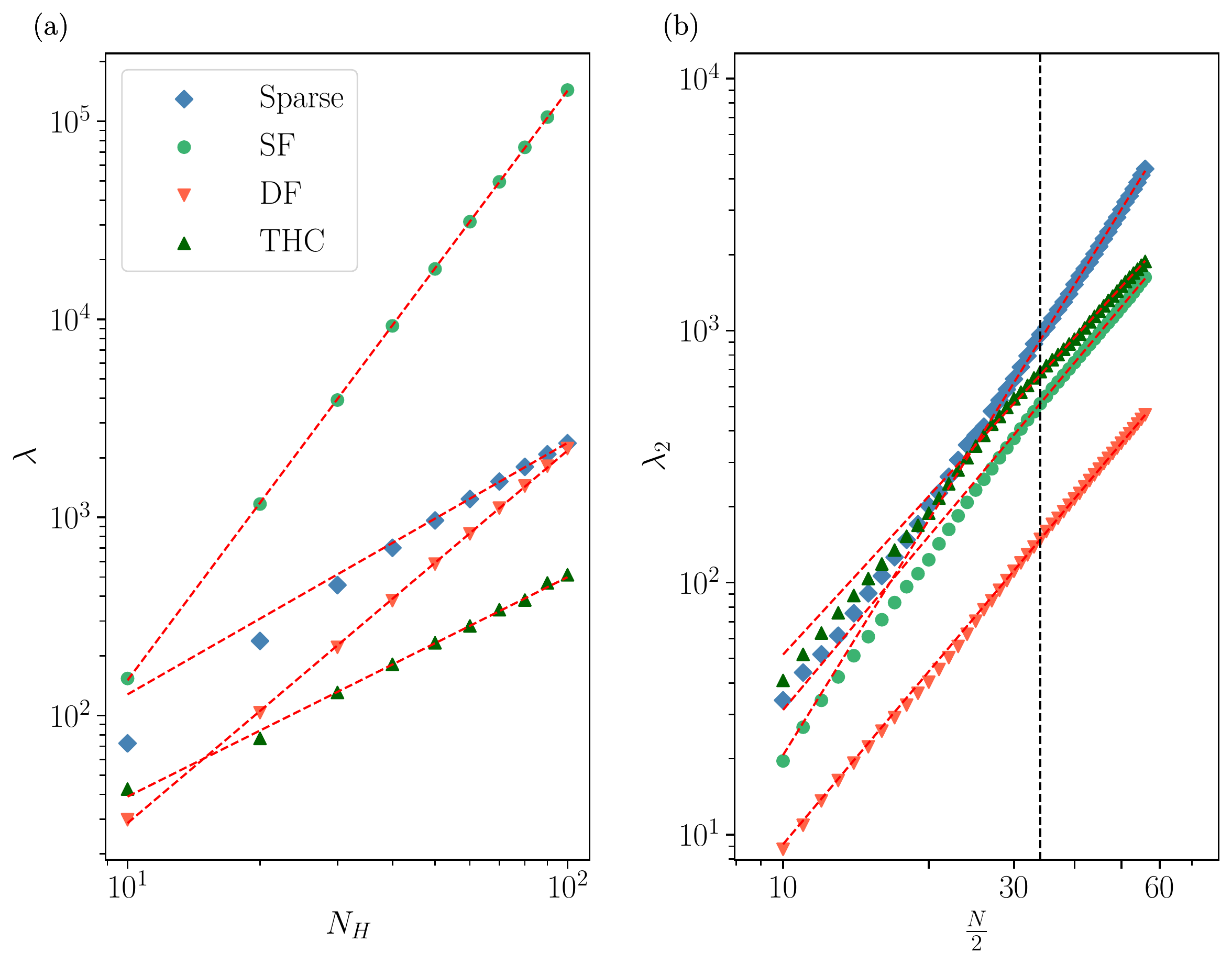}
\caption{\label{fig:h4hchain}
(a) The number of H atoms ($N_H$) versus total $\lambda$ values and (b) the number of orbitals ($N/2$) versus two-body $\lambda$ values for different approaches.
Note that these $\lambda$ values in (a) include both one-body and two-body $\lambda$ values relevant to each approach
and the two-body $\lambda_2$ values in (b) are also approach-specific values.
The vertical black dotted line in (b) is drawn at $N/2=36$ beyond which all lines behave linearly and
associated CCSD(T) correlation energy errors are roughly constant.
Red dotted lines represent the linear fits on a log scale for each data whose slopes are listed in 
\tab{hchainslopes} and \tab{h4slopes}. Those slopes are used to compute scalings listed in \tab{h4hchain}.
}
\end{figure}

Values of $\lambda$ for each method are presented in \fig{h4hchain}.
In (a), we have the hydrogen chain data
from which 
we can infer the thermodynamic size scaling of different approaches.
In (b), we have the \ce{H4} two-body $\lambda$ data from which we can find the continuum limit scaling of different approaches.
Since \ce{H4} is a very small system, the one-body contribution is comparable to the two-body contribution to $\lambda$. 
However, as we are only interested in asymptotes, we focus just on the two-body contribution because asymptotically the two-body term dominates $\lambda$. 
These scalings of $\lambda$ are used to generate the Toffoli complexity scaling data in \tab{h4hchain}.
For the space complexity scaling, we need
additional information for the sparse method (i.e., the number of non-zero elements) and
the DF method (i.e., the average number of eigenvectors in the second factorization).
These are provided in \app{sparse} and \app{double_low_rank}, respectively.
We perform linear fits to obtain the asymptotic scaling of each method appropriately.
The results of these linear fits are summarized in \tab{hchainslopes} and \tab{h4slopes} along with data range and $R^2$ values.

\begin{table*}
\begin{tabular}{|c|c|c|c|c|}
\hline
$N_\text{H}$ & Sparse & SF & DF & THC \\ \hline
10 & 3.9$\times 10^{-8}$ & 7.4$\times 10^{-6}$ & 3.5$\times 10^{-5}$ & 4.4$\times 10^{-6}$ \\ \hline
20 & 1.2$\times 10^{-6}$ & 8.3$\times 10^{-6}$ & 3.4$\times 10^{-5}$ & 1.7$\times 10^{-5}$ \\ \hline
30 & 5.3$\times 10^{-6}$ & 8.6$\times 10^{-6}$ & 4.6$\times 10^{-5}$ & -2.9$\times 10^{-7}$ \\ \hline
40 & 1.2$\times 10^{-5}$ & 8.8$\times 10^{-6}$ & 4.7$\times 10^{-5}$ & 1.2$\times 10^{-5}$ \\ \hline
50 & 1.7$\times 10^{-5}$ & 8.9$\times 10^{-6}$ & 3.6$\times 10^{-5}$ & 5.4$\times 10^{-6}$ \\ \hline
60 & 2.1$\times 10^{-5}$ & 9.0$\times 10^{-6}$ & 3.8$\times 10^{-5}$ & 5.1$\times 10^{-5}$ \\ \hline
70 & 2.5$\times 10^{-5}$ & 9.0$\times 10^{-6}$ & 4.1$\times 10^{-5}$ & 1.9$\times 10^{-5}$ \\ \hline
80 & 2.7$\times 10^{-5}$ & 9.0$\times 10^{-6}$ & 3.5$\times 10^{-5}$ & 4.9$\times 10^{-6}$ \\ \hline
90 & 2.2$\times 10^{-5}$ & 9.1$\times 10^{-6}$ & 3.5$\times 10^{-5}$ & 2.2$\times 10^{-5}$ \\ \hline
100 & 3.4$\times 10^{-6}$ & 9.1$\times 10^{-6}$ & 4.1$\times 10^{-5}$ & -1.6$\times 10^{-5}$ \\ \hline
\end{tabular}
\caption[Errors in representations of hydrogen chains]{
CCSD(T) correlation energy error per hydrogen in Hartree of each hydrogen chain system for various methods.
We aimed to reach 10--50 {\textmu}Hartrees for all methods but SF. 
For the SF method, the largest error below 50 {\textmu}Hartrees / H atom that we could find was about 9 {\textmu}Hartrees.
}
\label{tab:hchain}
\end{table*}

\begin{table*}
\begin{tabular}{|c|c|c|c|}
\hline
approach & slope & data range & $R^2$ \\ \hline
Sparse & 1.27 & the last five points & 0.9998 \\ \hline
SF & 2.98 & all & 1.0000 \\ \hline
DF & 1.88 & all & 0.9998 \\ \hline
THC & 1.11 & exclude the first two and the last two points & 0.9991\\\hline
\end{tabular}
\caption[Slope of hydrogen chain linear fits on a log scale in \fig{h4hchain}]{
Slopes of linear fits on a log scale to the hydrogen chain data in \figa{h4hchain}{(a)}.
}
\label{tab:hchainslopes}
\end{table*}

\begin{table*}
\begin{tabular}{|c|c|c|}
\hline
approach & slope & $R^2$ \\ \hline
Sparse & 3.10 & 0.9993 \\ \hline
SF & 2.29  & 0.9998 \\ \hline
DF & 2.28  & 0.9999 \\ \hline
THC & 2.09 &  0.9991\\\hline
\end{tabular}
\caption[Slopes of H$_4$ linear fits on a log scale in \fig{h4hchain}]{
Slopes of linear fits on a log scale to the \ce{H4} data in \figa{h4hchain}{(b)}.
All data points above $N/2 = 35$ are used in the linear fit. 
}
\label{tab:h4slopes}
\end{table*}

\subsection{Surface code compilation for the active space model of FeMoCo by Reiher~\emph{et al.}}
\label{sec:surface_code}

We now come to the problem of determining how much physical space and time will be used by our computations if we plan to realize them assisted by quantum error-correction within the surface code.
Producing these estimates requires making assumptions about the performance characteristics of physical qubits and the classical control system making up the quantum computer.
We will be assuming a physical gate error rate of 0.1\%, a control system reaction time of 10 microseconds, and a surface code cycle time of 1 microsecond.
We will also consider the more optimistic gate error rate of 0.01\% to give a point of comparison.

\subsubsection{Space and time constraints}

In surface code quantum computations, space-time tradeoffs are ubiquitous, meaning one can often shorten the length of a computation at the expense of a higher physical qubit overhead and vice versa.
A good first approximation is that a quantum computation is like a liquid that can be poured into any desired spacetime volume.
For example, often one can (roughly) halve the amount of time taken by a computation by doubling the number of magic state factories.
Of course, the ``uncompressible liquid computation volume'' approximation does break down when pushed enough. For example, as the space available is reduced, the contortions needed to fit the computation become more and more challenging.
The time has to increase by proportionally more, to accommodate the contortions.
There is also a minimum number of qubits needed to store the system being simulated, which no amount of contortion will get below.
Additionally, there are constraints that prevent the {\em time} of a quantum computation from being reduced arbitrarily far, such as the code distance $d$, the reaction time $t$, and details of the algorithm being run.
To give the reader a better understanding of the layout decisions we made, we will discuss these two constraints in more detail.

The code distance $d$ determines the level of protection against errors errors one has, and is chosen based on a combination of the physical error rate and number of operations in the computation.  For most practical code implementations, protection against time-like errors up to that distance $d$ dictates a local physical stabilizer has to be measured $O(d)$ times before it is possible to error correct those measurements to a desired level of reliability.
We will call the amount of time it takes to run $d$ rounds of the surface code cycle a ``beat''.
Any surface code construction that involves changing which stabilizers are being measured will (usually) take at least one beat to complete, since that is how long it takes to become confident about the values of the new stabilizers.

The reaction time $t$ measures how long it takes for the classical control system to receive the raw physical measurements corresponding to a logical measurement, error correct them to recover the logical measurement, and trigger a following logical measurement that is either in the X or Z basis depending on the value of the previous logical measurement.
This quantity becomes relevant when performing non-Clifford operations via magic state distillation and gate teleportation.
The gate teleportation will apply the desired operation, but will also randomly apply other Clifford operations that need to be undone by corrective Clifford operations.
If the corrective Clifford operations needed to finish a gate teleportation are applied ``in place'', by waiting for the correction to be known and then applying operations to the target qubits, then the correction process would have a time measured in beats.
This is a strategy we have used in the past when laying out surface code computations \cite{BabbushSpectraB}.
However, when using this strategy, it is difficult for the computation consuming magic states to keep pace with even a single magic state factory.
A more time efficient strategy is to precompute multiple world-lines, one for each possible Clifford correction, and selectively teleport the affected qubits through the appropriate world line \cite{Fowler2012aB,Gidney2019aB,  Litinski2018}.
The selective teleportation can be controlled by measuring certain qubits in either the X or Z basis.
Because logical X and Z measurements are implemented transversally, the time they take does not depend on the code distance, and so they are not beat limited.
Instead, the limiting time is how quickly the control system can process the deluge of data coming at it and iteratively figure out what the next basis to measure in is.

It is not always feasible to execute one non-Clifford operation per reaction time.
For example, there might not be enough magic state factories to produce one magic state per reaction time.
To account for this, we will introduce another time unit: the ``tick''.
A tick is either the reaction time of the control system or the average period between the production of magic states; whichever is slower.

Optimally packing a quantum computation often involves balancing tick-based constraints and beat-based constraints.
For example, the driving force of a QROM read is unary iteration sequentially preparing qubits flagging whether or not the address register was equal to each possible address value \cite{BabbushSpectraB}.
Unary iteration is made up of a series of interdependent Toffolis; it is primarily tick-constrained.
On the other hand, the flag qubits prepared by unary iteration are used as controls for huge multi-target CNOTs reaching into the target data registers.
These CNOTs are done via lattice surgery, which involves changing which stabilizers are being measured, and so the CNOTs are primarily beat-constrained.
Optimally packing a QROM computation into spacetime requires balancing the beat-based constraints from the multi-target CNOTs and the tick-based constraints from the unary iteration.

\subsubsection{QROM tradeoffs}

The performance in the surface code depends not only on the raw operations, but also on the layout design of the qubits.
Suppose we lay out data qubits into columns with every third column left empty as an access hallway.
In this layout, each data qubit has a single side exposed to a hallway.
We will call this layout ``single side storage''.
When using single side storage, a multi-target CNOT will monopolize the single side of its target qubits for one beat.
A basic QROM read performs one multi-target CNOT per possible address value, and therefore when using single side storage a basic QROM read with $n$ addresses will take at least $n$ beats to complete.

If $n$ beats is too slow (e.g.\ if it is significantly slower than the tick-limited unary iteration), we could instead use a storage layout where every second column was left empty as an access hallway.
In this ``double side storage'' layout, every qubit has two exposed sides.
This allows a second CNOT to start before the first has finished, by targeting the second side, lowering the minimum time required for an $n$-address QROM read from $n$ beats to $n/2$ beats. If $n/2$ beats is still too slow, alternative QROM circuits can reduce the number of multi-target CNOTs by an ``output expansion factor" $k < \sqrt{n}$ by using $k$ times more workspace qubits \cite{Lowpreparation,Berry2019B}.
These alternative circuits also lower the number of Toffoli gates required, reducing the tick-based lower bound from the unary iteration.

There are other possibilities for optimizing the packing of a QROM read, but for the physical parameter regime we are interested in there are three relevant design choices. (1) Should we use single or double sided storage? (2) How many magic state factories should there be? (3) What should the expansion factor $k$ be, for each QROM read?
After some experimentation and iteration we settled on using single side storage, four magic state factories, and an expansion factor of 16 for the QROM read during the $\prep$ subroutine and 1 for the QROM reads during the $\sel$ subroutine.
These decisions were based on looking at the algorithm's resource utilization at a global level, for various possible choices. We will now describe how that was analyzed.

\subsubsection{Diagram driven decisions}

\begin{figure}
    \centering
    
    \resizebox{.7\linewidth}{!}{
        \includegraphics{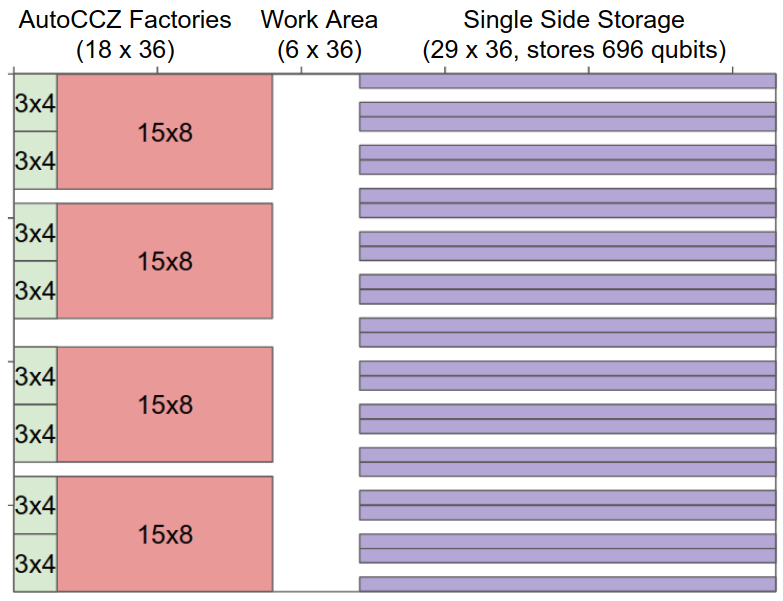}
    }
    \caption{Floorplan for the THC based FeMoCo qubitized phase estimation computation using the Reiher \emph{et al.}~Hamiltonian \cite{Reiher2017}. Covers $53 \times 36 = 1908$ logical qubits.
    The code distance is 31.
    The number of physical qubits is $1908 \cdot 2 \cdot 32^2 \approx 4 \cdot 10^6$.
    Red areas are the CCZ factory from \cite{Gidney2019aB}, with a level 1 code distance of 19 and a level 2 code distance of 31.
    Green areas are for the AutoCCZ fixup box from \cite{Gidney2019aB}.
    Purple areas are for data qubit storage.
    White areas are for routing, performing Clifford gates, and teleporting magic states into data qubits.
    }
    \label{fig:floorplan}
\end{figure}

When laying out a quantum algorithm in the surface code, there are two diagrams we recommend making.
The first diagram is a floorplan, or footprint, of where the various parts of the computation will live.
Our final floorplan diagram is in \fig{floorplan}.
The goal of the floorplan diagram is not to find a perfectly optimal layout, but rather to get a rough idea of how things will fit together and how much space will be needed to ensure it is possible to route data and magic states into operating areas quickly enough to keep the computation going.

\begin{figure}
    \centering
        \includegraphics[scale=0.5]{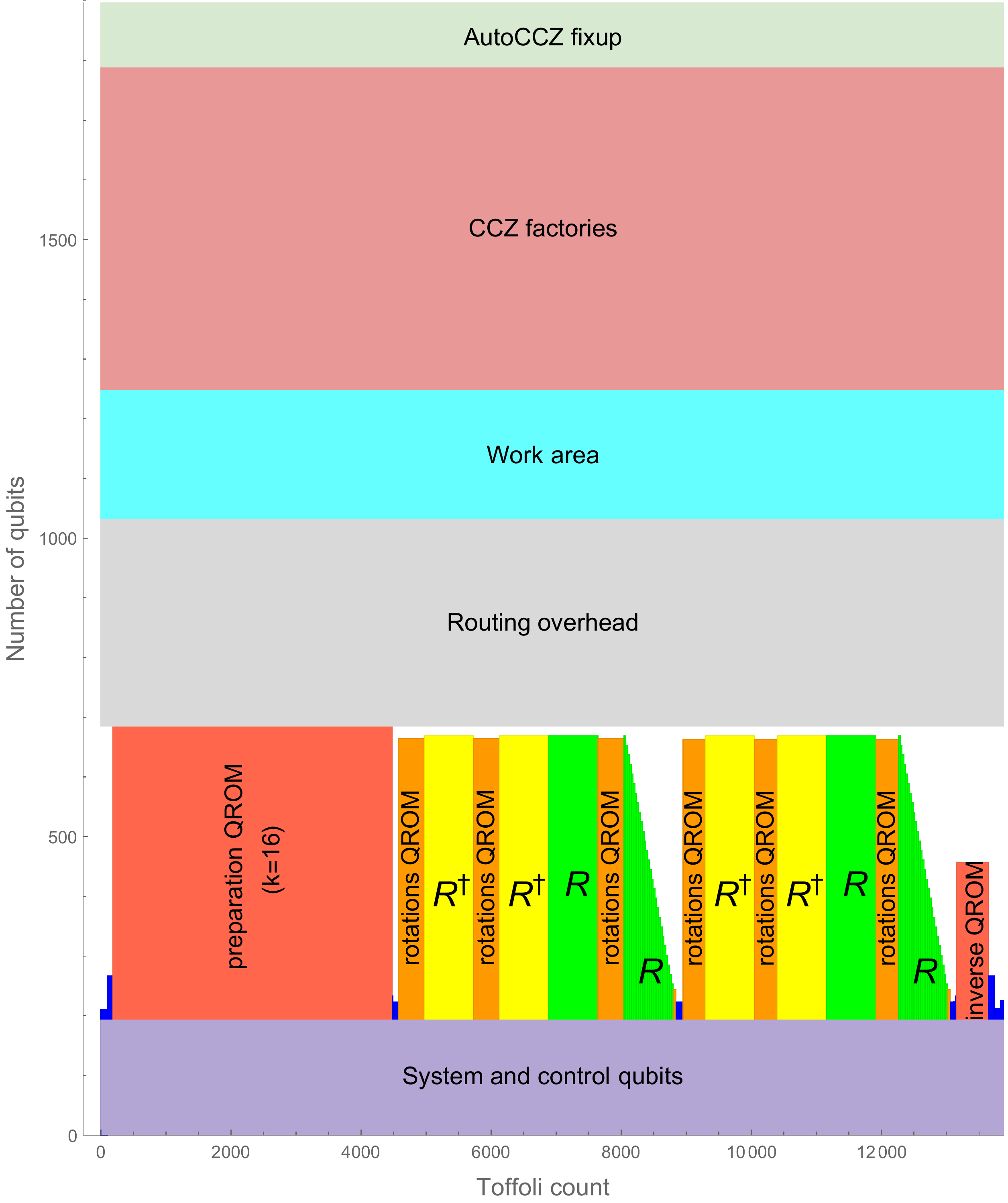}
    \caption{An improved strategy for the inner loop of the computation, where half of the phasing angles are loaded at a time.
    The output expansion factor during the first QROM read has correspondingly been decreased from 32 to 16.}
    \label{fig:qubittimelayout_new}
\end{figure}

The second useful diagram to make is an inventory of allocated qubits over time, which shows how many qubits are allocated as the algorithm progresses and also what they are being used for.
Our final diagram of this type is in \fig{qubittimelayout_new}.
This diagram can reveal spikes where too many qubits are being used, and holes where a lot of qubits are available for use (e.g.\ as additional magic state factories).

\begin{figure}
    \centering
        \includegraphics[scale=0.5]{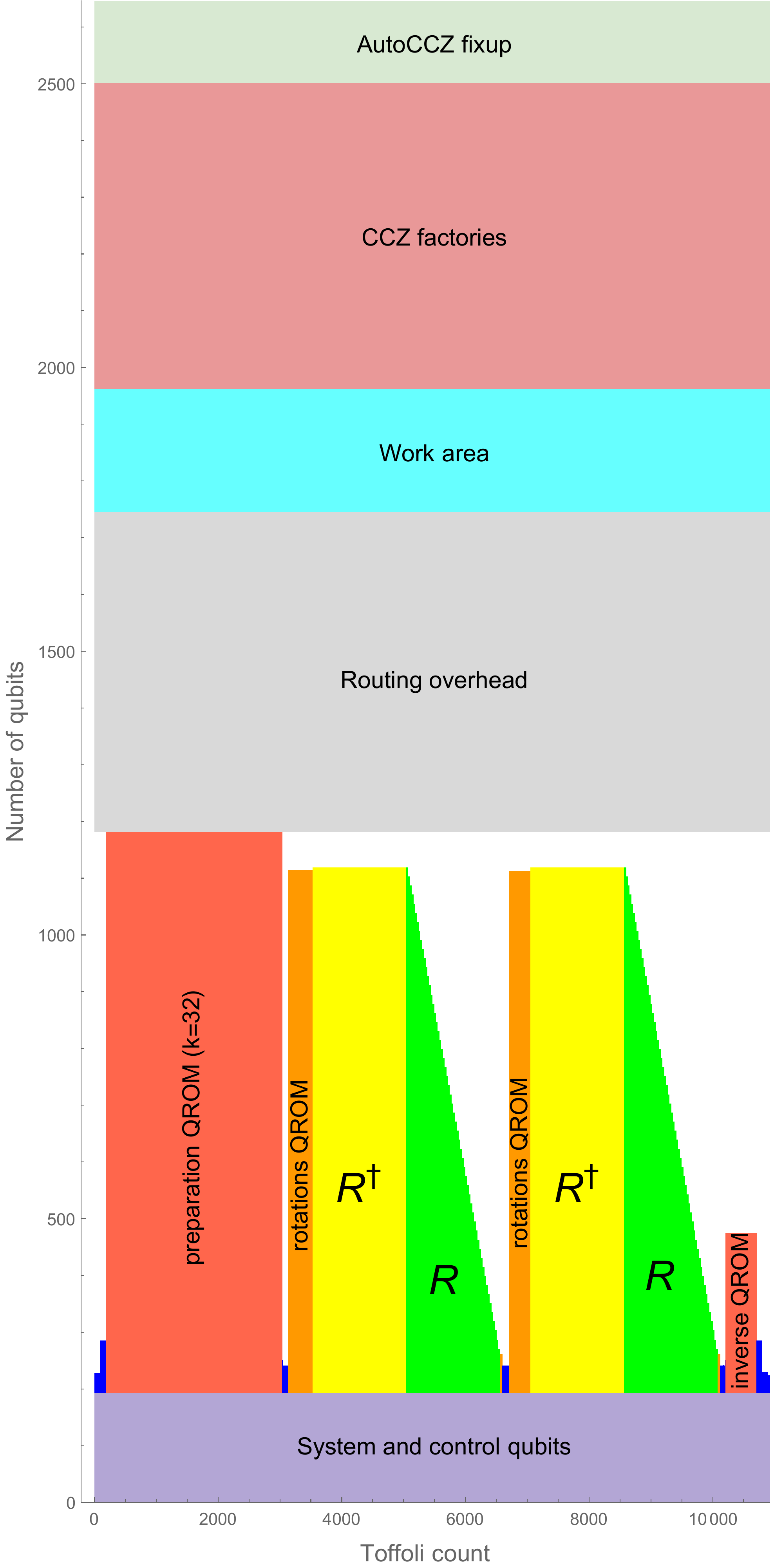}
    \caption{The original strategy we intended to use for the inner loop of the computation.
    The angles loaded from QROM in order to perform phasing operations are all loaded at the same time, to minimize the number of angle-loading QROM reads (in orange).
    It is clear from the diagram that the QROM reads are not the dominant cost; the diagram suggests performing redundant reads while storing fewer angles.}
    \label{fig:qubittimelayout_original}
\end{figure}

The reason these diagrams are useful to make is that understanding resource utilization allows one to make optimizations.
For example, elsewhere in this paper we note that the phasing operations during the $\sel$ subroutine will use angles loaded from QROM.
Originally, we intended for all the angles to be loaded simultaneously from one QROM read.
This avoids redundant QROM reads from the same address register and correspondingly minimizes the number of Toffolis.
However, when plotting allocated qubits over time as in \fig{qubittimelayout_original}, it became obvious that loading the angles is much faster than using them, and so loading half of the angles at a time would not cost much more overall.
The benefit of loading half of the angles at a time is that half as many output qubits are needed.
This is a substantial space savings for an acceptable time loss.

Because of the lower number of qubits available, we also had to adjust the output expansion factor of the QROM read during the prepare subroutine.
We reduced it from 32 to 16, which again increases the Toffoli count slightly but substantially reduces the workspace needed.
We also changed the reflections to be directly controlled by the control qubits used for phase estimation, instead of using the unary iteration circuit shown in \fig{controls} which needs extra ancillas.
Contrast the original allocation plan in \fig{qubittimelayout_original} with the more space efficient allocation plan in \fig{qubittimelayout_new}. In short, because we made a diagram showing allocated qubits over time, giving us a global view of our resource usage, we spotted a reasonably simple optimization that reduced the number of data qubits by 40\%.
This is why we recommend making these diagrams.

\subsubsection{Routing and distilling}

We now consider the routing of data during the phasing operations within the $\sel$ subroutine.
These phasing operations are implemented by adding qubits output from a QROM read into a phase gradient register, controlled by a target qubit.
This process creates an amount of phase kickback proportional to the amount read from QROM, and this phase kickback acts on the target qubit due to it being used as a control.
Because the phase gradient register is used for every single phasing operation, the phase gradient register should be moved out of the storage area and kept in the operating area.
The data being added into the phase gradient register then needs to be gradually streamed out of the storage area as the phasing operations are applied.
The majority of this data is the data that was produced by the QROM read.
In order to avoid traffic jams, it is important that data that will be needed at the same time be placed down separate hallways.
The QROM circuit construction gives us full control over where the target data lives, so this is not a difficult constraint to meet.

In the past, we have had trouble laying out quantum computations in a way that could consume incoming magic states quickly enough \cite{BabbushSpectraB}.
This is because the layout strategy we were using involved performing the probabilistic Clifford operations resulting from this process directly on the qubits being acted on by the magic states.
We are now using a different strategy, based on using AutoCCZ states which attach the corrections to the magic state instead of to the target qubits \cite{Fowler2012aB,Gidney2019aB,Litinski2018}.
When using AutoCCZs, the corrective operations can be packed into spacetime far away from the data qubits being operated on, allowing the computation to progress at a reaction-limited rate instead of a beat-limited rate.
This removes a significant constraint we were previously operating under.
Instead of struggling to keep pace with one magic state factory, we could now in principle keep pace with ten or more (if we were willing to pay the qubit cost of operating so many factories).

The magic state factory we will be using is the CCZ factory from \cite{Gidney2019aB}.
Note that results from \cite{Litinski2019B} suggest the factory can be reduced in size, due to the distillation process heralding topological errors.
We have not incorporated these results into our factory designs, but intend to do so in the future.

After exploring several cases using the estimation spreadsheet from the ancillary files of \cite{Gidney2018pub}, tweaked slightly to account for the improvements to the factory made in \cite{Gidney2019aB} and for the presence of multiple factories, we decided to use four factories with a level 1 code distance of 19 and a level 2 code distance of 31.
According to the ``logical\_factory\_dimensions" method from the ``estimate\_costs.py" script in the ancillary files of \cite{Gidney2019bB}, the large level 1 code distance results in the factory having a footprint of $15d \times 8d$ and a depth of $5d$ where $d=31$ is the level 2 code distance.
This choice gives a roughly 0.1\% chance of any distillation failure occurring when executing ten billion Toffolis, allocates nearly a million physical qubits for use as factories, and results in a Toffoli production rate of 25 kHz.
We give each factory a row of routing space to move the produced CCZ states to where they are needed, and two $3 \times 4$ AutoCCZ fixup areas for correcting the teleportation of the magic states being produced.

As shown in \fig{qubittimelayout_new}, the number of data qubits we store when running FeMoCo is less than 700.
The number of Toffoli operations is less than ten billion, and we perform Toffolis at a rate of 25 kHz.
According to the spreadsheet from \cite{Gidney2018pub}, a data code distance of 31 is sufficient for this regime while maintaining a 1\% total error budget.
We summarize this information in \fig{floorplan}, which shows one potential way to lay out the factories and the data qubits.
With this layout the FeMoCo computation would span four million physical qubits.

The total Toffoli count of the algorithm is approximately $483,\!000$ inner loops times $13,\!880$ Toffolis per inner loop, which equals $6.7$ billion Toffolis.
It would take 3 days to perform these Toffolis with four CCZ factories distilling at a total rate of 25 kHz.

\subsubsection{Estimate at optimistic error rates}

So far in this section, our estimates have been based on conservatively assuming a physical error rate of 0.1\% and a corresponding error suppression factor $\Lambda$ of 10 (meaning we assume the logical error rate per round goes down by a factor of 10 each time we increase the code distance by 2).
If we optimistically assume physical error rates of 0.01\%, and a corresponding $\Lambda$ of 100, all of the code distances can be cut in half while achieving the same reliability.

We again refer to the spreadsheet from Ref.~\cite{Gidney2018pub}.
We find that, under a total error budget of 1\% and a physical error rate of 0.01\%, ten billion Toffolis can be executed using a level 1 distillation code distance of 9, a level 2 distillation code distance of 15, and a data code distance of 15.
The resulting CCZ factory has a footprint of $14d \times 8d$, proportionally slightly better than the $15d \times 8d$ we had with a physical error rate of 0.1\%.
Therefore we can simply use the floorplan from \fig{floorplan} unchanged, but using $d=15$ instead of $d=31$.
Since the physical qubit count per logical qubit is $2(d+1)^2$, reducing the code distance from 31 to 15 reduces the physical qubit count by a factor of $32^2/16^2 = 4$.
The physical qubit count shrinks from four million to one million.

Because the computation is still beat limited at distance 15 when using four factories (as opposed to being reaction limited) the execution time at distance 15 is the execution time at distance 31 multiplied by $15/31$.
The execution time shrinks from 3.5 days to 1.5 days.

\section{Conclusions}
\label{sec:conclusion}

The primary contribution of this paper has been to introduce a new method for simulating electronic Hamiltonians in arbitrary basis on a quantum computer. In terms of both the asymptotic scaling and the finite resources required for complex benchmark molecules, our method appears to have lower cost for realization on a fault-tolerant quantum computer than any prior algorithm in the literature. We then compile these circuits and perform an analysis of error-correction overheads to arrive at the most accurate estimates yet of what would be required to realize an important and classically intractable quantum computation of chemistry within the surface code. 

The method we introduce uses a number of now standard techniques for quantum simulation including phase estimation of quantum walks, qubitization, QROM, coherent alias sampling and unary iteration. Our key innovation was to adapt this set of tools to the tensor hypercontraction representation of quantum chemistry, which had previously gone unexplored in the context of quantum computing. While the standard THC representation compresses the Hamiltonian considerably, enabling highly efficient quantum walks, using that representation directly leads to very large $\lambda$ values, requiring many repetitions of the quantum walk. To avoid this, we use tensors obtained from THC to transform the standard electronic Hamiltonian into a new representation which has a diagonal Coulomb operator but in a larger and non-orthogonal auxiliary basis. We then develop algorithms for realizing the associated qubitization oracles which work by rotating into the correct basis one tensor factor at a time, thus avoiding the need for a global non-orthogonal basis rotation, and allowing us to apply our algorithm without using any more system qubits than what is required by the original Hamiltonian.

The most efficient prior algorithms for simulating arbitrary basis quantum chemistry were the ``sparse'' method of Berry \emph{et al.}~\cite{Berry2019B} and the ``double low rank'' method of von Burg \emph{et al.}~\cite{vonBurg2020}. 
In both cases we correct errors in the 
original work (leading to reduced Toffoli complexity of the former approach). The sparse algorithm has Toffoli complexity $\widetilde{\cal O}((N + \sqrt{S})\lambda_V / \epsilon)$ and space complexity $\widetilde{\cal O}(N + \sqrt{S})$ where $S$ is the Hamiltonian sparsity. The double low rank algorithm has Toffoli complexity $\widetilde{\cal O}(N\lambda_{\rm DF} \sqrt{\Xi} / \epsilon)$ and space complexity $\widetilde{\cal O}(N\lambda_{\rm DF} \sqrt{\Xi})$ where $\Xi$ is the rank of the second tensor factorization discussed for quantum computing in \cite{Motta2018v2}. By contrast, the tensor hypercontraction approach introduced here has Toffoli complexity $\widetilde{\cal O}(N \lambda_\zeta / \epsilon)$ and space complexity $\widetilde{\cal O}(N)$. To contextualize the scaling of these $\lambda$ values as well as $S$ and $\Xi$ we analyze the asymptotic scaling of these methods applied to hydrogen systems growing towards both continuum and thermodynamic limits. 
Of the two prior algorithms, the double low rank method scales better towards the continuum limit with Toffoli complexity $\widetilde{\cal O}(N^{3.9}/\epsilon)$ and space complexity $\widetilde{\cal O}(N^{1.5})$. The THC algorithm has a favorable scaling with Toffoli complexity of $\widetilde{\cal O}(N^{3.1}/\epsilon)$ and space complexity of $\widetilde{\cal O}(N)$. When scaling towards the thermodynamic limit the sparse algorithm scales considerably better than the double low rank algorithm, having Toffoli complexity $\widetilde{\cal O}(N^{2.3} / \epsilon)$ and space complexity $\widetilde{\cal O}(N)$. Again, our THC algorithm still has even lower scaling with the same asymptotic space complexity but Toffoli complexity of $\widetilde{\cal O}(N^{2.1}/\epsilon)$. See \tab{hydrogen_scalings} for details.

Next, we compared the finite resources required to implement these methods for popular FeMoCo benchmarks. As can be seen in \tab{femoco_scalings}, the THC algorithm has both fewer logical qubits and less Toffoli complexity than any of the other competitive methods. We note that there might be more compact and accurate THC factors than those we found which could improve our results even further. Due to the difficulties associated with non-linear optimization, we expect that our solutions are suboptimal. In this sense, the estimates we report for the cost of the THC approach should be regarded as upper bounds on the cost of the most efficient possible implementations. We discuss a very detailed scheme for efficiently laying out the Reiher Hamiltonian FeMoCo computation within the surface code. Ultimately, we show that the computation could be realized with about four million physical qubits and under four days of runtime, assuming physical gate error rates of about $0.1\%$. With the more optimistic assumption of $0.01\%$ per gate error rates we could realize the same computation with about one million physical qubits and under two days of runtime.

It is interesting to note that, in our physical cost estimates, most qubits are used for routing and distillation.
Quantum circuits describing a quantum algorithm typically omit these qubits, which make up the majority of the cost.
For example, consider modifying a random access algorithm to use sequential access to save space by reducing routing overheads.
If avoiding random access forced a tradeoff versus the number of data qubits, the abstract circuit model would classify this improvement as a downgrade.
Because of this, we caution the reader that comparisons derived purely from quantum circuit metrics (such as counting Toffolis and data qubits) can be misleading.
They are good approximations, but these approximations can fail in known ways.

Despite the rapid progress detailed in \tab{big_table}, we expect that we are nearing the end of a series of asymptotic speedups for arbitrary basis quantum chemistry algorithms. At least for second quantized approaches based on linear combinations of unitaries, it would be difficult to imagine an algorithm that improves more than logarithmically on the $\widetilde{\cal O}(N \lambda / \epsilon)$ Toffoli complexity achieved by our approach (although it is possible that one could reduce $\lambda$). This is because it would seem that $\Omega(N)$ complexity should be required for any nontrivial quantum walk on $N$ qubits and phase estimation of LCU methods generically requires $\Omega(\lambda/\epsilon)$ repetitions. Indeed, we suspect that the near $\widetilde{\cal O}(N^2/ \epsilon)$ complexity obtained when scaling towards the thermodynamic limit for hydrogen chains would be the lowest possible scaling for simulation of the Coulomb operator as a consequence of its pairwise nature; also, this roughly matches the lowest scaling that has been achieved for simulating the Coulomb operator in special basis sets \cite{Childs2019BError}. However, perhaps there is room to improve over the $\widetilde{\cal O}(N^3/\epsilon)$ scaling we observe when scaling hydrogen systems towards their continuum limit. Still, there are several ways that we might hope to extend these methods and reduce constant factors. \insertrev{We emphasize that these asymptotic scalings should be further investigated in a more general set-up than hydrogenic systems in the future, though we expect that the same asymptotes will be observed in a much larger system.}

While the THC results presented in this work show remarkable improvements over previous qubitization approaches, the non-linear optimization associated in obtaining the THC factorization is challenging for broad applications. Future research and some ongoing effort in quantum chemistry may help to resolve this issue \cite{A.Matthews2020ImprovedHypercontraction,lee2019systematically}.
One natural question to ask is whether the THC representation (and in particular, our non-orthogonal THC Hamiltonian in \eq{larger_basis}) has utility for quantum computing beyond qubitization based methods; e.g., can one combine the THC representation with Trotter methods? Another question is whether one can leverage hierarchical matrix representations of the Coulomb operator \cite{xing2020linear} to compress the Hamiltonian in a way that is useful for qubitization either within the THC framework or within the sparse method. Another area to consider would be developing strategies of choosing active space orbitals with an aim towards reducing the value of $\lambda$ associated with qubitizing the resultant Hamiltonian. 

Similar to other prior papers on quantum computing for chemistry, the current work focuses on eigenstate preparation with the assumption of a sufficiently good initial guess state. The usual justification for this is that: (1) for most small molecules near their equilibrium geometry, simple initial states such as the Hartree-Fock state have reasonably good overlap with the ground state, (2) more complex state preparation procedures such as adiabatic state preparation may have a negligible additive cost to the cost of phase estimation which could produce a better overlapping initial state, and (3) as long as the overlap is not too small and especially when one has further knowledge of the gap there are methods for improving the ${\cal O}(1/a)$ scaling with the initial state overlap $a$ \cite{Berry2018,Lin2020b}. In \cite{McClean2014b}, authors argued that this state preparation cost should often increase exponentially as one grows systems towards a thermodynamic limit, and this might also be expected (at least in the worst case) from the QMA-Hardness of the electronic structure problem \cite{Verstraete2009}. However, in practice this asymptotic scaling may not matter because correlation lengths are finite and we tend to only need to perform very accurate correlated calculations on finite sized systems. Nevertheless, for very challenging systems such as FeMoCo, better methods of state preparation might be required. Future work should place more emphasis on analyzing and account for those costs.

Finally, we should continue to identify more concrete molecular benchmarks beyond the capabilities of classical electronic structure methods that could be solved on a quantum computer to provide  insights about chemistry. Noting that by combining qubitization with quantum signal processing \cite{Low2017} one can adapt our approach to perform highly efficient time evolutions rather than phase estimation, it is also worth exploring applications of quantum chemistry that would benefit from time evolution of the electronic Hamiltonian.

\subsection*{Data and code availability}

To maximize reproducibility, we share data and code used in this work on a public Zenodo repository \cite{zenodo}.
The repository includes all of the integral tensors (i.e., both exact and approximate) for systems studied in this work, python codes for generating the DF factors and computing $\lambda$ values as well as Mathematica notebooks for evaluating the cost of the SF, DF, sparse, and THC methods. Since THC factorization done in this work involves challenging non-linear optimization, it may be difficult for others to reproduce our numerical THC data (although it is easily verified). Thus, we recommend that interested readers use the available THC factors on \cite{zenodo} for future study.

\subsection*{Acknowledgements}

The authors thank Nicholas Rubin for helpful discussions, the authors of Ref.~\cite{vonBurg2020} and Ref.~\cite{Li2019} for sharing FeMoCo molecular integrals used in their work, and Michael Streif for pointing out a mistake in the one-body $\lambda$ value of hydrogen chains. DWB~worked on this project under a sponsored research agreement with Google Quantum AI.
DWB is also supported by Australian Research Council Discovery Projects DP190102633 and DP210101367.
NW was funded by a grant from Google Quantum AI, and his theoretical work on randomized simulation was  supported by the U.S. Department of Energy, Office of Science, National Quantum Information Science Research Centers, Co-Design Center for Quantum Advantage under contract number DE-SC0012704, support for NW's other theoretical contributions came from the Pacific Northwest National Laboratory LDRD program and the ``Embedding Quantum
Computing into Many-Body Frameworks for Strongly Correlated Molecular and Materials Systems'' project,
funded by the US Department of Energy (DOE). JL thanks David Reichman for encouragement and support.

\subsection*{Author contributions}

JL and RB proposed idea to combine tensor hypercontraction with qubitization. JL performed tensor hypercontraction analysis and all quantum chemistry numerics and proposed the non-orthogonal representation. DWB developed and analyzed quantum algorithm implementations with contribution from RB, including the qubitization technique for simulating the non-orthogonal representation. CG contributed the surface code layout analysis. NW wrote most of Appendix D. WJH, NW and JRM proposed and investigated a scheme for simulating the non-orthogonal representation by using quantum linear algebra tools, 
which was ultimately found to be less efficient than the one described. WJH and JRM also participated in many discussions and made useful suggestions. The paper was written by JL, DWB and RB with contributions from the others. Collaboration was organized and managed by RB.

\bibliography{other_references,Mendeley}

\begin{thebibliography}{105}%
\makeatletter
\providecommand \@ifxundefined [1]{%
 \@ifx{#1\undefined}
}%
\providecommand \@ifnum [1]{%
 \ifnum #1\expandafter \@firstoftwo
 \else \expandafter \@secondoftwo
 \fi
}%
\providecommand \@ifx [1]{%
 \ifx #1\expandafter \@firstoftwo
 \else \expandafter \@secondoftwo
 \fi
}%
\providecommand \natexlab [1]{#1}%
\providecommand \enquote  [1]{``#1''}%
\providecommand \bibnamefont  [1]{#1}%
\providecommand \bibfnamefont [1]{#1}%
\providecommand \citenamefont [1]{#1}%
\providecommand \href@noop [0]{\@secondoftwo}%
\providecommand \href [0]{\begingroup \@sanitize@url \@href}%
\providecommand \@href[1]{\@@startlink{#1}\@@href}%
\providecommand \@@href[1]{\endgroup#1\@@endlink}%
\providecommand \@sanitize@url [0]{\catcode `\\12\catcode `\$12\catcode
  `\&12\catcode `\#12\catcode `\^12\catcode `\_12\catcode `\%12\relax}%
\providecommand \@@startlink[1]{}%
\providecommand \@@endlink[0]{}%
\providecommand \url  [0]{\begingroup\@sanitize@url \@url }%
\providecommand \@url [1]{\endgroup\@href {#1}{\urlprefix }}%
\providecommand \urlprefix  [0]{URL }%
\providecommand \Eprint [0]{\href }%
\providecommand \doibase [0]{http://dx.doi.org/}%
\providecommand \selectlanguage [0]{\@gobble}%
\providecommand \bibinfo  [0]{\@secondoftwo}%
\providecommand \bibfield  [0]{\@secondoftwo}%
\providecommand \translation [1]{[#1]}%
\providecommand \BibitemOpen [0]{}%
\providecommand \bibitemStop [0]{}%
\providecommand \bibitemNoStop [0]{.\EOS\space}%
\providecommand \EOS [0]{\spacefactor3000\relax}%
\providecommand \BibitemShut  [1]{\csname bibitem#1\endcsname}%
\let\auto@bib@innerbib\@empty
\bibitem [{\citenamefont {Bauer}\ \emph {et~al.}(2020)\citenamefont {Bauer},
  \citenamefont {Bravyi}, \citenamefont {Motta},\ and\ \citenamefont
  {Kin-Lic~Chan}}]{Bauer2020QuantumScience}%
  \BibitemOpen
  \bibfield  {author} {\bibinfo {author} {\bibfnamefont {B.}~\bibnamefont
  {Bauer}}, \bibinfo {author} {\bibfnamefont {S.}~\bibnamefont {Bravyi}},
  \bibinfo {author} {\bibfnamefont {M.}~\bibnamefont {Motta}}, \ and\ \bibinfo
  {author} {\bibfnamefont {G.}~\bibnamefont {Kin-Lic~Chan}},\ }\href
  {https://arxiv.org/abs/2001.03685} {\bibfield  {journal} {\bibinfo  {journal}
  {arXiv:2001.03685}\ } (\bibinfo {year} {2020})}\BibitemShut {NoStop}%
\bibitem [{\citenamefont {Cao}\ \emph {et~al.}(2019)\citenamefont {Cao},
  \citenamefont {Romero}, \citenamefont {P.~Olson}, \citenamefont {Degroote},
  \citenamefont {D.~Johnson}, \citenamefont {Kieferov{\'{a}}}, \citenamefont
  {D.~Kivlichan}, \citenamefont {Menke}, \citenamefont {Peropadre},
  \citenamefont {P.~D.~Sawaya}, \citenamefont {Sim}, \citenamefont {Veis},\
  and\ \citenamefont {Aspuru-Guzik}}]{Cao2019QuantumComputing}%
  \BibitemOpen
  \bibfield  {author} {\bibinfo {author} {\bibfnamefont {Y.}~\bibnamefont
  {Cao}}, \bibinfo {author} {\bibfnamefont {J.}~\bibnamefont {Romero}},
  \bibinfo {author} {\bibfnamefont {J.}~\bibnamefont {P.~Olson}}, \bibinfo
  {author} {\bibfnamefont {M.}~\bibnamefont {Degroote}}, \bibinfo {author}
  {\bibfnamefont {P.}~\bibnamefont {D.~Johnson}}, \bibinfo {author}
  {\bibfnamefont {M.}~\bibnamefont {Kieferov{\'{a}}}}, \bibinfo {author}
  {\bibfnamefont {I.}~\bibnamefont {D.~Kivlichan}}, \bibinfo {author}
  {\bibfnamefont {T.}~\bibnamefont {Menke}}, \bibinfo {author} {\bibfnamefont
  {B.}~\bibnamefont {Peropadre}}, \bibinfo {author} {\bibfnamefont
  {N.}~\bibnamefont {P.~D.~Sawaya}}, \bibinfo {author} {\bibfnamefont
  {S.}~\bibnamefont {Sim}}, \bibinfo {author} {\bibfnamefont {L.}~\bibnamefont
  {Veis}}, \ and\ \bibinfo {author} {\bibfnamefont {A.}~\bibnamefont
  {Aspuru-Guzik}},\ }\href {\doibase 10.1021/acs.chemrev.8b00803} {\bibfield
  {journal} {\bibinfo  {journal} {Chemical Reviews}\ }\textbf {\bibinfo
  {volume} {119}},\ \bibinfo {pages} {10856} (\bibinfo {year}
  {2019})}\BibitemShut {NoStop}%
\bibitem [{\citenamefont {McArdle}\ \emph {et~al.}(2020)\citenamefont
  {McArdle}, \citenamefont {Endo}, \citenamefont {Aspuru-Guzik}, \citenamefont
  {Benjamin},\ and\ \citenamefont {Yuan}}]{McArdle2020QuantumChemistry}%
  \BibitemOpen
  \bibfield  {author} {\bibinfo {author} {\bibfnamefont {S.}~\bibnamefont
  {McArdle}}, \bibinfo {author} {\bibfnamefont {S.}~\bibnamefont {Endo}},
  \bibinfo {author} {\bibfnamefont {A.}~\bibnamefont {Aspuru-Guzik}}, \bibinfo
  {author} {\bibfnamefont {S.~C.}\ \bibnamefont {Benjamin}}, \ and\ \bibinfo
  {author} {\bibfnamefont {X.}~\bibnamefont {Yuan}},\ }\href {\doibase
  10.1103/RevModPhys.92.015003} {\bibfield  {journal} {\bibinfo  {journal}
  {Reviews of Modern Physics}\ }\textbf {\bibinfo {volume} {92}},\ \bibinfo
  {pages} {015003} (\bibinfo {year} {2020})}\BibitemShut {NoStop}%
\bibitem [{\citenamefont {Aspuru-Guzik}\ \emph {et~al.}(2005)\citenamefont
  {Aspuru-Guzik}, \citenamefont {Dutoi}, \citenamefont {Love},\ and\
  \citenamefont {Head-Gordon}}]{Aspuru-Guzik2005}%
  \BibitemOpen
  \bibfield  {author} {\bibinfo {author} {\bibfnamefont {A.}~\bibnamefont
  {Aspuru-Guzik}}, \bibinfo {author} {\bibfnamefont {A.~D.}\ \bibnamefont
  {Dutoi}}, \bibinfo {author} {\bibfnamefont {P.~J.}\ \bibnamefont {Love}}, \
  and\ \bibinfo {author} {\bibfnamefont {M.}~\bibnamefont {Head-Gordon}},\
  }\href {\doibase 10.1126/science.1113479} {\bibfield  {journal} {\bibinfo
  {journal} {Science}\ }\textbf {\bibinfo {volume} {309}},\ \bibinfo {pages}
  {1704} (\bibinfo {year} {2005})}\BibitemShut {NoStop}%
\bibitem [{\citenamefont {Mardirossian}\ and\ \citenamefont
  {Head-Gordon}(2017)}]{mardirossian2017thirty}%
  \BibitemOpen
  \bibfield  {author} {\bibinfo {author} {\bibfnamefont {N.}~\bibnamefont
  {Mardirossian}}\ and\ \bibinfo {author} {\bibfnamefont {M.}~\bibnamefont
  {Head-Gordon}},\ }\href {\doibase 10.1080/00268976.2017.1333644} {\bibfield
  {journal} {\bibinfo  {journal} {Molecular Physics}\ }\textbf {\bibinfo
  {volume} {115}},\ \bibinfo {pages} {2315} (\bibinfo {year}
  {2017})}\BibitemShut {NoStop}%
\bibitem [{\citenamefont {Ruedenberg}\ \emph {et~al.}(1973)\citenamefont
  {Ruedenberg}, \citenamefont {Raffenetti},\ and\ \citenamefont
  {Bardo}}]{ruedenberg1973energy}%
  \BibitemOpen
  \bibfield  {author} {\bibinfo {author} {\bibfnamefont {K.}~\bibnamefont
  {Ruedenberg}}, \bibinfo {author} {\bibfnamefont {R.}~\bibnamefont
  {Raffenetti}}, \ and\ \bibinfo {author} {\bibfnamefont {R.}~\bibnamefont
  {Bardo}},\ }in\ \href@noop {} {\emph {\bibinfo {booktitle} {Proceedings of
  the 1972 Boulder Seminar Research Conference on Theoretical Chemistry}}}\
  (\bibinfo {organization} {Wiley, New York},\ \bibinfo {year} {1973})\ p.\
  \bibinfo {pages} {164}\BibitemShut {NoStop}%
\bibitem [{\citenamefont {Kitaev}(1995)}]{Kitaev1995B}%
  \BibitemOpen
  \bibfield  {author} {\bibinfo {author} {\bibfnamefont {A.~Y.}\ \bibnamefont
  {Kitaev}},\ }\href {https://arxiv.org/abs/quant-ph/9511026} {\bibfield
  {journal} {\bibinfo  {journal} {arXiv:quant-ph/9511026}\ } (\bibinfo {year}
  {1995})}\BibitemShut {NoStop}%
\bibitem [{\citenamefont {Abrams}\ and\ \citenamefont
  {Lloyd}(1999)}]{Abrams1999}%
  \BibitemOpen
  \bibfield  {author} {\bibinfo {author} {\bibfnamefont {D.~S.}\ \bibnamefont
  {Abrams}}\ and\ \bibinfo {author} {\bibfnamefont {S.}~\bibnamefont {Lloyd}},\
  }\href {\doibase 10.1103/PhysRevLett.83.5162} {\bibfield  {journal} {\bibinfo
   {journal} {Physical Review Letters}\ }\textbf {\bibinfo {volume} {83}},\
  \bibinfo {pages} {5162} (\bibinfo {year} {1999})}\BibitemShut {NoStop}%
\bibitem [{\citenamefont {Berry}\ \emph {et~al.}(2019)\citenamefont {Berry},
  \citenamefont {Gidney}, \citenamefont {Motta}, \citenamefont {McClean},\ and\
  \citenamefont {Babbush}}]{Berry2019B}%
  \BibitemOpen
  \bibfield  {author} {\bibinfo {author} {\bibfnamefont {D.~W.}\ \bibnamefont
  {Berry}}, \bibinfo {author} {\bibfnamefont {C.}~\bibnamefont {Gidney}},
  \bibinfo {author} {\bibfnamefont {M.}~\bibnamefont {Motta}}, \bibinfo
  {author} {\bibfnamefont {J.}~\bibnamefont {McClean}}, \ and\ \bibinfo
  {author} {\bibfnamefont {R.}~\bibnamefont {Babbush}},\ }\href
  {https://quantum-journal.org/papers/q-2019-12-02-208/} {\bibfield  {journal}
  {\bibinfo  {journal} {Quantum}\ }\textbf {\bibinfo {volume} {3}},\ \bibinfo
  {pages} {208} (\bibinfo {year} {2019})}\BibitemShut {NoStop}%
\bibitem [{\citenamefont {von Burg}\ \emph {et~al.}(2020)\citenamefont {von
  Burg}, \citenamefont {Low}, \citenamefont {Haner}, \citenamefont {Steiger},
  \citenamefont {Reiher}, \citenamefont {Roetteler},\ and\ \citenamefont
  {Troyer}}]{vonBurg2020}%
  \BibitemOpen
  \bibfield  {author} {\bibinfo {author} {\bibfnamefont {V.}~\bibnamefont {von
  Burg}}, \bibinfo {author} {\bibfnamefont {G.~H.}\ \bibnamefont {Low}},
  \bibinfo {author} {\bibfnamefont {T.}~\bibnamefont {Haner}}, \bibinfo
  {author} {\bibfnamefont {D.}~\bibnamefont {Steiger}}, \bibinfo {author}
  {\bibfnamefont {M.}~\bibnamefont {Reiher}}, \bibinfo {author} {\bibfnamefont
  {M.}~\bibnamefont {Roetteler}}, \ and\ \bibinfo {author} {\bibfnamefont
  {M.}~\bibnamefont {Troyer}},\ }\href {https://arxiv.org/abs/2007.14460}
  {\bibfield  {journal} {\bibinfo  {journal} {arXiv:2007.14460}\ } (\bibinfo
  {year} {2020})}\BibitemShut {NoStop}%
\bibitem [{\citenamefont {Whitfield}\ \emph {et~al.}(2011)\citenamefont
  {Whitfield}, \citenamefont {Biamonte},\ and\ \citenamefont
  {Aspuru-Guzik}}]{Whitfield2010}%
  \BibitemOpen
  \bibfield  {author} {\bibinfo {author} {\bibfnamefont {J.~D.}\ \bibnamefont
  {Whitfield}}, \bibinfo {author} {\bibfnamefont {J.}~\bibnamefont {Biamonte}},
  \ and\ \bibinfo {author} {\bibfnamefont {A.}~\bibnamefont {Aspuru-Guzik}},\
  }\href {\doibase 10.1080/00268976.2011.552441} {\bibfield  {journal}
  {\bibinfo  {journal} {Molecular Physics}\ }\textbf {\bibinfo {volume}
  {109}},\ \bibinfo {pages} {735} (\bibinfo {year} {2011})}\BibitemShut
  {NoStop}%
\bibitem [{\citenamefont {Low}\ and\ \citenamefont {Chuang}(2019)}]{Low2016}%
  \BibitemOpen
  \bibfield  {author} {\bibinfo {author} {\bibfnamefont {G.~H.}\ \bibnamefont
  {Low}}\ and\ \bibinfo {author} {\bibfnamefont {I.~L.}\ \bibnamefont
  {Chuang}},\ }\href {https://doi.org/10.22331/q-2019-07-12-163} {\bibfield
  {journal} {\bibinfo  {journal} {Quantum}\ }\textbf {\bibinfo {volume} {3}},\
  \bibinfo {pages} {163} (\bibinfo {year} {2019})}\BibitemShut {NoStop}%
\bibitem [{\citenamefont {Poulin}\ \emph {et~al.}(2017)\citenamefont {Poulin},
  \citenamefont {Kitaev}, \citenamefont {Steiger}, \citenamefont {Hastings},\
  and\ \citenamefont {Troyer}}]{Poulin2017}%
  \BibitemOpen
  \bibfield  {author} {\bibinfo {author} {\bibfnamefont {D.}~\bibnamefont
  {Poulin}}, \bibinfo {author} {\bibfnamefont {A.~Y.}\ \bibnamefont {Kitaev}},
  \bibinfo {author} {\bibfnamefont {D.}~\bibnamefont {Steiger}}, \bibinfo
  {author} {\bibfnamefont {M.}~\bibnamefont {Hastings}}, \ and\ \bibinfo
  {author} {\bibfnamefont {M.}~\bibnamefont {Troyer}},\ }\href
  {https://journals.aps.org/prl/abstract/10.1103/PhysRevLett.121.010501}
  {\bibfield  {journal} {\bibinfo  {journal} {Physical Review Letters}\
  }\textbf {\bibinfo {volume} {121}},\ \bibinfo {pages} {010501} (\bibinfo
  {year} {2017})}\BibitemShut {NoStop}%
\bibitem [{\citenamefont {Berry}\ \emph {et~al.}(2018)\citenamefont {Berry},
  \citenamefont {Kieferov{\'{a}}}, \citenamefont {Scherer}, \citenamefont
  {Sanders}, \citenamefont {Low}, \citenamefont {Wiebe}, \citenamefont
  {Gidney},\ and\ \citenamefont {Babbush}}]{Berry2018}%
  \BibitemOpen
  \bibfield  {author} {\bibinfo {author} {\bibfnamefont {D.~W.}\ \bibnamefont
  {Berry}}, \bibinfo {author} {\bibfnamefont {M.}~\bibnamefont
  {Kieferov{\'{a}}}}, \bibinfo {author} {\bibfnamefont {A.}~\bibnamefont
  {Scherer}}, \bibinfo {author} {\bibfnamefont {Y.~R.}\ \bibnamefont
  {Sanders}}, \bibinfo {author} {\bibfnamefont {G.~H.}\ \bibnamefont {Low}},
  \bibinfo {author} {\bibfnamefont {N.}~\bibnamefont {Wiebe}}, \bibinfo
  {author} {\bibfnamefont {C.}~\bibnamefont {Gidney}}, \ and\ \bibinfo {author}
  {\bibfnamefont {R.}~\bibnamefont {Babbush}},\ }\href {\doibase
  10.1038/s41534-018-0071-5} {\bibfield  {journal} {\bibinfo  {journal} {npj
  Quantum Information}\ }\textbf {\bibinfo {volume} {4}},\ \bibinfo {pages}
  {22} (\bibinfo {year} {2018})}\BibitemShut {NoStop}%
\bibitem [{\citenamefont {Babbush}\ \emph
  {et~al.}(2018{\natexlab{a}})\citenamefont {Babbush}, \citenamefont {Wiebe},
  \citenamefont {McClean}, \citenamefont {McClain}, \citenamefont {Neven},\
  and\ \citenamefont {Chan}}]{BabbushLow}%
  \BibitemOpen
  \bibfield  {author} {\bibinfo {author} {\bibfnamefont {R.}~\bibnamefont
  {Babbush}}, \bibinfo {author} {\bibfnamefont {N.}~\bibnamefont {Wiebe}},
  \bibinfo {author} {\bibfnamefont {J.}~\bibnamefont {McClean}}, \bibinfo
  {author} {\bibfnamefont {J.}~\bibnamefont {McClain}}, \bibinfo {author}
  {\bibfnamefont {H.}~\bibnamefont {Neven}}, \ and\ \bibinfo {author}
  {\bibfnamefont {G.~K.-L.}\ \bibnamefont {Chan}},\ }\href
  {https://journals.aps.org/prx/abstract/10.1103/PhysRevX.8.011044} {\bibfield
  {journal} {\bibinfo  {journal} {Physical Review X}\ }\textbf {\bibinfo
  {volume} {8}},\ \bibinfo {pages} {011044} (\bibinfo {year}
  {2018}{\natexlab{a}})}\BibitemShut {NoStop}%
\bibitem [{\citenamefont {Babbush}\ \emph
  {et~al.}(2018{\natexlab{b}})\citenamefont {Babbush}, \citenamefont {Gidney},
  \citenamefont {Berry}, \citenamefont {Wiebe}, \citenamefont {McClean},
  \citenamefont {Paler}, \citenamefont {Fowler},\ and\ \citenamefont
  {Neven}}]{BabbushSpectraB}%
  \BibitemOpen
  \bibfield  {author} {\bibinfo {author} {\bibfnamefont {R.}~\bibnamefont
  {Babbush}}, \bibinfo {author} {\bibfnamefont {C.}~\bibnamefont {Gidney}},
  \bibinfo {author} {\bibfnamefont {D.~W.}\ \bibnamefont {Berry}}, \bibinfo
  {author} {\bibfnamefont {N.}~\bibnamefont {Wiebe}}, \bibinfo {author}
  {\bibfnamefont {J.}~\bibnamefont {McClean}}, \bibinfo {author} {\bibfnamefont
  {A.}~\bibnamefont {Paler}}, \bibinfo {author} {\bibfnamefont
  {A.}~\bibnamefont {Fowler}}, \ and\ \bibinfo {author} {\bibfnamefont
  {H.}~\bibnamefont {Neven}},\ }\href
  {https://journals.aps.org/prx/abstract/10.1103/PhysRevX.8.041015} {\bibfield
  {journal} {\bibinfo  {journal} {Physical Review X}\ }\textbf {\bibinfo
  {volume} {8}},\ \bibinfo {pages} {041015} (\bibinfo {year}
  {2018}{\natexlab{b}})}\BibitemShut {NoStop}%
\bibitem [{\citenamefont {Low}\ and\ \citenamefont {Wiebe}(2018)}]{Low2018}%
  \BibitemOpen
  \bibfield  {author} {\bibinfo {author} {\bibfnamefont {G.~H.}\ \bibnamefont
  {Low}}\ and\ \bibinfo {author} {\bibfnamefont {N.}~\bibnamefont {Wiebe}},\
  }\href {http://arxiv.org/abs/1805.00675} {\bibfield  {journal} {\bibinfo
  {journal} {arXiv:1805.00675}\ } (\bibinfo {year} {2018})}\BibitemShut
  {NoStop}%
\bibitem [{\citenamefont {Kivlichan}\ \emph {et~al.}(2020)\citenamefont
  {Kivlichan}, \citenamefont {Gidney}, \citenamefont {Berry}, \citenamefont
  {Wiebe}, \citenamefont {McClean}, \citenamefont {Sun}, \citenamefont {Jiang},
  \citenamefont {Rubin}, \citenamefont {Fowler}, \citenamefont {Aspuru-Guzik},
  \citenamefont {Neven},\ and\ \citenamefont {Babbush}}]{Kivlichan2019}%
  \BibitemOpen
  \bibfield  {author} {\bibinfo {author} {\bibfnamefont {I.~D.}\ \bibnamefont
  {Kivlichan}}, \bibinfo {author} {\bibfnamefont {C.}~\bibnamefont {Gidney}},
  \bibinfo {author} {\bibfnamefont {D.~W.}\ \bibnamefont {Berry}}, \bibinfo
  {author} {\bibfnamefont {N.}~\bibnamefont {Wiebe}}, \bibinfo {author}
  {\bibfnamefont {J.}~\bibnamefont {McClean}}, \bibinfo {author} {\bibfnamefont
  {W.}~\bibnamefont {Sun}}, \bibinfo {author} {\bibfnamefont {Z.}~\bibnamefont
  {Jiang}}, \bibinfo {author} {\bibfnamefont {N.}~\bibnamefont {Rubin}},
  \bibinfo {author} {\bibfnamefont {A.}~\bibnamefont {Fowler}}, \bibinfo
  {author} {\bibfnamefont {A.}~\bibnamefont {Aspuru-Guzik}}, \bibinfo {author}
  {\bibfnamefont {H.}~\bibnamefont {Neven}}, \ and\ \bibinfo {author}
  {\bibfnamefont {R.}~\bibnamefont {Babbush}},\ }\href
  {https://quantum-journal.org/papers/q-2020-07-16-296/} {\bibfield  {journal}
  {\bibinfo  {journal} {Quantum}\ }\textbf {\bibinfo {volume} {4}},\ \bibinfo
  {pages} {296} (\bibinfo {year} {2020})}\BibitemShut {NoStop}%
\bibitem [{\citenamefont {McClean}\ \emph {et~al.}(2020)\citenamefont
  {McClean}, \citenamefont {Faulstich}, \citenamefont {Zhu}, \citenamefont
  {O'Gorman}, \citenamefont {Qiu}, \citenamefont {White}, \citenamefont
  {Babbush},\ and\ \citenamefont {Lin}}]{McClean2020}%
  \BibitemOpen
  \bibfield  {author} {\bibinfo {author} {\bibfnamefont {J.~R.}\ \bibnamefont
  {McClean}}, \bibinfo {author} {\bibfnamefont {F.~M.}\ \bibnamefont
  {Faulstich}}, \bibinfo {author} {\bibfnamefont {Q.}~\bibnamefont {Zhu}},
  \bibinfo {author} {\bibfnamefont {B.}~\bibnamefont {O'Gorman}}, \bibinfo
  {author} {\bibfnamefont {Y.}~\bibnamefont {Qiu}}, \bibinfo {author}
  {\bibfnamefont {S.~R.}\ \bibnamefont {White}}, \bibinfo {author}
  {\bibfnamefont {R.}~\bibnamefont {Babbush}}, \ and\ \bibinfo {author}
  {\bibfnamefont {L.}~\bibnamefont {Lin}},\ }\href {\doibase
  10.1088/1367-2630/ab9d9f} {\bibfield  {journal} {\bibinfo  {journal} {New
  Journal of Physics}\ }\textbf {\bibinfo {volume} {22}},\ \bibinfo {pages}
  {093015} (\bibinfo {year} {2020})}\BibitemShut {NoStop}%
\bibitem [{\citenamefont {White}(2017)}]{white2017hybrid}%
  \BibitemOpen
  \bibfield  {author} {\bibinfo {author} {\bibfnamefont {S.~R.}\ \bibnamefont
  {White}},\ }\href {\doibase 10.1063/1.5007066} {\bibfield  {journal}
  {\bibinfo  {journal} {The Journal of Chemical Physics}\ }\textbf {\bibinfo
  {volume} {147}},\ \bibinfo {pages} {244102} (\bibinfo {year}
  {2017})}\BibitemShut {NoStop}%
\bibitem [{\citenamefont {Kassal}\ \emph {et~al.}(2008)\citenamefont {Kassal},
  \citenamefont {Jordan}, \citenamefont {Love}, \citenamefont {Mohseni},\ and\
  \citenamefont {Aspuru-Guzik}}]{Kassal2008}%
  \BibitemOpen
  \bibfield  {author} {\bibinfo {author} {\bibfnamefont {I.}~\bibnamefont
  {Kassal}}, \bibinfo {author} {\bibfnamefont {S.~P.}\ \bibnamefont {Jordan}},
  \bibinfo {author} {\bibfnamefont {P.~J.}\ \bibnamefont {Love}}, \bibinfo
  {author} {\bibfnamefont {M.}~\bibnamefont {Mohseni}}, \ and\ \bibinfo
  {author} {\bibfnamefont {A.}~\bibnamefont {Aspuru-Guzik}},\ }\href
  {http://www.pnas.org/content/105/48/18681.abstract} {\bibfield  {journal}
  {\bibinfo  {journal} {Proceedings of the National Academy of Sciences}\
  }\textbf {\bibinfo {volume} {105}},\ \bibinfo {pages} {18681} (\bibinfo
  {year} {2008})}\BibitemShut {NoStop}%
\bibitem [{\citenamefont {Babbush}\ \emph
  {et~al.}(2019{\natexlab{a}})\citenamefont {Babbush}, \citenamefont {Berry},
  \citenamefont {McClean},\ and\ \citenamefont {Neven}}]{BabbushContinuum}%
  \BibitemOpen
  \bibfield  {author} {\bibinfo {author} {\bibfnamefont {R.}~\bibnamefont
  {Babbush}}, \bibinfo {author} {\bibfnamefont {D.~W.}\ \bibnamefont {Berry}},
  \bibinfo {author} {\bibfnamefont {J.~R.}\ \bibnamefont {McClean}}, \ and\
  \bibinfo {author} {\bibfnamefont {H.}~\bibnamefont {Neven}},\ }\href
  {https://www.nature.com/articles/s41534-019-0199-y} {\bibfield  {journal}
  {\bibinfo  {journal} {npj Quantum Information}\ }\textbf {\bibinfo {volume}
  {5}},\ \bibinfo {pages} {92} (\bibinfo {year}
  {2019}{\natexlab{a}})}\BibitemShut {NoStop}%
\bibitem [{\citenamefont {Reiher}\ \emph {et~al.}(2017)\citenamefont {Reiher},
  \citenamefont {Wiebe}, \citenamefont {Svore}, \citenamefont {Wecker},\ and\
  \citenamefont {Troyer}}]{Reiher2017}%
  \BibitemOpen
  \bibfield  {author} {\bibinfo {author} {\bibfnamefont {M.}~\bibnamefont
  {Reiher}}, \bibinfo {author} {\bibfnamefont {N.}~\bibnamefont {Wiebe}},
  \bibinfo {author} {\bibfnamefont {K.~M.}\ \bibnamefont {Svore}}, \bibinfo
  {author} {\bibfnamefont {D.}~\bibnamefont {Wecker}}, \ and\ \bibinfo {author}
  {\bibfnamefont {M.}~\bibnamefont {Troyer}},\ }\href
  {http://www.pnas.org/content/114/29/7555.abstract} {\bibfield  {journal}
  {\bibinfo  {journal} {Proceedings of the National Academy of Sciences}\
  }\textbf {\bibinfo {volume} {114}},\ \bibinfo {pages} {7555} (\bibinfo {year}
  {2017})}\BibitemShut {NoStop}%
\bibitem [{\citenamefont {Beinert}\ \emph {et~al.}(1997)\citenamefont
  {Beinert}, \citenamefont {Holm},\ and\ \citenamefont {Munck}}]{Beinert1997}%
  \BibitemOpen
  \bibfield  {author} {\bibinfo {author} {\bibfnamefont {H.}~\bibnamefont
  {Beinert}}, \bibinfo {author} {\bibfnamefont {R.}~\bibnamefont {Holm}}, \
  and\ \bibinfo {author} {\bibfnamefont {E.}~\bibnamefont {Munck}},\ }\href
  {\doibase http://science.sciencemag.org/content/277/5326/653} {\bibfield
  {journal} {\bibinfo  {journal} {Science}\ }\textbf {\bibinfo {volume}
  {277}},\ \bibinfo {pages} {653} (\bibinfo {year} {1997})}\BibitemShut
  {NoStop}%
\bibitem [{\citenamefont {McClean}\ \emph
  {et~al.}(2014{\natexlab{a}})\citenamefont {McClean}, \citenamefont {Babbush},
  \citenamefont {Love},\ and\ \citenamefont {Aspuru-Guzik}}]{McClean2014}%
  \BibitemOpen
  \bibfield  {author} {\bibinfo {author} {\bibfnamefont {J.~R.}\ \bibnamefont
  {McClean}}, \bibinfo {author} {\bibfnamefont {R.}~\bibnamefont {Babbush}},
  \bibinfo {author} {\bibfnamefont {P.~J.}\ \bibnamefont {Love}}, \ and\
  \bibinfo {author} {\bibfnamefont {A.}~\bibnamefont {Aspuru-Guzik}},\ }\href
  {\doibase 10.1021/jz501649m} {\bibfield  {journal} {\bibinfo  {journal} {The
  Journal of Physical Chemistry Letters}\ }\textbf {\bibinfo {volume} {5}},\
  \bibinfo {pages} {4368} (\bibinfo {year} {2014}{\natexlab{a}})}\BibitemShut
  {NoStop}%
\bibitem [{\citenamefont {Fowler}\ \emph {et~al.}(2012)\citenamefont {Fowler},
  \citenamefont {Mariantoni}, \citenamefont {Martinis},\ and\ \citenamefont
  {Cleland}}]{Fowler2012}%
  \BibitemOpen
  \bibfield  {author} {\bibinfo {author} {\bibfnamefont {A.~G.}\ \bibnamefont
  {Fowler}}, \bibinfo {author} {\bibfnamefont {M.}~\bibnamefont {Mariantoni}},
  \bibinfo {author} {\bibfnamefont {J.~M.}\ \bibnamefont {Martinis}}, \ and\
  \bibinfo {author} {\bibfnamefont {A.~N.}\ \bibnamefont {Cleland}},\ }\href
  {\doibase 10.1103/PhysRevA.86.032324} {\bibfield  {journal} {\bibinfo
  {journal} {Physical Review A}\ }\textbf {\bibinfo {volume} {86}},\ \bibinfo
  {pages} {32324} (\bibinfo {year} {2012})}\BibitemShut {NoStop}%
\bibitem [{\citenamefont {Litinski}(2019{\natexlab{a}})}]{Litinski2018}%
  \BibitemOpen
  \bibfield  {author} {\bibinfo {author} {\bibfnamefont {D.}~\bibnamefont
  {Litinski}},\ }\href {https://quantum-journal.org/papers/q-2019-03-05-128/}
  {\bibfield  {journal} {\bibinfo  {journal} {Quantum}\ }\textbf {\bibinfo
  {volume} {3}},\ \bibinfo {pages} {128} (\bibinfo {year}
  {2019}{\natexlab{a}})}\BibitemShut {NoStop}%
\bibitem [{\citenamefont {Fowler}\ and\ \citenamefont
  {Gidney}(2018)}]{Fowler2018}%
  \BibitemOpen
  \bibfield  {author} {\bibinfo {author} {\bibfnamefont {A.~G.}\ \bibnamefont
  {Fowler}}\ and\ \bibinfo {author} {\bibfnamefont {C.}~\bibnamefont
  {Gidney}},\ }\href {http://arxiv.org/abs/1808.06709} {\bibfield  {journal}
  {\bibinfo  {journal} {arXiv:1808.06709}\ } (\bibinfo {year}
  {2018})}\BibitemShut {NoStop}%
\bibitem [{\citenamefont {Motta}\ \emph {et~al.}(2021)\citenamefont {Motta},
  \citenamefont {Ye}, \citenamefont {McClean}, \citenamefont {Li},
  \citenamefont {Minnich}, \citenamefont {Babbush},\ and\ \citenamefont
  {Chan}}]{Motta2018v2}%
  \BibitemOpen
  \bibfield  {author} {\bibinfo {author} {\bibfnamefont {M.}~\bibnamefont
  {Motta}}, \bibinfo {author} {\bibfnamefont {E.}~\bibnamefont {Ye}}, \bibinfo
  {author} {\bibfnamefont {J.~R.}\ \bibnamefont {McClean}}, \bibinfo {author}
  {\bibfnamefont {Z.}~\bibnamefont {Li}}, \bibinfo {author} {\bibfnamefont
  {A.~J.}\ \bibnamefont {Minnich}}, \bibinfo {author} {\bibfnamefont
  {R.}~\bibnamefont {Babbush}}, \ and\ \bibinfo {author} {\bibfnamefont
  {G.~K.-L.}\ \bibnamefont {Chan}},\ }\href {\doibase
  10.1038/s41534-021-00416-z} {\bibfield  {journal} {\bibinfo  {journal} {npj
  Quantum Inf.}\ }\textbf {\bibinfo {volume} {7}},\ \bibinfo {pages} {1}
  (\bibinfo {year} {2021})}\BibitemShut {NoStop}%
\bibitem [{\citenamefont {Aquilante}\ \emph {et~al.}(2011)\citenamefont
  {Aquilante}, \citenamefont {Boman}, \citenamefont {Bostr{\"o}m},
  \citenamefont {Koch}, \citenamefont {Lindh}, \citenamefont {de~Mer{\'a}s},\
  and\ \citenamefont {Pedersen}}]{aquilante2011cholesky}%
  \BibitemOpen
  \bibfield  {author} {\bibinfo {author} {\bibfnamefont {F.}~\bibnamefont
  {Aquilante}}, \bibinfo {author} {\bibfnamefont {L.}~\bibnamefont {Boman}},
  \bibinfo {author} {\bibfnamefont {J.}~\bibnamefont {Bostr{\"o}m}}, \bibinfo
  {author} {\bibfnamefont {H.}~\bibnamefont {Koch}}, \bibinfo {author}
  {\bibfnamefont {R.}~\bibnamefont {Lindh}}, \bibinfo {author} {\bibfnamefont
  {A.~S.}\ \bibnamefont {de~Mer{\'a}s}}, \ and\ \bibinfo {author}
  {\bibfnamefont {T.~B.}\ \bibnamefont {Pedersen}},\ }in\ \href@noop {} {\emph
  {\bibinfo {booktitle} {Linear-Scaling Techniques in Computational Chemistry
  and Physics}}}\ (\bibinfo  {publisher} {Springer},\ \bibinfo {year} {2011})\
  pp.\ \bibinfo {pages} {301--343}\BibitemShut {NoStop}%
\bibitem [{\citenamefont {Whitten}(1973)}]{Whitten1973}%
  \BibitemOpen
  \bibfield  {author} {\bibinfo {author} {\bibfnamefont {J.~L.}\ \bibnamefont
  {Whitten}},\ }\href {\doibase 10.1063/1.1679012} {\bibfield  {journal}
  {\bibinfo  {journal} {The Journal of Chemical Physics}\ }\textbf {\bibinfo
  {volume} {58}},\ \bibinfo {pages} {4496} (\bibinfo {year}
  {1973})}\BibitemShut {NoStop}%
\bibitem [{\citenamefont {Baerends}\ \emph {et~al.}(1973)\citenamefont
  {Baerends}, \citenamefont {Ellis},\ and\ \citenamefont {Ros}}]{Baerends1973}%
  \BibitemOpen
  \bibfield  {author} {\bibinfo {author} {\bibfnamefont {E.}~\bibnamefont
  {Baerends}}, \bibinfo {author} {\bibfnamefont {D.}~\bibnamefont {Ellis}}, \
  and\ \bibinfo {author} {\bibfnamefont {P.}~\bibnamefont {Ros}},\ }\href
  {\doibase 10.1016/0301-0104(73)80059-X} {\bibfield  {journal} {\bibinfo
  {journal} {Chemical Physics}\ }\textbf {\bibinfo {volume} {2}},\ \bibinfo
  {pages} {41} (\bibinfo {year} {1973})}\BibitemShut {NoStop}%
\bibitem [{\citenamefont {Jafri}\ and\ \citenamefont
  {Whitten}(1974)}]{Jafri1974}%
  \BibitemOpen
  \bibfield  {author} {\bibinfo {author} {\bibfnamefont {J.~A.}\ \bibnamefont
  {Jafri}}\ and\ \bibinfo {author} {\bibfnamefont {J.~L.}\ \bibnamefont
  {Whitten}},\ }\href {\doibase 10.1063/1.1682222} {\bibfield  {journal}
  {\bibinfo  {journal} {The Journal of Chemical Physics}\ }\textbf {\bibinfo
  {volume} {61}},\ \bibinfo {pages} {2116} (\bibinfo {year}
  {1974})}\BibitemShut {NoStop}%
\bibitem [{\citenamefont {Peng}\ and\ \citenamefont
  {Kowalski}(2017)}]{peng2017highly}%
  \BibitemOpen
  \bibfield  {author} {\bibinfo {author} {\bibfnamefont {B.}~\bibnamefont
  {Peng}}\ and\ \bibinfo {author} {\bibfnamefont {K.}~\bibnamefont
  {Kowalski}},\ }\href {\doibase 10.1021/acs.jctc.7b00605} {\bibfield
  {journal} {\bibinfo  {journal} {Journal of Chemical Theory and Computation}\
  }\textbf {\bibinfo {volume} {13}},\ \bibinfo {pages} {4179} (\bibinfo {year}
  {2017})}\BibitemShut {NoStop}%
\bibitem [{\citenamefont {Li}\ \emph {et~al.}(2019)\citenamefont {Li},
  \citenamefont {Li}, \citenamefont {Dattani}, \citenamefont {Umrigar},\ and\
  \citenamefont {Chan}}]{Li2019}%
  \BibitemOpen
  \bibfield  {author} {\bibinfo {author} {\bibfnamefont {Z.}~\bibnamefont
  {Li}}, \bibinfo {author} {\bibfnamefont {J.}~\bibnamefont {Li}}, \bibinfo
  {author} {\bibfnamefont {N.~S.}\ \bibnamefont {Dattani}}, \bibinfo {author}
  {\bibfnamefont {C.~J.}\ \bibnamefont {Umrigar}}, \ and\ \bibinfo {author}
  {\bibfnamefont {G.~K.-L.}\ \bibnamefont {Chan}},\ }\href {\doibase
  10.1063/1.5063376} {\bibfield  {journal} {\bibinfo  {journal} {The Journal of
  Chemical Physics}\ }\textbf {\bibinfo {volume} {150}},\ \bibinfo {pages}
  {024302} (\bibinfo {year} {2019})}\BibitemShut {NoStop}%
\bibitem [{\citenamefont {Low}\ \emph {et~al.}(2018)\citenamefont {Low},
  \citenamefont {Kliuchnikov},\ and\ \citenamefont
  {Schaeffer}}]{Lowpreparation}%
  \BibitemOpen
  \bibfield  {author} {\bibinfo {author} {\bibfnamefont {G.~H.}\ \bibnamefont
  {Low}}, \bibinfo {author} {\bibfnamefont {V.}~\bibnamefont {Kliuchnikov}}, \
  and\ \bibinfo {author} {\bibfnamefont {L.}~\bibnamefont {Schaeffer}},\ }\href
  {https://arxiv.org/abs/1812.00954} {\bibfield  {journal} {\bibinfo  {journal}
  {arXiv:1812.00954}\ } (\bibinfo {year} {2018})}\BibitemShut {NoStop}%
\bibitem [{\citenamefont {Hohenstein}\ \emph
  {et~al.}(2012{\natexlab{a}})\citenamefont {Hohenstein}, \citenamefont
  {Parrish},\ and\ \citenamefont {Mart{\'{i}}nez}}]{Hohenstein2012}%
  \BibitemOpen
  \bibfield  {author} {\bibinfo {author} {\bibfnamefont {E.~G.}\ \bibnamefont
  {Hohenstein}}, \bibinfo {author} {\bibfnamefont {R.~M.}\ \bibnamefont
  {Parrish}}, \ and\ \bibinfo {author} {\bibfnamefont {T.~J.}\ \bibnamefont
  {Mart{\'{i}}nez}},\ }\href {\doibase 10.1063/1.4732310} {\bibfield  {journal}
  {\bibinfo  {journal} {The Journal of Chemical Physics}\ }\textbf {\bibinfo
  {volume} {137}},\ \bibinfo {pages} {1085} (\bibinfo {year}
  {2012}{\natexlab{a}})}\BibitemShut {NoStop}%
\bibitem [{\citenamefont {Parrish}\ \emph {et~al.}(2012)\citenamefont
  {Parrish}, \citenamefont {Hohenstein}, \citenamefont {Mart{\'{i}}nez},\ and\
  \citenamefont {Sherrill}}]{Parrish2012}%
  \BibitemOpen
  \bibfield  {author} {\bibinfo {author} {\bibfnamefont {R.~M.}\ \bibnamefont
  {Parrish}}, \bibinfo {author} {\bibfnamefont {E.~G.}\ \bibnamefont
  {Hohenstein}}, \bibinfo {author} {\bibfnamefont {T.~J.}\ \bibnamefont
  {Mart{\'{i}}nez}}, \ and\ \bibinfo {author} {\bibfnamefont {C.~D.}\
  \bibnamefont {Sherrill}},\ }\href {\doibase 10.1063/1.4768233} {\bibfield
  {journal} {\bibinfo  {journal} {The Journal of Chemical Physics}\ }\textbf
  {\bibinfo {volume} {137}},\ \bibinfo {pages} {224106} (\bibinfo {year}
  {2012})}\BibitemShut {NoStop}%
\bibitem [{\citenamefont {Hohenstein}\ \emph
  {et~al.}(2012{\natexlab{b}})\citenamefont {Hohenstein}, \citenamefont
  {Parrish}, \citenamefont {Sherrill},\ and\ \citenamefont
  {Mart{\'{i}}nez}}]{Hohenstein2012a}%
  \BibitemOpen
  \bibfield  {author} {\bibinfo {author} {\bibfnamefont {E.~G.}\ \bibnamefont
  {Hohenstein}}, \bibinfo {author} {\bibfnamefont {R.~M.}\ \bibnamefont
  {Parrish}}, \bibinfo {author} {\bibfnamefont {C.~D.}\ \bibnamefont
  {Sherrill}}, \ and\ \bibinfo {author} {\bibfnamefont {T.~J.}\ \bibnamefont
  {Mart{\'{i}}nez}},\ }\bibfield  {booktitle} {\emph {\bibinfo {booktitle} {The
  Journal of Chemical Physics}},\ }\href {\doibase 10.1063/1.4768241}
  {\bibfield  {journal} {\bibinfo  {journal} {The Journal of Chemical Physics}\
  }\textbf {\bibinfo {volume} {137}},\ \bibinfo {pages} {221101} (\bibinfo
  {year} {2012}{\natexlab{b}})}\BibitemShut {NoStop}%
\bibitem [{\citenamefont {Kivlichan}\ \emph {et~al.}(2018)\citenamefont
  {Kivlichan}, \citenamefont {McClean}, \citenamefont {Wiebe}, \citenamefont
  {Gidney}, \citenamefont {Aspuru-Guzik}, \citenamefont {Chan},\ and\
  \citenamefont {Babbush}}]{kivlichan2018quantum}%
  \BibitemOpen
  \bibfield  {author} {\bibinfo {author} {\bibfnamefont {I.~D.}\ \bibnamefont
  {Kivlichan}}, \bibinfo {author} {\bibfnamefont {J.}~\bibnamefont {McClean}},
  \bibinfo {author} {\bibfnamefont {N.}~\bibnamefont {Wiebe}}, \bibinfo
  {author} {\bibfnamefont {C.}~\bibnamefont {Gidney}}, \bibinfo {author}
  {\bibfnamefont {A.}~\bibnamefont {Aspuru-Guzik}}, \bibinfo {author}
  {\bibfnamefont {G.~K.-L.}\ \bibnamefont {Chan}}, \ and\ \bibinfo {author}
  {\bibfnamefont {R.}~\bibnamefont {Babbush}},\ }\href {\doibase
  PhysRevLett.120.110501} {\bibfield  {journal} {\bibinfo  {journal} {Physical
  Review Letters}\ }\textbf {\bibinfo {volume} {120}},\ \bibinfo {pages}
  {110501} (\bibinfo {year} {2018})}\BibitemShut {NoStop}%
\bibitem [{\citenamefont {Seeley}\ \emph {et~al.}(2012)\citenamefont {Seeley},
  \citenamefont {Richard},\ and\ \citenamefont {Love}}]{Seeley2012}%
  \BibitemOpen
  \bibfield  {author} {\bibinfo {author} {\bibfnamefont {J.~T.}\ \bibnamefont
  {Seeley}}, \bibinfo {author} {\bibfnamefont {M.~J.}\ \bibnamefont {Richard}},
  \ and\ \bibinfo {author} {\bibfnamefont {P.~J.}\ \bibnamefont {Love}},\
  }\href {\doibase 10.1063/1.4768229} {\bibfield  {journal} {\bibinfo
  {journal} {Journal of Chemical Physics}\ }\textbf {\bibinfo {volume} {137}},\
  \bibinfo {pages} {224109} (\bibinfo {year} {2012})}\BibitemShut {NoStop}%
\bibitem [{\citenamefont {Wecker}\ \emph {et~al.}(2014)\citenamefont {Wecker},
  \citenamefont {Bauer}, \citenamefont {Clark}, \citenamefont {Hastings},\ and\
  \citenamefont {Troyer}}]{Wecker2014B}%
  \BibitemOpen
  \bibfield  {author} {\bibinfo {author} {\bibfnamefont {D.}~\bibnamefont
  {Wecker}}, \bibinfo {author} {\bibfnamefont {B.}~\bibnamefont {Bauer}},
  \bibinfo {author} {\bibfnamefont {B.~K.}\ \bibnamefont {Clark}}, \bibinfo
  {author} {\bibfnamefont {M.~B.}\ \bibnamefont {Hastings}}, \ and\ \bibinfo
  {author} {\bibfnamefont {M.}~\bibnamefont {Troyer}},\ }\href {\doibase
  10.1103/PhysRevA.90.022305} {\bibfield  {journal} {\bibinfo  {journal}
  {Physical Review A}\ }\textbf {\bibinfo {volume} {90}},\ \bibinfo {pages}
  {022305} (\bibinfo {year} {2014})}\BibitemShut {NoStop}%
\bibitem [{\citenamefont {Toloui}\ and\ \citenamefont
  {Love}(2013)}]{Toloui2013}%
  \BibitemOpen
  \bibfield  {author} {\bibinfo {author} {\bibfnamefont {B.}~\bibnamefont
  {Toloui}}\ and\ \bibinfo {author} {\bibfnamefont {P.~J.}\ \bibnamefont
  {Love}},\ }\href {http://arxiv.org/abs/1312.2579} {\bibfield  {journal}
  {\bibinfo  {journal} {arXiv:1312.2579}\ } (\bibinfo {year}
  {2013})}\BibitemShut {NoStop}%
\bibitem [{\citenamefont {Hastings}\ \emph {et~al.}(2015)\citenamefont
  {Hastings}, \citenamefont {Wecker}, \citenamefont {Bauer},\ and\
  \citenamefont {Troyer}}]{Hastings2015}%
  \BibitemOpen
  \bibfield  {author} {\bibinfo {author} {\bibfnamefont {M.~B.}\ \bibnamefont
  {Hastings}}, \bibinfo {author} {\bibfnamefont {D.}~\bibnamefont {Wecker}},
  \bibinfo {author} {\bibfnamefont {B.}~\bibnamefont {Bauer}}, \ and\ \bibinfo
  {author} {\bibfnamefont {M.}~\bibnamefont {Troyer}},\ }\href
  {http://arxiv.org/abs/1403.1539} {\bibfield  {journal} {\bibinfo  {journal}
  {Quantum Information {\&} Computation}\ }\textbf {\bibinfo {volume} {15}},\
  \bibinfo {pages} {1} (\bibinfo {year} {2015})}\BibitemShut {NoStop}%
\bibitem [{\citenamefont {Poulin}\ \emph {et~al.}(2015)\citenamefont {Poulin},
  \citenamefont {Hastings}, \citenamefont {Wecker}, \citenamefont {Wiebe},
  \citenamefont {Doherty},\ and\ \citenamefont {Troyer}}]{Poulin2014}%
  \BibitemOpen
  \bibfield  {author} {\bibinfo {author} {\bibfnamefont {D.}~\bibnamefont
  {Poulin}}, \bibinfo {author} {\bibfnamefont {M.~B.}\ \bibnamefont
  {Hastings}}, \bibinfo {author} {\bibfnamefont {D.}~\bibnamefont {Wecker}},
  \bibinfo {author} {\bibfnamefont {N.}~\bibnamefont {Wiebe}}, \bibinfo
  {author} {\bibfnamefont {A.~C.}\ \bibnamefont {Doherty}}, \ and\ \bibinfo
  {author} {\bibfnamefont {M.}~\bibnamefont {Troyer}},\ }\href
  {http://arxiv.org/abs/1406.4920} {\bibfield  {journal} {\bibinfo  {journal}
  {Quantum Information {\&} Computation}\ }\textbf {\bibinfo {volume} {15}},\
  \bibinfo {pages} {361} (\bibinfo {year} {2015})}\BibitemShut {NoStop}%
\bibitem [{\citenamefont {Babbush}\ \emph {et~al.}(2015)\citenamefont
  {Babbush}, \citenamefont {McClean}, \citenamefont {Wecker}, \citenamefont
  {Aspuru-Guzik},\ and\ \citenamefont {Wiebe}}]{BabbushTrotter}%
  \BibitemOpen
  \bibfield  {author} {\bibinfo {author} {\bibfnamefont {R.}~\bibnamefont
  {Babbush}}, \bibinfo {author} {\bibfnamefont {J.}~\bibnamefont {McClean}},
  \bibinfo {author} {\bibfnamefont {D.}~\bibnamefont {Wecker}}, \bibinfo
  {author} {\bibfnamefont {A.}~\bibnamefont {Aspuru-Guzik}}, \ and\ \bibinfo
  {author} {\bibfnamefont {N.}~\bibnamefont {Wiebe}},\ }\href
  {http://dx.doi.org/10.1103/PhysRevA.91.022311} {\bibfield  {journal}
  {\bibinfo  {journal} {Physical Review A}\ }\textbf {\bibinfo {volume} {91}},\
  \bibinfo {pages} {22311} (\bibinfo {year} {2015})}\BibitemShut {NoStop}%
\bibitem [{\citenamefont {Babbush}\ \emph {et~al.}(2016)\citenamefont
  {Babbush}, \citenamefont {Berry}, \citenamefont {Kivlichan}, \citenamefont
  {Wei}, \citenamefont {Love},\ and\ \citenamefont
  {Aspuru-Guzik}}]{BabbushSparse1}%
  \BibitemOpen
  \bibfield  {author} {\bibinfo {author} {\bibfnamefont {R.}~\bibnamefont
  {Babbush}}, \bibinfo {author} {\bibfnamefont {D.~W.}\ \bibnamefont {Berry}},
  \bibinfo {author} {\bibfnamefont {I.~D.}\ \bibnamefont {Kivlichan}}, \bibinfo
  {author} {\bibfnamefont {A.~Y.}\ \bibnamefont {Wei}}, \bibinfo {author}
  {\bibfnamefont {P.~J.}\ \bibnamefont {Love}}, \ and\ \bibinfo {author}
  {\bibfnamefont {A.}~\bibnamefont {Aspuru-Guzik}},\ }\href {\doibase
  10.1088/1367-2630/18/3/033032} {\bibfield  {journal} {\bibinfo  {journal}
  {New Journal of Physics}\ }\textbf {\bibinfo {volume} {18}},\ \bibinfo
  {pages} {33032} (\bibinfo {year} {2016})}\BibitemShut {NoStop}%
\bibitem [{\citenamefont {Babbush}\ \emph
  {et~al.}(2018{\natexlab{c}})\citenamefont {Babbush}, \citenamefont {Berry},
  \citenamefont {Sanders}, \citenamefont {Kivlichan}, \citenamefont {Scherer},
  \citenamefont {Wei}, \citenamefont {Love},\ and\ \citenamefont
  {Aspuru-Guzik}}]{BabbushSparse2}%
  \BibitemOpen
  \bibfield  {author} {\bibinfo {author} {\bibfnamefont {R.}~\bibnamefont
  {Babbush}}, \bibinfo {author} {\bibfnamefont {D.~W.}\ \bibnamefont {Berry}},
  \bibinfo {author} {\bibfnamefont {Y.~R.}\ \bibnamefont {Sanders}}, \bibinfo
  {author} {\bibfnamefont {I.~D.}\ \bibnamefont {Kivlichan}}, \bibinfo {author}
  {\bibfnamefont {A.}~\bibnamefont {Scherer}}, \bibinfo {author} {\bibfnamefont
  {A.~Y.}\ \bibnamefont {Wei}}, \bibinfo {author} {\bibfnamefont {P.~J.}\
  \bibnamefont {Love}}, \ and\ \bibinfo {author} {\bibfnamefont
  {A.}~\bibnamefont {Aspuru-Guzik}},\ }\href
  {http://iopscience.iop.org/article/10.1088/2058-9565/aa9463/meta} {\bibfield
  {journal} {\bibinfo  {journal} {Quantum Science and Technology}\ }\textbf
  {\bibinfo {volume} {3}},\ \bibinfo {pages} {015006} (\bibinfo {year}
  {2018}{\natexlab{c}})}\BibitemShut {NoStop}%
\bibitem [{\citenamefont {Campbell}(2019)}]{Campbell2018B}%
  \BibitemOpen
  \bibfield  {author} {\bibinfo {author} {\bibfnamefont {E.}~\bibnamefont
  {Campbell}},\ }\href {\doibase 10.1103/PhysRevLett.123.070503} {\bibfield
  {journal} {\bibinfo  {journal} {Physical Review Letters}\ }\textbf {\bibinfo
  {volume} {123}},\ \bibinfo {pages} {070503} (\bibinfo {year}
  {2019})}\BibitemShut {NoStop}%
\bibitem [{\citenamefont {Kivlichan}\ \emph {et~al.}(2019)\citenamefont
  {Kivlichan}, \citenamefont {Granade},\ and\ \citenamefont
  {Wiebe}}]{Kivlichan2019PhaseHamiltoniansB}%
  \BibitemOpen
  \bibfield  {author} {\bibinfo {author} {\bibfnamefont {I.~D.}\ \bibnamefont
  {Kivlichan}}, \bibinfo {author} {\bibfnamefont {C.~E.}\ \bibnamefont
  {Granade}}, \ and\ \bibinfo {author} {\bibfnamefont {N.}~\bibnamefont
  {Wiebe}},\ }\href {https://arxiv.org/abs/1907.10070} {\bibfield  {journal}
  {\bibinfo  {journal} {arXiv:1907.10070}\ } (\bibinfo {year}
  {2019})}\BibitemShut {NoStop}%
\bibitem [{\citenamefont {Kivlichan}\ \emph {et~al.}(2017)\citenamefont
  {Kivlichan}, \citenamefont {Wiebe}, \citenamefont {Babbush},\ and\
  \citenamefont {Aspuru-Guzik}}]{Kivlichan2016}%
  \BibitemOpen
  \bibfield  {author} {\bibinfo {author} {\bibfnamefont {I.~D.}\ \bibnamefont
  {Kivlichan}}, \bibinfo {author} {\bibfnamefont {N.}~\bibnamefont {Wiebe}},
  \bibinfo {author} {\bibfnamefont {R.}~\bibnamefont {Babbush}}, \ and\
  \bibinfo {author} {\bibfnamefont {A.}~\bibnamefont {Aspuru-Guzik}},\ }\href
  {http://iopscience.iop.org/article/10.1088/1751-8121/aa77b8} {\bibfield
  {journal} {\bibinfo  {journal} {Journal of Physics A: Mathematical and
  Theoretical}\ }\textbf {\bibinfo {volume} {50}},\ \bibinfo {pages} {305301}
  (\bibinfo {year} {2017})}\BibitemShut {NoStop}%
\bibitem [{\citenamefont {Gidney}\ and\ \citenamefont
  {Fowler}(2019{\natexlab{a}})}]{Gidney2018pub}%
  \BibitemOpen
  \bibfield  {author} {\bibinfo {author} {\bibfnamefont {C.}~\bibnamefont
  {Gidney}}\ and\ \bibinfo {author} {\bibfnamefont {A.~G.}\ \bibnamefont
  {Fowler}},\ }\href {https://doi.org/10.22331/q-2019-04-30-135} {\bibfield
  {journal} {\bibinfo  {journal} {Quantum}\ }\textbf {\bibinfo {volume} {3}},\
  \bibinfo {pages} {135} (\bibinfo {year} {2019}{\natexlab{a}})}\BibitemShut
  {NoStop}%
\bibitem [{\citenamefont {Parrish}\ \emph {et~al.}(2013)\citenamefont
  {Parrish}, \citenamefont {Hohenstein}, \citenamefont {Mart{\'{i}}nez},\ and\
  \citenamefont {Sherrill}}]{Parrish2013a}%
  \BibitemOpen
  \bibfield  {author} {\bibinfo {author} {\bibfnamefont {R.~M.}\ \bibnamefont
  {Parrish}}, \bibinfo {author} {\bibfnamefont {E.~G.}\ \bibnamefont
  {Hohenstein}}, \bibinfo {author} {\bibfnamefont {T.~J.}\ \bibnamefont
  {Mart{\'{i}}nez}}, \ and\ \bibinfo {author} {\bibfnamefont {C.~D.}\
  \bibnamefont {Sherrill}},\ }\href {\doibase 10.1063/1.4802773} {\bibfield
  {journal} {\bibinfo  {journal} {The Journal of Chemical Physics}\ }\textbf
  {\bibinfo {volume} {138}},\ \bibinfo {pages} {194107} (\bibinfo {year}
  {2013})}\BibitemShut {NoStop}%
\bibitem [{\citenamefont {Hohenstein}\ \emph
  {et~al.}(2013{\natexlab{a}})\citenamefont {Hohenstein}, \citenamefont
  {Kokkila}, \citenamefont {Parrish},\ and\ \citenamefont
  {Mart{\'{i}}nez}}]{Hohenstein2013a}%
  \BibitemOpen
  \bibfield  {author} {\bibinfo {author} {\bibfnamefont {E.~G.}\ \bibnamefont
  {Hohenstein}}, \bibinfo {author} {\bibfnamefont {S.~I.}\ \bibnamefont
  {Kokkila}}, \bibinfo {author} {\bibfnamefont {R.~M.}\ \bibnamefont
  {Parrish}}, \ and\ \bibinfo {author} {\bibfnamefont {T.~J.}\ \bibnamefont
  {Mart{\'{i}}nez}},\ }\href {\doibase 10.1063/1.4795514} {\bibfield  {journal}
  {\bibinfo  {journal} {The Journal of Chemical Physics}\ }\textbf {\bibinfo
  {volume} {138}},\ \bibinfo {pages} {124111} (\bibinfo {year}
  {2013}{\natexlab{a}})}\BibitemShut {NoStop}%
\bibitem [{\citenamefont {Hohenstein}\ \emph
  {et~al.}(2013{\natexlab{b}})\citenamefont {Hohenstein}, \citenamefont
  {Kokkila}, \citenamefont {Parrish},\ and\ \citenamefont
  {Mart{\'{i}}nez}}]{Hohenstein2013}%
  \BibitemOpen
  \bibfield  {author} {\bibinfo {author} {\bibfnamefont {E.~G.}\ \bibnamefont
  {Hohenstein}}, \bibinfo {author} {\bibfnamefont {S.~I.}\ \bibnamefont
  {Kokkila}}, \bibinfo {author} {\bibfnamefont {R.~M.}\ \bibnamefont
  {Parrish}}, \ and\ \bibinfo {author} {\bibfnamefont {T.~J.}\ \bibnamefont
  {Mart{\'{i}}nez}},\ }\href {\doibase 10.1021/jp4021905} {\bibfield  {journal}
  {\bibinfo  {journal} {The Journal of Physical Chemistry B}\ }\textbf
  {\bibinfo {volume} {117}},\ \bibinfo {pages} {12972} (\bibinfo {year}
  {2013}{\natexlab{b}})}\BibitemShut {NoStop}%
\bibitem [{\citenamefont {Benedikt}\ \emph {et~al.}(2013)\citenamefont
  {Benedikt}, \citenamefont {B{\"o}hm},\ and\ \citenamefont
  {Auer}}]{benedikt2013tensor}%
  \BibitemOpen
  \bibfield  {author} {\bibinfo {author} {\bibfnamefont {U.}~\bibnamefont
  {Benedikt}}, \bibinfo {author} {\bibfnamefont {K.-H.}\ \bibnamefont
  {B{\"o}hm}}, \ and\ \bibinfo {author} {\bibfnamefont {A.~A.}\ \bibnamefont
  {Auer}},\ }\href {\doibase 10.1063/1.4833565} {\bibfield  {journal} {\bibinfo
   {journal} {The Journal of Chemical Physics}\ }\textbf {\bibinfo {volume}
  {139}},\ \bibinfo {pages} {224101} (\bibinfo {year} {2013})}\BibitemShut
  {NoStop}%
\bibitem [{\citenamefont {Parrish}\ \emph {et~al.}(2014)\citenamefont
  {Parrish}, \citenamefont {Sherrill}, \citenamefont {Hohenstein},
  \citenamefont {Kokkila},\ and\ \citenamefont {Mart{\'{i}}nez}}]{Parrish2014}%
  \BibitemOpen
  \bibfield  {author} {\bibinfo {author} {\bibfnamefont {R.~M.}\ \bibnamefont
  {Parrish}}, \bibinfo {author} {\bibfnamefont {C.~D.}\ \bibnamefont
  {Sherrill}}, \bibinfo {author} {\bibfnamefont {E.~G.}\ \bibnamefont
  {Hohenstein}}, \bibinfo {author} {\bibfnamefont {S.~I.}\ \bibnamefont
  {Kokkila}}, \ and\ \bibinfo {author} {\bibfnamefont {T.~J.}\ \bibnamefont
  {Mart{\'{i}}nez}},\ }\href {\doibase 10.1063/1.4876016} {\bibfield  {journal}
  {\bibinfo  {journal} {The Journal of Chemical Physics}\ }\textbf {\bibinfo
  {volume} {140}},\ \bibinfo {pages} {181102} (\bibinfo {year}
  {2014})}\BibitemShut {NoStop}%
\bibitem [{\citenamefont {Lu}\ and\ \citenamefont {Ying}(2015)}]{Lu2015}%
  \BibitemOpen
  \bibfield  {author} {\bibinfo {author} {\bibfnamefont {J.}~\bibnamefont
  {Lu}}\ and\ \bibinfo {author} {\bibfnamefont {L.}~\bibnamefont {Ying}},\
  }\href {\doibase 10.1016/j.jcp.2015.09.014} {\bibfield  {journal} {\bibinfo
  {journal} {Journal of Computational Physics}\ }\textbf {\bibinfo {volume}
  {302}},\ \bibinfo {pages} {329} (\bibinfo {year} {2015})}\BibitemShut
  {NoStop}%
\bibitem [{\citenamefont {{Kokkila Schumacher}}\ \emph
  {et~al.}(2015)\citenamefont {{Kokkila Schumacher}}, \citenamefont
  {Hohenstein}, \citenamefont {Parrish}, \citenamefont {Wang},\ and\
  \citenamefont {Mart{\'{i}}nez}}]{KokkilaSchumacher2015}%
  \BibitemOpen
  \bibfield  {author} {\bibinfo {author} {\bibfnamefont {S.~I.}\ \bibnamefont
  {{Kokkila Schumacher}}}, \bibinfo {author} {\bibfnamefont {E.~G.}\
  \bibnamefont {Hohenstein}}, \bibinfo {author} {\bibfnamefont {R.~M.}\
  \bibnamefont {Parrish}}, \bibinfo {author} {\bibfnamefont {L.~P.}\
  \bibnamefont {Wang}}, \ and\ \bibinfo {author} {\bibfnamefont {T.~J.}\
  \bibnamefont {Mart{\'{i}}nez}},\ }\href {\doibase 10.1021/acs.jctc.5b00272}
  {\bibfield  {journal} {\bibinfo  {journal} {Journal of Chemical Theory and
  Computation}\ }\textbf {\bibinfo {volume} {11}},\ \bibinfo {pages} {3042}
  (\bibinfo {year} {2015})}\BibitemShut {NoStop}%
\bibitem [{\citenamefont {Song}\ and\ \citenamefont
  {Mart{\'{i}}nez}(2016)}]{Song2016}%
  \BibitemOpen
  \bibfield  {author} {\bibinfo {author} {\bibfnamefont {C.}~\bibnamefont
  {Song}}\ and\ \bibinfo {author} {\bibfnamefont {T.~J.}\ \bibnamefont
  {Mart{\'{i}}nez}},\ }\href {\doibase 10.1063/1.4948438} {\bibfield  {journal}
  {\bibinfo  {journal} {The Journal of Chemical Physics}\ }\textbf {\bibinfo
  {volume} {144}},\ \bibinfo {pages} {174111} (\bibinfo {year}
  {2016})}\BibitemShut {NoStop}%
\bibitem [{\citenamefont {Lu}\ and\ \citenamefont {Ying}(2016)}]{Lu2016}%
  \BibitemOpen
  \bibfield  {author} {\bibinfo {author} {\bibfnamefont {J.}~\bibnamefont
  {Lu}}\ and\ \bibinfo {author} {\bibfnamefont {L.}~\bibnamefont {Ying}},\
  }\href {\doibase 10.4310/AMSA.2016.v1.n2.a3} {\bibfield  {journal} {\bibinfo
  {journal} {Annals of Mathematical Sciences and Applications}\ }\textbf
  {\bibinfo {volume} {1}},\ \bibinfo {pages} {321} (\bibinfo {year}
  {2016})}\BibitemShut {NoStop}%
\bibitem [{\citenamefont {Hu}\ \emph {et~al.}(2017)\citenamefont {Hu},
  \citenamefont {Lin},\ and\ \citenamefont {Yang}}]{Hu2017}%
  \BibitemOpen
  \bibfield  {author} {\bibinfo {author} {\bibfnamefont {W.}~\bibnamefont
  {Hu}}, \bibinfo {author} {\bibfnamefont {L.}~\bibnamefont {Lin}}, \ and\
  \bibinfo {author} {\bibfnamefont {C.}~\bibnamefont {Yang}},\ }\href {\doibase
  10.1021/acs.jctc.7b00807} {\bibfield  {journal} {\bibinfo  {journal} {Journal
  of Chemical Theory and Computation}\ }\textbf {\bibinfo {volume} {13}},\
  \bibinfo {pages} {5420} (\bibinfo {year} {2017})}\BibitemShut {NoStop}%
\bibitem [{\citenamefont {Song}\ and\ \citenamefont
  {Mart{\'{i}}nez}(2017{\natexlab{a}})}]{Song2017}%
  \BibitemOpen
  \bibfield  {author} {\bibinfo {author} {\bibfnamefont {C.}~\bibnamefont
  {Song}}\ and\ \bibinfo {author} {\bibfnamefont {T.~J.}\ \bibnamefont
  {Mart{\'{i}}nez}},\ }\href {\doibase 10.1063/1.4973840} {\bibfield  {journal}
  {\bibinfo  {journal} {The Journal of Chemical Physics}\ }\textbf {\bibinfo
  {volume} {146}},\ \bibinfo {pages} {034104} (\bibinfo {year}
  {2017}{\natexlab{a}})}\BibitemShut {NoStop}%
\bibitem [{\citenamefont {Lu}\ and\ \citenamefont {Thicke}(2017)}]{Lu2017}%
  \BibitemOpen
  \bibfield  {author} {\bibinfo {author} {\bibfnamefont {J.}~\bibnamefont
  {Lu}}\ and\ \bibinfo {author} {\bibfnamefont {K.}~\bibnamefont {Thicke}},\
  }\href {\doibase 10.1016/j.jcp.2017.09.012} {\bibfield  {journal} {\bibinfo
  {journal} {Journal of Computational Physics}\ }\textbf {\bibinfo {volume}
  {351}},\ \bibinfo {pages} {187} (\bibinfo {year} {2017})}\BibitemShut
  {NoStop}%
\bibitem [{\citenamefont {Song}\ and\ \citenamefont
  {Mart{\'{i}}nez}(2017{\natexlab{b}})}]{Song2017a}%
  \BibitemOpen
  \bibfield  {author} {\bibinfo {author} {\bibfnamefont {C.}~\bibnamefont
  {Song}}\ and\ \bibinfo {author} {\bibfnamefont {T.~J.}\ \bibnamefont
  {Mart{\'{i}}nez}},\ }\href {\doibase 10.1063/1.4997997} {\bibfield  {journal}
  {\bibinfo  {journal} {The Journal of Chemical Physics}\ }\textbf {\bibinfo
  {volume} {147}},\ \bibinfo {pages} {161723} (\bibinfo {year}
  {2017}{\natexlab{b}})}\BibitemShut {NoStop}%
\bibitem [{\citenamefont {Hummel}\ \emph {et~al.}(2017)\citenamefont {Hummel},
  \citenamefont {Tsatsoulis},\ and\ \citenamefont
  {Gr{\"u}neis}}]{hummel2017low}%
  \BibitemOpen
  \bibfield  {author} {\bibinfo {author} {\bibfnamefont {F.}~\bibnamefont
  {Hummel}}, \bibinfo {author} {\bibfnamefont {T.}~\bibnamefont {Tsatsoulis}},
  \ and\ \bibinfo {author} {\bibfnamefont {A.}~\bibnamefont {Gr{\"u}neis}},\
  }\href {\doibase 10.1063/1.4977994} {\bibfield  {journal} {\bibinfo
  {journal} {The Journal of Chemical Physics}\ }\textbf {\bibinfo {volume}
  {146}},\ \bibinfo {pages} {124105} (\bibinfo {year} {2017})}\BibitemShut
  {NoStop}%
\bibitem [{\citenamefont {Schutski}\ \emph {et~al.}(2017)\citenamefont
  {Schutski}, \citenamefont {Zhao}, \citenamefont {Henderson},\ and\
  \citenamefont {Scuseria}}]{schutski2017tensor}%
  \BibitemOpen
  \bibfield  {author} {\bibinfo {author} {\bibfnamefont {R.}~\bibnamefont
  {Schutski}}, \bibinfo {author} {\bibfnamefont {J.}~\bibnamefont {Zhao}},
  \bibinfo {author} {\bibfnamefont {T.~M.}\ \bibnamefont {Henderson}}, \ and\
  \bibinfo {author} {\bibfnamefont {G.~E.}\ \bibnamefont {Scuseria}},\ }\href
  {\doibase 10.1063/1.4996988} {\bibfield  {journal} {\bibinfo  {journal} {The
  Journal of Chemical Physics}\ }\textbf {\bibinfo {volume} {147}},\ \bibinfo
  {pages} {184113} (\bibinfo {year} {2017})}\BibitemShut {NoStop}%
\bibitem [{\citenamefont {Dong}\ \emph {et~al.}(2018)\citenamefont {Dong},
  \citenamefont {Hu},\ and\ \citenamefont {Lin}}]{Dong2018}%
  \BibitemOpen
  \bibfield  {author} {\bibinfo {author} {\bibfnamefont {K.}~\bibnamefont
  {Dong}}, \bibinfo {author} {\bibfnamefont {W.}~\bibnamefont {Hu}}, \ and\
  \bibinfo {author} {\bibfnamefont {L.}~\bibnamefont {Lin}},\ }\href {\doibase
  10.1021/acs.jctc.7b01113} {\bibfield  {journal} {\bibinfo  {journal} {Journal
  of Chemical Theory and Computation}\ }\textbf {\bibinfo {volume} {14}},\
  \bibinfo {pages} {1311} (\bibinfo {year} {2018})}\BibitemShut {NoStop}%
\bibitem [{\citenamefont {Hu}\ \emph {et~al.}(2018)\citenamefont {Hu},
  \citenamefont {Shao}, \citenamefont {Cepellotti}, \citenamefont {da~Jornada},
  \citenamefont {Lin}, \citenamefont {Thicke}, \citenamefont {Yang},\ and\
  \citenamefont {Louie}}]{Hu2018}%
  \BibitemOpen
  \bibfield  {author} {\bibinfo {author} {\bibfnamefont {W.}~\bibnamefont
  {Hu}}, \bibinfo {author} {\bibfnamefont {M.}~\bibnamefont {Shao}}, \bibinfo
  {author} {\bibfnamefont {A.}~\bibnamefont {Cepellotti}}, \bibinfo {author}
  {\bibfnamefont {F.~H.}\ \bibnamefont {da~Jornada}}, \bibinfo {author}
  {\bibfnamefont {L.}~\bibnamefont {Lin}}, \bibinfo {author} {\bibfnamefont
  {K.}~\bibnamefont {Thicke}}, \bibinfo {author} {\bibfnamefont
  {C.}~\bibnamefont {Yang}}, \ and\ \bibinfo {author} {\bibfnamefont {S.~G.}\
  \bibnamefont {Louie}},\ }in\ \href {\doibase 10.1007/978-3-319-93701-4_48}
  {\emph {\bibinfo {booktitle} {Lecture Notes in Computer Science (including
  subseries Lecture Notes in Artificial Intelligence and Lecture Notes in
  Bioinformatics)}}},\ Vol.\ \bibinfo {volume} {10861 LNCS}\ (\bibinfo
  {publisher} {Springer, Cham},\ \bibinfo {year} {2018})\ pp.\ \bibinfo {pages}
  {604--617}\BibitemShut {NoStop}%
\bibitem [{\citenamefont {Duchemin}\ and\ \citenamefont
  {Blase}(2019)}]{Duchemin2019}%
  \BibitemOpen
  \bibfield  {author} {\bibinfo {author} {\bibfnamefont {I.}~\bibnamefont
  {Duchemin}}\ and\ \bibinfo {author} {\bibfnamefont {X.}~\bibnamefont
  {Blase}},\ }\href {\doibase 10.1063/1.5090605} {\bibfield  {journal}
  {\bibinfo  {journal} {The Journal of Chemical Physics}\ }\textbf {\bibinfo
  {volume} {150}},\ \bibinfo {pages} {174120} (\bibinfo {year}
  {2019})}\BibitemShut {NoStop}%
\bibitem [{\citenamefont {Lee}\ \emph {et~al.}(2019)\citenamefont {Lee},
  \citenamefont {Lin},\ and\ \citenamefont
  {Head-Gordon}}]{lee2019systematically}%
  \BibitemOpen
  \bibfield  {author} {\bibinfo {author} {\bibfnamefont {J.}~\bibnamefont
  {Lee}}, \bibinfo {author} {\bibfnamefont {L.}~\bibnamefont {Lin}}, \ and\
  \bibinfo {author} {\bibfnamefont {M.}~\bibnamefont {Head-Gordon}},\ }\href
  {\doibase 10.1021/acs.jctc.9b00820} {\bibfield  {journal} {\bibinfo
  {journal} {Journal of Chemical Theory and Computation}\ }\textbf {\bibinfo
  {volume} {16}},\ \bibinfo {pages} {243} (\bibinfo {year} {2019})}\BibitemShut
  {NoStop}%
\bibitem [{\citenamefont
  {A.~Matthews}(2020)}]{A.Matthews2020ImprovedHypercontraction}%
  \BibitemOpen
  \bibfield  {author} {\bibinfo {author} {\bibfnamefont {D.}~\bibnamefont
  {A.~Matthews}},\ }\href {\doibase 10.1021/acs.jctc.9b01205} {\bibfield
  {journal} {\bibinfo  {journal} {Journal of Chemical Theory and Computation}\
  }\textbf {\bibinfo {volume} {16}},\ \bibinfo {pages} {1382} (\bibinfo {year}
  {2020})}\BibitemShut {NoStop}%
\bibitem [{\citenamefont {Huggins}\ \emph {et~al.}(2021)\citenamefont
  {Huggins}, \citenamefont {McClean}, \citenamefont {Rubin}, \citenamefont
  {Jiang}, \citenamefont {Wiebe}, \citenamefont {Whaley},\ and\ \citenamefont
  {Babbush}}]{Huggins2019EfficientComputersNew}%
  \BibitemOpen
  \bibfield  {author} {\bibinfo {author} {\bibfnamefont {W.~J.}\ \bibnamefont
  {Huggins}}, \bibinfo {author} {\bibfnamefont {J.~R.}\ \bibnamefont
  {McClean}}, \bibinfo {author} {\bibfnamefont {N.~C.}\ \bibnamefont {Rubin}},
  \bibinfo {author} {\bibfnamefont {Z.}~\bibnamefont {Jiang}}, \bibinfo
  {author} {\bibfnamefont {N.}~\bibnamefont {Wiebe}}, \bibinfo {author}
  {\bibfnamefont {K.~B.}\ \bibnamefont {Whaley}}, \ and\ \bibinfo {author}
  {\bibfnamefont {R.}~\bibnamefont {Babbush}},\ }\href {\doibase
  10.1038/s41534-020-00341-7} {\bibfield  {journal} {\bibinfo  {journal} {npj
  Quantum Information}\ }\textbf {\bibinfo {volume} {7}},\ \bibinfo {pages}
  {23} (\bibinfo {year} {2021})}\BibitemShut {NoStop}%
\bibitem [{zen(2020)}]{zenodo}%
  \BibitemOpen
  \href {\doibase 10.5281/zenodo.4248322} {\enquote {\bibinfo {title} {Data and
  code repository for `even more efficient quantum computations of chemistry
  through tensor hypercontraction'},}\ } (\bibinfo {year} {2020})\BibitemShut
  {NoStop}%
\bibitem [{\citenamefont {{Virtanen}}\ \emph {et~al.}(2020)\citenamefont
  {{Virtanen}}, \citenamefont {{Gommers}}, \citenamefont {{Oliphant}},
  \citenamefont {{Haberland}}, \citenamefont {{Reddy}}, \citenamefont
  {{Cournapeau}}, \citenamefont {{Burovski}}, \citenamefont {{Peterson}},
  \citenamefont {{Weckesser}}, \citenamefont {{Bright}}, \citenamefont {{van
  der Walt}}, \citenamefont {{Brett}}, \citenamefont {{Wilson}}, \citenamefont
  {{Jarrod Millman}}, \citenamefont {{Mayorov}}, \citenamefont {{Nelson}},
  \citenamefont {{Jones}}, \citenamefont {{Kern}}, \citenamefont {{Larson}},
  \citenamefont {{Carey}}, \citenamefont {{Polat}}, \citenamefont {{Feng}},
  \citenamefont {{Moore}}, \citenamefont {{VanderPlas}}, \citenamefont
  {{Laxalde}}, \citenamefont {{Perktold}}, \citenamefont {{Cimrman}},
  \citenamefont {{Henriksen}}, \citenamefont {{Quintero}}, \citenamefont
  {{Harris}}, \citenamefont {{Archibald}}, \citenamefont {{Ribeiro}},
  \citenamefont {{Pedregosa}}, \citenamefont {{van Mulbregt}},\ and\
  \citenamefont {{SciPy 1.0 Contributors}}}]{2020SciPy-NMeth}%
  \BibitemOpen
  \bibfield  {author} {\bibinfo {author} {\bibfnamefont {P.}~\bibnamefont
  {{Virtanen}}}, \bibinfo {author} {\bibfnamefont {R.}~\bibnamefont
  {{Gommers}}}, \bibinfo {author} {\bibfnamefont {T.~E.}\ \bibnamefont
  {{Oliphant}}}, \bibinfo {author} {\bibfnamefont {M.}~\bibnamefont
  {{Haberland}}}, \bibinfo {author} {\bibfnamefont {T.}~\bibnamefont
  {{Reddy}}}, \bibinfo {author} {\bibfnamefont {D.}~\bibnamefont
  {{Cournapeau}}}, \bibinfo {author} {\bibfnamefont {E.}~\bibnamefont
  {{Burovski}}}, \bibinfo {author} {\bibfnamefont {P.}~\bibnamefont
  {{Peterson}}}, \bibinfo {author} {\bibfnamefont {W.}~\bibnamefont
  {{Weckesser}}}, \bibinfo {author} {\bibfnamefont {J.}~\bibnamefont
  {{Bright}}}, \bibinfo {author} {\bibfnamefont {S.~J.}\ \bibnamefont {{van der
  Walt}}}, \bibinfo {author} {\bibfnamefont {M.}~\bibnamefont {{Brett}}},
  \bibinfo {author} {\bibfnamefont {J.}~\bibnamefont {{Wilson}}}, \bibinfo
  {author} {\bibfnamefont {K.}~\bibnamefont {{Jarrod Millman}}}, \bibinfo
  {author} {\bibfnamefont {N.}~\bibnamefont {{Mayorov}}}, \bibinfo {author}
  {\bibfnamefont {A.~R.~J.}\ \bibnamefont {{Nelson}}}, \bibinfo {author}
  {\bibfnamefont {E.}~\bibnamefont {{Jones}}}, \bibinfo {author} {\bibfnamefont
  {R.}~\bibnamefont {{Kern}}}, \bibinfo {author} {\bibfnamefont
  {E.}~\bibnamefont {{Larson}}}, \bibinfo {author} {\bibfnamefont
  {C.}~\bibnamefont {{Carey}}}, \bibinfo {author} {\bibfnamefont
  {{\.I}.}~\bibnamefont {{Polat}}}, \bibinfo {author} {\bibfnamefont
  {Y.}~\bibnamefont {{Feng}}}, \bibinfo {author} {\bibfnamefont {E.~W.}\
  \bibnamefont {{Moore}}}, \bibinfo {author} {\bibfnamefont {J.}~\bibnamefont
  {{VanderPlas}}}, \bibinfo {author} {\bibfnamefont {D.}~\bibnamefont
  {{Laxalde}}}, \bibinfo {author} {\bibfnamefont {J.}~\bibnamefont
  {{Perktold}}}, \bibinfo {author} {\bibfnamefont {R.}~\bibnamefont
  {{Cimrman}}}, \bibinfo {author} {\bibfnamefont {I.}~\bibnamefont
  {{Henriksen}}}, \bibinfo {author} {\bibfnamefont {E.~A.}\ \bibnamefont
  {{Quintero}}}, \bibinfo {author} {\bibfnamefont {C.~R.}\ \bibnamefont
  {{Harris}}}, \bibinfo {author} {\bibfnamefont {A.~M.}\ \bibnamefont
  {{Archibald}}}, \bibinfo {author} {\bibfnamefont {A.~H.}\ \bibnamefont
  {{Ribeiro}}}, \bibinfo {author} {\bibfnamefont {F.}~\bibnamefont
  {{Pedregosa}}}, \bibinfo {author} {\bibfnamefont {P.}~\bibnamefont {{van
  Mulbregt}}}, \ and\ \bibinfo {author} {\bibnamefont {{SciPy 1.0
  Contributors}}},\ }\href {\doibase 10.1038/s41592-019-0686-2} {\bibfield
  {journal} {\bibinfo  {journal} {Nature Methods}\ }\textbf {\bibinfo {volume}
  {17}},\ \bibinfo {pages} {261} (\bibinfo {year} {2020})}\BibitemShut
  {NoStop}%
\bibitem [{\citenamefont {Bradbury}\ \emph {et~al.}(2018)\citenamefont
  {Bradbury}, \citenamefont {Frostig}, \citenamefont {Hawkins}, \citenamefont
  {Johnson}, \citenamefont {Leary}, \citenamefont {Maclaurin},\ and\
  \citenamefont {Wanderman-Milne}}]{jax2018github}%
  \BibitemOpen
  \bibfield  {author} {\bibinfo {author} {\bibfnamefont {J.}~\bibnamefont
  {Bradbury}}, \bibinfo {author} {\bibfnamefont {R.}~\bibnamefont {Frostig}},
  \bibinfo {author} {\bibfnamefont {P.}~\bibnamefont {Hawkins}}, \bibinfo
  {author} {\bibfnamefont {M.~J.}\ \bibnamefont {Johnson}}, \bibinfo {author}
  {\bibfnamefont {C.}~\bibnamefont {Leary}}, \bibinfo {author} {\bibfnamefont
  {D.}~\bibnamefont {Maclaurin}}, \ and\ \bibinfo {author} {\bibfnamefont
  {S.}~\bibnamefont {Wanderman-Milne}},\ }\href {http://github.com/google/jax}
  {\enquote {\bibinfo {title} {{JAX}: composable transformations of
  {P}ython+{N}um{P}y programs},}\ } (\bibinfo {year} {2018})\BibitemShut
  {NoStop}%
\bibitem [{\citenamefont {Duchi}\ \emph {et~al.}(2011)\citenamefont {Duchi},
  \citenamefont {Hazan},\ and\ \citenamefont {Singer}}]{duchi2011adaptive}%
  \BibitemOpen
  \bibfield  {author} {\bibinfo {author} {\bibfnamefont {J.}~\bibnamefont
  {Duchi}}, \bibinfo {author} {\bibfnamefont {E.}~\bibnamefont {Hazan}}, \ and\
  \bibinfo {author} {\bibfnamefont {Y.}~\bibnamefont {Singer}},\ }\href
  {https://dl.acm.org/doi/10.5555/1953048.2021068} {\bibfield  {journal}
  {\bibinfo  {journal} {Journal of Machine Learning Research}\ }\textbf
  {\bibinfo {volume} {12}},\ \bibinfo {pages} {2121} (\bibinfo {year}
  {2011})}\BibitemShut {NoStop}%
\bibitem [{\citenamefont {Childs}\ \emph {et~al.}(2021)\citenamefont {Childs},
  \citenamefont {Su}, \citenamefont {Tran}, \citenamefont {Wiebe},\ and\
  \citenamefont {Zhu}}]{Childs2019BError}%
  \BibitemOpen
  \bibfield  {author} {\bibinfo {author} {\bibfnamefont {A.~M.}\ \bibnamefont
  {Childs}}, \bibinfo {author} {\bibfnamefont {Y.}~\bibnamefont {Su}}, \bibinfo
  {author} {\bibfnamefont {M.~C.}\ \bibnamefont {Tran}}, \bibinfo {author}
  {\bibfnamefont {N.}~\bibnamefont {Wiebe}}, \ and\ \bibinfo {author}
  {\bibfnamefont {S.}~\bibnamefont {Zhu}},\ }\href {\doibase
  10.1103/PhysRevX.11.011020} {\bibfield  {journal} {\bibinfo  {journal}
  {Physical Review X}\ }\textbf {\bibinfo {volume} {11}},\ \bibinfo {pages}
  {011020} (\bibinfo {year} {2021})}\BibitemShut {NoStop}%
\bibitem [{\citenamefont {Childs}\ and\ \citenamefont
  {Wiebe}(2012)}]{Childs2012}%
  \BibitemOpen
  \bibfield  {author} {\bibinfo {author} {\bibfnamefont {A.~M.}\ \bibnamefont
  {Childs}}\ and\ \bibinfo {author} {\bibfnamefont {N.}~\bibnamefont {Wiebe}},\
  }\href {https://dl.acm.org/citation.cfm?id=2481570} {\bibfield  {journal}
  {\bibinfo  {journal} {Quantum Information {\&} Computation}\ }\textbf
  {\bibinfo {volume} {12}},\ \bibinfo {pages} {901} (\bibinfo {year}
  {2012})}\BibitemShut {NoStop}%
\bibitem [{\citenamefont {Gily{\'{e}}n}\ \emph {et~al.}(2019)\citenamefont
  {Gily{\'{e}}n}, \citenamefont {Su}, \citenamefont {Low},\ and\ \citenamefont
  {Wiebe}}]{Gilyen2019QuantumArithmetics}%
  \BibitemOpen
  \bibfield  {author} {\bibinfo {author} {\bibfnamefont {A.}~\bibnamefont
  {Gily{\'{e}}n}}, \bibinfo {author} {\bibfnamefont {Y.}~\bibnamefont {Su}},
  \bibinfo {author} {\bibfnamefont {G.~H.}\ \bibnamefont {Low}}, \ and\
  \bibinfo {author} {\bibfnamefont {N.}~\bibnamefont {Wiebe}},\ }in\ \href
  {\doibase 10.1145/3313276.3316366} {\emph {\bibinfo {booktitle} {Proceedings
  of the 51st Annual ACM SIGACT Symposium on Theory of Computing}}},\ \bibinfo
  {series and number} {STOC 2019}\ (\bibinfo  {publisher} {Association for
  Computing Machinery},\ \bibinfo {address} {New York, NY, USA},\ \bibinfo
  {year} {2019})\ pp.\ \bibinfo {pages} {193--204}\BibitemShut {NoStop}%
\bibitem [{\citenamefont {Szegedy}(2004)}]{Szegedy2004}%
  \BibitemOpen
  \bibfield  {author} {\bibinfo {author} {\bibfnamefont {M.}~\bibnamefont
  {Szegedy}},\ }in\ \href {\doibase 10.1109/FOCS.2004.53} {\emph {\bibinfo
  {booktitle} {45th Annual IEEE Symposium on Foundations of Computer
  Science}}}\ (\bibinfo  {publisher} {IEEE},\ \bibinfo {year} {2004})\ pp.\
  \bibinfo {pages} {32--41}\BibitemShut {NoStop}%
\bibitem [{\citenamefont {Sanders}\ \emph {et~al.}(2020)\citenamefont
  {Sanders}, \citenamefont {Berry}, \citenamefont {Costa}, \citenamefont
  {Tessler}, \citenamefont {Wiebe}, \citenamefont {Gidney}, \citenamefont
  {Neven},\ and\ \citenamefont
  {Babbush}}]{Sanders2020CompilationOptimizationB}%
  \BibitemOpen
  \bibfield  {author} {\bibinfo {author} {\bibfnamefont {Y.~R.}\ \bibnamefont
  {Sanders}}, \bibinfo {author} {\bibfnamefont {D.~W.}\ \bibnamefont {Berry}},
  \bibinfo {author} {\bibfnamefont {P.~C.}\ \bibnamefont {Costa}}, \bibinfo
  {author} {\bibfnamefont {L.~W.}\ \bibnamefont {Tessler}}, \bibinfo {author}
  {\bibfnamefont {N.}~\bibnamefont {Wiebe}}, \bibinfo {author} {\bibfnamefont
  {C.}~\bibnamefont {Gidney}}, \bibinfo {author} {\bibfnamefont
  {H.}~\bibnamefont {Neven}}, \ and\ \bibinfo {author} {\bibfnamefont
  {R.}~\bibnamefont {Babbush}},\ }\href {\doibase 10.1103/PRXQuantum.1.020312}
  {\bibfield  {journal} {\bibinfo  {journal} {PRX Quantum}\ }\textbf {\bibinfo
  {volume} {1}},\ \bibinfo {pages} {020312} (\bibinfo {year}
  {2020})}\BibitemShut {NoStop}%
\bibitem [{\citenamefont {Gidney}(2018)}]{GidneyAdder}%
  \BibitemOpen
  \bibfield  {author} {\bibinfo {author} {\bibfnamefont {C.}~\bibnamefont
  {Gidney}},\ }\href {\doibase 10.22331/q-2018-06-18-74} {\bibfield  {journal}
  {\bibinfo  {journal} {Quantum}\ }\textbf {\bibinfo {volume} {2}},\ \bibinfo
  {pages} {74} (\bibinfo {year} {2018})}\BibitemShut {NoStop}%
\bibitem [{\citenamefont {Szabo}\ and\ \citenamefont
  {Ostlund}(1996)}]{szabo1996modern}%
  \BibitemOpen
  \bibfield  {author} {\bibinfo {author} {\bibfnamefont {A.}~\bibnamefont
  {Szabo}}\ and\ \bibinfo {author} {\bibfnamefont {N.~S.}\ \bibnamefont
  {Ostlund}},\ }\href@noop {} {\emph {\bibinfo {title} {Modern quantum
  chemistry: introduction to advanced electronic structure theory}}}\ (\bibinfo
   {publisher} {Courier Corporation},\ \bibinfo {year} {1996})\BibitemShut
  {NoStop}%
\bibitem [{\citenamefont {Shavitt}\ and\ \citenamefont
  {Bartlett}(2009)}]{shavitt2009many}%
  \BibitemOpen
  \bibfield  {author} {\bibinfo {author} {\bibfnamefont {I.}~\bibnamefont
  {Shavitt}}\ and\ \bibinfo {author} {\bibfnamefont {R.~J.}\ \bibnamefont
  {Bartlett}},\ }\href@noop {} {\emph {\bibinfo {title} {Many-body methods in
  chemistry and physics: MBPT and coupled-cluster theory}}}\ (\bibinfo
  {publisher} {Cambridge University Press},\ \bibinfo {year}
  {2009})\BibitemShut {NoStop}%
\bibitem [{\citenamefont {Lee}\ and\ \citenamefont
  {Head-Gordon}(2018)}]{lee2018regularized}%
  \BibitemOpen
  \bibfield  {author} {\bibinfo {author} {\bibfnamefont {J.}~\bibnamefont
  {Lee}}\ and\ \bibinfo {author} {\bibfnamefont {M.}~\bibnamefont
  {Head-Gordon}},\ }\href {\doibase 10.1021/acs.jctc.8b00731} {\bibfield
  {journal} {\bibinfo  {journal} {Journal of Chemical Theory and Computation}\
  }\textbf {\bibinfo {volume} {14}},\ \bibinfo {pages} {5203} (\bibinfo {year}
  {2018})}\BibitemShut {NoStop}%
\bibitem [{\citenamefont {Raghavachari}\ \emph {et~al.}(1989)\citenamefont
  {Raghavachari}, \citenamefont {Trucks}, \citenamefont {Pople},\ and\
  \citenamefont {Head-Gordon}}]{raghavachari1989fifth}%
  \BibitemOpen
  \bibfield  {author} {\bibinfo {author} {\bibfnamefont {K.}~\bibnamefont
  {Raghavachari}}, \bibinfo {author} {\bibfnamefont {G.~W.}\ \bibnamefont
  {Trucks}}, \bibinfo {author} {\bibfnamefont {J.~A.}\ \bibnamefont {Pople}}, \
  and\ \bibinfo {author} {\bibfnamefont {M.}~\bibnamefont {Head-Gordon}},\
  }\href {\doibase S0009-2614(89)87395-6} {\bibfield  {journal} {\bibinfo
  {journal} {Chemical Physics Letters}\ }\textbf {\bibinfo {volume} {157}},\
  \bibinfo {pages} {479} (\bibinfo {year} {1989})}\BibitemShut {NoStop}%
\bibitem [{\citenamefont {Weigend}(2006)}]{weigend2006accurate}%
  \BibitemOpen
  \bibfield  {author} {\bibinfo {author} {\bibfnamefont {F.}~\bibnamefont
  {Weigend}},\ }\href {\doibase 10.1039/B515623H} {\bibfield  {journal}
  {\bibinfo  {journal} {Physical Chemistry Chemical Physics}\ }\textbf
  {\bibinfo {volume} {8}},\ \bibinfo {pages} {1057} (\bibinfo {year}
  {2006})}\BibitemShut {NoStop}%
\bibitem [{\citenamefont {Aquilante}\ \emph {et~al.}(2010)\citenamefont
  {Aquilante}, \citenamefont {De~Vico}, \citenamefont {Ferr{\'e}},
  \citenamefont {Ghigo}, \citenamefont {Malmqvist}, \citenamefont
  {Neogr{\'a}dy}, \citenamefont {Pedersen}, \citenamefont {Pito{\v{n}}{\'a}k},
  \citenamefont {Reiher}, \citenamefont {Roos}, \citenamefont
  {Serrano-Andr\'es}, \citenamefont {Urban}, \citenamefont {Veryazov},\ and\
  \citenamefont {Lindh}}]{aquilante2010molcas}%
  \BibitemOpen
  \bibfield  {author} {\bibinfo {author} {\bibfnamefont {F.}~\bibnamefont
  {Aquilante}}, \bibinfo {author} {\bibfnamefont {L.}~\bibnamefont {De~Vico}},
  \bibinfo {author} {\bibfnamefont {N.}~\bibnamefont {Ferr{\'e}}}, \bibinfo
  {author} {\bibfnamefont {G.}~\bibnamefont {Ghigo}}, \bibinfo {author}
  {\bibfnamefont {P.-{\aa}.}\ \bibnamefont {Malmqvist}}, \bibinfo {author}
  {\bibfnamefont {P.}~\bibnamefont {Neogr{\'a}dy}}, \bibinfo {author}
  {\bibfnamefont {T.~B.}\ \bibnamefont {Pedersen}}, \bibinfo {author}
  {\bibfnamefont {M.}~\bibnamefont {Pito{\v{n}}{\'a}k}}, \bibinfo {author}
  {\bibfnamefont {M.}~\bibnamefont {Reiher}}, \bibinfo {author} {\bibfnamefont
  {B.~O.}\ \bibnamefont {Roos}}, \bibinfo {author} {\bibfnamefont
  {L.}~\bibnamefont {Serrano-Andr\'es}}, \bibinfo {author} {\bibfnamefont
  {M.}~\bibnamefont {Urban}}, \bibinfo {author} {\bibfnamefont
  {V.}~\bibnamefont {Veryazov}}, \ and\ \bibinfo {author} {\bibfnamefont
  {R.}~\bibnamefont {Lindh}},\ }\href {\doibase 10.1002/jcc.21318} {\bibfield
  {journal} {\bibinfo  {journal} {Journal of Computational Chemistry}\ }\textbf
  {\bibinfo {volume} {31}},\ \bibinfo {pages} {224} (\bibinfo {year}
  {2010})}\BibitemShut {NoStop}%
\bibitem [{\citenamefont {Dunning}(1989)}]{Dunning1989}%
  \BibitemOpen
  \bibfield  {author} {\bibinfo {author} {\bibfnamefont {T.~H.}\ \bibnamefont
  {Dunning}},\ }\href {\doibase 10.1063/1.456153} {\bibfield  {journal}
  {\bibinfo  {journal} {The Journal of Chemical Physics}\ }\textbf {\bibinfo
  {volume} {90}},\ \bibinfo {pages} {1007} (\bibinfo {year}
  {1989})}\BibitemShut {NoStop}%
\bibitem [{\citenamefont {Edmiston}\ and\ \citenamefont
  {Ruedenberg}(1963)}]{edmiston1963localized}%
  \BibitemOpen
  \bibfield  {author} {\bibinfo {author} {\bibfnamefont {C.}~\bibnamefont
  {Edmiston}}\ and\ \bibinfo {author} {\bibfnamefont {K.}~\bibnamefont
  {Ruedenberg}},\ }\href {\doibase 10.1103/RevModPhys.35.457} {\bibfield
  {journal} {\bibinfo  {journal} {Reviews of Modern Physics}\ }\textbf
  {\bibinfo {volume} {35}},\ \bibinfo {pages} {457} (\bibinfo {year}
  {1963})}\BibitemShut {NoStop}%
\bibitem [{\citenamefont {Fowler}(2012)}]{Fowler2012aB}%
  \BibitemOpen
  \bibfield  {author} {\bibinfo {author} {\bibfnamefont {A.~G.}\ \bibnamefont
  {Fowler}},\ }\href {http://arxiv.org/abs/1210.4626} {\bibfield  {journal}
  {\bibinfo  {journal} {arXiv:1210.4626}\ } (\bibinfo {year}
  {2012})}\BibitemShut {NoStop}%
\bibitem [{\citenamefont {Gidney}\ and\ \citenamefont
  {Fowler}(2019{\natexlab{b}})}]{Gidney2019aB}%
  \BibitemOpen
  \bibfield  {author} {\bibinfo {author} {\bibfnamefont {C.}~\bibnamefont
  {Gidney}}\ and\ \bibinfo {author} {\bibfnamefont {A.~G.}\ \bibnamefont
  {Fowler}},\ }\href {http://arxiv.org/abs/1905.08916} {\bibfield  {journal}
  {\bibinfo  {journal} {arXiv:1905.08916}\ } (\bibinfo {year}
  {2019}{\natexlab{b}})}\BibitemShut {NoStop}%
\bibitem [{\citenamefont {Litinski}(2019{\natexlab{b}})}]{Litinski2019B}%
  \BibitemOpen
  \bibfield  {author} {\bibinfo {author} {\bibfnamefont {D.}~\bibnamefont
  {Litinski}},\ }\href {\doibase 10.22331/q-2019-12-02-205} {\bibfield
  {journal} {\bibinfo  {journal} {{Quantum}}\ }\textbf {\bibinfo {volume}
  {3}},\ \bibinfo {pages} {205} (\bibinfo {year}
  {2019}{\natexlab{b}})}\BibitemShut {NoStop}%
\bibitem [{\citenamefont {Gidney}\ and\ \citenamefont
  {Eker{\aa}}(2019)}]{Gidney2019bB}%
  \BibitemOpen
  \bibfield  {author} {\bibinfo {author} {\bibfnamefont {C.}~\bibnamefont
  {Gidney}}\ and\ \bibinfo {author} {\bibfnamefont {M.}~\bibnamefont
  {Eker{\aa}}},\ }\href {http://arxiv.org/abs/1905.09749} {\bibfield  {journal}
  {\bibinfo  {journal} {arXiv:1905.09749}\ } (\bibinfo {year}
  {2019})}\BibitemShut {NoStop}%
\bibitem [{\citenamefont {Childs}\ and\ \citenamefont
  {Su}(2019)}]{Childs2019c}%
  \BibitemOpen
  \bibfield  {author} {\bibinfo {author} {\bibfnamefont {A.~M.}\ \bibnamefont
  {Childs}}\ and\ \bibinfo {author} {\bibfnamefont {Y.}~\bibnamefont {Su}},\
  }\href {\doibase 10.1103/PhysRevLett.123.050503} {\bibfield  {journal}
  {\bibinfo  {journal} {Physical Review Letters}\ }\textbf {\bibinfo {volume}
  {123}},\ \bibinfo {pages} {050503} (\bibinfo {year} {2019})}\BibitemShut
  {NoStop}%
\bibitem [{\citenamefont {Xing}\ \emph {et~al.}(2020)\citenamefont {Xing},
  \citenamefont {Huang},\ and\ \citenamefont {Chow}}]{xing2020linear}%
  \BibitemOpen
  \bibfield  {author} {\bibinfo {author} {\bibfnamefont {X.}~\bibnamefont
  {Xing}}, \bibinfo {author} {\bibfnamefont {H.}~\bibnamefont {Huang}}, \ and\
  \bibinfo {author} {\bibfnamefont {E.}~\bibnamefont {Chow}},\ }\href {\doibase
  10.1063/5.0010732} {\bibfield  {journal} {\bibinfo  {journal} {The Journal of
  Chemical Physics}\ }\textbf {\bibinfo {volume} {153}},\ \bibinfo {pages}
  {084119} (\bibinfo {year} {2020})}\BibitemShut {NoStop}%
\bibitem [{\citenamefont {Lin}\ and\ \citenamefont {Tong}(2020)}]{Lin2020b}%
  \BibitemOpen
  \bibfield  {author} {\bibinfo {author} {\bibfnamefont {L.}~\bibnamefont
  {Lin}}\ and\ \bibinfo {author} {\bibfnamefont {Y.}~\bibnamefont {Tong}},\
  }\href {\doibase 10.22331/q-2020-12-14-372} {\bibfield  {journal} {\bibinfo
  {journal} {Quantum}\ }\textbf {\bibinfo {volume} {4}},\ \bibinfo {pages}
  {372} (\bibinfo {year} {2020})}\BibitemShut {NoStop}%
\bibitem [{\citenamefont {McClean}\ \emph
  {et~al.}(2014{\natexlab{b}})\citenamefont {McClean}, \citenamefont {Babbush},
  \citenamefont {Love},\ and\ \citenamefont {Aspuru-Guzik}}]{McClean2014b}%
  \BibitemOpen
  \bibfield  {author} {\bibinfo {author} {\bibfnamefont {J.}~\bibnamefont
  {McClean}}, \bibinfo {author} {\bibfnamefont {R.}~\bibnamefont {Babbush}},
  \bibinfo {author} {\bibfnamefont {P.}~\bibnamefont {Love}}, \ and\ \bibinfo
  {author} {\bibfnamefont {A.}~\bibnamefont {Aspuru-Guzik}},\ }\href {\doibase
  10.1021/jz501649m} {\bibfield  {journal} {\bibinfo  {journal} {The Journal of
  Physical Chemistry Letters}\ }\textbf {\bibinfo {volume} {5}},\ \bibinfo
  {pages} {4368 } (\bibinfo {year} {2014}{\natexlab{b}})}\BibitemShut {NoStop}%
\bibitem [{\citenamefont {Verstraete}\ \emph {et~al.}(2009)\citenamefont
  {Verstraete}, \citenamefont {Cirac},\ and\ \citenamefont
  {Latorre}}]{Verstraete2009}%
  \BibitemOpen
  \bibfield  {author} {\bibinfo {author} {\bibfnamefont {F.}~\bibnamefont
  {Verstraete}}, \bibinfo {author} {\bibfnamefont {J.~I.}\ \bibnamefont
  {Cirac}}, \ and\ \bibinfo {author} {\bibfnamefont {J.~I.}\ \bibnamefont
  {Latorre}},\ }\href {\doibase 10.1103/PhysRevA.79.032316} {\bibfield
  {journal} {\bibinfo  {journal} {Physical Review A}\ }\textbf {\bibinfo
  {volume} {79}},\ \bibinfo {pages} {32316} (\bibinfo {year}
  {2009})}\BibitemShut {NoStop}%
\bibitem [{\citenamefont {Low}\ and\ \citenamefont {Chuang}(2017)}]{Low2017}%
  \BibitemOpen
  \bibfield  {author} {\bibinfo {author} {\bibfnamefont {G.~H.}\ \bibnamefont
  {Low}}\ and\ \bibinfo {author} {\bibfnamefont {I.~L.}\ \bibnamefont
  {Chuang}},\ }\href {\doibase 10.1103/PhysRevLett.118.010501} {\bibfield
  {journal} {\bibinfo  {journal} {Physical Review Letters}\ }\textbf {\bibinfo
  {volume} {118}},\ \bibinfo {pages} {010501} (\bibinfo {year}
  {2017})}\BibitemShut {NoStop}%
\bibitem [{\citenamefont {Babbush}\ \emph
  {et~al.}(2019{\natexlab{b}})\citenamefont {Babbush}, \citenamefont {Berry},\
  and\ \citenamefont {Neven}}]{BabbushSYK}%
  \BibitemOpen
  \bibfield  {author} {\bibinfo {author} {\bibfnamefont {R.}~\bibnamefont
  {Babbush}}, \bibinfo {author} {\bibfnamefont {D.~W.}\ \bibnamefont {Berry}},
  \ and\ \bibinfo {author} {\bibfnamefont {H.}~\bibnamefont {Neven}},\ }\href
  {\doibase 10.1103/PhysRevA.99.040301} {\bibfield  {journal} {\bibinfo
  {journal} {Physical Review A}\ }\textbf {\bibinfo {volume} {99}},\ \bibinfo
  {pages} {040301} (\bibinfo {year} {2019}{\natexlab{b}})}\BibitemShut
  {NoStop}%
\bibitem [{\citenamefont {Berry}\ \emph {et~al.}(2020)\citenamefont {Berry},
  \citenamefont {Childs}, \citenamefont {Su}, \citenamefont {Wang},\ and\
  \citenamefont {Wiebe}}]{Berry2020}%
  \BibitemOpen
  \bibfield  {author} {\bibinfo {author} {\bibfnamefont {D.~W.}\ \bibnamefont
  {Berry}}, \bibinfo {author} {\bibfnamefont {A.~M.}\ \bibnamefont {Childs}},
  \bibinfo {author} {\bibfnamefont {Y.}~\bibnamefont {Su}}, \bibinfo {author}
  {\bibfnamefont {X.}~\bibnamefont {Wang}}, \ and\ \bibinfo {author}
  {\bibfnamefont {N.}~\bibnamefont {Wiebe}},\ }\href {\doibase
  10.22331/q-2020-04-20-254} {\bibfield  {journal} {\bibinfo  {journal}
  {Quantum}\ }\textbf {\bibinfo {volume} {4}},\ \bibinfo {pages} {254}
  (\bibinfo {year} {2020})}\BibitemShut {NoStop}%
\bibitem [{\citenamefont {Hodges}\ and\ \citenamefont
  {Lehmann}(1963)}]{hodges1963}%
  \BibitemOpen
  \bibfield  {author} {\bibinfo {author} {\bibfnamefont {J.~L.}\ \bibnamefont
  {Hodges}}\ and\ \bibinfo {author} {\bibfnamefont {E.~L.}\ \bibnamefont
  {Lehmann}},\ }\href {\doibase 10.1214/aoms/1177704172} {\bibfield  {journal}
  {\bibinfo  {journal} {Annals of Mathematical Statistics}\ }\textbf {\bibinfo
  {volume} {34}},\ \bibinfo {pages} {598} (\bibinfo {year} {1963})}\BibitemShut
  {NoStop}%
\bibitem [{\citenamefont {Lehmann}(1999)}]{lehmann99}%
  \BibitemOpen
  \bibfield  {author} {\bibinfo {author} {\bibfnamefont {E.~L.}\ \bibnamefont
  {Lehmann}},\ }\href@noop {} {\emph {\bibinfo {title} {Elements of
  Large-Sample Theory}}}\ (\bibinfo  {publisher} {Springer-Verlag},\ \bibinfo
  {year} {1999})\BibitemShut {NoStop}%
\end{thebibliography}%

\appendix

\section{The ``sparse'' algorithm of Berry \emph{et al.}}
\label{app:sparse}

\subsection{Representing the sparse Hamiltonian as a linear combination of unitaries}

Here we review the method of \cite{Berry2019B}, with some further optimizations of the method.
The first step in simulating a Hamiltonian using qubitization is to represent it as a linear combination of unitaries. How one chooses this linear combination of unitaries has significant ramifications for the $\lambda$-factor which will scale the cost of the qubitized simulation. We start by expressing the electronic Hamiltonian in an arbitrary second-quantized basis as in \eq{full}. We can map these operators to qubits as
\begin{equation}
T = \frac{1}{2}\sum_{\sigma\in\{\uparrow,\downarrow\}} \sum_{p,q=1}^{N/2} T_{p q} \left(a_{p,\sigma}^\dagger a_{q,\sigma} + a_{q,\sigma}^\dagger a_{p,\sigma}\right) = \frac{1}{2} \sum_{\sigma\in\{\uparrow,\downarrow\}} \sum_{p,q=1}^{N/2} T_{p q} Q'_{pq\sigma}
\end{equation}
and
\begin{equation}
\label{eq:V_expanded}
V = \frac 18 \sum_{\alpha,\beta\in \{\uparrow, \downarrow\}} \sum_{p,q,r,s=1}^{N/2} V_{pqrs}\left(a^\dagger_{p,\alpha} a_{q,\alpha}+a^\dagger_{q,\alpha} a_{p,\alpha}\right)\left(a^\dagger_{r,\beta} a_{s,\beta}+a^\dagger_{s,\beta} a_{r,\beta}\right) =  \frac 18 \sum_{\alpha,\beta\in \{\uparrow, \downarrow\}} \sum_{p,q,r,s=1}^{N/2} V_{pqrs} Q'_{pq\alpha} Q'_{rs\beta} ,
\end{equation}
where the Jordan-Wigner transform that is used is expressed as
\begin{equation}
\label{eq:jw}
Q'_{pq\sigma}=\begin{cases}
    X_{p,\sigma} \vec Z X_{q,\sigma}, & p< q, \\
    Y_{p,\sigma} \vec Z Y_{q,\sigma}, & p> q, \\
\openone - Z_{p,\sigma} , & p=q,
\end{cases} \qquad 
a^\dagger_{p,\sigma} a_{q,\sigma} + a^\dagger_{q,\sigma} a_{p, \sigma} \mapsto \frac{X_{p,\sigma} \vec Z X_{q,\sigma} + Y_{p,\sigma} \vec Z Y_{q,\sigma}}{2}, \qquad
a^\dagger_{p,\sigma} a_{p,\sigma} \mapsto \frac{\openone - Z_{p,\sigma}}{2},
\end{equation}
where $X$, $Y$ and $Z$ are the Pauli operators, the subscripts indicate the qubits these operators act on, and $A_p \vec Z A_q$ is shorthand for $A_p Z_{p+1} \cdots Z_{q-1} A_q$. Note that we get a factor of $1/8$ in front of \eq{V_expanded} from multiplying the original factor of $1/2$ by a factor of $1/4$ that comes from expanding the $pq$ and $rs$ terms as the sum of their Hermitian conjugate for each index.
Note that we have made a correction of a factor of 2 from $Q_{pq\sigma}$ as defined in \cite{Berry2019B}.

An improvement we can make is to remove the identity from the case where $p=q$, so that $Q_{pq\sigma}$ has the same weightings with on-diagonal and off-diagonal terms, as
\begin{equation}\label{eq:jw2}
    Q_{pq\sigma}=\begin{cases}
    X_{p,\sigma} \vec Z X_{q,\sigma}, & p< q, \\
    Y_{p,\sigma} \vec Z Y_{q,\sigma}, & p> q, \\
- Z_{p,\sigma} , & p=q.
\end{cases}
\end{equation}
Then the expansion of $T$ can be written as
\begin{equation}
T = \frac 12 \sum_{\sigma\in\{\uparrow,\downarrow\}} \sum_{p,q=1}^{N/2} T_{p q} Q_{pq\sigma} + \sum_{p=1}^{N/2} T_{p p} \openone,
\end{equation}
and the expansion of $V$ can then be written as
\begin{equation}
V = \frac 18 \sum_{\alpha,\beta\in \{\uparrow, \downarrow\}} \left(\sum_{p,q,r,s=1}^{N/2} V_{pqrs} Q_{pq\alpha} Q_{rs\beta} + \sum_{p,r,s=1}^{N/2} V_{pprs} Q_{rs\beta}+ \sum_{p,q,r=1}^{N/2} V_{pqrr} Q_{pq\alpha}+ \sum_{p,r=1}^{N/2} V_{pprr}\openone \right).
\end{equation}
Here the middle two terms correspond to one-body terms that can be combined with $T$, and the fourth is a constant offset that can be omitted.
We can therefore write the Hamiltonian in terms of a new $T'$ and $V'$, given by
\begin{align}
T' &= \frac 12 \sum_{\sigma\in\{\uparrow,\downarrow\}} \sum_{p,q=1}^{N/2} T'_{p q} Q_{pq\sigma}\, , \\
V' &= \frac 18 \sum_{\alpha,\beta\in \{\uparrow, \downarrow\}} \sum_{p,q,r,s=1}^{N/2} V_{pqrs} Q_{pq\alpha} Q_{rs\beta}\, ,
\end{align}
with
\begin{equation}\label{eq:Tpdef}
T'_{pq} = T_{pq} + \sum_{r=1}^{N/2} V_{pqrr}.
\end{equation}
Then $H=T'+V'$ plus a term proportional to the identity, which can be omitted because it gives a constant shift to the eigenvalues.

Because the operator $Q_{pq\sigma}$ as well as products $Q_{pq\alpha} Q_{rs\beta}$ are all unitary operators we now see that $H = T' + V'$ is a linear combination of unitaries. In this representation the associated $\lambda$ values are
\begin{align}
    \label{eq:lambda_V}
    \lambda = \lambda_T + \lambda_V \qquad \qquad
    \lambda_T = \sum_{p,q=1}^{N/2} \left | T_{pq} + \sum_{r=1}^{N/2} V_{pqrr} \right |,
    \qquad \qquad
    \lambda_V = \frac 12  \sum_{p,q,r,s=1}^{N/2} \left | V_{pqrs} \right|.
\end{align}
For $T'$ the multiplying the factor of $1/2$ cancels with a factor of 2 for summing the spin degree of freedom.
The factor of $1/2$ in front of the $V$ term comes from multiplying the factor of 1/8 by 4 for summing the spin degrees of freedom.

Note that these $\lambda$ definitions differ from those given in \cite{Berry2019B}. At first glance $\lambda_V$ appears to be a factor of 8 smaller. The first reason for this is because we deviated from the convention of absorbing the factor of 1/2 into $V$ (since this is just a difference in convention, it is consistent with prior work). However, there is roughly another factor of 4 difference that comes in because the work of \cite{Berry2019B} accidentally left out a factor of 1/2 in \eq{jw}. This leads to a $\lambda_V$ that is reduced by a factor of 4. For the Reiher Hamiltonian \cite{Reiher2017}, we previously reported 
$\lambda_T = 1{,}490 \, {\rm  a.u.}$, $\lambda_V = 8{,}373 \, {\rm a.u.}$ so
$\lambda=9{,}863\,{\rm a.u.}$ with a truncation threshold of $2\times10^{-4}$. 
This should be updated to
$\lambda_T = 90 \, {\rm  a.u.}$, $\lambda_V = 2,045 \, {\rm a.u.}$ so
$\lambda=2{,}135\,{\rm a.u.}$\ with a truncation threshold of $7.5\times10^{-5}$.
For the Li Hamiltonian \cite{Li2019} we previously gave $\lambda_T = 3{,}446 \, {\rm  a.u.}$, $\lambda_V = 4{,}168 \, {\rm a.u.}$ so
$\lambda=7{,}614\,{\rm a.u.}$\ with a truncation threshold of $1\times10^{-4}$.
This should be updated to $\lambda_T = 561 \, {\rm  a.u.}$, $\lambda_V = 986 \, {\rm a.u.}$ so
$\lambda=1{,}547\,{\rm a.u.}$\ with a truncation threshold of $3.5\times10^{-5}$. The truncation threshold became tighter in this work because the metric we chose (i.e., CCSD(T) correlation energy error) is more conservative than what was used in our previous work \cite{Berry2019B}.

\begin{table*}
\begin{tabular}{|c|c|c|c|}
\hline
threshold & $d$  & $\lambda$ & CCSD(T) error (m$E_h$) \\ \hline
0.001 & 122980 & 1445.0 & -190.61 \\ \hline
0.0005 & 233787 & 1736.8 & -16.76 \\ \hline
0.00025 & 391732 & 1954.4 & 0.84 \\ \hline
0.0002 & 449699 & 2005.4 & 2.04 \\ \hline
1.00$\times10^{-4}$ & 633943 & 2110.5 & 0.73 \\ \hline
8.75$\times10^{-5}$ & 668079 & 2123.2 & 0.63 \\ \hline
\color{blue}7.50$\times10^{-5}$ & \color{blue}705831 & \color{blue}2135.3 & \color{blue}0.33 \\ \hline
5.00$\times10^{-5}$ & 796197 & 2157.6 & 0.22 \\ \hline
2.50$\times10^{-5}$ & 916649 & 2175.2 & 0.05 \\ \hline
1.00$\times10^{-5}$ & 1014792 & 2181.9 & 0.0027 \\ \hline
5.00$\times10^{-6}$ & 1055829 & 2183.2 & 0.0003 \\ \hline
2.50$\times10^{-6}$ & 1078952 & 2183.5 & 0.0011 \\ \hline
1.00$\times10^{-6}$ & 1094260 & 2183.6 & 0.0004 \\ \hline
\end{tabular}
\caption[Sparse algorithm overheads for Reiher FeMoCo Hamiltonian]{Sparse method data for the Reiher Hamiltonian. Here $d$ is the number of permutation-unique non-zero elements above a given threshold, and $\lambda_T$ is 90.4 Hartree. The entry in blue corresponds to the threshold used in our resource estimates.}
\label{tab:sparseR}
\end{table*}

\begin{table*}
\begin{tabular}{|c|c|c|c|}
\hline
threshold & $d$ & $\lambda$ & CCSD(T) error (m$E_h$) \\ \hline
0.001 & 45201 & 1345.9 & -637.18 \\ \hline
0.0005 & 78643 & 1403.5 & -239.20 \\ \hline
0.00025 & 131232 & 1452.8 & -64.44 \\ \hline
0.0001 & 239085 & 1504.9 & -8.05 \\ \hline
5.0$\times10^{-5}$ & 359942 & 1534.7 & -1.49 \\ \hline
4.5$\times10^{-5}$ & 382082 & 1538.6 & -1.18 \\ \hline
4.0$\times10^{-5}$ & 408391 & 1542.7 & -0.88 \\ \hline
\color{blue}3.5$\times10^{-5}$ & \color{blue}440501 & \color{blue}1547.3 & \color{blue}-0.47 \\ \hline
3.0$\times10^{-5}$ & 480468 & 1552.2 & -0.26 \\ \hline
2.5$\times10^{-5}$ & 532212 & 1557.6 & -0.16 \\ \hline
1.0$\times10^{-5}$ & 881193 & 1579.2 & 0.03 \\ \hline
5.0$\times10^{-6}$ & 1259007 & 1589.8 & 0.01 \\ \hline
2.5$\times10^{-6}$ & 1722770 & 1596.4 & 0.01 \\ \hline
1.0$\times10^{-6}$ & 2410637 & 1600.9 & 0.005 \\ \hline
\end{tabular}
\caption[Sparse algorithm overheads for Li FeMoCo Hamiltonian]{Sparse method data for the Li Hamiltonian.
Here $d$ is the number of permutation-unique non-zero elements above a given threshold, and
$\lambda_T$ is 561.5 Hartree. The entry in blue corresponds to the threshold used in our resource estimates.}
\label{tab:sparseL}
\end{table*}

\subsection{The cost of qubitization of the sparse chemistry Hamiltonian}

The state to be prepared is similar to that in Eq.~(48) of \cite{Berry2019B}, except $T_{pq}$ is replaced with $T'_{pq}$ and $V_{pqrs}$ is replaced with $V_{pqrs}/2$ (due to the factor of 2 in the definition of $V$), so the state to be prepared is
\begin{equation}\label{eq:sparsestate}
\ket{0}\ket{+}\ket{0}\ket{0}\! \sum_{\sigma\in\{\uparrow,\downarrow\}}\sum_{p,q=1}^{N/2} \sqrt{\frac{|T'_{pq}|}\lambda} \ket{\theta^T_{pq}}\ket{p,q,\sigma}\ket{0,0,0} + \ket{1}\ket{+}\ket{+}\ket{+}\!\sum_{\alpha,\beta\in \{\uparrow, \downarrow\}}\sum_{p,q,r,s=1}^{N/2}\sqrt{\frac{|\widetilde V_{pqrs}|}{2\lambda}} \ket{\theta^V_{pqrs}}\ket{p,q,\alpha}\ket{r,s,\beta} .
\end{equation}
This state can be prepared as described in \cite{Berry2019B}, using controlled swaps to generate the symmetries of the state.

\begin{figure}[tbh]
\centering
\includegraphics[scale=0.925]{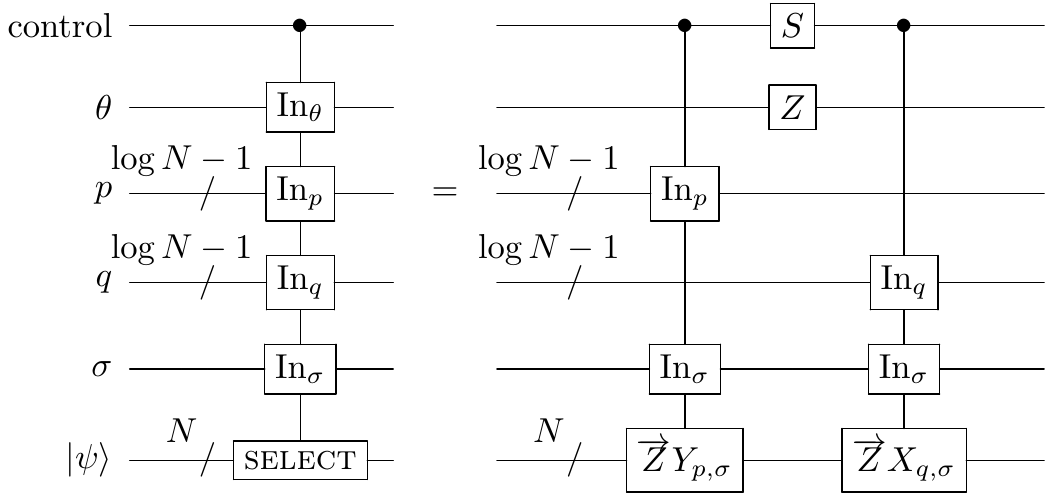}
\caption{\label{fig:selecth} The circuit needed to perform a controlled $\textsc{select}$ operation. 
This is similar to that in \cite{Berry2019B}, except it is not necessary to perform the equality test $p=q$.
The unitaries labeled as $\protect\overrightarrow{Z} A_j$ apply the operation $Z_0 \cdots Z_{j-1} A_j$ to the target register, depending on the value from the input register, using the technique shown in Figure 9 of \cite{BabbushSpectraB}.
The sign is shown as being obtained via a $Z$ gate on the sign qubit, but in practice this would be obtained as part of the state preparation.
}
\end{figure}

Because we are here not including the identity in $Q_{pq\sigma}$, the controlled unitaries can be performed in a simpler way than shown in Figure 1 of \cite{Berry2019B}.
In that work there are inequality tests between the $p$ and $q$ registers which are needed to produce the sum of the identity and $Z$ operations.
When the identity is not included, then the circuit needed simplifies to just two applications of the circuit for Majorana operators shown in Figure 9 of \cite{BabbushSpectraB} and Figure 1 of \cite{BabbushSYK}.
The simplified circuit is shown in \fig{selecth}.
When the \sel\ is controlled as shown, then the complexity is $2(N-1)$.
If it does not need to be controlled, then the complexity is only $2(N-2)$.
We will require only one of these two \sel\ operations to be controlled, because for the case of the one-body term in the Hamiltonian only one \sel\ should be applied, so one of the \sel\ needs to be controlled on the qubit flagging one- and two-body terms in the Hamiltonian.
The total complexity of these \sel\ operations is therefore $4N-6$.

In the sparse simulation method, the relevant parameters are the number of orbitals $N$, the $\lambda$ value, and the number of unique nonzero entries $d$.
If we are allowing error in the energy due to the state preparation of $\epsilon_{\textsc{prep}}$,
the output size for the keep probabilities for the QROM is
\begin{equation}
\zetabits = \left\lceil \log\left(\frac{\lambda}{2\epsilon_{\textsc{prep}}}\right)\right\rceil .
\end{equation}
There are eight registers of size $n_N=\lceil \log  (N/2) \rceil$, because the sparse preparation scheme needs to output ind values and alt values of $p,q,r,s$.
There are also $2$ qubits needed for the two output values of $\theta$ (one for the ind and one for the alt values of $p,q,r,s$), as well as $2$ qubits used for ind and alt values of the first register which distinguishes between $T$ and $V$.
As a result the QROM output size is
\begin{equation}
\wid = \zetabits +8n_N + 4.
\end{equation}
The cost of the preparation is then
\begin{equation}
\lceil\dm/\chunk_1\rceil+\wid(\chunk_1-1)
\end{equation}
and of the inverse preparation is
\begin{equation}
\lceil\dm/\chunk_2\rceil+\chunk_2 .
\end{equation}

To begin the state preparation, we need to prepare an equal superposition state over $d$ basis states.
Given that $2^\factor$ is a factor of $d$, the procedure and its costs are as follows.
\begin{enumerate}
\item Perform an inequality test on $\lceil \log d\rceil-\factor$ bits with $\lceil \log d\rceil-\factor-1$ Toffolis.
\item Rotate an ancilla qubit to obtain overall amplitude for success of $1/2$ using $\rotprec$ bits of precision. This has cost $\rotprec-3$ Toffolis (since the rotation angle is given classically).
\item Reflect on success for both, which may be performed via a controlled phase gate (no Toffolis).
\item Invert the rotation with cost $\rotprec-3$.
\item Invert the inequality test, which may be performed with Cliffords provided the first inequality test was performed using the out-of-place adder.
\item Perform a reflection on $\lceil \log d\rceil-\factor+1$ qubits with cost $\lceil \log d\rceil-\factor-1$.
\item Perform the inequality test again, with cost $\lceil \log d\rceil-\factor-1$.
\end{enumerate}
The total cost is then $3\lceil \log d\rceil-3\factor+2\rotprec-9$.  This is a cost paid both for the preparation and for the inverse preparation. 

Other minor Toffoli costs are as follows.
In the following we can use extra ancillas to save cost, because a large number of ancillas were used for the QROM, and can be reused here without increasing the maximum number of ancillas used.
\begin{enumerate}
\item Perform \sel\ as shown in \fig{selecth} twice, with only one being controlled.
As discussed above, this has complexity $4N-6$.
\item The state preparation needs an inequality test on $\zetabits$ qubits, as well as controlled swaps.
The cost of the inequality test on $\zetabits$ qubits is $\zetabits$, and by performing an inequality test with the out-of-place adder we can invert it in the inverse preparation with no additional Toffoli cost.
The controlled swaps are on $4n_N+1$ qubits.
Here the $4\lceil\log(N/2)\rceil$ are for the values of $p$, $q$, $r$, and $s$, and the $+1$ is for the qubit which distinguished between the one- and two-electron terms.
It is not necessary to perform a controlled swap on the sign qubits output, because the correct phase can be applied with Clifford gates.
It is possible to invert the controlled swaps in the inverse preparation with Cliffords.
The method is to copy all values being swapped before they are swapped.
Then to invert the controlled swap, perform measurements on the swapped values in the $X$ basis.
We can perform phase fixups using controlled-phase operations, where the control is the control qubit for the controlled swaps, and the targets are the copies of the registers.
That means we can eliminate the non-Clifford cost of the inverse preparation, giving a Toffoli cost of $\zetabits+4n_N+1$.
\item The controlled swaps used to generate the symmetries have a cost of $4n_N$.
Again these controlled swaps can be inverted for the inverse preparation with measurements and Clifford gates.
\item A reflection on the ancilla is needed as well.
The qubits that need to be reflected on are the $\lceil \log d \rceil$ qubits needed for preparing the state, the $\zetabits$ qubits used for the equal superposition state in the coherent alias sampling, the three qubits that are used for controlled swaps to generate the symmetries of the state, and the ancilla qubit that is rotated to produce the equal superposition state.
That gives a Toffoli cost of $\lceil \log d\rceil+\zetabits+2$.
\item For each step one more Toffoli is needed for the unary iteration used for the phase estimation, and one more Toffoli is needed to make the reflection controlled.
\end{enumerate}

Adding all these minor costs together gives
\begin{align}
&2(3\lceil\log d\rceil - 3\factor+2\rotprec-9) + (4N-6) + (\zetabits + 4n_N+1)+4n_N+\lceil\log d\rceil+\zetabits+4 \nn
&= 4N + 8n_N + 2\zetabits + 7\lceil\log d\rceil - 6\factor +4\rotprec - 19.
\end{align}
The total cost for a single step is then
\begin{equation}\label{eq:sparsetoffoli}
\left\lceil\frac{\dm}{\chunk_1}\right\rceil+\wid(\chunk_1-1) + \left\lceil\frac{\dm}{\chunk_2}\right\rceil+\chunk_2 + 4N + 8n_N + 2\zetabits + 7\lceil\log d\rceil - 6\factor +4\rotprec - 19 ,
\end{equation}
with $\wid = \zetabits +8n_N + 4$, $n_N=\lceil\log(N/2)\rceil$, and $\factor$ an integer such that $2^\factor$ is a factor of $d$.

The logical qubits used are as follows.
\begin{enumerate}
    \item The control register for the phase estimation uses $\lceil\log(\mathcal{I}+1)\rceil$ qubits, and there are $\lceil\log(\mathcal{I}+1)\rceil-1$ qubits for the unary iteration.
    \item The system uses $N$ qubits.
    \item The QROM uses a state with $\lceil \log d \rceil$ qubits.
    \item A qubit is needed to flag success of the equal superposition state preparation.
    \item An ancilla qubit is rotated in the preparation of the equal superposition state.
    \item The phase gradient state uses $\rotprec$ qubits.
    \item The equal superposition state used for the coherent alias sampling, which has $\zetabits$ qubits.
    \item The QROM uses qubits (including the output) $m\chunk_1+\lceil \log(d/\chunk_1)\rceil$.
\end{enumerate}
That gives a total number of logical qubits
\begin{equation}\label{eq:sparsequbits}
2\lceil\log(\mathcal{I}+1)\rceil + N + \lceil \log d \rceil + \rotprec + \zetabits + m\chunk_1+\lceil \log(d/\chunk_1)\rceil+1,
\end{equation}
with $\wid = \zetabits +8n_N + 4$.

\subsection{Counting the number of permutation-unique elements}
The two-electron integral tensor $V_{pqrs}$ exhibits a 8-fold permutational symmetry:
\begin{equation}
V_{pqrs}
=
V_{pqsr}
=
V_{qprs}
=
V_{qpsr}
=
V_{rspq}
=
V_{rsqp}
=
V_{srpq}
=
V_{srqp}
\end{equation}
To evaluate the cost of the sparse method presented above, 
we need to count the number of permutation-unique elements above a given truncation threshold.
We provide more details about how this counting was performed so that others can easily reproduce the numerical data
presented here. 
We note that the number of permutation-unique elements in $V_{pqrs}$ is given as \cite{szabo1996modern}
\begin{equation}
\frac18
\left(\frac N2\right)
\left(\frac N2 + 1\right)
\left(\left(\frac N2\right)^2 + \frac N2 + 2\right) .
\end{equation}
This expression can be obtained by considering the following four different classes of $V_{pqrs}$:
\begin{enumerate}
\item
$p$, $q$, $r$, and $s$ are all unique indices.
We loop over a total of $\binom{N/2}{4}$ unique combination of quartets $(p,q,r,s)$ and
count $V_{pqrs}$, $V_{prqs}$, and $V_{psrq}$ in this category.
\item
Two indices are redundant (i.e., only three unique indices).
We loop over a total of $\binom{N/2}{3}$ unique combination of triplets $(p,q,r)$
and count
$V_{ppqr}$,
$V_{pqpr}$,
$V_{qqpr}$,
$V_{qrpq}$,
$V_{rrpq}$, and
$V_{rpqr}$
in this category.
\item
Three indices are redundant (i..e, only two unique indices).
We loop over a total of $\binom{N/2}{2}$ unique combination of doublets $(p,q)$
and count
$V_{ppqq}$,
$V_{pqpq}$,
$V_{pppq}$, and
$V_{qqqp}$ here.
\item
All four indices are redundant. 
We loop over $p$ and count $V_{pppp}$ in this category.
\end{enumerate}
In order to obtain $d$, we 
enumerate each category
and count
the number of elements above a threshold.
Adding $(1/2)(N/2)(N/2+1)$ to account for the symmetry-unique part of the one-body operator to this directly gives us the numerical data in \tab{sparseR} and \tab{sparseL}.

\subsection{Numerical data for hydrogen chains and \texorpdfstring{\ce{H4}}{H4}}
\begin{figure}[ht!]
\includegraphics[scale=0.55]{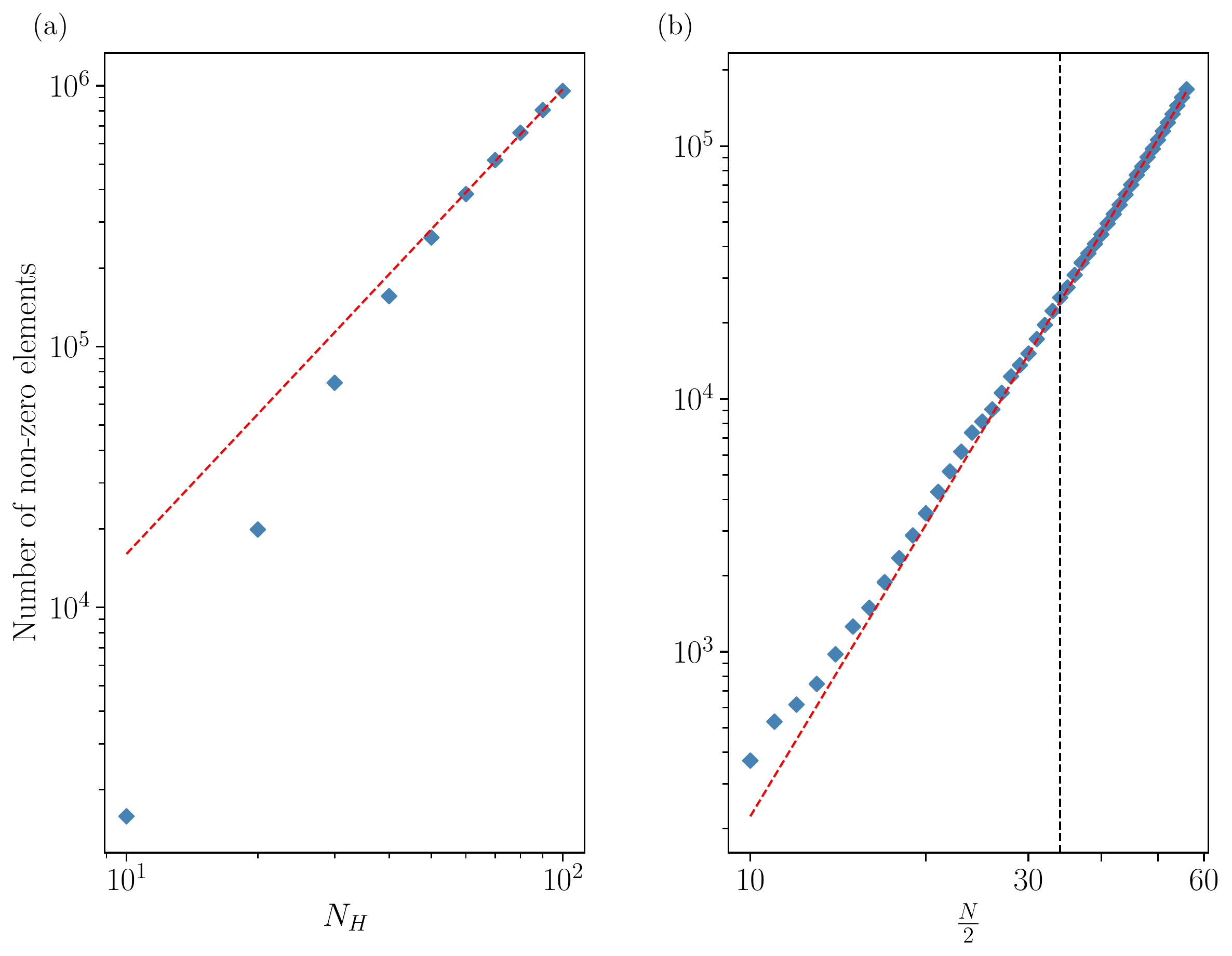}
\caption{\label{fig:h4hchainnnz}
(a) the number of H atom ($N_H$) versus the number of non-zero elements in $V$ and (b) the number of orbitals ($N/2$) versus the number of non-zero elements in $V$.
The last five points in (a) are used in the linear fit on a log scale where the slope is given as 1.78 with $R^2 = 0.9985$ (where $R$ is the correlation coefficient).
In (b), every point beyond $N/2 = 35$ is used for the linear fit on a log scale, which yields a slope of 3.83 with $R^2 = 0.9993$.
}
\end{figure}
In \fig{h4hchainnnz}, we present the number of non-zero values of $V$
with a truncation threshold, $5\times10^{-5}$, for hydrogen chains and \ce{H4}.

\section{The ``single low rank factorization'' algorithm of Berry \emph{et al.}}
\label{app:low_rank}

\subsection{Representing the single low rank factorized Hamiltonian as a linear combination of unitaries}

In Berry \textit{et al.}~the approach focuses on simulating a truncated low rank decomposition of the Coulomb operator which expresses the two-body terms in \eq{full} as a sum of squared one-body terms in a form that was first described for quantum computation in \cite{Poulin2014}. The first work to suggest exploiting the low rank properties of this decomposition was \cite{Motta2018v2}. Specifically the factorization of the Coulomb operator used in that work is
\begin{equation}
\label{eq:low_rank}
V \approx W = \frac{1}{2}\sum_{\ell=1}^{L} \left(\sum_{\sigma \in \{\uparrow,\downarrow\}}\sum_{p,q=1}^{N/2} W^{(\ell)}_{pq}  a^\dagger_{p,\sigma} a_{q,\sigma}\right)^2 = \frac{1}{8}\sum_{\ell=1}^{L} \left(\sum_{\sigma \in \{\uparrow,\downarrow\}}\sum_{p,q=1}^{N/2} W^{(\ell)}_{pq} \left( a^\dagger_{p,\sigma} a_{q,\sigma} + a^\dagger_{q,\sigma} a_{p,\sigma} \right)\right)^2
\end{equation}
where the $W_{pq}^{(\ell)}$ are scalars obtained by performing a Cholesky decomposition on a flattened version of the $V_{pqrs}$ tensor and $L = \widetilde{\cal O}(N)$. Note that the factor of 1/2 becomes a factor of 1/8 because there is a factor of 1/2 that is then squared for adding the Hermitian conjugates. Using the Jordan-Wigner transform as in \eq{jw} our Hamiltonian can be represented as the linear combination of unitaries:
\begin{align}\label{eq:firstham}
W = 
\frac{1}{8} \sum_{\ell=1}^L \left(\sum_{\sigma \in \{\uparrow,\downarrow\}}\sum_{p,q=1}^{N/2} W^{(\ell)}_{pq} Q'_{pq\sigma} \right)^2.
\end{align}
As before, when we separate out the identity for $p=q$, we can express this in terms of $Q'_{pq\sigma}$ from \eq{jw2} as
\begin{align}
\label{eq:singleW}
W &= 
\frac{1}{8} \sum_{\ell=1}^L \left(\sum_{\sigma \in \{\uparrow,\downarrow\}}\sum_{p,q=1}^{N/2} W^{(\ell)}_{pq} Q_{pq\sigma} + 2\sum_{r=1}^{N/2} W^{(\ell)}_{rr}\openone\right)^2\nn
&=\frac{1}{8} \sum_{\ell=1}^L \left(\sum_{\sigma \in \{\uparrow,\downarrow\}}\sum_{p,q=1}^{N/2} W^{(\ell)}_{pq} Q_{pq\sigma}\right)^2
+\frac{1}{2} \sum_{\ell=1}^L \sum_{\sigma \in \{\uparrow,\downarrow\}}\sum_{p,q,r=1}^{N/2} W^{(\ell)}_{pq} W^{(\ell)}_{rr} Q_{pq\sigma}
+\frac 12 \sum_{\ell=1}^L \left(\sum_{r=1}^{N/2} W^{(\ell)}_{rr}\openone\right)^2\nn
&=\frac{1}{8} \sum_{\ell=1}^L \left(\sum_{\sigma \in \{\uparrow,\downarrow\}}\sum_{p,q=1}^{N/2} W^{(\ell)}_{pq} Q_{pq\sigma}\right)^2
+\frac{1}{2} \sum_{\sigma \in \{\uparrow,\downarrow\}}\sum_{p,q,r=1}^{N/2} V_{pqrr} Q_{pq\sigma}
+\frac 12 \sum_{\ell=1}^L \left(\sum_{r=1}^{N/2} W^{(\ell)}_{rr}\openone\right)^2 \, . 
\end{align}
In the last line we have used that $\sum_\ell W^{(\ell)}_{pq} W^{(\ell)}_{rs}=V_{pqrs}$, which is exact when $L$ is taken to be sufficiently large.
In this expression, the second term can be combined with $T$ to give a one-body operator $T'$ that is identical to that in \app{sparse},
\begin{equation}
T'=\frac 12 \sum_{\sigma \in \{\uparrow,\downarrow\}}\sum_{p,q=1}^{N/2} \left( T_{pq} + \sum_{r=1}^{N/2} V_{pqrr} \right) Q_{pq\sigma} \, . 
\end{equation}
The third term in \eq{singleW} is proportional to the identity, and can be omitted.
The fundamental two-body term in $W$ is then just the first term of \eq{singleW},
\begin{equation}
W' = \frac{1}{8} \sum_{\ell=1}^L \left(\sum_{\sigma \in \{\uparrow,\downarrow\}}\sum_{p,q=1}^{N/2} W^{(\ell)}_{pq} Q_{pq\sigma}\right)^2.
\end{equation}
The Hamiltonian to be simulated is $T'+W'$.

We can now give the expression for $\lambda = \lambda_T + \lambda_{\rm SF}$.
Because $T'$ is the same as in \app{sparse}, $\lambda_T$ is as in \eq{lambda_V}.
For $\lambda_{\rm SF}$ we take the sum of the weightings in $W'$ to give
\begin{align}
    \label{eq:lambdas}
    \lambda_{\rm SF} = \frac 14 \sum_{\ell=1}^L \left(\sum_{p,q=1}^{N/2} \left |W_{pq}^{(\ell)} \right |\right)^2.
\end{align}
There is a factor of 4 due to summing over the spins, which would give a factor out the front of $1/2$.
However, then this $\lambda$ can be effectively divided by 2 if we choose to realize the squared operator by performing oblivious amplitude amplification as they do in \cite{vonBurg2020}. After oblivious amplitude amplification of the operator $A$ we effectively implement $2 A^2 - \openone$ where we can ignore the identity.
Since this gives us a factor of 2 we can divide the $\lambda$ by two, giving $\lambda_{\rm SF}$ as in \eq{lambdas}.

\subsection{The cost of qubitization of the single low rank factorized Hamiltonian}

Next we describe the costing of the implementation of the single low rank factorised Hamiltonian as in \cite{Berry2019B}, except taking into account the amplitude amplification approach for reducing $\lambda$.
The complete Hamiltonian written as a linear combination of unitaries is (ignoring an additive term proportional to the identity)
\begin{equation}
H \approx \frac{1}{2} \sum_{\sigma\in\{\uparrow,\downarrow\}} \sum_{p,q=1}^{N/2} T'_{p q}  Q_{pq\sigma} + \frac{1}{8} \sum_{\ell=1}^L \left(\sum_{\sigma \in \{\uparrow,\downarrow\}}\sum_{p,q=1}^{N/2} W^{(\ell)}_{pq} Q_{pq\sigma} \right)^2 ,
\end{equation}
where $T'_{pq}$ is as in \eq{Tpdef}.
The procedure as given in \cite{Berry2019B} corresponds to preparing a state, performing controlled operations, then inverting the preparation.
Now, we need to perform the state preparation separately on two registers.
First, prepare a superposition over the first register, which selects between the terms in the factorisation, as well as the one-body term.
Next, prepare a superposition on the second register with weights corresponding to the square roots of $W_{pq}^{(\ell)}$ and $T_{pq}$.
Perform the \sel\ operation by using the circuit shown in Figure 1 of \cite{Berry2019B}.
Then invert the state preparation on the second register, reflect on that register, then perform the preparation, \sel, and inverse preparation again.

\begin{figure}[tbh]
    \centering
\includegraphics[scale=0.914]{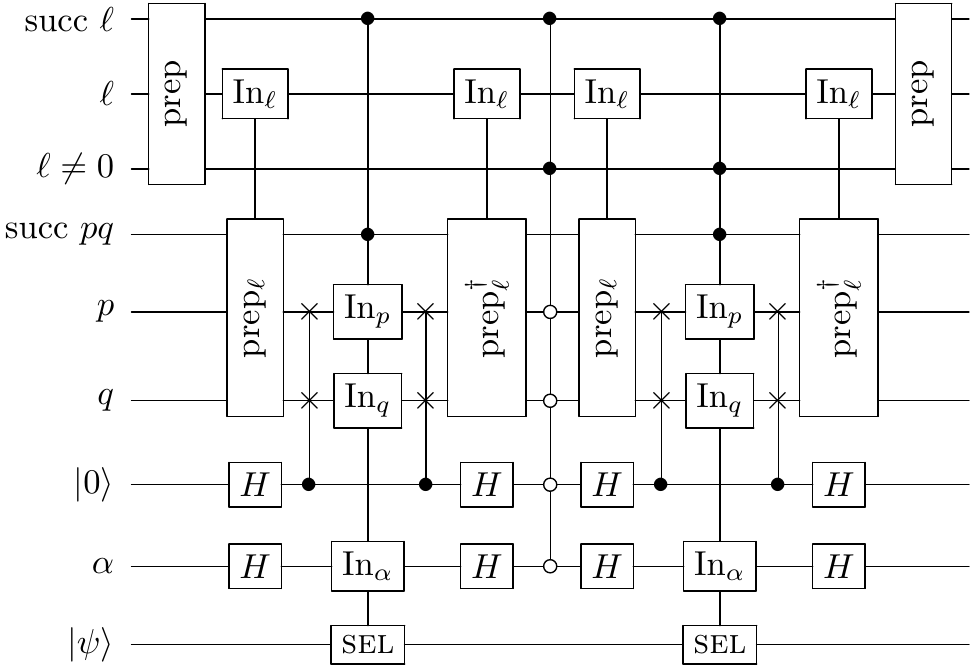}\\
\caption{The circuit for performing the state preparation and controlled operations for the single factorisation approach.
The register labelled $\ell$ is the first control register containing $\ell$, and $p,q$ labels the second and third control registers.
The registers labelled succ $\ell$ and succ $pq$ are the registers that flag success of the preparation of the equal superposition state for the $\ell$ and $p,q$ registers.
The register labelled $\ell\ne 0$ is a temporary register used to keep the result of an inequality test checking that $\ell\ne 0$, which is used to ensure that the second half of the circuit has no effect for that value of $\ell$, which is used to label the one-electron term.
The register labelled $\ket{0}$ is used to control swaps of $p$ and $q$.
The register labelled $\alpha$ is used to select the spin (and is initialised as $\ket{0}$).
The bottom register labelled $\ket{\psi}$ is the target system.
The operation labelled \textsc{sel} is the \sel\ as shown in Figure 1 of \cite{Berry2019B}.
}
    \label{fig:selectsf}
\end{figure}

The steps are shown in \fig{selectsf}, and in detail are as follows.
\begin{enumerate}
\item We first prepare a state on the first register as
\begin{equation}
\frac 1{\sqrt{\lambda}} \left(\ket{0}\sqrt{2\sum_{p\le q}|T'_{pq}|} + \sum_{\ell=1}^L\ket{\ell}\sum_{p\le q}|W_{pq}^{(\ell)}|  \right).
\end{equation}
This has complexity as follows.
\begin{enumerate}
\item Preparing an equal superposition on $L+1$ basis states has complexity $3n_L-3\factor+2b_r-9$, where $b_r$ is the number of bits used for the rotation on the ancilla,
\begin{equation}
n_L = \lceil\log(L+1)\rceil,
\end{equation}
and $\factor$ is a number such that $L+1$ is divisible by $2^\factor$.
\item A QROM is applied with output size $b_L=n_L+\zetabits_1+2$, where the extra 2 qubits are for outputting the qubit showing if $\ell=0$.
The complexity is
\begin{equation}
\left\lceil \frac{L+1}{k_L} \right\rceil + b_L(k_L-1).
\end{equation}
\item An inequality test is performed with complexity $\zetabits_1$.
\item A controlled swap is performed with complexity $n_L+1$.
\end{enumerate}
\item Next, we prepare a state on the second register as
\begin{equation}
\frac 1{\sqrt{\lambda}} \left(\ket{0}\sum_{p\le q}\sqrt{2|T'_{pq}|}\ket{\theta_{pq}^{(0)}}\ket{p,q} + \sum_{\ell=1}^L\ket{\ell}\sqrt{\sum_{r\le s}|W_{rs}^{(\ell)}|}
\sum_{p\le q}\sqrt{|W_{pq}^{(\ell)}|}\ket{\theta_{pq}^{(\ell)}}\ket{p,q}\right)\ket{+}\ket{+},
\end{equation}
where $\theta_{pq}^{(\ell)}$ are used to obtain the correct signs on the terms, and the $\ket{+}$ states at the end are used to select the spin and control the swap between the $p$ and $q$ registers.
The complexity of this state preparation is as follows.
\begin{enumerate}
\item First, an equal superposition over $p$ and $q$ is prepared with $p\le q\le N/2$.
Letting $n_N=\lceil \log(N/2)\rceil$ be the numbers of bits used for the $p$ and $q$ registers, the cost of preparing this equal superposition is as follows.
\begin{enumerate}
\item For testing the two inequalities the cost is $n_N-1$ for $q\le N/2$ and $n_N$ for $p\le q$.
\item An ancilla is rotated with cost $b_r-3$ for $b_r$ bits of accuracy.
\item A reflection is performed controlled on the result of the two inequality tests and the ancilla qubit, with cost 1.
\item The inequality tests are inverted with Cliffords, and there is a cost of $b_r-3$ for another qubit rotation.
\item There is a reflection on the $p$ and $q$ registers and the ancilla with cost $2n_N-1$.
\item The inequality tests are performed again with cost $2n_N-1$.
\item Another Toffoli is used to give a single qubit flagging success for both inequality tests.
\end{enumerate}
That gives a cost of $6n_N+2b_r-7$.
\item Next, a contiguous register is computed from $p$, and $q$ as
\begin{equation}\label{eq:qromreg}
s = p(p-1)/2 + q .
\end{equation}
In \app{contig} we show that the complexity of computing $p(p-1)/2 + q$ is $n_N^2+n_N-1$.
We are making an improvement over the method in \cite{Berry2019B} by not computing a contiguous register including $\ell$ as well.
This is because it is possible to apply the QROM to two registers (provided there are not restrictions like $\ell\le s$), as we show in \app{qrom}.
\item Perform the QROM on the two registers as described in \app{qrom} with output of size $b_{p}=2n_N+\zetabits_2+2$, with complexity
\begin{equation}\label{eq:qromprep2a}
\left\lceil \frac{L+1}{k_{p1}} \right\rceil 
\left\lceil \frac{N^2/8+N/2}{k_{p2}} \right\rceil + b_p(k_{p1}k_{p2}-1).
\end{equation}
\item Perform the inequality test with cost $\zetabits_2$.
\item Perform the controlled swap with the alt values with cost $2n_N$.
The sign required for the sign qubits can be implemented with Cliffords as before.
\end{enumerate}
\item Perform a controlled swap between the $p$ and $q$ registers, with cost $n_N$.
\item Perform \sel\ as shown in Figure 1 of \cite{Berry2019B}.
As discussed in \app{sparse} the cost is $2(N-2)$ Toffolis when it does not need to be controlled.
\item Reverse steps 2 and 3, where the complexities are the same except the QROM complexity which is changed to
\begin{equation}\label{eq:qromprep2b}
\left\lceil \frac{L+1}{k'_{p1}} \right\rceil 
\left\lceil \frac{N^2/8+N/2}{k'_{p2}} \right\rceil
+ k'_{p1}k'_{p2}.
\end{equation}
\item Reflect on the qubits that were prepared in step 2.
The qubits we need to reflect on are as follows.
\begin{enumerate}
    \item The $2n_N$ qubits for the $p$ and $q$ registers.
    \item We need to reflect on the $\zetabits_2$ registers that are used for the equal superposition state for the state preparation.
    \item One that is rotated for the amplitude amplification.
    \item One for the spin.
    \item One for controlling the swap between the $p$ and $q$ registers.
\end{enumerate}
That gives a total of $2n_N+\zetabits_2+3$ qubits.
The reflection needs to be controlled on the success of the preparation on the $\ell$ register, and $\ell\ne 0$, making the total cost
$2n_N+\zetabits_2+3$ Toffolis.
\item Perform steps 2 to 5 again, but this time $L+1$ is replaced with $L$ in \eq{qromprep2a} and \eq{qromprep2b}.
Also, the \sel\ operation needs to be controlled on $\ell\ne 0$, which flags the one-body term.
That requires another Toffoli.
\item Invert the state preparation on the $\ell$ register, where the complexity of the QROM is reduced to
\begin{equation}
\left\lceil \frac{L+1}{k'_L} \right\rceil + k'_L.
\end{equation}
\item To complete the step of the quantum walk, perform a reflection on the ancillas used for the state preparation.
There are $n_L+2n_N+\zetabits_1+\zetabits_2+4$, where the qubits we need to reflect on are as follows.
\begin{enumerate}
    \item The $n_L$ qubits for the $\ell$ register.
    \item The $2n_N$ qubits for the $p$ and $q$ registers.
    \item The $\zetabits_1$ qubits for the equal superposition state used for preparing the state on the $\ell$ register using the coherent alias sampling.
    \item The $\zetabits_2$ qubits for the equal superposition state for preparing the state on the $p,q$ registers.
    \item Two qubits rotated for the amplitude amplification.
    \item One qubit for the spin.
    \item One qubit for controlling the swap of the $p$ and $q$ registers.
\end{enumerate}
This reflection has cost $n_L+2n_N+\zetabits_1+\zetabits_2+2$.
\item The steps of the walk are made controlled by using unary iteration on an ancilla used for the phase estimation.
Each step requires another two Toffolis for the unary iteration and making the reflection controlled.
\end{enumerate}
Adding all these complexities together gives
\begin{align}\label{eq:lowranktof}
&2(3n_L-3\factor+2b_r-9) 
+\left\lceil \frac{L+1}{k_L} \right\rceil 
+ b_L(k_L-1) + \left\lceil \frac{L+1}{k'_L} \right\rceil + k'_L
+ 2b_L-2 
+ 4(6n_N+2b_r-7) 
+ 4(n_N^2+n_N-1) 
\nn
& \quad + \left\lceil \frac{L+1}{k_{p1}} \right\rceil 
\left\lceil \frac{N^2/8+N/2}{k_{p2}} \right\rceil + b_p(k_{p1}k_{p2}-1) 
+ \left\lceil \frac{L+1}{k'_{p1}} \right\rceil 
\left\lceil \frac{N^2/8+N/2}{k'_{p2}} \right\rceil + k'_{p1}k'_{p2} 
+ \left\lceil \frac{L}{k_{p1}} \right\rceil 
\left\lceil \frac{N^2/8+N/2}{k_{p2}} \right\rceil \nn & \quad + b_p(k_{p1}k_{p2}-1) 
+ \left\lceil \frac{L}{k'_{p1}} \right\rceil 
\left\lceil \frac{N^2/8+N/2}{k'_{p2}} \right\rceil + k'_{p1}k'_{p2} 
+ 4\zetabits_2 
+ 8n_N 
+ 4n_N 
+ 4(N-2) 
+ 1 
+ 2n_N+\zetabits_2+3 
\nn & \quad + n_L+2n_N +\zetabits_1+\zetabits_2 +4 
\nn
&=7n_L+ 4 n_N^2+ 40 n_N
-6\factor+12b_r
+\left\lceil \frac{L+1}{k_L} \right\rceil
+ b_L(k_L+1) + \left\lceil \frac{L+1}{k'_L} \right\rceil + k'_L
\nn
&\quad + \left\lceil \frac{L+1}{k_{p1}} \right\rceil 
\left\lceil \frac{N^2/8+N/2}{k_{p2}} \right\rceil + 2b_pk_{p1}k_{p2}
+ \left\lceil \frac{L+1}{k'_{p1}} \right\rceil 
\left\lceil \frac{N^2/8+N/2}{k'_{p2}} \right\rceil + 2k'_{p1}k'_{p2} 
+ \left\lceil \frac{L}{k_{p1}} \right\rceil 
\left\lceil \frac{N^2/8+N/2}{k_{p2}} \right\rceil 
\nn
&\quad + \left\lceil \frac{L}{k'_{p1}} \right\rceil 
\left\lceil \frac{N^2/8+N/2}{k'_{p2}} \right\rceil
+ \zetabits_1+4\zetabits_2
+ 4N
-56
\end{align}
with $b_L=n_L+\zetabits_1+2$, $n_L=\lceil\log(L+1)\rceil$, $b_{p}=2n_N+\zetabits_2+2$, and $n_N=\lceil \log(N/2)\rceil$.

Next we consider the total number of logical qubits needed for the simulation via this method.
\begin{enumerate}
\item The control register for the phase estimation, and the ancillas for the unary iteration, together need $2\lceil\log\mathcal{I}\rceil-1$ qubits.
\item There are $N$ qubits for the target system.
\item There are $n_L+2$ qubits for the $\ell$ register, the qubit rotated in preparing the equal superposition, and the qubit flagging success of preparing the equal superposition.
\item The state preparation on the $\ell$ register uses $n_L+2\zetabits_1+3$ qubits.
Here $n_L$ is for the alt values, $\zetabits_1$ are for keep values, $\zetabits_1$ are for the equal superposition state, 1 is for the output of the inequality test, and 2 are for the qubit flagging $\ell\ne 0$ and its alternate value.
\item There are $2n_N+2$ qubits needed for the $p$ and $q$ registers, a qubit that is rotated for the equal superposition, and a qubit flagging success of preparing the equal superposition.
\item The contiguous register needs $\lceil \log(N^2/8+N/4)\rceil$ qubits.
\item The equal superposition state used for the second preparation uses $\zetabits_2$ qubits.
\item The phase gradient register uses $b_r$ qubits.
\item The QROM needs a number of qubits $b_pk_{p1}k_{p2}+\lceil\log[(L+1)/k_{p1}]\rceil+\lceil\log[(N^2/8+N/2)/k_{p2}]\rceil$.
\end{enumerate}
The QROM for the state preparation on the second register uses a large number of temporary ancillas, which can be reused by later parts of the algorithm, so those later parts of the algorithm do not need the number of qubits counted.
The total number of qubits used is then
\begin{equation}\label{eq:lowranklog}
2\lceil\log\mathcal{I}\rceil+N+2n_L+2\zetabits_1+\zetabits_2+b_r+2n_N+6 +\lceil \log(N^2/8+N/4)\rceil+b_pk_{p1}k_{p2}+\lceil\log[(L+1)/k_{p1}]\rceil+\lceil\log[(N^2/8+N/2)/k_{p2}]\rceil
\end{equation}
with $b_{p}=2n_N+\zetabits_2+2$, $n_N=\lceil \log(N/2)\rceil$, $n_L=\lceil\log(L+1)\rceil$.
This completes the costing of the low rank factorization method.

\begin{table*}
\begin{tabular}{|c|c|c|}
\hline
$L$ & $\lambda$ & CCSD(T) error (m$E_h$) \\ \hline
50 & 3367.6 & 4.79 \\ \hline
75 & 3657.9 & 0.21 \\ \hline
100 & 3854.3 & 1.55 \\ \hline
125 & 3997.4 & 3.08 \\ \hline
150 & 4112.7 & 2.07 \\ \hline
175 & 4199.2 & 1.63 \\ \hline
\color{blue}200 & \color{blue}4258.0 & \color{blue}0.10 \\ \hline
225 & 4300.7 & 0.38 \\ \hline
250 & 4331.9 & 0.16 \\ \hline
275 & 4354.9 & 0.29 \\ \hline
300 & 4372.0 & 0.18 \\ \hline
325 & 4385.3 & 0.13 \\ \hline
350 & 4395.6 & 0.06 \\ \hline
375 & 4403.4 & 0.03 \\ \hline
400 & 4409.3 & 0.02 \\ \hline
\end{tabular}
\caption[Singe low rank algorithm overheads for Reiher FeMoCo Hamiltonian]{Single low rank factorization data for the Reiher Hamiltonian \cite{Reiher2017}.
Here $\lambda_T$ is 90.4 Hartree. The entry in blue is the one used for our resource estimates.}
\end{table*}

\begin{table*}
\begin{tabular}{|c|c|c|}
\hline
$L$ & $\lambda$ & CCSD(T) error (m$E_h$) \\ \hline
50 & 2233.7 & -93.07 \\ \hline
75 & 2484.5 & -57.08 \\ \hline
100 & 2664.5 & -23.77 \\ \hline
125 & 2743.5 & -9.37 \\ \hline
150 & 2786.9 & -11.55 \\ \hline
175 & 2835.5 & -8.17 \\ \hline
200 & 2906.9 & 0.69 \\ \hline
225 & 2986.9 & 1.52 \\ \hline
250 & 3035.9 & 0.90 \\ \hline
\color{blue}275 & \color{blue}3071.8 & \color{blue}0.48 \\ \hline
300 & 3099.2 & -0.07 \\ \hline
325 & 3119.3 & -0.18 \\ \hline
350 & 3134.2 & -0.11 \\ \hline
375 & 3146.0 & -0.08 \\ \hline
400 & 3154.8 & -0.10 \\ \hline
\end{tabular}
\caption[Singe low rank algorithm overheads for Li FeMoCo Hamiltonian]{Single low rank factorization data for the Li Hamiltonian \cite{Li2019}.
Here $\lambda_T$ is 561.5 Hartree. The entry in blue is the one used for our resource estimates.}
\end{table*}

\section{The ``double low rank factorization'' algorithm of von Burg \emph{et al.}}
\label{app:double_low_rank}

\subsection{Representing the double low rank factorized Hamiltonian as a linear combination of unitaries}

The approach of \cite{vonBurg2020} is a modification of the approach of \cite{Berry2019B}. It is also based on applying qubitization to the representations discussed in \cite{Motta2018v2}, but the difference is that it uses a second factorization of the Coulomb operator. The idea to combine this second factorization with qubitization was suggested but not implemented in \cite{Berry2019B}. The idea is to further factorize the low rank operator as
\begin{equation}
V \approx F = \frac{1}{2}\sum_{\ell=1}^{\rankone} \left(\sum_{\sigma \in \{\uparrow,\downarrow\}}\sum_{p,q=1}^{N/2} W^{(\ell)}_{pq}  a^\dagger_{p,\sigma} a_{q,\sigma}\right)^2 = \frac{1}{2} \sum_{\ell=1}^{\rankone} U_{\ell} \left(\sum_{\sigma \in \{\uparrow, \downarrow\}} \sum_{p=1}^{N/2} f_{p}^{(\ell)} n_{p,\sigma} \right)^2 U_{\ell}^\dagger
\end{equation}
where $n_{p,\sigma} = a^\dagger_{p, \sigma} a_{p, \sigma}$, $W_{pq}^{(\ell)}$ are scalars obtained by performing a Cholesky decomposition on a flattened version of the $V_{pqrs}$ tensor, $f_{p}^{(\ell)}$ are scalars obtained from the diagonalization of the squared one-body operator, and $U_\ell$ are the eigenvector matrices of that diagonalization.
The sum over $p$ can be truncated at $\ranktwo^{(\ell)}$ to a good approximation.
Generally, we have that $\rankone = \widetilde{\cal O}(N)$ and $\ranktwo^{(\ell)} < N$ but in some rather special cases $\ranktwo^{(\ell)}$ can be asymptotically less than this (e.g., for very large systems scaling towards the thermodynamic limit).
For the moment we will keep the sum at $N/2$ to show how part of this two-body operator can be combined with the one-body operator.
We will also assume for the moment that $L=N^2/4$ so there is no approximation involved in summing to $L$.

Using the Jordan-Wigner transformation of \eq{jw} one can map the operator $F$ to qubits as
\begin{equation}
F = \frac{1}{2} \sum_{\ell=1}^{\rankone} U_{\ell} \left(\sum_{\sigma \in \{\uparrow, \downarrow\}} \sum_{p=1}^{N/2} f_{p}^{(\ell)} \left(\frac{\openone - Z_{p,\sigma}}{2}\right) \right)^2 U_{\ell}^\dagger = \frac{1}{8} \sum_{\ell=1}^{\rankone} U_{\ell} \left(\sum_{\sigma \in \{\uparrow, \downarrow\}} \sum_{p=1}^{N/2} f_{p}^{(\ell)} \left(\openone - Z_{p,\sigma}\right) \right)^2 U_{\ell}^\dagger .
\end{equation}
The factor of 1/2 becomes a factor of 1/8 because a factor of 1/2 comes from the Jordan-Wigner transformation which is then squared. This operator may then be expressed as
\begin{equation}
F = \frac{1}{8} \sum_{\ell=1}^{\rankone} U_{\ell} \left({\cal I}_\ell - {\cal Z}_\ell \right)^2 U_{\ell}^\dagger =  \frac{1}{8} \sum_{\ell=1}^{\rankone} U_{\ell} \left(2 \, {\cal I}_\ell \left({\cal I}_\ell - {\cal Z}_\ell\right) - {\cal I}_\ell^2 + {\cal Z}_\ell^2 \right) U_{\ell}^\dagger \equiv \frac{1}{8}\sum_{\ell=1}^{\rankone} \left(2\, {\cal I}_\ell U_\ell {\cal N}_\ell  U_\ell^\dagger-{\cal I}_\ell^2 + U_\ell {\cal Z}_\ell^2 U_\ell^\dagger\right)
\end{equation}
where we used the definitions
\begin{equation}
{\cal I}_\ell = 2 \sum_{p=1}^{N/2} f_{p}^{(\ell)} \openone, \qquad {\cal Z}_\ell = \sum_{\sigma \in \{\uparrow, \downarrow\}} \sum_{p=1}^{N/2} f_{p}^{(\ell)} Z_{p,\sigma}, \qquad {\cal N}_\ell = {\cal I}_\ell - {\cal Z}_\ell = \sum_{\sigma \in \{\uparrow, \downarrow\}} \sum_{p=1}^{N/2} f_{p}^{(\ell)} n_{p,\sigma}.
\end{equation}
The authors of \cite{vonBurg2020} add the one-body part of this expression into the one-body operator to provide a new one-body operator
\begin{equation}
\frac{1}{2} \sum_{\sigma\in\{\uparrow,\downarrow\}} \sum_{p,q=1}^{N/2} T_{p q} a^\dagger_{p,\sigma}a_{q,\sigma} + \frac{1}{4} \sum_{\ell=1}^{\rankone} {\cal I}_\ell U_\ell  {\cal N}_\ell U_\ell^\dagger = \frac{1}{2} \sum_{\sigma\in\{\uparrow,\downarrow\}} \sum_{p,q=1}^{N/2} \left(T_{p q} + \frac{1}{2}\sum_{\ell=1}^{\rankone} {\cal I}_\ell W_{pq}^{(\ell)}\right) a^\dagger_{p,\sigma}a_{q,\sigma}
\end{equation}
where we have used the relation that
\begin{equation}
U_\ell {\cal N}_\ell U_\ell^\dagger = 
U_\ell \left(\sum_{\sigma \in \{\uparrow, \downarrow\}} \sum_{p=1}^{N/2} f_{p}^{(\ell)} n_{p,\sigma}\right) U_\ell^\dagger =
\sum_{\sigma \in \{\uparrow,\downarrow\}} \sum_{pq=1}^{N/2} W_{pq}^{(\ell)} a^\dagger_{p,\sigma}a_{q,\sigma}.
\end{equation}
The expression for the one-body term can be simplified by noting that 
the trace is unchanged under $U_\ell$, so
\begin{equation}
{\cal I}_\ell =
2 \sum_{p=1}^{N/2} f_{p}^{(\ell)} \openone  = 2 \sum_{r=1}^{N/2} W_{rr}^{(\ell)} \openone .
\end{equation}
Thus, the additional term becomes
\begin{align}
\frac{1}{2} \sum_{\ell=1}^{\rankone} {\cal I}_\ell W_{pq}^{(\ell)} &= \sum_{\ell=1}^{\rankone} \sum_{r=1}^{N/2} W_{rr}^{(\ell)} W_{pq}^{(\ell)}
=\sum_{r=1}^{N/2}\sum_{\ell=1}^{\rankone} W_{rr}^{(\ell)} W_{pq}^{(\ell)}
=\sum_{r=1}^{N/2}V_{pqrr}\, .
\end{align}
The last equality is exact when $L$ is large enough that there is no truncation of the rank.
Then the one-body term becomes, with no approximation due to truncation of the rank,
\begin{equation}
\frac{1}{2} \sum_{\sigma\in\{\uparrow,\downarrow\}} \sum_{p,q=1}^{N/2} T'_{p q} a^\dagger_{p,\sigma}a_{q,\sigma},
\end{equation}
where $T'_{p q}$ is as in \eq{Tpdef}.
Thus the one-body operator can be taken to be $T'$, the same as for the sparse case in \app{sparse} and the single factorization in \app{low_rank}.

For the two-body operator $F$, we can omit $-{\cal I}_\ell^2$, which only gives a uniform shift in the energy, so in $F'$ only $U_\ell {\cal Z}_\ell^2 U_\ell^\dagger$ is retained.
Moreover, in this two-body operator one can now truncate the sum over $p$ at $\ranktwo^{(\ell)}$, so
\begin{equation}\label{eq:Fprime}
F' = \frac{1}{8}\sum_{\ell=1}^{\rankone} U_\ell \left( \sum_{\sigma \in \{\uparrow, \downarrow\}} \sum_{p=1}^{\ranktwo^{(\ell)}} f_{p}^{(\ell)} Z_{p,\sigma} \right)^2 U_\ell^\dagger.
\end{equation}
In this expression one can also take a truncation with $L<N^2/4$, so there is a truncation of the rank for the two-body operator but not the one-body operator.

The key insight of \cite{vonBurg2020} is a method for implementing the $U_\ell$ such that one obtains a lambda scaling as $\lambda = \lambda_{T'} + \lambda_{\rm DF}$ where
\begin{equation}
\lambda_{\rm DF}= \frac{1}{4} \sum_{\ell=1}^{\rankone} \left(\sum_{p=1}^{\ranktwo^{(\ell)}} \left |f_p^{(\ell)} \right|\right)^2.
\end{equation}
The factor of 1/8 becomes a factor of 1/4 because we must multiply by $2^2 = 4$ due to the sum over the spin degree of freedom but we can then divide by two by using the oblivious amplitude amplification, Chebyshev polynomial technique. The contribution to $\lambda$ from $T'$ can be determined as follows.
One can perform a rotation of the basis to diagonalise $T'$ as well, so the value of $\lambda_{T'}$ is the Schatten 1-norm (sum of the absolute values of the eigenvalues, or trace norm) of the matrix $T'_{pq}=T_{p q} + \sum_{rr=1}^{N/2} V_{pqrr}$.
The main reason that this double-factorization method is more efficient than the one in \cite{Berry2019B} is because it has reduced $\lambda$. While the authors of \cite{vonBurg2020} write that their method is more efficient by more than an order of magnitude for various systems studied, this turns out not to be the case comparing to the most efficient method in \cite{Berry2019B} and correcting the estimate of $\lambda$.

\subsection{Cost of qubitization of the double low rank factorized Hamiltonian}

The primary cost of the double-factorisation method from \cite{vonBurg2020} comes from the circuit shown in Eq.~(78) of their Supplementary Material.
The costs there are given by the following parts.
\begin{enumerate}
\item A state preparation as shown in their Eq.~(79). This needs to be performed and inverted twice.
\item A QROM to output the angles needed in their Lemma 8.  This QROM needs to be called and erased twice.
\item The $Z$ rotations as per their Eq.~(68) need to be performed $4N$ times.
Those have complexity $\rotbits-2$ each, with
\begin{equation}\label{eq:beta}
\rotbits=\lceil 5.652+\log(\lambda N/\epsilon)\rceil
\end{equation}
the bits of precision of the rotation.
\item There are $N/2$ controlled swaps performed before and after the rotations as shown in Eq.~(62) of \cite{vonBurg2020}.
Since that is repeated, the cost is $2N$.
\item The reflection in Eq.~(78) of \cite{vonBurg2020} has a complexity depending on how the reflection is performed.
\end{enumerate}

First note that there seems to be a problem in using the circuit exactly as shown in Eq.~(79) of \cite{vonBurg2020}.
It is using a joint state preparation on two registers, but the qubits before the preparation cannot be cleanly divided into those for the first register and those for the second register.
That would be needed in order to reflect on only one, as is needed in the procedure in Eq.~(79) of \cite{vonBurg2020}.
We will first discuss the costing for the procedure as shown in Eq.~(79), then describe how to perform the separate state preparations on the two registers.

In comparing the complexity as presented here to that in \cite{vonBurg2020}, it must be taken into account that they are defining $N$ in a different way.
In that work $N$ is the number of orbitals, whereas here it is the number of \emph{spin-}orbitals, which is twice as large, because it is multiplied by the spin degree of freedom.
Therefore, \cite{vonBurg2020} takes $N=54$ for the FeMoCo orbitals of \cite{Reiher2017}, whereas here we take $N=108$.
Also, in the expression for $\rotbits$, $\epsilon$ is the error in synthesising a single step, which is taken to be 0.0001 Hartree in \cite{vonBurg2020}.

The dominant costs are from the QROM and the rotations.
In using our THC representation we are able to reduce the QROM costs below those in \cite{vonBurg2020}, though the rotation costs are unchanged.
To understand the QROM costs, we first list the main variables used.
\begin{itemize}
    \item The main rank $\rankone$.
    \item $\ranktwo^{(\ell)}$ is the rank of each term.
    \item $(1/\rankone)\ranktwo =\sum_{\ell=1}^{\rankone} \ranktwo^{(\ell)}$ is the average rank of the second factorization; thus, the total number of coefficients is $\rankone \ranktwo$.
\end{itemize}
Before you perform the state preparation, you need to prepare an equal superposition over $\rankone\ranktwo$ basis states.
That has cost $3\lceil \log \rankone\ranktwo\rceil-3\factor+2\rotprec-9$ Toffolis, as per the analysis in \app{sparse}, where $\rankone\ranktwo$ is divisible by $2^\factor$.
The QROM used for the state preparation has an output size
\begin{equation}
b= 2n_{\ranktwo} + 2n_\rankone + \zetabits + 2,
\end{equation}
where 
\begin{align}\label{eq:numbersms}
n_{\ranktwo} = \lceil \log (\max_\ell \ranktwo^{(\ell)})\rceil, \qquad \qquad
n_\rankone = \lceil \log(\rankone+1)\rceil, \qquad \qquad
\zetabits = \lceil 2.5 + \log(\lambda/\epsilon) \rceil,
\end{align}
are the bits for the two registers and the bits of precision $\zetabits$ used for the keep values.
We take $n_\rankone = \lceil \log(\rankone+1)\rceil$ rather than $n_\rankone = \lceil \log\rankone\rceil$ in order to use an additional value of $\ell$ to account for the one-electron term.
There is also a size for the contiguous register
\begin{equation}
    n_{\rankone,\ranktwo} = \lceil \log(\rankone\ranktwo + N/2) \rceil.
\end{equation}
Here the $N/2$ will be to account for the one-electron term in the simulation.
In the expression for $b$ the factor of 2 on $n_{\ranktwo}$ and $n_\rankone$ accounts for the ind and alt values in sparse state preparation, and the $+2$ is for the sign bit and its alternate value.
(In \cite{vonBurg2020} only one bit is assumed for the sign.)
The complexity for the state preparation is that of the QROM, plus $\zetabits$ for the inequality test and $n_{\ranktwo}+n_\rankone$ for the controlled swaps.

For the state preparation, since the number of items of data is $\rankone\ranktwo$, and each has size $b$, the complexity of the QROM computation is
\begin{equation}
\left\lceil\frac {\rankone\ranktwo}{k_p}\right\rceil + b(k_p-1),
\end{equation}
where $k_p$ must be chosen as a power of $2$.
The uncomputation cost is then
\begin{equation}
\left\lceil\frac {\rankone\ranktwo}{k'_p}\right\rceil + k'_p .
\end{equation}
There is cost $n_\ranktwo+n_\rankone$ for the controlled swaps and $\zetabits$ for the inequality test, which are inverted as well for the inverse preparation, giving cost $2n_\ranktwo+2n_\rankone+2\zetabits=b+\zetabits-2$.
Taking account of the need to perform the state preparation and inverse state preparation twice,
we give the cost of the state preparation for the two-body term as
\begin{equation}\label{eq:qrom1}
2\left\lceil\frac {\rankone \ranktwo}{k_p}\right\rceil + 2bk_p + 2\left\lceil\frac {\rankone\ranktwo}{k'_p}\right\rceil + 2k'_p + 2\zetabits -4.
\end{equation}

Before performing the QROM for the rotations, it is necessary to use QROM to generate a register with the offsets in order to produce a contiguous register.
This procedure is described in Eq.~(39) and Lemma 7 of \cite{vonBurg2020}.
The complexity of the QROM for the offsets is
\begin{equation}
\left\lceil\frac \rankone{k_\rankone}\right\rceil + n_{\rankone,\ranktwo} (k_\rankone-1) .
\end{equation}
In uncomputing the complexity is
\begin{equation}
\left\lceil\frac \rankone{k'_\rankone}\right\rceil + k'_\rankone.
\end{equation}
We also need to perform addition of registers twice and invert addition of registers twice, which has cost $4(n_{\rankone,\ranktwo}-1)$.
That gives a total cost of
\begin{equation}\label{eq:qrom0}
    2\left\lceil\frac {\rankone}{k_\rankone}\right\rceil + 2n_{\rankone,\ranktwo} (k_\rankone+1) + 2\left\lceil\frac {\rankone}{k'_\rankone}\right\rceil + 2k'_\rankone - 4.
\end{equation}

Once the contiguous register is produced, one can apply the QROM to output the rotation angles.
Because there are $\rankone\ranktwo$ sets of $N\rotbits/2$ bits for the rotation angles,
the cost of the QROM used for the rotation angles can be given as
\begin{equation}
\left\lceil\frac {\rankone\ranktwo}{k_r}\right\rceil + N\rotbits(k_r-1)/2,
\end{equation}
and the cost of the inverse QROM as
\begin{equation}
\left\lceil\frac {\rankone\ranktwo}{k'_r}\right\rceil + k'_r.
\end{equation}
In \cite{vonBurg2020}, $\rotbits$ is taken to be
\begin{equation}
\rotbits=\lceil 5.652 + \log(N\lambda/2\epsilon) \rceil .
\end{equation}
Note that we are using our convention for $N$, which is double that used in \cite{vonBurg2020}.
Accounting for two steps with preparation and inverse preparation the cost is
\begin{equation}\label{eq:qrom2}
2\left\lceil\frac {\rankone\ranktwo}{k_r}\right\rceil + N\rotbits(k_r-1) + 2\left\lceil\frac {\rankone \ranktwo}{k'_r}\right\rceil + 2k'_r .
\end{equation}
For the total cost of the two-body term in the double-factorisation method of \cite{vonBurg2020}, we add the QROM costs in \eq{qrom1}, \eq{qrom0}, and \eq{qrom2}, plus the $4N(\rotbits-2)$ cost for the controlled rotations, plus the $2N$ cost for the controlled swaps (for the spin), plus twice the $3n_{\rankone,\ranktwo}-3\factor+2\rotprec-9$ cost for preparing the equal superposition state, $n_{\rankone,\ranktwo}-1$ for reflection and $2$ for the unary iteration for the phase estimation and making the reflection controlled.
There is also a cost of $n_\ranktwo-1$ for the reflection on the second register, if this were possible as claimed in \cite{vonBurg2020}.
The total cost is therefore
\begin{align}
&2\left\lceil\frac {\rankone \ranktwo}{k_p}\right\rceil + 2bk_p + 2\left\lceil\frac {\rankone\ranktwo}{k'_p}\right\rceil + 2k'_p + 2\zetabits
+2\left\lceil\frac {\rankone}{k_\rankone}\right\rceil + 2n_{\rankone,\ranktwo} (k_\rankone+1) +2\left\lceil\frac {\rankone}{k'_\rankone}\right\rceil + 2k'_\rankone\nn
&+2\left\lceil\frac {\rankone\ranktwo}{k_r}\right\rceil + N\rotbits(k_r-1) + 2\left\lceil\frac {\rankone \ranktwo}{k'_r}\right\rceil + 2k'_r +
4N\rotbits-6N+7n_{\rankone,\ranktwo}-6\factor+4\rotprec+n_\ranktwo-26,
\end{align}
where $b=2n_{\ranktwo} + 2n_\rankone + \zetabits + 2$ with the numbers of qubits as given in \eq{numbersms}.
The values of $k$ should be chosen as powers of 2 that minimise the cost.

As mentioned above, the procedure needs to be made more complicated, because the state preparation should be performed separately on the two registers, rather than jointly on both registers at once, because a reflection needs to be made on one register.
An alternative approach is to prepare the state on the first register, then prepare the state on the second register controlled on the first register, rather than preparing the state on both registers together.

It is also necessary to account for the one-electron term.
It is possible to add that in an explicit way as proposed in \cite{vonBurg2020}, but that significantly increases the complexity because of the need to perform high accuracy rotations again.
Instead, one can combine it with the two-electron term in a similar way as we have described above for the single low rank factorization.
The principle is to use an additional basis state on the first register to flag the one-electron term.
Then the second register will have a state preparation corresponding to the one-electron term for this basis state on the first register, but \emph{only} for the first part, not the second, because we are not applying the oblivious amplitude amplification within a single step for the one-electron term.

\begin{figure}[tbh]
    \centering
\includegraphics[scale=0.925]{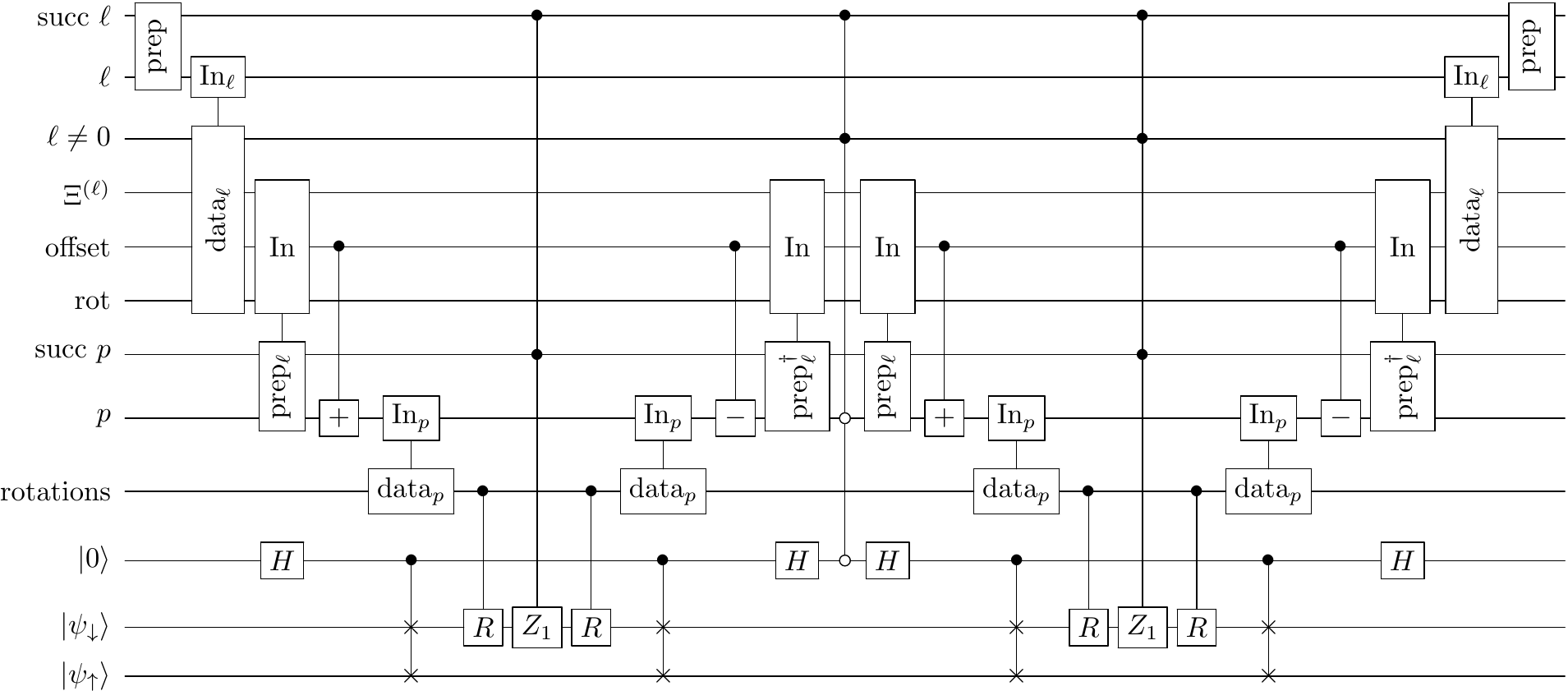}
\caption{The circuit for performing the state preparation and controlled operations (\sel) for the double factorisation approach.
The register labelled $\ell$ is the first control register containing $\ell$, and $p$ labels the second control register.
The registers labelled succ $\ell$ and succ $p$ are the registers that flag success of the preparation of the equal superposition state for the $\ell$ and $p$ registers.
The register labelled $\ell\ne 0$ is a temporary register used to keep the result of an inequality test checking that $\ell\ne 0$, which is used to ensure that the second half of the circuit has no effect for that value of $\ell$, which is used to label the one-electron term.
The registers labelled $\ranktwo^{(\ell)}$, ``offset'' and ``rot'' are the outputs of the QROM on the $\ell$ register that are mainly used for the controlled preparation of the state on the $p$ register.
The register labelled ``rotations'' is the data needed for the basis rotations for implementing $Z_{p,\sigma}$.
The register labelled $\ket{0}$ is used to select the spin, and $\ket{\psi_\downarrow}$ and $\ket{\psi_\uparrow}$ are the spin down and up components of the target system.
}
    \label{fig:selectdf}
\end{figure}

\begin{figure}[tbh]
\centering
\includegraphics[scale=0.925]{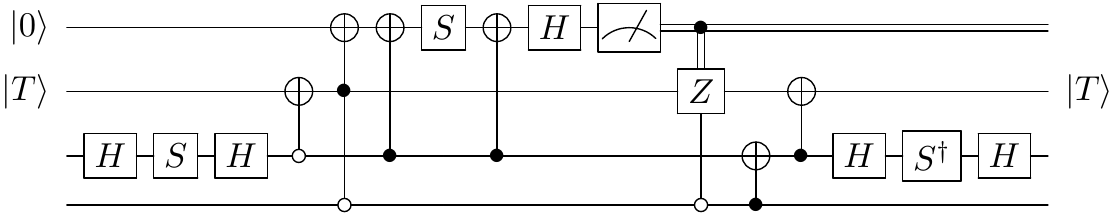}
\caption{A quantum circuit for catalytically implementing a controlled Hadamard using a single Toffoli gate, Clifford gates, and a catalytic state $\ket{T}=T\ket{+}$.
 The bottom qubit is the control, and the third qubit (second from the bottom) is the target.}
\label{fig:chad}
\end{figure}

The detailed procedure is shown in \fig{selectdf}.
The steps are as follows.
\begin{enumerate}
\item First we need to prepare the appropriate superposition state on the $\ell$ register, which is indicated as the box labelled ``prep'' in \fig{selectdf}.
The steps to perform that are as follows.
\begin{enumerate}
\item Prepare an equal superposition over $\rankone+1$ basis states, with cost $3n_\rankone-3\factor+2\rotprec-9$ Toffolis, as per the analysis in \app{sparse}, where $\rankone$ is divisible by $2^\factor$.
\item Use a QROM for state preparation on the first register outputting an alt value of size $n_\rankone$ and a keep value of size $\zetabits_1$.
The output size is
\begin{equation}
b_{p1} = n_\rankone+\zetabits_1,
\end{equation}
so the QROM cost is
\begin{equation}
\left\lceil\frac{\rankone+1}{k_{p1}}\right\rceil + b_{p1}(k_{p1}-1),
\end{equation}
\item Perform an inequality test with cost $\zetabits_1$.
\item Perform the controlled swap with cost $n_\rankone$, completing the state preparation on the first register.
\end{enumerate}
\item Before we prepare the state on the $p$ register, we need to output data from the $\ell$ register.
This is indicated by the ``In$_\ell$'' and ``data$_\ell$'' in \fig{selectdf}.
For each value on the first register we need to output $\ranktwo^{(\ell)}$ (with $n_\ranktwo$ bits), and $\rotprec$ bits for a rotation on an ancilla qubit.
It is also convenient to output the offset (with $n_{\rankone,\ranktwo}$ bits) at this stage, and determine if $\ell=0$, giving output size
\begin{equation}
b_o = n_\ranktwo 
+n_{\rankone,\ranktwo}+\rotprec +1,
\end{equation}
and QROM cost
\begin{equation}
\left\lceil\frac{\rankone+1}{k_o}\right\rceil + b_o(k_o-1).
\end{equation}
\item Next we need to prepare the state on the $p$ register controlled on the $\ell$ register.
This step is indicated by the ``In'' and ``prep$_\ell$'' in \fig{selectdf}, and uses the data output by the QROM in the previous step.
The steps needed for the preparation are as follows.
\begin{enumerate}
\item The controlled preparation of an equal superposition state may be performed in the following way with complexity $7n_\ranktwo
+2\rotprec-6$. The steps are as follows.
\begin{enumerate}
\item Use a string of $n_\ranktwo-1$ Toffolis (and Clifford gates) to produce a new $n_\ranktwo$-qubit register that has zeros matching the leading zeros in the binary representation of $\ranktwo^{(\ell)}$, and ones after that.
These qubits are used to control the Hadamards in the next step.
\item Perform $n_\ranktwo$ controlled Hadamards with cost $n_\ranktwo$. A controlled Hadamard can be performed with a single Toffoli by catalytically using a T state, as shown in \fig{chad}.
Therefore we count the cost of each controlled Hadamard as 1.
\item Perform an inequality test with the register containing $\ranktwo^{(\ell)}$ with cost $n_\ranktwo$.
\item Rotate the ancilla qubit with cost $\rotprec-2$ based on the rotation angle given by the output of the QROM.
\item The reflection on the result of the inequality test and ancilla qubit is a controlled phase which is a Clifford gate.
\item Invert the rotation with cost $\rotprec-2$ and the inequality test with Cliffords.
\item Perform the $n_\ranktwo$ controlled Hadamards again.
\item Reflect about the zero state on $n_\ranktwo+1$ qubits (the qubits where the state preparation is being performed and the rotated ancilla) with cost $n_\ranktwo-1$.
\item Perform the $n_\ranktwo$ controlled Hadamards again. Now the binary-to-unary conversion can be inverted with Cliffords.
\item Perform an inequality test again with cost $n_\ranktwo$.
Now, flagged on the success of the inequality test, we have prepared an equal superposition state on the second register that can be used for state preparation on the second register.
\end{enumerate}
\item We now need to add the offset to the second register to provide a contiguous register to apply QROM to for state preparation. This addition has cost $n_{\rankone,\ranktwo}-1$.
\item Next, apply the QROM to output the alt and keep values with size
\begin{equation}
b_{p2} = n_\ranktwo +\zetabits_2+2,
\end{equation}
where the $+2$ is for the sign bit and its alt value.
In this part there are $\rankone \ranktwo+N/2$ outputs to give, because the rotations for the one-electron term are needed as well, then in the next part there are only $\rankone \ranktwo$.
Therefore the two QROMs have cost
\begin{equation}\label{eq:prepqr1}
\left\lceil \frac {\rankone \ranktwo+N/2}{k_{p2}}\right\rceil + b_{p2}(k_{p2}-1),
\end{equation}
\item The remainder of the state preparation uses an inequality test with cost $\zetabits_2$, and a controlled swap with cost $n_\ranktwo$.
\end{enumerate}
\item Apply the number operators via rotations and QROM with the following steps.
\begin{enumerate}
\item To apply the QROM for the rotation angles, we need to again add an offset to the second register to provide a contiguous register, with cost again $n_{\rankone,\ranktwo}-1$.
This operation is indicated by the controlled box with $+$ in it in \fig{selectdf}, and a box with $-$ in it for its inversion.
\item Use QROM to output the rotation angles, which is shown as ``In$_p$'' and ``data$_p$'' in \fig{selectdf}. The QROM cost is
\begin{equation}
\left\lceil\frac {\rankone\ranktwo+N/2}{k_r}\right\rceil + N\rotbits(k_r-1)/2 .
\end{equation}
\item Perform controlled swaps with the spin qubit as control, with cost $N/2$.
\item Apply the controlled rotations to rotate the basis with cost $N(\rotbits-2)$.
\item Apply the $Z_1$ operation, controlled on success of the preparation of the $\ell$ and $p$ registers.
This control gives a cost of 1.
\item Reverse the controlled rotations and controlled swaps, with cost $N(\rotbits-2)$ and $N/2$.
\item Reverse the QROM with cost
\begin{equation}
\left\lceil\frac {\rankone \ranktwo+N/2}{k'_r}\right\rceil + k'_r .
\end{equation}
\item Reverse the addition onto the contiguous register, with cost again $n_{\rankone,\ranktwo}-1$.
\end{enumerate}
\item Invert the state preparation on the $p$ register, which is shown as ``In'' and ``prep$_p^\dagger$'' in \fig{selectdf}.
The cost is the same as given in step 2, except the cost of the QROM is reduced to
\begin{equation}\label{eq:prepiqr}
\left\lceil\frac {\rankone \ranktwo+N/2}{k'_{p2}}\right\rceil + k'_{p2}.
\end{equation}
\item The reflection on the second register \emph{only} (which is what caused the procedure to be this complicated) is performed.
The qubits that need to be reflected on are as follows.
\begin{enumerate}
\item The $n_\ranktwo$ qubits that the state is prepared on.
\item The $\zetabits_2$ that are used for the superposition state for the coherent alias sampling.
\item One that is rotated for the amplitude amplification.
\item One for the spin.
\end{enumerate}
That gives a total of $n_\ranktwo+\zetabits_2+2$.
This reflection also needs to be controlled on success for preparation of the $\ell$ register, and $\ell\ne 0$, so the complexity of the reflection is $n_\ranktwo+\zetabits_2+2$.
\item Repeat steps 2 to 5. This time, because we do not need to output values for the one-electron terms, the costs of the QROM and inverse QROM for state preparation are reduced to
\begin{equation}
\label{eq:prepqr2}
\left\lceil \frac {\rankone \ranktwo}{k_{p2}}\right\rceil + b_{p2}(k_{p2}-1), \qquad \left\lceil\frac {\rankone \ranktwo}{k'_{p2}}\right\rceil + k'_{p2}.
\end{equation}
The costs for the QROM for the rotations are reduced to
\begin{equation}
\left\lceil\frac {\rankone\ranktwo}{k_r}\right\rceil + N\rotbits(k_r-1)/2, \qquad \left\lceil\frac {\rankone \ranktwo}{k'_r}\right\rceil + k'_r .
\end{equation}
This time $Z_1$ must be controlled by the qubit flagging $\ell\ne 0$, which introduces a cost of another Toffoli.
\item Invert the QROM in step 2 and the preparation in step 1, where the costs of the QROMs are reduced to
\begin{equation}
\left\lceil\frac{\rankone+1}{k'_{p1}}\right\rceil + k'_{p1}, \qquad \left\lceil\frac{\rankone+1}{k'_o}\right\rceil + k'_o.
\end{equation}
\item There is also a cost for the reflection needed for constructing the quantum walk.
The qubits that need to be reflected on are as follows.
\begin{enumerate}
\item The $n_\rankone$ qubits for the $\ell$ register.
\item The $n_\ranktwo$ qubits for the $p$ register.
\item The $\zetabits_1$ qubits for the equal superposition state for the alias sampling for the $\ell$ register.
\item The $\zetabits_2$ qubits for the equal superposition state for the $p$ register.\
\item The two qubits that are rotated.
\item One qubit for the spin.
\end{enumerate}
That gives a total of $n_\rankone+n_\ranktwo+\zetabits_1+\zetabits_2+3$ so the cost is $n_\rankone+n_\ranktwo+\zetabits_1+\zetabits_2+1$.
\item The steps of the quantum walk are made controlled using unary iteration on an ancilla.
This introduces a cost of another two Toffolis for the unary iteration and making the reflection controlled.
\end{enumerate}
Adding together all these costs, then gives
\begin{align}
& 2(3n_\rankone-3\factor+2\rotprec-9) 
+\left\lceil\frac{\rankone+1}{k_{p1}}\right\rceil + b_{p1}(k_{p1}-1) 
+2\zetabits_1+2n_\rankone 
+\left\lceil\frac{\rankone+1}{k_o}\right\rceil + b_o(k_o-1) 
+\left\lceil\frac{\rankone+1}{k'_{p1}}\right\rceil + k'_{p1}\nn &\quad + \left\lceil\frac{\rankone+1}{k'_o}\right\rceil + k'_o 
+\left\lceil\frac {\rankone\ranktwo+N/2}{k_r}\right\rceil+\left\lceil\frac {\rankone\ranktwo}{k_r}\right\rceil + N\rotbits(k_r-1) 
+ \left\lceil\frac {\rankone \ranktwo+N/2}{k'_r}\right\rceil+ \left\lceil\frac {\rankone \ranktwo}{k'_r}\right\rceil + 2k'_r 
\nn &\quad+4(7n_\ranktwo+2\rotprec-6) 
+8(n_{\rankone,\ranktwo}-1) 
  +\left\lceil \frac {\rankone \ranktwo+N/2}{k_{p2}}\right\rceil+\left\lceil \frac {\rankone \ranktwo}{k_{p2}}\right\rceil + 2b_{p2}(k_{p2}-1) + \left\lceil\frac {\rankone \ranktwo+N/2}{k'_{p2}}\right\rceil \nn &\quad + \left\lceil\frac {\rankone \ranktwo}{k'_{p2}}\right\rceil + 2k'_{p2} 
+4\zetabits_2+4n_\ranktwo 
+4N(\rotbits-2)+2N+2 
+n_\ranktwo+\zetabits_2+2 
+n_\rankone+n_\ranktwo+\zetabits_1+\zetabits_2+3 
\nn &=
9n_\rankone-6\factor+12\rotprec
+\left\lceil\frac{\rankone+1}{k_{p1}}\right\rceil + b_{p1}(k_{p1}-1)
+\left\lceil\frac{\rankone+1}{k_o}\right\rceil + b_o(k_o-1)
+\left\lceil\frac{\rankone+1}{k'_{p1}}\right\rceil + k'_{p1}+ \left\lceil\frac{\rankone+1}{k'_o}\right\rceil + k'_o
\nn &
\quad +\left\lceil\frac {\rankone\ranktwo+N/2}{k_r}\right\rceil+\left\lceil\frac {\rankone\ranktwo}{k_r}\right\rceil + N\rotbits k_r + \left\lceil\frac {\rankone \ranktwo+N/2}{k'_r}\right\rceil+ \left\lceil\frac {\rankone \ranktwo}{k'_r}\right\rceil + 2k'_r
+34n_\ranktwo 
+8 n_{\rankone,\ranktwo}
\nn &\quad +\left\lceil \frac {\rankone \ranktwo+N/2}{k_{p2}}\right\rceil+\left\lceil \frac {\rankone \ranktwo}{k_{p2}}\right\rceil + 2b_{p2}(k_{p2}-1)  + \left\lceil\frac {\rankone \ranktwo+N/2}{k'_{p2}}\right\rceil+ \left\lceil\frac {\rankone \ranktwo}{k'_{p2}}\right\rceil + 2k'_{p2}
+3\zetabits_1+6\zetabits_2
+3N\rotbits-6N
-43,\label{eq:MStoffolis}
\end{align}
with $b_{p1} = n_\rankone+\zetabits_1$, $b_o = n_\ranktwo+n_{\rankone,\ranktwo}+p$, and $b_{p2} = n_\ranktwo +\zetabits_2+2$, and the numbers of qubits as defined in \eq{numbersms}.

Next consider the logical qubit count.
The qubits used are as follows.
\begin{enumerate}
\item The $\lceil\log(\mathcal{I}+1)\rceil$ qubits for the control register for the state preparation and $\lceil\log(\mathcal{I}+1)\rceil-1$ for the unary iteration on this register.
\item There are $N$ qubits used for the system.
\item The first register that is prepared needs $n_\rankone$ qubits, plus one to flag success of the state preparation and one that is rotated as part of the state preparation.
\item The output of the QROM needs $n_\rankone+\zetabits_1$, another $\zetabits_1$ are used for the equal superposition state to take the inequality with, and another qubit is needed as the output of the inequality test. We can ignore temporary ancillas used as more will be needed at a later step.
\item The second QROM on the first register used to output the data needed for the equal superposition state on the second register uses $b_o$ output bits.
\item The second register needs $n_\ranktwo$ bits, plus another bit for flagging success of preparing the equal superposition and another that is rotated.
\item A register with size $n_{\rankone,\ranktwo}$ for outputting the contiguous register for the state preparation on the second register.
This is a temporary ancilla that can be erased after the QROM with Cliffords by using the out-of-place adder.
It then is computed again for the QROM for the rotation angles.  There it can be computed in place, so does not add to the qubit count.
\item The QROM used for preparing the state on the second register uses $b_{p2}$ qubits output, as well as a number of temporary qubits that are less than those needed for the rotation angles output later.
\item The state preparation needs another $\zetabits_2$ qubits in a superposition, as well as another qubit flagging success of the inequality test.
\item The angles for the rotations need $k_r N\rotbits/2$.
\item The phase gradient state which needs $\rotbits$ qubits.
\item Another control qubit is needed for the spin.
\item A T state on a single qubit is used to perform the controlled Hadamards with Toffoli gates.
\end{enumerate}
Adding together these ancilla costs gives
\begin{align}\label{eq:MSqubits}
N+2n_\rankone
+n_\ranktwo
+2\zetabits_1
+\zetabits_2
+\rotbits
+b_o
+b_{p2}
+k_r N\rotbits/2
+2\lceil\log(\mathcal{I}+1)\rceil
+7.
\end{align}

\begin{table*}
\begin{tabular}{|c|c|c|c|c|}
\hline
threshold & $L$ & Eigenvectors & $\lambda$ & CCSD(T) error (m$E_h$) \\ \hline
0.1 & 53 & 765 & 253.6 & -87.91 \\ \hline
0.05 & 95 & 1274 & 262.1 & -12.39 \\ \hline
0.025 & 135 & 2319 & 272.0 & -2.78 \\ \hline
0.0125 & 182 & 3983 & 280.8 & -1.10 \\ \hline
0.01 & 195 & 4700 & 283.4 & -0.18 \\ \hline
0.0075 & 216 & 5678 & 286.3 & 1.33 \\ \hline
0.005 & 242 & 7181 & 289.5 & 2.27 \\ \hline
0.0025 & 300 & 9930 & 292.9 & 1.02 \\ \hline
\color{blue}0.00125 & \color{blue}360 & \color{blue}13031 & \color{blue}294.8 & \color{blue}0.44 \\ \hline
0.001 & 384 & 14062 & 295.2 & 0.25 \\ \hline
0.00075 & 414 & 15419 & 295.6 & 0.20 \\ \hline
0.0005 & 444 & 17346 & 296.1 & 0.09 \\ \hline
0.000125 & 567 & 24005 & 296.7 & -0.01 \\ \hline
0.0001 & 581 & 25054 & 296.7 & -0.02 \\ \hline
0.00005 & 645 & 28469 & 296.8 & 0.00 \\ \hline
\end{tabular}
\caption[Double low rank algorithm overheads for Reiher FeMoCo Hamiltonian]{Double low rank factorization data for the Reiher Hamiltonian \cite{Reiher2017}.
Here $\lambda_{T'}$ is 38.6 Hartree, which is the Schatten-norm of $T'$.
A threshold is used to truncate eigenvectors based on the truncation strategy described in \cite{vonBurg2020} and \eq{dfthresh}. The entry in blue is the one used for our resource estimates.
}
\label{tab:dfR}
\end{table*}

\begin{table*}
\begin{tabular}{|c|c|c|c|c|}
\hline
threshold & $L$ & Eigenvectors & $\lambda$ & CCSD(T) error (m$E_h$) \\ \hline
0.1 & 94 & 1998 & 1076.5 & -287.44 \\ \hline
0.05 & 184 & 3765 & 1108.9 & -91.76 \\ \hline
0.025 & 205 & 5992 & 1136.0 & -20.52 \\ \hline
0.0125 & 247 & 8450 & 1152.1 & -12.74 \\ \hline
0.01 & 261 & 9302 & 1155.5 & -6.73 \\ \hline
0.0075 & 278 & 10493 & 1159.1 & -5.77 \\ \hline
0.005 & 312 & 12508 & 1163.4 & -2.30 \\ \hline
0.0025 & 344 & 16355 & 1168.6 & 1.53 \\ \hline
\color{blue}0.00125 & \color{blue}394 & \color{blue}20115 & \color{blue}1171.2 & \color{blue}0.07 \\ \hline
0.001 & 413 & 21407 & 1171.7 & 0.42 \\ \hline
0.00075 & 434 & 23145 & 1172.2 & 0.40 \\ \hline
0.0005 & 470 & 25751 & 1172.8 & 0.42 \\ \hline
0.000125 & 589 & 35006 & 1173.7 & -0.04 \\ \hline
0.0001 & 614 & 36557 & 1173.7 & -0.02 \\ \hline
0.00005 & 679 & 41563 & 1173.9 & -0.01 \\ \hline
\end{tabular}
\caption[Double low rank algorithm overheads for Li FeMoCo Hamiltonian]{Double low rank factorization data for the Li Hamiltonian \cite{Li2019}.
Here $\lambda_{T'}$ is 478.1 Hartree, which is the Schatten-norm of $T'$.
A threshold is used to truncate eigenvectors based on the truncation strategy described in \cite{vonBurg2020} and \eq{dfthresh}. The entry in blue is the one used for our resource estimates.
}
\label{tab:dfL}
\end{table*}

\subsection{Numerical determination of double low rank factorization}

The numerical determination of double low rank factorization 
can be ambiguous without further details.
In this subsection, 
we aim to provide full details of 
how the factorization is obtained in this work.
The original approach proposed by Peng and Kowalski \cite{peng2017highly}
worked with separate thresholds 
for the first and the second factorizations.
Instead, von Burg \emph{et al.}~\cite{vonBurg2020}
proposed a truncation scheme just based on a single threshold
which we further elaborate here.
\begin{enumerate}
\item First, we perform either
the eigendecomposition or the Cholesky decomposition of
$V$,
$V_{pqrs} = \sum_{\ell=1}^L W_{pq}^{(\ell)} W_{rs}^{(\ell)}$
and sort $\ell$
in the ascending order of
the corresponding eigenvalues
such that the magnitude of $W_{pq}^{(\ell)}$ for small $\ell$'s are
larger than that of
large $\ell$'s
.
\item The second factorization was originally proposed to be done with the singular value decomposition of $W^{(\ell)}$ \cite{peng2017highly}, but
following von Burg \emph{et al.}~\cite{vonBurg2020} we use eigendecomposition:
$W_{pq}^{(\ell)} = \sum_{m=1}^{N/2} f_m^{(\ell)} U_{pm}^{(\ell)} U_{qm}^{(\ell)}$.
\item For a given threshold and looping over the first factorization index $\ell$ from 1 to $L$, 
we discard the $m$-th eigenvector in the second factorization that satisfies
\begin{equation}
\left(\sum_{p=1}^{N/2} |f_p^{(\ell)}|\right)
|f_m^{(\ell)}| < \text{threshold}\, .
\label{eq:dfthresh}
\end{equation}
We note that if no eigenvectors are kept at $\ell_0$ then we discard the rest of vectors for $\ell>\ell_0$ without going through them.
\end{enumerate}
This scheme was used to generate the numerical data presented in \tab{dfR} and \tab{dfL}.

\subsection{Numerical data for hydrogen chains and \texorpdfstring{\ce{H4}}{H4}}
\begin{figure}[ht!]
\includegraphics[scale=0.50]{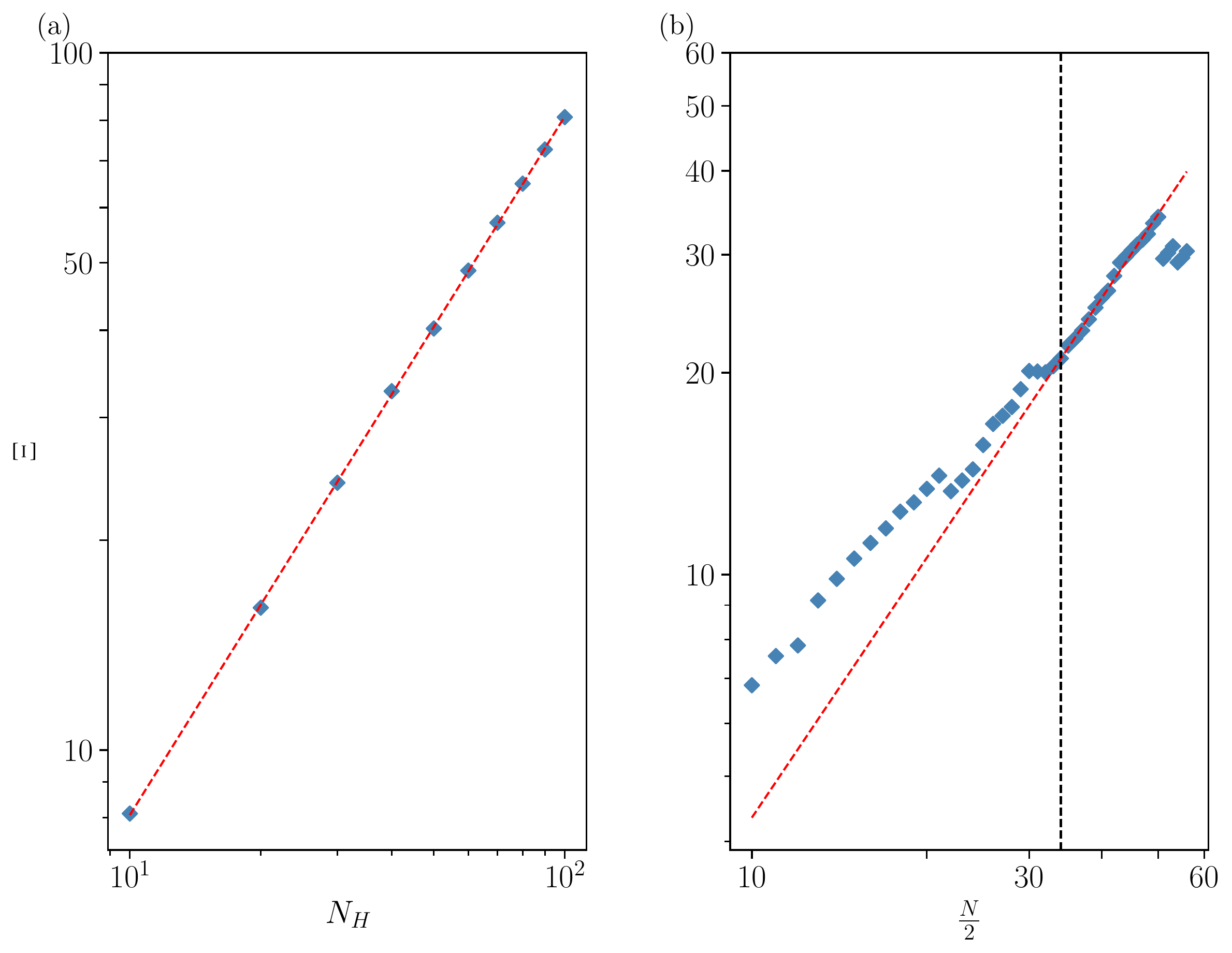}
\caption{\label{fig:h4hchainmu}
(a) The number of H atom ($N_H$) versus $\ranktwo$ and (b) the number of orbitals ($N/2$) versus $\ranktwo$.
All data points in (a) are used in the linear fit on a log scale where the slope is given as 1.00 with $R^2 = 0.9999$ (where $R$ is the correlation coefficient).
In (b), every point beyond $N/2 = 35$ except the last six points is used for the linear fit, which yields a slope of 1.29 with $R^2 = 0.9917$.
The DF threshold that was used is 0.01 for (a) and 1$\times10^{-4}$ for (b).
}
\end{figure}

The average rank of the second factorization for hydrogen chain and \ce{H4} is shown in \fig{h4hchainmu}.
Theoretically, $\ranktwo$ cannot scale worse than $O(N)$.
However, we obtained $O(N^{1.29})$ in \fig{h4hchainmu} (b).
Therefore, in \tab{hydrogen_scalings} we assumed that $\ranktwo$ scales as $O(N)$ for the \ce{H4} case.

\section{The randomized Trotter simulation algorithms of Campbell and Kivlichan \emph{et al.}}
\label{app:drift}

Here we discuss in more detail a recent randomized approach, often called qDRIFT~\cite{Campbell2018B}, that approximates time evolution using a randomized evolution.  
The randomized nature of qDRIFT leads to a subtle point arising between the cases considered here and the analysis provided in~\cite{Campbell2018B,Kivlichan2019PhaseHamiltoniansB} in that the results in both cases discuss the probability of failure for the algorithm rather than the mean-square error. The analysis below focuses on the mean-square error in phase estimation that arises from qDRIFT and we see that because of the possibility of large deviations, the worst case scaling of the complexity can be substantially worse than the $\mathcal{O}(\lambda^2/\epsilon^2)$ scaling anticipated from such results.

Before jumping into the analysis, we will review the fundamentals of the qDRIFT algorithm.  The idea behind the evolution is to implement a channel, for the Hamiltonian $H=\sum_j H_j$
\begin{equation} 
    U(\tau,0): \rho \mapsto \sum_j p_j e^{-iH_j \tau/p_j}\rho e^{iH_j \tau/p_j},\label{eq:qdrift}
\end{equation}
where $p_j = \|H_j\| / \sum_j \|H_j\|$ where $\|\cdot\|$ is the Schatten infinity norm (spectral norm).  In our case we will assume that each $H_j$ is proportional to a unitary so then $p_j = \|H_j\|/ \lambda$. This channel can be implemented by randomly selecting a term from the Hamiltonian, simulating the evolution of just that term for a duration that scales inversely with the probability of drawing the Hamiltonian term.  From~\cite{Campbell2018B} we have that if we define $\mathcal{E}(\tau,0): \rho \mapsto e^{-iH\tau} \rho e^{iH\tau}$
\begin{equation}
    \|U(\tau,0) - \mathcal{E}(\tau,0)\|_\diamond \le 2 \lambda^2 \tau^2.
\end{equation}

Note that this exact same bound holds if we consider a controlled evolution using qDRIFT, which can be thought of as simulating the Hamiltonian $H_{\rm ctrl} = \ketbra{1}{1} \otimes H = \sum_j \ketbra{1}{1}\otimes H_j$.
As discussed in \cite{Campbell2018B}, the value of $\lambda$ is unchanged so the bound on the error in the controlled Hamiltonian is the same as for the Hamiltonian without control.
For improved performance of the phase estimation, one can also use control between positive and negative Hamiltonian evolution \cite{BabbushSpectraB}.
That corresponds to $H_{\rm ctrl} = -Z \otimes H$, and again the analysis of the error for the controlled Hamiltonian is the same as without control, because the value of $\lambda$ is unchanged.
This implies that the same bound holds in the controlled and the uncontrolled case.  When used in conjunction with qDRIFT, the variance in the estimate is not the only source of error.

\subsection{qDRIFT mean-square error}
First, let us examine the scaling if we were to try to apply phase estimation using qDRIFT in order to achieve a fixed root-mean-square error in the energy (for example $1.6$ mHa).  Existing work in~\cite{Campbell2018B} discusses this case; however, the failure probability scaling of $O(1/(\delta^3\epsilon^2))$, where $\delta$ is the failure probability and $\epsilon$ is the uncertainty in the phase upon success, in that work seemingly precludes that analysis from yielding an inexpensive simulation that guarantees a fixed root-mean-square error.  Here we provide a tighter analysis and demonstrate that the scaling is worse than one would na\"{i}vely expect.

We begin by using the fact that the diamond norm bounds the worst-case trace norm between outputs that can be attained over all inputs to the channels.  Further, it can be seen using the arguments from~\cite{Campbell2018B,Berry2020} that for any positive integer $R$
\begin{equation}
    \|U^R(\tau,0) - \mathcal{E}^R(\tau,0)\|_\diamond \le 2 R\lambda^2 \tau^2.
\end{equation}
Thus by the von-Neumann trace inequality it follows that for any bounded observable $O$
\begin{equation}
    \max_{\rho}|{\rm Tr}(O U^R(\tau,0)(\rho)) - {\rm Tr}(O \mathcal{E}^R(\tau,0)(\rho)| \le 2 \|O\| R\lambda^2 \tau^2.
\end{equation}
Let $\phi$ be the exact phase for the exact evolution, $\hat\phi$ be the observable for the phase, and take $O = (\hat\phi-\phi)^2$ to be the observable for the squared error in the phase.  The quantity we wish to bound is then for initial state $\rho_0$
\begin{align}
    \langle (\hat\phi-\phi)^2 \rangle := {\rm Tr}((\hat{\phi} - \phi)^2\mathcal{E}^R(\tau,0)(\rho_0)) \le \left|{\rm Tr}((\hat{\phi} - \phi)^2U^R(\tau,0)(\rho_0))\right| +\left|{\rm Tr}(O U^R(\tau,0)(\rho)) - {\rm Tr}(O \mathcal{E}^R(\tau,0)(\rho) \right| , \label{eq:expansion}
\end{align}
where the expectation value on the left is defined to be that for the qDRIFT evolution.

Then $\|O\|\le \max|\pi\pm\phi|^2$, and so
\begin{equation}
    |{\rm Tr}(O U^R(\tau,0)) - {\rm Tr}(O \mathcal{E}^R(\tau,0)| \le 2 \max|\pi\pm\phi|^2 R\lambda^2 \tau^2.
\end{equation}
That gives the maximum increase in the mean-square error in the phase.
Now consider the case that there is a different time interval for the qDRIFT $\tau$ versus the time interval for the phase estimation $t$, with $\tau=t/k$ for some integer $k$.
Then the total number of intervals for the qDRIFT $R$ is related to the total number for phase estimation $r$ as $R=kr$.
Then the mean-square error in the phase for the exact evolution is $\pi^2/r^2$, and so
\begin{equation}
    \langle (\hat\phi-\phi)^2 \rangle \le \frac{\pi^2}{r^2} + \frac{2 \max|\pi\pm\phi|^2 r\lambda^2 t^2}k.
\end{equation}

This expression corresponds to the uncertainty in the phase that we would achieve through the use of traditional phase estimation; however, we can improve the phase achieved per unit time by controlling the direction of the phase evolution as discussed in~\cite{Berry2020,Wecker2014B}.  Using this observation we can define an energy operator $\hat{E}=\hat{\phi} /2t$ so that
\begin{equation}
    \langle (\hat{E} - E)^2\rangle \le \frac{\pi^2}{4r^2 t^2} + \frac{ \max|\pi\pm\phi|^2 r\lambda^2}{2k}.
\end{equation}
Then writing this in terms of $N_{\exp} =  kr$,
\begin{equation}\label{eq:eerrbnd}
    \langle (\hat{E} - E)^2\rangle \le \frac{\pi^2 k^2}{4N_{\exp}^2 t^2} + \frac{ \max|\pi\pm\phi|^2 N_{\exp}\lambda^2}{2k^2}.
\end{equation}
We aim to find the smallest $N_{\exp}$ such that the right-hand side is bounded above by $\epsilon^2$, where $\epsilon$ is as before the bound on the allowable root-mean square error in the energy estimate.
Minimising $N_{\exp}$ for constant $\epsilon$ is the same as minimising $\epsilon$ for constant $N_{\exp}$.
To minimise over $k$, take the derivative of the right-hand side to give
\begin{equation}
\frac{\partial }{\partial k}\left(\frac{\pi^2 k^2}{4N_{\exp}^2 t^2} + \frac{ \max|\pi\pm\phi|^2 N_{\exp}\lambda^2}{2k^2} \right) =  \frac{\pi^2 k}{2N_{\exp}^2 t^2} - \frac{ \max|\pi\pm\phi|^2 N_{\exp}\lambda^2}{k^3}.
\end{equation}
The turning point is for
\begin{equation}
k^2 = \sqrt{2N_{\exp}^3} \max|1\pm\phi/\pi| \lambda t.
\end{equation}
Putting that into \eq{eerrbnd} gives
\begin{equation}
    \langle (\hat{E} - E)^2\rangle \le \frac{ \pi\max|\pi\pm\phi| \lambda}{\sqrt{2N_{\exp}}t}.
\end{equation}
Setting the right-hand side equal to $\epsilon^2$ then gives
\begin{equation}
    N_{\exp} = \frac{ \pi^2\max|\pi\pm\phi|^2 \lambda^2}{2\epsilon^4 t^2}.
\end{equation}
Because $E=\phi/2t$ and $E$ can, in principle, be as large as $\lambda$, $\phi=\pi$ should correspond to $E\ge\lambda$ in order to resolve all possible values of $E$.
That implies that the maximum allowable value of $t$ is $\pi/2\lambda$, which gives
\begin{equation}
    N_{\exp} = \frac{ 4\max|\pi\pm\phi|^2 \lambda^4}{2\epsilon^4}\le \frac{8\pi^2 \lambda^4}{\epsilon^4}.
\end{equation}
This scaling is as $\mathcal{O}(\lambda^4/\epsilon^4)$ and will therefore yield large gate counts for the problems that we focus on.
The reason for the large scaling, is that the diamond norm bounds the difference in the probabilities of measurement results, which can allow an increase in probability concentrated at the maximum error leading to large mean-square error.

The gate complexity for the unamplified qDRIFT subroutine can be found by multiplying the number of exponentials by the number of Toffoli gates needed per exponential.
The Hamiltonian can be expressed such that each $H_j$ is a weighted Pauli operator.  In this case $e^{-iH_j \tau /p_j}$ can be implemented using at most $\mathcal{O}(N)$ one and two qubit Clifford operations and a single $R_z$ gate~\cite{Whitfield2010}.
We implement these rotations using a state that is constructed using a $q+1$ qubit phase gradient state:
\begin{equation}
    \ket{\psi_{q+1}} = \frac{1}{\sqrt{2^{q+1}}} \sum_{j=0}^{2^{q+1}-1} e^{\pi ij / 2^{q}} \ket{j} . \label{eq:resource}
\end{equation}
We can then use this as a resource state to implement a single qubit rotation using incrementer gates as described in Appendix A of~\cite{Sanders2020CompilationOptimizationB} which requires $q-1$ Toffoli gates to implement $e^{-iZ \pi/2^q}$ using the $q+1$ qubit resource state in~\eqref{eq:resource}. 

Next note that for any value of $\lambda t/k$ we can round the desired rotation angle to $\lambda t/k \ge \pi / 2^{\lceil \log (\pi k/\lambda t)\rceil}$.  Thus we can choose the evolution time $t$ such that we will, at worst, need to halve the evolution time to ensure that the rotation angle coincides with the angle returned by the phase gradient state.  This will necessitate choosing $t\in [\pi/4\lambda, \pi/2\lambda]$ which corresponds in the best case scenario to taking $q= \log(2 k)\approx \frac{1}{2}\log(32\pi^4 \lambda^6 / \epsilon^6) +1$. 
The number of Toffoli gates needed is
\begin{equation}
    N_{\rm Toff} = N_{\exp} (q-1) \approx \frac{4\pi^2 \lambda^4 \log(32 \pi^4 \lambda^6/\epsilon^6)}{\epsilon^4} .\label{eq:ToffqDRIFT} 
\end{equation}

The total number of ancillae needed by the algorithm is given by the sum of the ancillae needed to store the resource state $\ket{\psi_{q+1}}$, the ancillae needed for the phase estimation step using the method of~\cite{BabbushSpectraB} and any ancillae needed to implement the carry logic incrementer circuit~\cite{GidneyAdder} and finally the $N$ target qubits.  These spatial overheads sum to
\begin{align}
    N_{\rm qubits} &= N+ (q+1) + 2\lceil \log(r+1) \rceil -1 + q \lesssim N+2\log(N_{\exp}r)+2\nonumber\\
    &\lesssim N +2\log\left(\frac{2\lambda^2}{\epsilon^2} \right)+2 . \label{eq:ancqDrift}
\end{align}

The scaling of this algorithm is quartically worse than the cost of applying qubitized phase estimation.  These comparisons suggest at first glance that randomized simulations are not a competitive technique for simulating such dynamical systems.  However, the results in~\cite{Kivlichan2019} suggest that incorporating randomness to simulate low importance terms and using conventional methods to simulate high-importance terms may be a much more profitable approach to using randomized simulation methods for challenge problems in chemistry.  

The specific numbers for the Reiher \emph{et al.}\ and Li \emph{et al.}\ Hamiltonians can be found using~\eqref{eq:ToffqDRIFT} and~\eqref{eq:ancqDrift} together with the $\lambda$ values from the sparse Hamiltonian simulation method.  Since qDRIFT does not have quantum costs that depend at all on the number of terms in the Hamiltonian we take the smallest thresholds considered in Tables~\ref{tab:sparseR} and~\ref{tab:sparseL} ($5\times 10^{-5}$) for the Reiher \emph{et al.}\ and Li \emph{et al.}\ Hamiltonians respectively.  This corresponds to $\lambda = 2183.6$ and $\lambda = 1600.9$ respectively.  Taking chemical accuracy of 0.0016 Hartree as our target accuracy then yields $N_{\rm Toff} = 1.8\times 10^{28}$ for the Reiher Hamiltonian and $N_{\rm Toff} = 5.2\times 10^{27}$ for the Li Hamiltonian.  Further, the Reiher Hamiltonian requires $168$ logical qubits to simulate within this accuracy whereas the Li Hamiltonian would require approximately $211$ logical qubits.

\subsection{Confidence intervals for the eigenphases}

Although the presence of (possibly) fat tails for the phase estimation distribution using qDRIFT prevents our analysis from yielding low cost estimates for achieving a fixed mean-squared-error in the estimate of the ground state energy for FeMoCo, such an estimate may not strictly speaking be necessary.
The reason is that one may instead aim to obtain a confidence interval of small size, rather than a small mean-square error.
Alternatively one may wish to consider repeated application of the quantum algorithm with sampling of the measurement results.
In that case the root-mean square error is only important if one is taking the average of the measurement results, but one can instead use the Hodges-Lehmann estimator \cite{hodges1963} which can have small root-mean square error despite a large root-mean square error in the individual samples.
One could also use classically obtained prior knowledge about the ground state energy to reject values that are outside the range of possible values, thereby reducing the impact of the fat tails.

In the following we will consider the complexity when aiming to obtain a confidence interval of small size, then the Hodges-Lehmann estimator.
An innovation we introduce here is a new form of phase estimation that uses a Kaiser window to optimally yield a sample that is with $95\%$ probability within chemical accuracy of the ground state (although other levels of confidence can also be found numerically using the same strategy).
To explain this method, we first review the formalism of phase estimation.

In canonical phase estimation (as opposed to iterative phase estimation) we have a state of the form
\begin{equation}
\ket{\psi(\phi)} = \sum_{j=0}^{r-1} e^{ij\phi} \psi_j \ket{j} .
\end{equation}
The inner product with the phase state
\begin{equation}
\ket{\hat\phi} = \frac 1{\sqrt r}\sum_{j=0}^{r-1} e^{ij\hat\phi} \ket{j}
\end{equation}
where $\hat\phi$ is the estimator for $\phi$ (rather than an operator)
gives
\begin{equation}
\braket{\hat\phi}{\psi(\phi)} = \frac 1{\sqrt r}\sum_{j=0}^{r-1} e^{ij(\phi-\hat\phi)} \psi_j \, .
\end{equation}
Let us put
\begin{equation}
x_j = 2j/r - 1 +1/r
\end{equation}
so $x_0=- 1 +1/r$ and $x_{r-1} = 1 - 1/r$ which gives
\begin{equation}
\braket{\hat\phi}{\psi(\phi)} = \frac 1{\sqrt r}\sum_{j=0}^{r-1} e^{i(x_j +1-1/r)(\phi-\hat\phi)r/2} \psi_j \, .
\end{equation}
Putting $\omega=(\phi-\hat\phi)r/2$, we have
\begin{equation}
\braket{\hat\phi}{\psi(\phi)}e^{-i(1-1/r)\omega} = \frac 1{\sqrt r}\sum_{j=0}^{r-1} e^{ix_j \omega} \psi_j \, .
\end{equation}
Now you can define
\begin{equation}
\psi(x) = \sum_j \psi_j \delta(x+1-(2j+1)/r) \, ,
\end{equation}
which gives
\begin{equation}
\braket{\hat\phi}{\psi(\phi)}e^{-i(1-1/r)\omega} = \frac 1{\sqrt r}\int dx \, e^{ix \omega} \psi(x) \, .
\end{equation}
Because $\psi(x)$ can be obtained by multiplying by a comb function with spacing $2/r$, $|\braket{\hat\phi}{\psi(\phi)}|$ as a function of $\omega$ must repeat with spacing $\pi r$ (that is, it is effectively convolved with a comb).

To consider the scaling in the limit of large $r$, it is convenient to consider a continuous $\psi$ that is nonzero in the interval $[-1,1]$, which would give $|\braket{\hat\phi}{\psi(\phi)}|$ that is non-repeating in $\omega$, and consider the variance of $\omega$ when using $|\braket{\hat\phi}{\psi(\phi)}|^2$ as a probability distribution.
The quantity $\braket{\hat\phi}{\psi(\phi)}$ is equivalent to a Fourier transform of $\psi(x)$, and
the standard function on the interval $[-1,1]$ that minimises the variance of its Fourier transform is
\begin{equation}
\psi(x) = \begin{cases} \cos(\pi x/2) & |x|\le 1, \\
0 & |x|>1.
\end{cases}
\end{equation}
The Fourier transform is
\begin{equation}
\frac{\sqrt{8\pi} \cos{\omega}}{\pi^2-4\omega^2} .
\end{equation}
The variance of $\omega$ can be found as
\begin{equation}
\int d\omega \, \omega^2 \frac{8\pi \cos^2{\omega}}{(\pi^2-4\omega^2)^2} = \frac{\pi^2}4 .
\end{equation}
In the case of finite $r$ we note that $\omega=(\phi-\hat\phi)r/2$, so the variance in the phase estimate is divided by $r^2/4$, giving
$\pi^2/r^2$.

If we want to find the 95\% confidence interval, we need to solve
\begin{equation}
\int_{-a}^a d\omega \, \omega^2 \frac{8\pi \cos^2{\omega}}{(\pi^2-4\omega^2)^2} = 0.95 \, .
\end{equation}
Numerically solving gives $a=2.863325$, 
so the ratio of the size of the 95\% confidence interval to the standard deviation is $1.822849$, 
which is a bit less than the ratio for Gaussians of $1.959964$. 

If we want to do better, then we can use the Kaiser window.
The square of the Fourier transform of the Kaiser window on $[-1,1]$ is proportional to
\begin{equation}
\frac{\sinh^2\sqrt{\alpha^2-\omega^2}}{\alpha^2-\omega^2},
\end{equation}
where we can adjust $\alpha$ to optimise the size of the confidence interval.
Numerically solving gives the narrowest 95\% confidence interval as $a=2.542853$ 
with $\alpha=2.179411$. 

Now, what we are interested in is the minimum cost when we perform qDRIFT, and allow some probability of error from the qDRIFT and some probability from the phase estimation.
With $r$ the number of repetitions, the error from qDRIFT is
\begin{equation}
2r\lambda^2 t^2.
\end{equation}
Note that this is equivalent to taking $k=1$ in the notation of the previous subsection, which means that the qDRIFT interval is the same as that used for phase estimation.
Therefore, for smaller $t$, the error due to qDRIFT is smaller, but the error in estimating the \emph{energy} is scaled by $t$, so smaller $t$ leads to larger error in the energy.
The total confidence interval of 95\% can be obtained with
\begin{equation}
\int_0^{\epsilon t r} d\omega \frac{\sinh^2\sqrt{\alpha^2-\omega^2}}{\alpha^2-\omega^2} = \left( 0.95 + r\lambda^2 t^2\right)
\int_0^\infty d\omega \frac{\sinh^2\sqrt{\alpha^2-\omega^2}}{\alpha^2-\omega^2} .
\end{equation}
The reason for the upper limit on the integral of $\epsilon t r$ is that the allowable error in the energy is $\epsilon$.
Since $E = \phi/2t$, that means the allowable error in the phase is $2\epsilon t$.
Since $\omega=(\phi-\hat\phi)r/2$, the allowable error in $\omega$ is $\epsilon t r$.
The integral on the right-hand side is for normalisation.
The expression $0.95 + r\lambda^2 t^2$ indicates that the confidence interval due to the phase estimation needs an additional $r\lambda^2 t^2$ to account for the additional probability of error from qDRIFT.
The probability of error outside the interval due to qDRIFT can only increase by half of $2r\lambda^2 t^2$, because increasing probability outside the interval by $r\lambda^2 t^2$ means that the probability elsewhere needs to be increased by $r\lambda^2 t^2$, resulting in the total error in the probability distribution of $2r\lambda^2 t^2$.

We then aim to solve for $r=N_{\exp}$ for any given $t$ and $\alpha$, and minimise it.
We set $a=\epsilon t r$ and $\delta=r\lambda^2 t^2$, and find
\begin{equation}
\frac{\epsilon^2 r}{\lambda^2}= \frac{a^2}{\delta}
\end{equation}
so
\begin{equation}\label{eq:CIest}
N_{\exp}=\frac{\lambda^2 a^2}{\epsilon^2\delta} .
\end{equation}
Note that this shows $N_{\exp}$ is proportional to $\lambda^2/\epsilon^2$, which gives the scaling of the complexity up to logarithmic factors (from synthesising rotations).
Minimising, we get the minimum value of $a^2/\delta$ as $304.744$, 
with $\alpha=3.05961$, 
$a=3.37625$, and 
$\delta=0.0374053$. 
We take $\epsilon=0.0016$, so for Reiher with $\lambda=2183.6$ we get $N_{\exp}=5.67598\times 10^{14}$,
and for Li with $\lambda=1600.9$ we get $N_{\exp}=3.05087\times 10^{14}$.

Next, note that the rotations for the exponentials are by an angle $\lambda t$, which can be performed more efficiently if 
$\lambda t$ is equal to $\pi$ divided by a power of 2.
From the above $\lambda t=\epsilon\delta/\lambda a$, so
for the two cases we get $\lambda\tau$ equal to $\pi/2^{28}$ times $0.693643$ and $0.946117$.
We can round the first to $11/16$, and the second to $7/8$.
That means the first would need 31 Toffolis per step, and the second would need 30 Toffolis per step for the exponential.
There will also be a single Toffoli per step for the unary iteration on the control register for the phase estimation, and another Toffoli for using that control register to control between forward and reverse evolution.
Then we can minimise $r$ over $\alpha$ for the constant product $\lambda t$ to give the following.
\begin{enumerate}
\item For Reiher, the minimum $N_{\exp}$ is $5.67874\times 10^{14}$ for $\alpha=3.03125$ with $\lambda\tau=(11/16)\pi/2^{28}$, giving a number of Toffolis $1.9\times 10^{16}$ and a number of logical qubits at most equal to $270$. 
\item For Li, the minimum $N_{\exp}$ is $3.12299\times 10^{14}$ for $\alpha=2.8794$ with $\lambda\tau=(7/8)\pi/2^{28}$, giving a number of Toffolis $1.0\times 10^{16}$ and a number of logical qubits at most equal to $310$. 
\end{enumerate}

These results are substantially better than those predicted using the bounds in Campbell's work.  Assuming a $95\%$ confidence interval, the number of exponentials is given for a probability of failure $P_0 = 0.05$ (see Eq.~(45) in the Supplemental Material of \cite{Campbell2018B})
\begin{equation}
    N_{\exp} \le \frac{27\pi^2\lambda^2}{2\epsilon^2 P_0^3} ,
\end{equation}
which yields $4.0\times 10^{18}$ exponentials for the Reiher Hamiltonian and $2.1\times 10^{18}$ exponentials for the Li Hamiltonian.

An alternative approach is to consider the case that the Hodges-Lehmann estimator \cite{hodges1963} is used instead of the mean when combining results of multiple runs of the quantum algorithm.
The Hodges-Lehmann estimator is based on taking all pairwise means of the samples, and finding the median of those pairwise means.
For $M$ samples, $M$ times the mean-square error of the estimate is asymptotically given by (see p.~245 of \cite{lehmann99})
\begin{equation}
    \frac 1{12\left(\int d\omega\, p^2(\omega)\right)^2}
\end{equation}
for symmetric probability distribution $p(\omega)$.
A reasonable goal in the case of qDRIFT is for this quantity to be equal to $\epsilon^2$, rather than the variance, because this quantity will correspond to the asymptotic variance under multiple samples.
For the case of the probability distribution
\begin{equation}
    p(\omega) = \frac{8\pi\cos^2\omega}{(\pi^2-4\omega^2)^2},
\end{equation}
the asymptotic value for the Hodges-Lehmann mean-square error is 2.38991, which is about 3\% below the mean-square error for an individual sample of $\pi^2/4$.

In order to bound the effect of qDRIFT error on the Hodges-Lehmann estimator, when modifying the probability distribution the maximum reduction in $\int d\omega\, p^2(\omega)$ will be reducing the maximum probability, while increasing the probability for large values of $\omega$ by very small amounts over a wide region.
It should be noted that this assumes that the error is symmetric.
Calling the modified probability distribution $q$, we therefore need
\begin{equation}
    \frac 1{12r^2 t^2\left(\int d\omega\, q^2(\omega)\right)^2} \le \epsilon^2 ,
\end{equation}
while
\begin{equation}
    \int d\omega \, |p(\omega)-q(\omega)| \le 2r\lambda^2 t^2.
\end{equation}
The choice for $r$ is then
\begin{equation}
    r = \frac {\lambda^2}{6\epsilon^2 \left(\int d\omega\, q^2(\omega)\right)^2\int d\omega \, |p(\omega)-q(\omega)|}.
\end{equation}
The optimal choice of $q$ is to truncate the probability distribution at $0.648057$ times the maximum of $p$, which gives the multiplying factor on $\lambda^2/\epsilon^2$ as $35.5192$; that is,
\begin{equation}\label{eq:HLest}
    N_{\exp} = 35.5192 \frac{\lambda^2}{\epsilon^2}.
\end{equation}
Therefore, with this estimate, we have the same $\lambda^2/\epsilon^2$ scaling as for a confidence interval, but a considerably smaller constant factor.
In a similar way as for the confidence interval, we can adjust the parameters so that $\lambda t$ is a convenient number for the rotations.
\begin{enumerate}
    \item For the Reiher case we find the truncation can be adjusted to $0.683740$, giving $\lambda t=\pi/2^{26}$ and $r=6.7\times 10^{13}$. The rotations take 25 Toffolis, so there are $1.8\times 10^{15}$ Toffolis for the simulation and 250 qubits.
    \item For the Li orbitals, choosing a truncation $0.645979$ gives $\lambda t=3\pi/2^{27}$ and $r=3.6\times 10^{13}$.  The rotation takes 26 Toffolis, so the total cost is $1.0\times 10^{15}$ Toffolis for the simulation and 296 qubits.
\end{enumerate}
In this approach there is about a further order of magnitude improvement over the resource estimate based on the confidence interval.
This is because it is allowing considerably larger error due to the qDRIFT, on the assumption that it can be eliminated by sampling.
This estimate of the complexity does assume that the qDRIFT error is symmetric, so is not completely rigorous.

\section{Direct qubitization of the standard tensor hypercontraction representation}
\label{app:standard_thc}

In this appendix we will discuss the cost of quantum simulation when qubitization is applied directly to the tensor hypercontraction representation of \eq{thc}. While the result is also an algorithm with gate complexity $\widetilde{\cal O}(N \lambda/\epsilon)$, the associated $\lambda$ is much larger than the $\lambda$ for the method described in the main body of the paper. We include this appendix for completeness. Our algorithm will use qubitization as described in \sec{qubitization} of the main paper. Here, we describe how one would implement \sel\, and \prep\, oracles. As we show, each of these primitives has complexity no more than $\widetilde{\cal O}(N)$; thus, consistent with \eq{lcu_cost}, the overall complexity is no more than $\widetilde{\cal O}(N \lambda /\epsilon)$.

\subsection{Specification of oracles for qubitization and implementation of Hamiltonian selection oracle}

To specify the requirements for the qubitization oracles that we will query to block encode the THC Hamiltonian, we can use a similar application of the Jordan-Wigner transformation as in \app{sparse} to give
\begin{equation}
H = \frac 12 \sum_{\sigma\in\{\uparrow,\downarrow\}} \sum_{p,q=1}^{N/2} T'_{p q} Q_{pq\sigma} + \frac 18 \sum_{\alpha,\beta\in \{\uparrow, \downarrow\}} \sum_{p,q,r,s=1}^{N/2} V_{pqrs} Q_{pq\alpha} Q_{rs\beta}\, ,
\end{equation}
where $Q_{pq\alpha}$ is as defined in \eq{jw2}.
For THC, we the replace $V_{pqrs}$ with the approximation
\begin{equation}
G_{pqrs} =
\sum_{\mu, \nu = 1}^{M} \chi_{p}^{(\mu)} \chi_{q}^{(\mu)} \zeta_{\mu\nu} \chi_{r}^{(\nu)}  \chi_{s}^{(\nu)} .
\end{equation}
That gives a $\lambda$ value of
\begin{equation}
\lambda = \sum_{p,q=1}^{N/2} \left| T_{pq}+\sum_{r=1}^{N/2} V_{pqrr} \right| + \frac 12 \sum_{p,q,r,s=1}^{N/2} \sum_{\mu,\nu=1}^M \left| \chi_{p}^{(\mu)} \chi_{q}^{(\mu)} \zeta_{\mu\nu} \chi_{r}^{(\nu)}  \chi_{s}^{(\nu)} \right| .
\end{equation}
The state to be prepared is of the form
\begin{align}\label{eq:complete_state}
 &\ket{0}\sum_{p,q=1}^{N/2} \sqrt{\frac{|T'_{pq}|}\lambda}\ket{\theta^T_{pq}}\ket{p} \ket{q}\ket{0}\ket{0}\ket{0}\ket{0}\ket{0}\ket{0}\ket{0}\ket{0}\nn
&+\ket{1}\sum_{p,q,r,s=1}^{N/2} \sum_{\mu,\nu=1}^M \sqrt{\frac{\left| \chi_{p}^{(\mu)} \chi_{q}^{(\mu)} \zeta_{\mu\nu} \chi_{r}^{(\nu)}  \chi_{s}^{(\nu)}\right|}{2\lambda}}
\ket{\theta^{\zeta}_{\mu\nu}}\ket{\mu}\ket{\nu}\ket{\theta^{(\mu)}_p}\ket{p}\ket{\theta^{(\mu)}_q}\ket{q}\ket{\theta^{(\nu)}_r}\ket{r}\ket{\theta^{(\nu)}_s}\ket{s}.
\end{align}
Here $\theta^T_{pq}$ gives the sign of $T'$ as before, $\theta^{(\mu)}_p$ gives the sign of $\chi^{(\mu)}_{p}$, and $\theta^{\zeta}_{\mu\nu}$ gives the sign of $\zeta_{\mu\nu}$.
The key thing to note is that the second part of this state, corresponding to the two-body terms, can be factorised as
\begin{align}\label{eq:complete_state2}
& \frac 1{\sqrt{2\lambda}} \ket{1}\sum_{\mu,\nu=1}^M \sqrt{|\zeta_{\mu\nu}|}\ket{\theta^{\zeta}_{\mu\nu}}\ket{\mu}\ket{\nu}
\left(\sum_{p=1}^{N/2}\sqrt{|\chi_{p}^{(\mu)}|}\ket{\theta^{(\mu)}_p}\ket{p}\right)
\left(\sum_{q=1}^{N/2}\sqrt{|\chi^{(\mu)}_{q}|}\ket{\theta^{(\mu)}_q}\ket{q}\right)\nn &\qquad \otimes
\left(\sum_{r=1}^{N/2}\sqrt{|\chi_{r}^{(\nu)}|}\ket{\theta^{(\nu)}_r}\ket{r}\right)
\left(\sum_{s=1}^{N/2}\sqrt{|\chi_{s}^{(\nu)}|}\ket{\theta^{(\nu)}_s}\ket{s}\right).
\end{align}
The fey feature of this state that makes it easier to prepare is that it factorises.
One may therefore start by preparing a state on $\mu$ and $\nu$, then controlled on $\mu$ and $\nu$ prepare the states in brackets.

\subsection{Using the structure of tensor hypercontraction to implement the qubitization state preparation}

In preparing the overall state in \eq{complete_state}, one can first prepare the state on the first 4 registers, with amplitudes proportional to $\sqrt{|T'_{pq}|}$ for the one-body term and $\sqrt{|\zeta_{\mu\nu}|}$ for the two-body-term.
That is, we first prepare the state
\begin{equation}
\ket{0}\sum_{p,q=1}^{N/2} \sqrt{\frac{|T'_{pq}|}{\lambda}} \ket{\theta^T_{pq}}\ket{p}\ket{q} + \ket{1}\sum_{\mu,\nu=1}^M \sqrt{\frac{|\zeta_{\mu\nu}|}{2\lambda}}\ket{\theta^\zeta_{\mu\nu}}\ket{\mu}\ket{\nu}.
\end{equation}
After this preparation, for the two-body term flagged by 1 on the first qubit, we need to perform four applications of the mapping
\begin{equation}
\label{eq:prepare_chi}
\textsc{prepare}_\chi \ket{0}^{\log (N/2)} \ket{\mu} \mapsto \sum_{p=1}^{N/2} \sqrt{|\chi_{p}^{(\mu)}|} \ket{\theta^{(\mu)}_p} \ket{p} \ket{\mu}.
\end{equation}
This is a preparation on two registers, but this time there is not symmetry.

For the first preparation, the steps to be performed are as follows.
\begin{enumerate}
\item Rotate the first qubit to give the appropriate relative weighting between the one- and two-body terms.
\item For 1 on the first qubit, prepare an equal superposition over $\mu$ and $\nu$ for $1\le \mu\le\nu\le M$, or for 0 on the first qubit prepare an equal superposition with the restrictions $1\le \mu\le\nu\le N/2$.
That can be performed using inequality tests and amplitude amplification in a similar way to that explained in \app{low_rank}.
The difference is that instead of just using the inequality test $\nu\le M$, you use a controlled inequality test of $\nu\le N/2$ for 0 on the first qubit, or $\nu\le M$ for 1 on the first qubit.
The extra inequality tests increase the Toffoli cost to $8n_M+2b_r+\mathcal{O}(1)$, where $b_r$ are the bits of precision for rotation on an ancilla and $n_M=\lceil\log M\rceil$.
\item Compute $\mu(\mu-1)/2+\nu$, which can be performed with complexity $n_M^2+n_M-1$.
In the case of 0 on the first qubit (for the one-body term), add $M(M+1)/2$ to yield a contiguous register.
That addition has complexity $2n_M+\mathcal{O}(1)$.
\item Perform a QROM using this contiguous register to alt values of $\mu$ and $\nu$, keep probabilities, and two values (one is the alt value) of the sign $\theta^\zeta_{\mu\nu}$.
The output size for the QROM is
\begin{equation}
b_\zeta = 2n_M+\zetabits+2,
\end{equation}
where $\zetabits$ are the bits of precision for $\zeta_{\mu\nu}$.
The complexity of the QROM is
\begin{equation}\label{eq:rankqr1}
\left\lceil \frac{L_\zeta}{k_\zeta} \right\rceil + b_\zeta (k_\zeta-1),
\end{equation}
where
\begin{equation}
    L_\zeta=M(M+1)/2+N^2/8+N/4
\end{equation}
and $k_\zeta$ must be a power of 2.
\item Perform an inequality test between the keep value and another register in an equal superposition, with cost $\zetabits$.
\item Perform a swap between $\mu,\nu$ and the alt values controlled on the result of the inequality test, with cost $2n_M$.
The signs can be applied with Clifford gates.
\item Apply a swap between the $\mu$ and $\nu$ registers controlled by a qubit in a $\ket{+}$ state, with cost $n_M$.
\end{enumerate}
The total complexity is then
\begin{equation}
C_{P_\zeta}= \left\lceil \frac{L_\zeta}{k_\zeta} \right\rceil + b_\zeta k_\zeta+n_M^2+12n_M+b_r+\mathcal{O}(\log(1/\epsilon_{\rm rot})),
\end{equation}
where $\epsilon_{\rm rot}$ is the accuracy required for the rotation on the first qubit.
In inverting the state preparation, all costs are the same except the QROM cost, which is reduced to give a total cost
\begin{equation}
C_{P^\dagger_\zeta}= \left\lceil \frac{L_\zeta}{k'_\zeta} \right\rceil + k'_\zeta +n_M^2+14n_M+b_r+\zetabits+\mathcal{O}(\log(1/\epsilon_{\rm rot})) .
\end{equation}

For the preparation given in \eq{prepare_chi}, the considerations are similar, except we do not take advantage of symmetry, so $L_\chi=MN/2$.
There the preparation needs a QROM to run through all values of $p$ and $\mu$, and it is convenient to use the QROM for two registers as in \app{qrom}.
The total Toffoli costs are
\begin{enumerate}
    \item Prepare an equal superposition over $N/2$ basis states in the $p$ register, with cost $3n_N-3\factor+2b_r-9$, where $\factor$ is the largest number such that $2^\factor$ is a factor of $N/2$, and $b_r$ is bit of precision for rotation on an ancilla.
    \item Because we only need to output alternate values of $p$ (and not $\mu$), the output size for the QROM is 
\begin{equation}
    b_\chi=n_N + \zetabits.
\end{equation}
    Therefore we can apply the QROM with cost (using the method for separate registers in \app{qrom})
\begin{equation}
    \left\lceil \frac{M}{k_{\chi 1}}\right\rceil \left\lceil \frac{N}{2 k_{\chi 2}}\right\rceil + b_\chi (k_{\chi 1}k_{\chi 2}-1).
\end{equation}
    \item Perform the inequality test with cost $\zetabits$.
    \item The controlled swap does not need to touch the register with $\mu$, because we are just preparing a superposition over $p$ for a given $\mu$.
    Therefore the controlled swap has cost $n_N$.
\end{enumerate}
The total costs are then
\begin{equation}
C_{P_\chi} = \left\lceil \frac{M}{k_{\chi 1}}\right\rceil \left\lceil \frac{N}{2 k_{\chi 2}}\right\rceil + b_\chi k_{\chi 1} k_{\chi 2} +
3n_N-3\factor+2b_r+
\mathcal{O}(1)
\end{equation}
for preparation and
\begin{equation}
C_{P_\chi^\dagger} = \left\lceil \frac{M}{k'_{\chi 1}}\right\rceil \left\lceil \frac{N}{2 k'_{\chi 2}}\right\rceil+ k'_{\chi 1} k'_{\chi 2} +b_\chi +
3n_N-3\factor+2b_r+\mathcal{O}(1)
\end{equation}
for inverse preparation.
It is also possible to combine the preparation of some of the equal superposition states together, which would give slightly different costs.
We will not analyse that here, because the method that is best tends to depend on the particular example.

We can now combine all of the Toffoli costs. Clearly, we will need to prepare and unprepare the $\chi$ state four times and prepare and unprepare the $\zeta$ state once. Thus the overall cost of preparation and unpreparation is
\begin{align}
C_P + C_{P^\dagger} & = C_{P_\zeta} + C_{P^\dagger_\zeta} +  4 C_{P_\chi} + 4 C_{P^\dagger_\chi}.
\label{eq:prep_cost}
\end{align}
For the total cost, there will also be $4N$ for the cost of implementing the \sel\ operation, and $2n_M + 4n_N + 5\zetabits+\mathcal{O}(1)$ for reflections on the control registers to construct the overall step of the quantum walk.
There are $2n_M + 4n_N+\mathcal{O}(1)$ qubits that the state is prepared on, and we need to reflect on the $5\zetabits$ qubits used for equal superposition states as well.

The logical qubit costs are as follows, where we omit a number of single ancillas (about 20) for simplicity.
\begin{enumerate}
    \item The $2\lceil\log(\mathcal{I}+1)\rceil-1$ qubits for the control register for state preparation and unary iteration on that register.
    \item The $N$ system registers.
    \item For storing $\mu,\nu,p,q,r,s$ there are $2n_M+4n_N$ qubits needed.
    \item The $5\zetabits$ from registers that are used as comparison registers for the inequality tests in the coherent alias sampling.
    \item The phase gradient state has $b_r$ bits.
    \item The contiguous register for the $\zeta$ state preparation which has $\lceil\log L_\zeta\rceil$ qubits.
    \item The output of the QROM in the $\zeta$ state preparation uses $b_\zeta$ bits.
    \item The QROM in the $\zeta$ state preparation uses another $b_\zeta (k_\zeta-1) +\lceil \log(L_\zeta/k_\zeta) \rceil$ temporary qubits.
    At this point the number of qubits used would normally be at a maximum, and one can find the number of qubits needed by adding this number of qubits to the numbers in the other items listed above.
    Otherwise, one would ignore this cost, and add the costs given below.
    \item There are $4b_\chi$ qubits needed for the outputs of the 4 QROMs for the 4 $\chi$ state preparations.
    \item There are another $b_\chi (k_{\chi 1}k_{\chi 2}-1)+ \lceil \log (M/k_{\chi 1}) \rceil \lceil \log (N/2k_{\chi 2}) \rceil$ temporary qubits used in the final QROM for $\chi$ state preparation.
\end{enumerate}
Adding these numbers of qubits gives us
\begin{align}
& 2\lceil\log(\mathcal{I}+1)\rceil + N + 2n_M+4n_N + 5\zetabits + b_r + \lceil\log L_\zeta\rceil + b_\zeta + \mathcal{O}(1) \nn & + \max \big( 
b_\zeta (k_\zeta-1) +\lceil \log(L_\zeta/k_\zeta) \rceil ~,~
3b_\chi + b_\chi k_{\chi 1}k_{\chi 2} + \lceil \log (M/k_{\chi 1}) \rceil \lceil \log (N/2k_{\chi 2}) \rceil
\big).
\end{align}
For realistic parameters, the first expression in the max should be taken, corresponding to the largest number of qubits being used in the $\zeta$ state preparation.

\subsection{Comparison of \texorpdfstring{$\lambda$}{lambda} associated with direct qubitization and non-orthogonal qubitization}

\begin{figure}[ht!]
\includegraphics[scale=0.45]{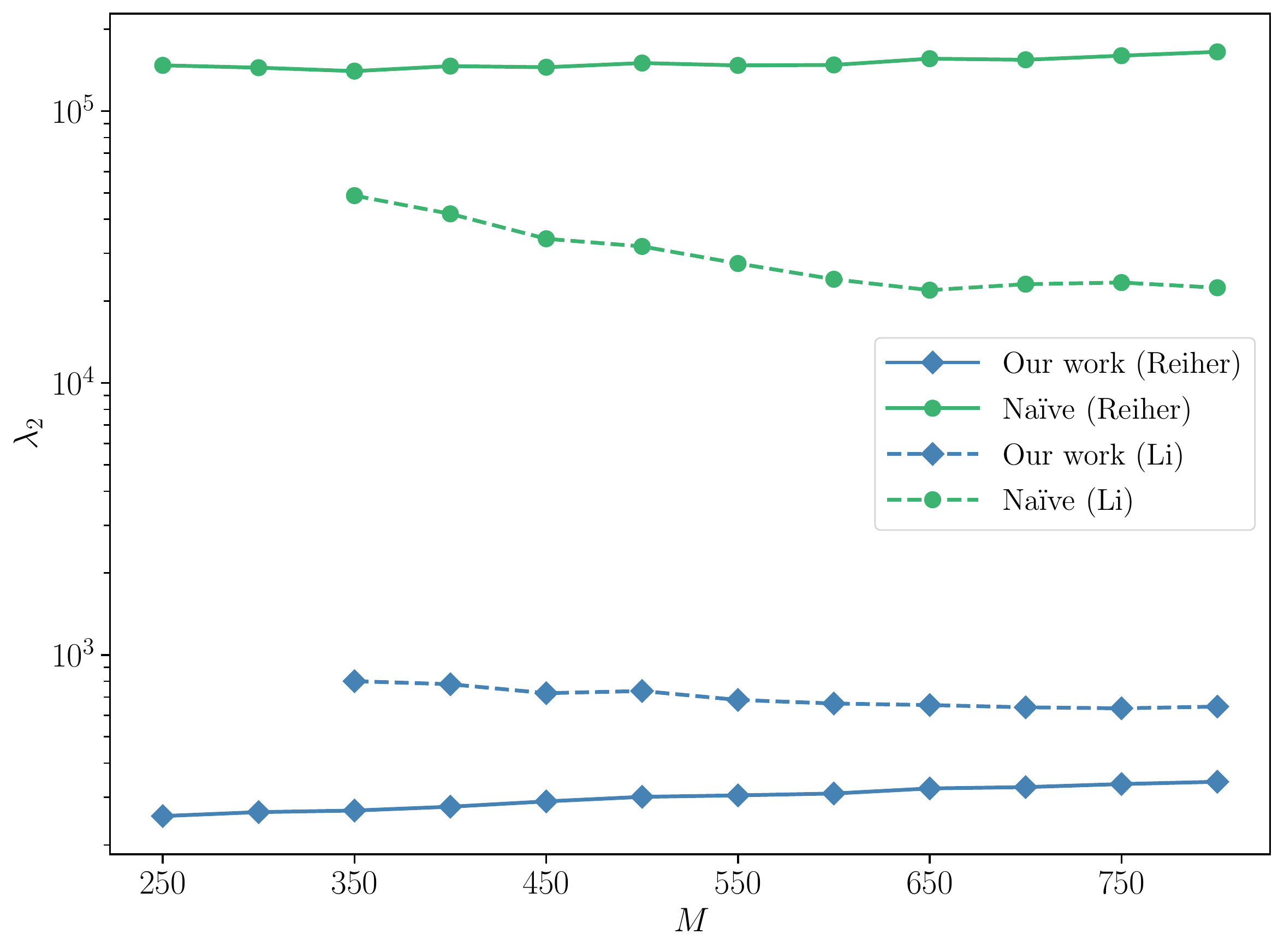}
\caption{\label{fig:naive}
Plots of $\lambda_2$ versus $M$ for two FeMoCo Hamiltonians (Reiher and Li) where $\lambda_2$ is computed by \eq{thc_lambda} (labeled ``Na{\"i}ve'') and
\eq{thirdlambda} (labeled ``Our work'').
These numerical values are based on the THC factors that produced the numerical data in \tab{femocothcR} and \tab{femocothcL}.
}
\end{figure}

\begin{figure}[ht!]
\includegraphics[scale=0.45]{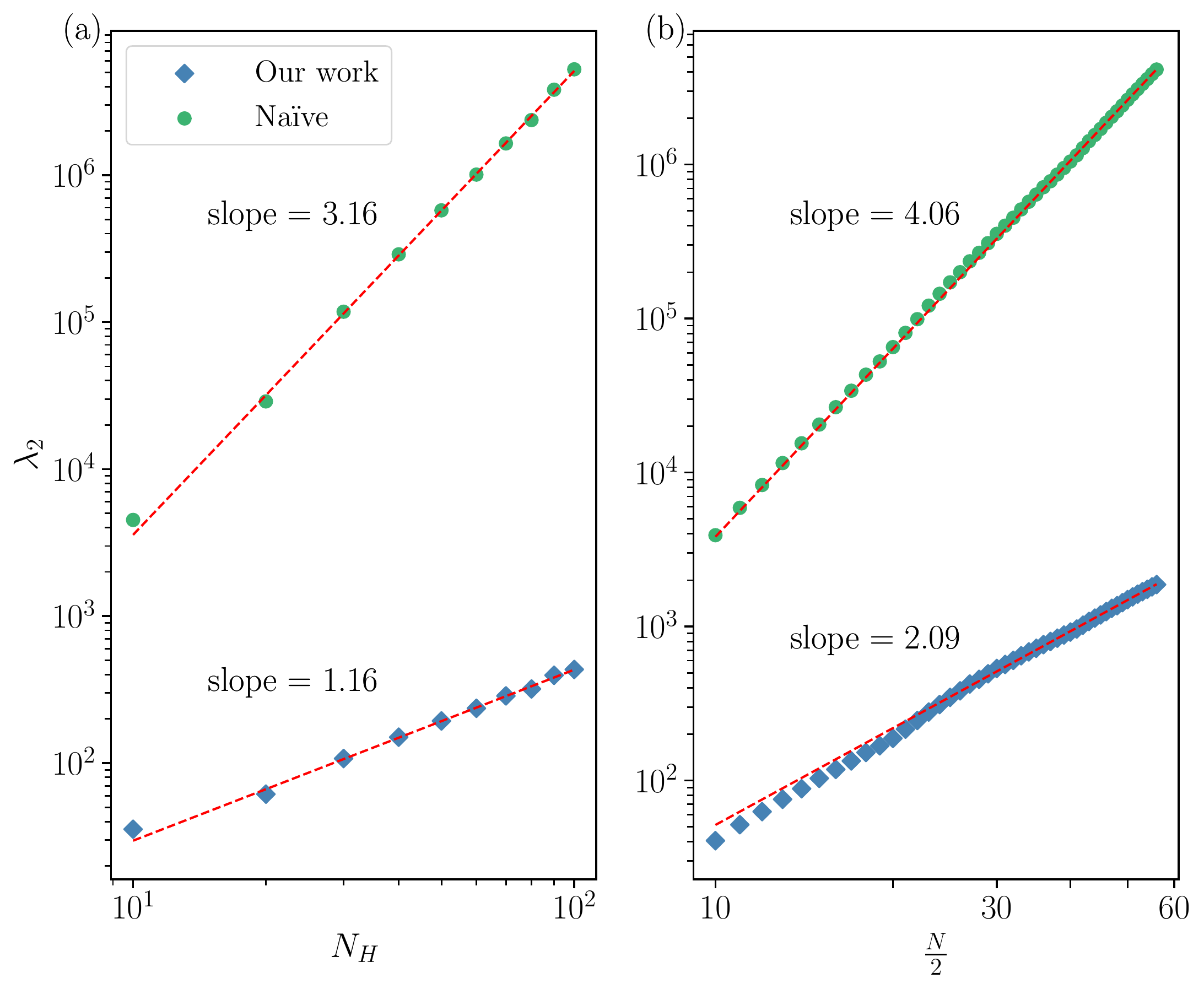}
\caption{\label{fig:naive2}
Plot of $\lambda_2$ versus $N_H$ for hydrogen chain and versus $N/2$ for \ce{H4}, where $\lambda_2$ is computed by \eq{thc_lambda} (labeled ``Na{\"i}ve'') and
\eq{thirdlambda} (labeled ``Our work'').
These numerical values are based on the THC factors that produced the numerical data in \fig{h4hchain}.
$R^2$ is greater than 0.99 in all fits.
Note that the slope of ``Our work'' in (a) is slightly different from what is reported in \tab{hchainslopes} because
(a) uses only $\lambda_2$ whereas \tab{hchainslopes} is based on total $\lambda$.
}
\end{figure}

Here we present numerical data for 
the two FeMoCo Hamiltonians
which compare our proposed use of THC (\eq{thirdlambda})
against the na{\"i}ve use of THC as defined in \eq{thc_lambda} and described in this appendix.
As mentioned in the main text, the use of \eq{thc_lambda} results in the two-body $\lambda$, i.e., $\lambda_2$, that is a few orders of magnitude greater
than
that of \eq{thirdlambda}.
In particular, we compare the values of \eq{thirdlambda} and \eq{thc_lambda} as a function of $M$.
As shown in \fig{naive}, significant reduction in $\lambda_2$ is observed going from \eq{thc_lambda} to \eq{thirdlambda}.
Furthermore, in \fig{naive2}, we illustrate the same point for hydrogen chain and \ce{H4}. Compared to the na{\"i}ve approach, we not only achieve more than an order of magnitude
reduction in $\lambda_2$ but also achieve a reduced scaling of $\lambda_2$ with respect to system size and the number of basis functions.
Specifically, we note that the empirical scaling based on the linear fits on a log scale suggests that we achieve
$N_H^{3.16}$ to $N_H^{1.16}$ size scaling reduction in hydrogen chain and $N^{4.06}$ to $N^{2.09}$ basis scaling reduction in \ce{H4}.
We argue that the use of THC in qubitization is only successful when utilizing its non-orthogonal form as proposed in this work.
This was one of the key observations in our work.


\section{Computing contiguous registers}
\label{app:contig}

In this appendix we show the formula for computing contiguous registers. This step is used in state preparations for both the main algorithm of this paper as well as a modification we introduce to the double factorization of von Burg \emph{et al.}~\cite{vonBurg2020} which is not discussed in their paper but which we believe is required for their algorithm to execute correctly. The need for this is first discussed in \eq{contiguous} of this paper. For contiguous registers we need to compute
\begin{equation}
p(p+1)/2+q
\end{equation}
where $p$ and $q$ have the same number of bits, which we will denote $n$.
We will show that this formula can be computed using $n^2+n-1$ Toffolis, considerably simplifying the analysis of the complexity.
In this appendix we take $p$ and $q$ to start from 0 rather than 1 as is used in the discussion of the Hamiltonians.
That is why the formula is $p(p+1)/2+q$ rather than $p(p-1)/2+q$.

Writing $p$ and $q$ in terms of their bits
\begin{equation}
p = \sum_{j=0}^{n-1} p_j 2^j, \qquad q = \sum_{j=0}^{n-1} q_j 2^j,
\end{equation}
we have
\begin{align}
p^2 &= \sum_{j=0}^{n-1}\sum_{k=0}^{n-1} p_j p_k 2^{j+k}\nn
&=  \sum_{j=0}^{n-1} p_j 2^{2j} + 2\sum_{j=0}^{n-1}\sum_{k=0}^{j-1} p_j p_k 2^{j+k}\nn
&= \sum_{j=0}^{n-1} p_j 2^{2j} +
\sum_{\ell=0}^{2n-3} 2^{\ell+1} \sum_{j=\max(0,\ell-n+1)}^{\lfloor (\ell-1)/2\rfloor} p_{\ell-j} p_j ,
\end{align}
where $\lfloor (\ell-1)/2\rfloor$ is to ensure that $\ell-j>j$.
Breaking this up into odd and even gives
\begin{equation}
p^2 = p_0 + \sum_{\ell=1}^{n-2} 2^{2\ell+1}  \sum_{j=\max(0,2\ell-n+1)}^{\ell-1} p_{2\ell-j} p_j +
\sum_{\ell=1}^{n-1} 2^{2\ell} \left( p_\ell+\sum_{j=\max(0,2\ell-n)}^{\ell-1} p_{2\ell-1-j} p_j\right) .
\end{equation}

That gives
\begin{equation}
p^2+p = 2p_0 + \sum_{\ell=1}^{n-1} 2^\ell p_\ell + \sum_{\ell=1}^{n-2} 2^{2\ell+1}   \sum_{j=\max(0,2\ell-n+1)}^{\ell-1} p_{2\ell-j} p_j +
\sum_{\ell=1}^{n-1} 2^{2\ell} \left( p_\ell +\sum_{j=\max(0,2\ell-n)}^{\ell-1} p_{2\ell-1-j} p_j \right) .
\end{equation}
Dividing by 2 then gives
\begin{equation}
p(p+1)/2 = p_0 + \sum_{\ell=1}^{n-1} 2^{\ell-1} p_\ell + \sum_{\ell=1}^{n-2} 2^{2\ell}  \sum_{j=\max(0,2\ell-n+1)}^{\ell-1} p_{2\ell-j} p_j +
\sum_{\ell=1}^{n-1} 2^{2\ell-1} \left( p_{\ell} +\sum_{j=\max(0,2\ell-n)}^{\ell-1} p_{2\ell-1-j} p_j \right) .
\end{equation}
Adding $q$ gives
\begin{align}
p(p+1)/2+q &= \sum_{\ell=0}^{n-1} 2^\ell q_\ell +p_0 + \sum_{\ell=1}^{n-1} 2^{\ell-1} p_\ell + \sum_{\ell=1}^{n-2} 2^{2\ell}  \!\!\!\sum_{j=\max(0,2\ell-n+1)}^{\ell-1} \! p_{2\ell-j} p_j +
\sum_{\ell=1}^{n-1} 2^{2\ell-1} \! \left( p_{\ell} +\!\!\!\sum_{j=\max(0,2\ell-n)}^{\ell-1}\!  p_{2\ell-1-j} p_j \right) \nn
&=q_0 + p_0 + p_1 + \sum_{\ell=1}^{n-2} 2^{\ell} (q_\ell+p_{\ell+1}) + 2^{n-1}q_{n-1}  
\nn & \quad +
\sum_{\ell=1}^{n-2} 2^{2\ell}  \sum_{j=\max(0,2\ell-n+1)}^{\ell-1} p_{2\ell-j} p_j +
\sum_{\ell=1}^{n-1} 2^{2\ell-1} \left( p_{\ell} +\sum_{j=\max(0,2\ell-n)}^{\ell-1} p_{2\ell-1-j} p_j \right) .
\end{align}
The next steps depend on whether $n$ is odd or even.
First, for $n$ even, $n-1=2\ell-1$ for $\ell=n/2$.
Then we can write
\begin{align}
p(p+1)/2+q &=q_0 + p_0 + p_1 + \sum_{\ell=1}^{n-2} 2^{\ell} (q_\ell+p_{\ell+1}) + 2^{n-1}q_{n-1} + \sum_{\ell=1}^{n/2-1} 2^{2\ell} \sum_{j=0}^{\ell-1} p_{2\ell-j} p_j\nn & \quad 
+ \sum_{\ell=n/2}^{n-2} 2^{2\ell} \sum_{j=2\ell-n+1}^{\ell-1} p_{2\ell-j} p_j
+\sum_{\ell=1}^{n/2-1} 2^{2\ell-1} \left( p_\ell+\sum_{j=0}^{\ell-1} p_{2\ell-1-j} p_j \right)\nn & \quad
+ 2^{n-1}\left( p_{n/2}+ \sum_{j=0}^{n/2-1} p_{n-1-j} p_j\right)
+\sum_{\ell=n/2+1}^{n-1} 2^{2\ell-1} \left( p_\ell+\sum_{j=2\ell-n}^{\ell-1} p_{2\ell-1-j} p_j \right).
\end{align}
This can be written as
\begin{subequations}
\begin{align}\label{eq:levelsa}
p(p+1)/2+q  &=q_0 + p_0 + p_1 \\ \label{eq:levelsb} & \quad 
+\sum_{\ell=1}^{n/2-1} 2^{2\ell-1} \left( q_{2\ell-1}+p_{2\ell} + p_\ell+\sum_{j=0}^{\ell-1} p_{2\ell-1-j} p_j \right)\\ \label{eq:levelsc} & \quad
+ \sum_{\ell=1}^{n/2-1} 2^{2\ell} \left( q_{2\ell}+p_{2\ell+1}+\sum_{j=0}^{\ell-1} p_{2\ell-j} p_j\right) \\ \label{eq:levelsd} & \quad 
+ 2^{n-1}\left( q_{n-1}+ p_{n/2}+\sum_{j=0}^{n/2-1} p_{n-1-j} p_j\right) \\ \label{eq:levelse} & \quad 
+ \sum_{\ell=n/2}^{n-2} 2^{2\ell} \sum_{j=2\ell-n+1}^{\ell-1} p_{2\ell-j} p_j \\ \label{eq:levelsf} & \quad 
+\sum_{\ell=n/2}^{n-2} 2^{2\ell+1}\left( p_{\ell+1}+ \sum_{j=2\ell-n+1}^{\ell-1} p_{2\ell-j} p_{j+1}\right).
\end{align}
\end{subequations}
The last line has shifted $\ell$ to simplify the following discussion.
This expression has split into lines as follows.
\begin{enumerate}
\item In \eq{levelsa} three bits which are added together.
\item In \eq{levelsb} bits which are added together and multiplied by odd powers $2^{2\ell-1}$ below $2^{n-1}$.
There are $\ell+3$ of these bits to be added together.
\item In \eq{levelsc} bits which are added together and multiplied by \emph{even} powers of $2^{2\ell}$ below $2^{n-1}$.
There are $\ell+2$ of these bits to be added together.
\item In \eq{levelsd} there are $n/2+2$ bits to be added together and multiplied by $2^{n-1}$.
\item In \eq{levelse} there are $n-\ell-1$ bits to be added together and multiplied by even powers $2^{2\ell}$ above $2^{n-1}$ up to $2^{2n-4}$.
\item In \eq{levelsf} there are $n-\ell$ bits to be added together and multiplied by odd powers $2^{2\ell+1}$ above $2^{n-1}$ up to $2^{2n-3}$.
\end{enumerate}

Now we can apply the approach used for bit sums in Appendix A of \cite{Kivlichan2019}.
We separate the bits into those that are multiplied by 1, 2, 4, 8, and so on.
We add triples of bits at each level using the adder in Figure 4(b) from \cite{GidneyAdder}.
Adding each triple of bits takes one Toffoli, reduces the number of bits at that level by 2, and increases the number of bits at the next level by 1.
So, if there are $m$ bits at one level, then the number of Toffolis needed for that level is $\lfloor m/2\rfloor$, and there are an additional $\lfloor m/2\rfloor$ carry bits at the next higher level.

Here the first level, those bits multiplied by 1, has only three bits, meaning one Toffoli is needed, and there is one more carry bit at level 2.
For level 2, $\ell=1$, and there are $\ell+3=4$ bits.
Adding the carry bit, there are now $5$ bits, which take 2 Toffolis to sum together, and give another 2 carry bits for the next level, 3, where the bits are multiplied by $2^2$.
For level 3 with $\ell=1$, there are $\ell+2=3$ bits, and incluing the 2 carry bits gives 5.
These 5 bits can again be added together with 2 Toffolis, giving 2 Toffolis for the next level.

In general, for the lines in \eq{levelsa} to \eq{levelsf} we have the following costings.
\begin{enumerate}
\item In \eq{levelsa} the three bits are added with one Toffoli, giving one carry bit.
\item In \eq{levelsb} For term $\ell$, we have $\ell$ carry bits from the previous level.
We can see this for the case $\ell=1$ because there was one carry bit from adding the triple of bits from level 1.
For $\ell>1$, we have the number of carry bits from \eq{levelsc} with term $\ell-1$, which gave $\ell$ carry bits.
The number of bits to be added together is therefore $\ell+3+\ell$.
These bits can be added together with cost $\lfloor(2\ell+3)/2\rfloor=\ell+1$ Toffolis and the same number of carry bits.
\item In \eq{levelsc} for term $\ell$, we have $\ell+1$ carry bits coming from \eq{levelsb} with the same value of $\ell$.
The number of bits is $\ell+2$ plus the $\ell+1$ carry bits, giving a total of $2\ell+3$.
These can again be added with cost $\ell+1$, giving $\ell+1$ carry bits.
\item In \eq{levelsd} there are $n/2+2$ bits, plus $n/2$ carry bits from \eq{levelsc}, since the last term there has $\ell=n/2-1$.
That gives a total of $n+2$ bits to add, which can be added with $n/2+1$ Toffolis and giving the same number of carry bits.
\item In \eq{levelse} there are $n-\ell+1$ carry bits to be included.
This can be seen for $\ell=n/2$, because there were $n/2+1$ carry bits from \eq{levelsd}.
For $\ell>n/2$, it can be seen because there are $n-\ell+1$ carry bits from the sixth line by using $n-\ell'$ with $\ell'=\ell-1$.
That gives a total number of bits $2(n-\ell)$, which can be added together with $n-\ell$ Toffolis and giving $n-\ell$ carry bits.
\item In \eq{levelsf} there are $n-\ell$ bits coming from the fifth line.
That gives a total number of bits $2(n-\ell)$, which can be added together with $n-\ell$ Toffolis and giving $n-\ell$ carry bits.
\item Note that for the final term from \eq{levelsf} with $\ell=n-2$, the number of carry bits is $n-(n-2)=2$.
These two bits can be added together with one more Toffoli gate.
\end{enumerate}
Now that we have quantified the number of Toffolis at each level, we can add them together to give
\begin{align}
1+2\sum_{\ell=1}^{n/2-1} (\ell+1) + \sum_{\ell=1}^{n/2-1} (\ell+1) + n/2+1 + 2\sum_{\ell=n/2}^{n-2} (n-\ell) +1 = n(n+3)/2-1
\end{align}
Toffolis.
There are also $n(n-1)/2$ Toffolis needed to compute all of the products $p_j p_k$ for $j\ne k$.
Adding those Toffolis gives the total $n^2+n-1$.
An illustration of the case of even $n=6$ is given in Table \ref{tab:n6}.

\begin{table*}[tbh]
\begin{tabular}{|c | c | c | c | c | c|c|c|c|c|c|c|c|c|c|}
\hline
level& factor & line of equation & $\ell$ & $q$   & $p$          & $p_0$               & $p_1$    & $p_2$    & $p_3$    & $p_4$    & $p_5$    &&  & Toffolis \\  \hline
 12  &$2^{11}$&                  &        &       &              &                     &          &          &          &          & $\times$ &  & & \\  
 11  &$2^{10}$&                  &        &       &              &                     &          &          &          &          & $\times$ & $\times$ &  & 1 \\  
 10  & $2^9$ & \eqref{eq:levelsf}& 4      &       &              &                     &          &          &          & $p_4p_5$ & $p_5$    & $\times$ & $\times$& 2 \\  
 9   & $2^8$ & \eqref{eq:levelse}& 4      &       &              &                     &          &          & $p_3p_5$ & $\times$ & $\times$ & $\times$ & & 2 \\  
 8   & $2^7$ & \eqref{eq:levelsf}& 3      &       &              &                     &          & $p_2p_5$ & $p_3p_4$ & $p_4$    & $\times$ & $\times$ & $\times$ & 3 \\
 7   & $2^6$ & \eqref{eq:levelse}& 3      &       &              &                     & $p_1p_5$ & $p_2p_4$ & $\times$ & $\times$ & $\times$ & $\times$ & &3 \\  
 6   & $2^5$ & \eqref{eq:levelsd}&        & $q_5$ &              & $p_0p_5$            & $p_1p_4$ & $p_2p_3$ & $p_3$    & $\times$ & $\times$ & $\times$ && 4  \\
 5   & $2^4$ & \eqref{eq:levelsc}& 2      & $q_4$ & $p_5$        & $p_0p_4$            & $p_1p_3$ & $\times$ & $\times$ & $\times$ &          & && 3 \\  
 4   & $2^3$ & \eqref{eq:levelsb}& 2      & $q_3$ & $p_4$        & $p_0p_3$            & $p_1p_2$ & $p_2$    & $\times$ & $\times$ &          && & 3 \\  
 3   & $2^2$ & \eqref{eq:levelsc}& 1      & $q_2$ & $p_3$        & $p_0p_2$            & $\times$ & $\times$ &          &          &          && & 2 \\  
 2   & 2     & \eqref{eq:levelsb}& 1      & $q_1$ & $p_2$        & $p_0p_1$            & $p_1$    &$\times$  &          &          &          &&& 2  \\  
 1   & 1     & \eqref{eq:levelsa}&        & $q_0$ & $p_1$        & \color{blue}{$p_0$} &          &          &          &          &          &&& 1 \\  
 0   & NA    &                   &        &       &\cancel{$p_0$}& \cancel{$p_0$}        &          &          &          &          &          && & \\  \hline
\end{tabular}
\caption[Computing a contiguous register for even number of bits]{\label{tab:n6}
Here we show how to compute a contiguous register for an even number of bits $n=6$.
The entries $p_0$, $p_0p_1$, and so forth show entries with different bits.
The entries with $\times$ show bits that are carried from lower levels.
The first column shows the level number, the second column shows the multiplying factor (power of 2), the third column shows which line from \eq{levelsa} to \eq{levelsf} this corresponds to, the fourth column shows the $\ell$ value from that equation, and the last column shows the number of Toffolis.
The $\cancel{p_0}$ shows identical bits that are added with no cost, and the resulting bit is shown in blue in the next level as \color{blue}{$p_0$}.}
\end{table*}

Now we consider the case for $n$ odd.  We can write
\begin{align}
p(p+1)/2+q &=q_0 + p_0 + p_1 + \sum_{\ell=1}^{n-2} 2^{\ell} (q_\ell+p_{\ell+1}) + 2^{n-1}q_{n-1} + 
\nn & \quad +
\sum_{\ell=1}^{(n-3)/2} 2^{2\ell}  \sum_{j=0}^{\ell-1} p_{2\ell-j} p_j 
+2^{n-1}  \sum_{j=0}^{(n-3)/2} p_{n-1-j} p_j
+\sum_{\ell=(n+1)/2}^{n-2} 2^{2\ell}  \sum_{j=2\ell-n+1}^{\ell-1} p_{2\ell-j} p_j\nn & \quad
+\sum_{\ell=1}^{(n-1)/2} 2^{2\ell-1} \left( p_{\ell} +\sum_{j=0}^{\ell-1} p_{2\ell-1-j} p_j \right) 
+\sum_{\ell=(n+1)/2}^{n-1} 2^{2\ell-1} \left( p_{\ell} +\sum_{j=2\ell-n}^{\ell-1} p_{2\ell-1-j} p_j \right).
\end{align}
As before, we will rewrite giving
\begin{subequations}
\begin{align}\label{eq:levodda}
p(p+1)/2+q &=q_0 + p_0 + p_1 \\ \label{eq:levoddb}
& \quad +\sum_{\ell=1}^{(n-1)/2} 2^{2\ell-1} \left( p_{2\ell-1}+p_{2\ell}+p_{\ell} +\sum_{j=0}^{\ell-1} p_{2\ell-1-j} p_j \right) \\ \label{eq:levoddc}
& \quad + \sum_{\ell=1}^{(n-3)/2} 2^{2\ell} \left( q_{2\ell}+p_{2\ell+1}+ \sum_{j=0}^{\ell-1} p_{2\ell-j} p_j \right)\\ \label{eq:levoddd}
& \quad +2^{n-1} \left( q_{n-1}+ \sum_{j=0}^{(n-3)/2} p_{n-1-j} p_j\right)\\ \label{eq:levodde}
& \quad +\sum_{\ell=(n+1)/2}^{n-1} 2^{2\ell-1} \left( p_{\ell} +\sum_{j=2\ell-n}^{\ell-1} p_{2\ell-1-j} p_j \right)\\ \label{eq:levoddf}
& \quad +\sum_{\ell=(n+1)/2}^{n-2} 2^{2\ell}  \sum_{j=2\ell-n+1}^{\ell-1} p_{2\ell-j} p_j.
\end{align}
\end{subequations}
Now the lines are as follows.
\begin{enumerate}
\item In \eq{levodda} three bits are added together.
\item In \eq{levoddb} bits which are added together and multiplied by odd powers $2^{2\ell-1}$ below $2^{n-1}$.
There are $\ell+3$ of these bits to be added together.
\item In \eq{levoddc} bits which are added together and multiplied by \textit{even} powers of $2^{2\ell}$ below $2^{n-1}$.
There are $\ell+2$ of these bits to be added together.
\item In \eq{levoddd} there are $(n+1)/2$ bits to be added together and multiplied by $2^{n-1}$.
\item In \eq{levodde} there are $n-\ell+1$ bits to be added together and multiplied by odd powers $2^{2\ell-1}$ above $2^{n-1}$ up to $2^{2n-3}$.
\item In \eq{levoddf} there are $n-\ell-1$ bits to be added together and multiplied by even powers $2^{2\ell}$ above $2^{n-1}$ up to $2^{2n-4}$.
\end{enumerate}
The addition works as follows.
\begin{enumerate}
\item In \eq{levodda} the three bits are added with one Toffoli, giving one carry bit.
\item As before, in \eq{levoddb} there are $\ell$ carry bits from the previous level, giving $2\ell+3$ terms, which can be added together with $\ell+1$ Toffolis and giving $\ell+1$ carry bits.
\item As before, in \eq{levoddc} for term $\ell$ there are $\ell+1$ carry bits from the previous level and $\ell+2$ bits giving a total of $2\ell+3$, so the number of Toffolis and carry bits are $\ell+1$.
\item In \eq{levoddd}, this time there is carry from \eq{levoddb} with $\ell=(n-1)/2$, so there are $(n+1)/2$ carry bits combined with the $(n+1)/2$ bits already in this term, for a total of $n+1$ bits to sum.  the number of Toffolis and carry bits to the next level is therefore $(n+1)/2$.
\item In \eq{levodde}, this time there are $n-\ell+1$ bits carried from the previous level.
This can be seen for the first term with $\ell=(n+1)/2$ because there are $(n+1)/2$ bits carried from \eq{levoddd}.
For $\ell=(n+1)/2$ there are $n-\ell'$ bits carried from \eq{levoddf} with $\ell'=\ell-1$.
Thus the total number of bits to be summed is $2(n-\ell+1)$, which can be done with $n-\ell+1$ Toffolis and carry bits.
\item In \eq{levoddf}, there are $n-\ell+1$ carry bits from \eq{levodde}, plus $n-\ell-1$, for a total of $2(n-\ell)$.
These may be summed with $n-\ell$ Toffolis and carry bits.
\item For the last term on \eq{levodde} with $\ell=n-1$, there are $n-\ell+1=n-(n-1)+1=2$ bits to sum, which takes one more Toffoli.
\end{enumerate}
The total number of Toffolis is therefore
\begin{equation}
1+\sum_{\ell=1}^{(n-1)/2}(\ell+1) +\sum_{\ell=1}^{(n-3)/2}(\ell+1) + (n+1)/2+\sum_{\ell=(n+1)/2}^{n-1}(n-\ell+1)+\sum_{\ell=(n+1)/2}^{n-2}(n-\ell)+1 =n(n+3)/2-1.
\end{equation}
This is the same formula as before, so adding the $n(n-1)/2$ Toffolis needed to compute the products $p_j p_k$ for $j\ne k$ again gives a total cost of $n^2+n-1$.
Thus we see that, regardless of whether the number of bits is odd or even, the number of Toffolis is $n^2+n-1$.
An example for odd $n=7$ is given in \tab{n7}.

\begin{table*}[tbh]
\begin{tabular}{|c | c | c | c | c | c|c|c|c|c|c|c|c|c|c|c|}
\hline
level& factor & line of equation  & $\ell$ & $q$   & $p$          & $p_0$               & $p_1$    & $p_2$    & $p_3$    & $p_4$    & $p_5$    & $p_6$    & &  & Toffolis \\  \hline
 14  &$2^{13}$&                   &        &       &              &                     &          &          &          &          &          & $\times$ &&  &   \\  
 13  &$2^{12}$&                   &        &       &              &                     &          &          &          &          &          & $\times$ & $\times$ &&  1 \\  
 12  &$2^{11}$& \eqref{eq:levodde}& 6      &       &              &                     &          &          &          &          & $p_5p_6$ & $p_6$    & $\times$ & $\times$ &  2 \\  
 11  &$2^{10}$& \eqref{eq:levoddf}& 5      &       &              &                     &          &          &          & $p_4p_6$ & $\times$ & $\times$ & $\times$ &&   2 \\  
 10  & $2^9$  & \eqref{eq:levodde}& 5      &       &              &                     &          &          & $p_3p_6$ & $p_4p_5$ & $p_5$    & $\times$ & $\times$ & $\times$ & 3 \\  
  9  & $2^8$  & \eqref{eq:levoddf}& 4      &       &              &                     &          & $p_2p_6$ & $p_3p_5$ & $\times$ & $\times$ & $\times$ & $\times$ &&  3 \\  
  8  & $2^7$  & \eqref{eq:levodde}& 4      &       &              &                     & $p_1p_6$ & $p_2p_5$ & $p_3p_4$ & $p_4$    & $\times$ & $\times$ & $\times$ & $\times$ & 4 \\  
  7  & $2^6$  & \eqref{eq:levoddd}&        & $q_6$ &              & $p_0p_6$            & $p_1p_5$ & $p_2p_4$ & $\times$ & $\times$ & $\times$ & $\times$ && &4 \\  
  6  & $2^5$  & \eqref{eq:levoddb}& 3      & $q_5$ & $p_6$        & $p_0p_5$            & $p_1p_4$ & $p_2p_3$ & $p_3$    & $\times$ & $\times$ & $\times$ &&& 4  \\
  5  & $2^4$  & \eqref{eq:levoddc}& 2      & $q_4$ & $p_5$        & $p_0p_4$            & $p_1p_3$ & $\times$ & $\times$ & $\times$ &          &          & && 3 \\  
  4  & $2^3$  & \eqref{eq:levoddb}& 2      & $q_3$ & $p_4$        & $p_0p_3$            & $p_1p_2$ & $p_2$    & $\times$ & $\times$ &          &          && & 3 \\  
  3  & $2^2$  & \eqref{eq:levoddc}& 1      & $q_2$ & $p_3$        & $p_0p_2$            & $\times$ & $\times$ &          &          &          &          && & 2 \\  
  2  & 2      & \eqref{eq:levoddb}& 1      & $q_1$ & $p_2$        & $p_0p_1$            & $p_1$    & $\times$ &          &          &          &          &  && 2  \\  
  1  & 1      & \eqref{eq:levodda}&        & $q_0$ & $p_1$        & \color{blue}{$p_0$} &          &          &          &          &          &          &&& 1 \\  
  0  & NA     &                   &        &       &\cancel{$p_0$}& \cancel{$p_0$}      &          &          &          &          &          &          && & \\  \hline
\end{tabular}
\caption[Computing a contiguous register for odd number of bits]{\label{tab:n7}
Here we show how to compute a contiguous register for an odd number of bits $n=7$.
The entries $p_0$, $p_0p_1$, and so forth show entries with different bits.
The entries with $\times$ show bits that are carried from lower levels.
The first column shows the level number, the second column shows the multiplying factor (power of 2), the third column shows which line from \eq{levodda} to \eq{levoddf} this corresponds to, the fourth column shows the $\ell$ value from that equation, and the last column shows the number of Toffolis.
The $\cancel{p_0}$ shows identical bits that are added with no cost, and the resulting bit is shown in blue in the next level as  \color{blue}{$p_0$}.}
\end{table*}

\section{QROM applied to two registers}
\label{app:qrom}

When performing QROM on two registers where one is iterating through all values of both registers, it is possible to perform the QROM efficiently without computing a contiguous register as was done in prior work.
Say we have registers with variables $x$ and $y$, which take $N_1$ and $N_2$ different values, and so can be represented on $n_1=\lceil \log N_1\rceil$ and $n_2=\lceil \log N_2\rceil$ bits.
We will choose $k_1$ and $k_2$ to be powers of 2.

Then for the QROM, we will iterate through $\lceil N_1/k_1\rceil$ values on the most significant $n_1-\log k_1$ bits of $x$, and $\lceil N_2/k_2\rceil$ values on the most significant $n_2-\log k_2$ bits of $y$.
For each of those values, we will give the QROM output for all combinations of values on the $\log k_1$ less-significant bits of $x$ and the $\log k_2$ less-significant bits of $y$.
Then the last step is to perform swaps controlled by the less-significant bits of $x$ and $y$ to move the correct data to the output.

For the complexity of this procedure, the complexity of the iteration through the $\lceil N_1/k_1\rceil$ values on the most significant bits of $x$ is $\lceil N_1/k_1\rceil-2$, or $\lceil N_1/k_1\rceil-1$ when it needs to be controlled by another qubit.
For each of those values on $x$, it is used as a control for iterating through $\lceil N_2/k_2\rceil$ values on the most significant bits of $y$.
Since that iteration is done in a controlled way, its cost is $\lceil N_2/k_2\rceil-1$.
Since it is done $\lceil N_1/k_1\rceil$ times, the overall cost is
\begin{equation}
    \left\lceil \frac{N_1}{k_1}\right\rceil 
    \left( 
    \left\lceil \frac{N_2}{k_2}\right\rceil -1
    \right).
\end{equation}
Adding to that the cost of the iteration on $x$ gives a cost of
\begin{equation}
     \left\lceil \frac{N_1}{k_1}\right\rceil 
    \left( 
    \left\lceil \frac{N_2}{k_2}\right\rceil -1
    \right)
    +\left\lceil \frac{N_1}{k_1}\right\rceil -2
    =\left\lceil \frac{N_1}{k_1}\right\rceil \left\lceil \frac{N_2}{k_2}\right\rceil -2,
\end{equation}
where the $-2$ is if the overall QROM does not need to be made controlled.
We will omit the $-2$ for simplicity, and consistency with the way these costs for QROM are usually quoted.

Then the cost of the controlled swaps at the end is identical to what it usually is for the QROM, so for output size $b$ it will be $b(k_1 k_2-1)$.
That gives a total cost (omitting the $-2$) of
\begin{equation}
    \left\lceil \frac{N_1}{k_1}\right\rceil \left\lceil \frac{N_2}{k_2}\right\rceil + b(k_1 k_2-1).
\end{equation}
The cost here should be compared to a cost for the case of a contiguous register
\begin{equation}
    \left\lceil \frac{N_1N_2}{k}\right\rceil + b(k-1),
\end{equation}
with $k=k_1k_2$.
The cost using a contiguous register will be slightly less, but the difference will often be less than the cost of computing a contiguous register.
For example, say $N_1=350$, $N_2=72$, and $b=20$.
Then with $k_1=8$ and $k_2=4$, the cost of the QROM is decreased by only 4 using a contiguous register, but the cost of computing the contiguous register is 17.

The number of ancillas is increased to
\begin{equation}
\left\lceil \log(N_1/k_1)\right\rceil +\left\lceil \log(N_2/k_2)\right\rceil+ bk_1 k_2.
\end{equation}
That is because there are $\left\lceil \log(N_1/k_1)\right\rceil$ required for iteration on the first register and $\left\lceil \log(N_2/k_2)\right\rceil$ for iteration on the second register, as well as $bk_1 k_2$ for the outputs.
In practice this usually only needs one more ancilla than using a contiguous register, which is less than would be needed to store the contiguous register itself.

In the same way, in uncomputing the QROM where it is necessary to perform a phase fixup, there is the same change to the cost for outputting the data, so the cost becomes
\begin{equation}
    \left\lceil \frac{N_1}{k_1}\right\rceil \left\lceil \frac{N_2}{k_2}\right\rceil + k_1 k_2.
\end{equation}
In the example with $N_1=350$, $N_2=72$, but taking $k_1=16$, $k_2=8$, the cost is only increased by one Toffoli over the cost for a contiguous register.
The number of ancillas needed is
\begin{equation}
\left\lceil \log(N_1/k_1)\right\rceil +\left\lceil \log(N_2/k_2)\right\rceil+ k_1 k_2.
\end{equation}

\end{document}